\documentclass[11pt,letterpaper,parskip=half]{scrartcl}

\usepackage[T1]{fontenc}
\usepackage[utf8]{inputenc}
\usepackage{mathpazo}          
\usepackage{microtype}         
\usepackage{amsmath}
\usepackage{amssymb}
\usepackage{amsthm}
\usepackage{bm}                
\usepackage{mathtools}         

\usepackage[dvipsnames]{xcolor}

\usepackage[margin=1in]{geometry}
\usepackage{setspace}
\usepackage{array}
\usepackage{booktabs}
\usepackage{longtable}
\usepackage{graphicx}
\usepackage{flafter}                   
\usepackage{needspace}                 

\usepackage[round]{natbib}
\usepackage[colorlinks=true,linkcolor=MidnightBlue,citecolor=OliveGreen,urlcolor=BrickRed]{hyperref}
\usepackage{doi}               
\usepackage{cleveref}

\theoremstyle{definition}

\theoremstyle{remark}
\newtheorem{remark}{Remark}[section]

\addtokomafont{disposition}{\rmfamily}     
\setkomafont{subtitle}{\Large}             
\setkomafont{author}{\normalsize}          
\setkomafont{date}{\small}                 

\newcommand{\M}{\mathcal{M}}          
\newcommand{\G}{\mathcal{G}}           
\newcommand{\R}{\mathbb{R}}            
\newcommand{\bS}{\bm{S}}              
\newcommand{\bA}{\bm{A}}              
\newcommand{\bC}{\bm{C}}              
\newcommand{\bT}{\bm{T}}              
\newcommand{\fs}{\mathfrak{s}}         
\newcommand{\fa}{\mathfrak{a}}         
\newcommand{\fc}{\mathfrak{c}}         
\newcommand{\ft}{\mathfrak{t}}         
\newcommand{\qx}{{}_1q_x}             
\newcommand{\logit}{\operatorname{logit}}
\newcommand{\expit}{\operatorname{expit}}
\newcommand{\ezero}{e_0}              
\newcommand{\nmode}[1]{\times_{#1}}    

\newcommand{\yobs}{\mathbf{y}_{\mathrm{obs}}}  
\newcommand{\zk}{\mathbf{z}_k}                
\newcommand{\tk}{\mathbf{t}_k}                
\newcommand{\dd}{\boldsymbol{\delta}_d}         
\newcommand{\eref}{e_0^{\mathrm{ref}}}         
\newcommand{\estar}{e_0^{*}}                   
\newcommand{\lhat}{\hat{\lambda}}              
\newcommand{\De}{\Delta e_0}                   

\newcommand{\includefigifexists}[2][width=\textwidth]{%
  \IfFileExists{#2}{%
    \includegraphics[#1]{#2}%
  }{%
    \fbox{\parbox{0.9\textwidth}{\centering\vspace{2cm}%
      \texttt{#2}\\[0.5em]%
      \textit{(Figure not yet generated -- run the QMD pipeline)}%
      \vspace{2cm}}}%
  }%
}

\title{Multi-dimensional Mortality}
\subtitle{Sex-Age-Specific Model Life Tables, Fitting, Prediction from\\
          Summary Mortality Indicators, and Forecasting}
\author{Samuel J.\ Clark\\
        \medskip
        {\small Department of Sociology, The Ohio State University}}
\date{\today}

\begin{document}
\maketitle

\begin{abstract}

Demographers rely on several distinct tools to work with mortality
schedules: model life table systems, relational fitting methods,
summary-indicator prediction models, and forecasting frameworks.  These
have been developed largely independently, each with its own statistical
machinery, and none provides structurally coherent sex-specific
schedules across all four tasks.  This paper develops a single framework
-- MDMx -- that unifies all four within one Tucker tensor decomposition
of the Human Mortality Database.

The HMD's period life tables (roughly 50 populations, up to 274 years)
are organized as a four-dimensional tensor of $\logit(q_x)$ indexed by
sex, age, country, and year.  A Tucker decomposition extracts shared
factor matrices for sex ($\bS$) and age ($\bA$); because these are
common to all populations and years, every schedule the system produces
is structurally coherent -- female and male mortality are linked through
the same basis by construction, not fitted independently or reconciled
post hoc.

The decomposition compresses the data into a low-rank representation
capturing $>$99.99\% of per-mode variance.  From this single
decomposition, four capabilities emerge.  \emph{Model life tables}:
clustering on level-controlled Tucker core features identifies canonical
mortality regimes, and smooth within-regime trajectories capture the
rotation of the age pattern of mortality decline as $\ezero$ rises.
\emph{Life table fitting}: a three-stage algorithm estimates cluster,
$\ezero$, and optional disruption parameters from an observed schedule,
with Laplace-approximated Bayes factors for disruption detection.
\emph{Summary-indicator prediction}: a neural network maps child
mortality ($_5q_0$) or both child and adult mortality to complete
schedules via truncated Tucker core weights, reformulating SVD-Comp
within the tensor framework.  \emph{Forecasting}: PCA-reduced core
matrices are projected forward through a damped local linear trend
Kalman filter with a two-level hierarchical drift target (80\% HMD-wide,
20\% country), producing probabilistic sex-age mortality surfaces.

In rolling-origin cross-validation (six origins, 15-year horizon), the
forecasting system achieves an $\ezero$ MAE of 1.44~years -- matching
the R \texttt{demography} package's Hyndman--Ullah implementation and
17\% better than Lee--Carter -- with a sex-gap MAE of 0.60~years (vs.\
0.84 for Hyndman--Ullah and 1.11 for Lee--Carter) and 93.7\% coverage
at nominal 95\%.  The architecture addresses the three persistent
challenges of the Lee--Carter family: age-pattern rotation emerges from
the joint evolution of multiple PCA scores rather than a fixed $b_x$;
sex coherence is structural rather than imposed post hoc; and prediction
intervals are calibrated from cross-validation performance.

\end{abstract}

\newpage
\tableofcontents
\listoffigures
\listoftables
\newpage


\section{Introduction}
\label{sec:introduction}

The age pattern of human mortality is remarkably regular.  Despite vast
differences in overall mortality levels across populations and historical
periods, the shape of the mortality curve -- high infant mortality declining
through childhood, a minimum in the pre-teen years, a modest rise through
young adulthood, and an approximately exponential increase through old
age -- recurs with striking consistency
\citep{Gompertz1825,Makeham1860,Heligman1980}.  Demographers have
exploited this regularity for over a century through model life table
systems that represent the diversity of human mortality experience using a
small number of parameters or components.

Classical approaches to model life tables fall into two broad families.
The first is \emph{regional}: a collection of mortality schedules is
grouped into discrete families, each indexed by a single parameter such
as life expectancy, and a user selects the family that best matches the
population of interest.  The Coale--Demeny regional model life tables
\citep{CoaleDemeny1966,CoaleDemeny1983} -- four families (North, South,
East, West) derived from 326 life tables -- and the United Nations model
tables \citep{UNModelLifeTables1982} are the most widely used systems of
this type.  The second family is \emph{relational}: a standard mortality
schedule is transformed through a parametric function to produce any
target schedule.  The Brass logit system \citep{Brass1971}, in which the
logit of survivorship in a target population is a linear function of the
logit of survivorship in a standard, is the canonical example.

A powerful third approach treats the problem as one of
\emph{dimensionality reduction}: represent a large collection of
mortality schedules as linear combinations of a small number of
age-varying basis vectors, so that any schedule can be compactly
described by its weights on those vectors.  This idea was introduced to
demography by \citet{LedermannBreas1959}, who applied factor analysis to
a set of 154 abridged life tables and found that four factors explained
approximately 95\% of the variance in age-specific probabilities of
dying -- demonstrating for the first time that the space of human
mortality schedules is intrinsically low-dimensional.  Building on these
factors, \citet{BourgeoisPichat1962} constructed a five-dimensional
system of model life tables, and \citet{Ledermann1969} formalized the
approach into one- and two-parameter model life tables that saw
operational use alongside the Coale--Demeny system, particularly in
Francophone demography.

The modern resurgence of dimensionality reduction in mortality modeling
began with the singular value decomposition (SVD) and principal component
analysis (PCA).  The Lee--Carter model \citep{LeeCarter1992} expresses
the log of age-specific mortality as a mean age schedule plus a single
time-varying component modulated by an age-specific sensitivity vector --
essentially a rank-one SVD of the mortality surface.  Higher-rank
extensions retain additional components to capture finer structure: the
modified logit system of \citet{MurrayEtAl2003} uses the same
mean-plus-residual architecture with an SVD-derived correction; the
log-quadratic model of \citet{Wilmoth2012} applies a quadratic mean
model and identifies its correction factor $v_x$ as the leading left
singular vector of the residual matrix; and \citet{FosdickHoff2014}
developed a Bayesian separable factor analysis model for the
four-dimensional HMD mortality array that parameterizes mode-specific
covariance through low-rank-plus-diagonal structure.

A parallel strand of work has pursued the multi-way generalization
directly, applying tensor decomposition to mortality arrays with three
or more dimensions.  \citet{RussolilloGiordanoHaberman2011} introduced
the Tucker3 model to mortality analysis, decomposing a three-way array
of log death rates indexed by age, year, and country for ten European
populations; the resulting time-mode components served as a synthesized
mortality index that could be projected forward via standard ARIMA
methods.  \citet{GiordanoHabermanRussolillo2019} extended this to
coherent subgroup modeling, and \citet{BergeronBoucherEtAl2018NAAJ}
adapted the Tucker3 framework to compositional mortality data -- life
table death distributions rather than log rates -- for forecasting
Canadian provincial mortality.  \citet{DongEtAl2020} conducted the
most systematic comparison to date, applying both canonical polyadic
(CP) decomposition and several Tucker variants to a three-way array
(age, year, country/gender) for ten European populations and two
genders, and demonstrated that multi-way decompositions consistently
outperform single-population Lee--Carter on out-of-sample forecasts.
Most recently, \citet{CardilloEtAl2024} moved to a full four-way
Tucker4 model -- cause of death, age, year, and country -- using WHO
data, illustrating the capacity of higher-order decompositions to
isolate cause-specific mortality patterns across populations.
All of these applications treat tensor decomposition as a
forecasting tool: the decomposition extracts latent temporal factors,
and time-series models project them forward.  None attempts to use the
decomposition as a generative model for producing mortality schedules
at specified parameters, none develops an inverse mapping for life
table fitting or classification, and none incorporates summary-indicator
prediction.  Moreover, sex is handled in these frameworks either by
separate analysis or by folding it into the country/population
dimension -- an approach that treats the sex differential as an
attribute of the population rather than as a structurally distinct mode
of the decomposition.  The question of how to produce sex-coherent
forecasts -- avoiding the divergent female--male trajectories that
independent fits invariably produce -- has been addressed through a
variety of post-hoc mechanisms: common-factor augmentation
\citep{LiLee2005}, product-ratio functional PCA
\citep{HyndmanBoothYasmeen2013}, complex-valued Lee--Carter
\citep{deJongTickleLi2016}, sex-ratio models
\citep{BergeronBoucherEtAl2018SR}, and compositional-data coherence
\citep{BergeronBoucherEtAl2017}, among others.  Each of these
achieves non-divergence between the sexes, but through constraints
imposed after the decomposition rather than through the structure of
the decomposition itself.

Clark developed a \emph{component model} of
mortality using the SVD in a new way
\citep{clark2001investigation,Clark2019}.  The model represents any
age-specific mortality schedule as a weighted sum of a small number of
orthogonal, age-varying vectors (components) identified by the SVD of a
matrix of observed schedules:
$\bm{m} \approx \sum_{i=1}^{c} w_i \, s_i \bm{u}_i$,
where $\bm{u}_i$ are the left singular vectors, $s_i$ the singular
values, and $w_i$ scalar weights that vary across schedules.  The
conceptual distinction from Wilmoth's earlier work is important.
Wilmoth developed an age--period--cohort mortality model
\citep{WilmothCaselli1987,Wilmoth1988,WilmothVallinCaselli1989,Wilmoth1990}
whose basic structure is $\log(m_x) = (\text{mean model}) +
(\text{residual model})$: age and period main effects are subtracted
first, and the SVD operates on the residuals to capture the remaining
age--period interaction -- a mean-plus-residual architecture that the
Lee--Carter model simplifies to rank one
\citep[\emph{cf.}][]{LeeCarter1992} and that the Log-Quad model of
\citet{Wilmoth2012} later adapted for generating model life tables.  In
all of these formulations the SVD term models what is left over after
a mean has been removed.  The component model has no separate mean:
the first component $s_1 \bm{u}_1$ captures the average age shape of
mortality and its weight encodes overall level, while higher-order
components encode age-specific deviations from that shape, independent
of level.  Because the SVD operates on the entire matrix of mortality
schedules rather than on residuals, the components jointly represent
everything -- level, shape, and their interaction -- in a single
decomposition \citep{Clark2019}.

In the first iteration of this work, \citet{clark2001investigation} applied PCA to mortality data
from 19 sites in the INDEPTH demographic surveillance network
\citep{INDEPTHNetwork2002} in sub-Saharan Africa -- a region not
represented in any existing model life table system.  Using the
component model and hierarchical clustering of the PCA coefficients
(with the level-controlling constant removed so that clusters reflect
age-pattern shape rather than overall mortality level), that work
identified seven characteristic mortality patterns for Africa, two of
which exhibited the distinctive young-adult excess associated with
HIV/AIDS.  The INDEPTH components reproduced all eight
Coale--Demeny and UN model patterns with near-perfect fidelity ($R^2
> 0.99$ using 15 components), confirming that the African data span a
space at least as rich as the one underlying the classical systems.  Superimposing
the HIV-related components at varying intensities onto a baseline pattern
produced a graded set of demonstration model life tables that
illustrated the demographic impact of the epidemic.

\citet{Clark2019} formalized and extended this approach as the
\emph{SVD-Comp} model, calibrated to 4{,}486 single-year-of-age life
tables from the Human Mortality Database.  The SVD of the HMD
mortality matrix yields age-varying components whose first four terms
capture more than 99.9\% of the total variance in the data -- a
dramatic compression from thousands of schedules to four vectors.  To
make the components operationally useful, Clark modeled each SVD weight
$w_i$ as a function of child mortality ($_5q_0$) or both child and
adult mortality ($_5q_0$, ${}_{45}q_{15}$) through polynomial
regression on logit-transformed inputs, with separate models for each
sex and each of the four retained components.  Given an input value
of $_5q_0$ (one-parameter model) or both $_5q_0$ and ${}_{45}q_{15}$
(two-parameter model), the regression predicts the weights, and the
weighted sum of scaled components produces a complete single-year-of-age
mortality schedule.  Cross-validation confirmed that the model is
robust to the selection of calibration schedules, and direct comparison
showed that SVD-Comp outperforms the Log-Quad model of
\citet{Wilmoth2012} across the HMD, particularly at younger ages where
the single-year resolution provides finer detail than the five-year age
groups of Log-Quad.

All of these matrix methods share a fundamental limitation when mortality
data are naturally indexed by more than two dimensions.  The Human
Mortality Database \citep[HMD;][]{HMD2024} provides annual,
single-year-of-age mortality rates for approximately 50 populations
across some 40 countries -- predominantly in Europe, but also including the United
States, Canada, Australia, New Zealand, Japan, South Korea, Taiwan, Hong
Kong, Chile, and Israel -- separately for females and males, spanning in
some cases more than two and a half centuries.  This produces a
four-dimensional data structure indexed by sex, age, country, and year.
Any matrix decomposition must first flatten at least two of these
dimensions together, sacrificing the ability to model their interactions
directly.  A matrix approach that concatenates female and male schedules,
for instance, treats the sex differential as part of the age pattern
rather than as a structurally distinct dimension.  Similarly, stacking
countries forces the decomposition to find components that are shared
across all populations or to treat country effects as noise.

Tensor decomposition provides a natural generalization.  Just as the
singular value decomposition \citep[SVD;][]{Eckart1936,GolubVanLoan2013}
factors a matrix into orthogonal components along its two dimensions, the
Tucker decomposition \citep{Tucker1966,DeLathauwer2000} factors a
higher-order tensor into orthogonal components along each of its
dimensions, linked by a core tensor that encodes their interactions
\citep[for a comprehensive review, see][]{KoldaBader2009}.
Applied to the four-dimensional mortality tensor $\M \in \R^{S \times A
\times C \times T}$ -- where $S = 2$ indexes sex (female, male), $A$
indexes single-year ages, $C$ indexes countries, and $T$ indexes calendar
years -- the Tucker decomposition simultaneously identifies a
small number of sex components (here exactly two, preserving the
female--male distinction); a set of age basis functions that capture the
characteristic shapes of mortality across the lifespan; a set of country
loadings that locate each population in a low-dimensional space; and a set
of temporal loadings that trace the trajectory of mortality change over
time.  The core tensor specifies precisely how these components combine to
produce the observed mortality at any given sex, age, country, and year.

The dimensionality-reduction approach has also been especially
influential in mortality forecasting -- and the Tucker framework provides
a natural forecasting target, because projecting the country-year core
matrices forward in time is equivalent to forecasting the complete
sex-age mortality surface.  Lee and Carter's original
contribution was as much a forecasting method as a decomposition: the
random walk with drift on the temporal index $k_t$ provided the first
widely-adopted probabilistic forecast of national mortality
\citep{LeeCarter1992}.  Subsequent work has extended the forecasting
architecture in several directions.  \citet{LeeMillerLeeCarter2001}
evaluated forecast performance and identified a systematic tendency
to underpredict life expectancy arising from the fixed age pattern.
\citet{BoothMaindonaldSmith2002} addressed this by optimising the
fitting period to better represent recent trends.
\citet{HyndmanUllah2007} elevated the rank from 1 to 6, treating
log mortality as a functional time series and forecasting each
component via random walk.  \citet{LiLee2005} introduced coherent
multi-population forecasting by augmenting the single-population
model with a common factor, and
\citet{HyndmanBoothYasmeen2013} achieved coherence through a
product-ratio approach.  \citet{deJongTickle2006} reformulated the
model in state-space form, enabling Kalman-filter estimation.  Recent
reviews \citep{BoothTickle2008,ShangBoothHyndman2011,
BaselliniCamardaBooth2023} document three persistent challenges: the
fixed age pattern in Lee--Carter tends to underpredict life expectancy,
independent sex-specific fits produce divergent forecasts, and
prediction intervals are sensitive to the time-series specification.

This paper extends the matrix-based component model of
\citet{clark2001investigation} and \citet{Clark2019} to a
multidimensional Tucker tensor decomposition and develops a complete
framework that addresses all four tasks -- model life tables,
fitting, summary-indicator prediction, and forecasting -- within
a single coherent architecture.  The framework delivers three
products:
\begin{enumerate}
    \item \textbf{A low-rank representation of human mortality.}  The
    Tucker decomposition of the HMD mortality tensor produces a compact set
    of basis functions and loadings from which any observed life table can
    be reconstructed with high accuracy using only a few components per
    dimension.  Because the decomposition operates on all four dimensions
    jointly, it captures the structure of sex differentials, age patterns,
    cross-country variation, and secular trends in a single coherent model.

    \item \textbf{A clustering of mortality regimes.}  The Tucker
    decomposition encodes each country-year's mortality schedule through
    an effective core matrix that combines country and year loadings with
    the core tensor.  By removing the first age component -- which captures
    overall mortality level -- the remaining features describe the
    \emph{shape} of the age schedule independent of level.  Clustering
    in this level-controlled feature space identifies canonical mortality
    regimes: recurring age-pattern signatures that capture systematic
    differences in the shape of the mortality curve.  Because the unit
    of clustering is the country-year rather than the country, a single
    population can occupy different regimes at different points in its
    history -- for instance, transitioning from one age-pattern family
    to another as its epidemiological profile evolves.  The regimes
    that emerge are canonical age-pattern families; populations from
    diverse geographies and time periods may share the same regime if
    their underlying age-pattern shapes are similar
    \citep{Omran1971,Mesle2004}.  The historical mortality
    transition itself is captured not by the cluster structure but by the
    within-cluster trajectories that trace how the age-sex pattern evolves
    as $\ezero$ rises.  Each cluster is characterized by a representative
    age-pattern shape and a typical sex differential.

    \item \textbf{A separable model for exceptional mortality.}  Wars and
    pandemics produce mortality patterns that are qualitatively different
    from the secular trends captured by the baseline decomposition.
    Combat mortality concentrates in young-adult males;
    pandemic mortality varies by pathogen type, with respiratory pandemics
    and enteric epidemics producing distinct age profiles.  The framework
    isolates these exceptional patterns, estimates canonical disruption
    profiles by type, and provides a linear model that adds exceptional
    mortality at arbitrary intensity to any baseline schedule.
\end{enumerate}

The combination of these three products yields a practical tool: given a
target mortality level (expressed as life expectancy at birth, $\ezero$),
a mortality regime (identified by cluster membership), and an optional
exceptional mortality scenario (specified by disruption type and
intensity), the framework produces a complete, sex-specific mortality
schedule across all ages.  Three additional capabilities extend this
generative core:
\begin{enumerate}
\setcounter{enumi}{3}

    \item \textbf{A life table fitter.}  The generative model maps
    parameters (cluster, $\ezero$, disruption type and intensity) to a
    mortality schedule; the fitter inverts this mapping, estimating all
    parameters from a single observed schedule.  A three-stage algorithm
    first selects the cluster by minimum residual norm, then estimates
    $\ezero$ via linearized projection along the cluster trajectory, and
    finally tests each disruption type via penalized projection with
    Laplace-approximated Bayes factors for model selection.  The fitter
    operates without any temporal context -- it needs only one
    age-sex-specific mortality schedule as input -- making it applicable
    to isolated surveys, census-derived estimates, or any setting where
    a complete schedule is available but the underlying regime and
    disruption status are unknown.

    \item \textbf{Summary-indicator prediction.}  In many applied settings,
    only summary mortality indicators are available rather than a full
    age schedule: child mortality (${}_5q_0$), or child and adult
    mortality (${}_5q_0$ and ${}_{45}q_{15}$) together.  This pathway
    maps these summary inputs to complete sex-specific schedules via a
    neural network trained with a reconstruction loss that enforces
    consistency with the Tucker basis.  The architecture exploits the
    same low-dimensional structure as the generative model: the network
    predicts Tucker core weights from the summary indicators, and the
    Tucker reconstruction delivers the full schedule.  This reformulates
    the classical SVD-Comp approach \citep{Clark2019} within the tensor
    framework, gaining structural sex coherence and the richer basis
    provided by the four-dimensional decomposition.

    \item \textbf{Mortality forecasting.}  The Tucker decomposition
    provides a natural forecasting target: projecting the effective core
    matrix $G_{ct}$ forward in time is equivalent to forecasting the
    complete sex-age mortality surface.  A PCA reduction compresses the
    84-dimensional $G_{ct}$ to five scores, and a damped local linear
    trend Kalman filter projects these scores forward with the drift
    state constrained toward a two-level hierarchical target that blends
    the HMD-wide consensus with country-specific trends.
    An empirical search over the full three-level simplex (HMD-wide,
    cluster, country) demonstrated that the cluster level adds no signal
    beyond what the HMD-wide and country levels already provide --
    a finding that simplifies the architecture while marginally improving
    accuracy.  Sex coherence is structural through the shared Tucker
    factor matrices rather than imposed post hoc.  Delta-method
    prediction intervals, validated against Monte Carlo propagation, are
    calibrated from cross-validation z-scores.
\end{enumerate}

The remainder of this paper is organized as follows.
\Cref{sec:data} describes the Human Mortality Database and the structure
of the data.
\Cref{sec:preprocessing} details the preprocessing steps applied to the
raw data, including the logit transformation, treatment of missing and
irregular values, and adaptive temporal pooling to address Poisson noise.
\Cref{sec:tucker} develops the Tucker decomposition in detail, building
intuition from the familiar SVD and emphasizing the geometric
interpretation of projections and subspaces.
\Cref{sec:application} applies the decomposition to the HMD mortality
tensor and interprets the resulting components demographically.
\Cref{sec:clustering} describes the clustering model used to identify
mortality regimes, using level-controlled age-structure features
derived from the Tucker core tensor, and the complementary epoch
classification that categorizes the time trend in mortality level.
\Cref{sec:reconstruction} develops the reconstruction model that produces
mortality schedules at arbitrary mortality levels within each cluster.
\Cref{sec:exceptional} presents the exceptional mortality model.
Each of these sections develops the relevant theory and then
immediately applies it to the current HMD, reporting the concrete
dimensions, selected ranks, cluster structure, reconstruction accuracy,
disruption profiles, and neural extensions that result.
\Cref{sec:fitting} develops the life table fitter.
\Cref{sec:svdcomp} develops the summary-indicator prediction pathway.
\Cref{sec:forecasting} presents the forecasting framework.
\Cref{sec:discussion} discusses the framework as a whole, its
limitations, and directions for further work.
\Cref{sec:notation} collects the principal notation in a single
reference table; notation is introduced where each symbol is first
needed, and the consolidated table serves as a reference rather than
prerequisite reading.
\Cref{sec:computational} documents the computational environment
and software.
Appendix~\ref{app:events} provides the complete event dictionaries used to
identify exceptional years.

\section{Data}
\label{sec:data}

\subsection{The Human Mortality Database}
\label{sec:data:hmd}

The data for this study come from the Human Mortality Database
\citep[HMD;][]{HMD2024}, a joint project of the University of California,
Berkeley and the Max Planck Institute for Demographic Research.  The HMD
provides detailed mortality data computed from official death counts and
population estimates using a uniform methodology documented in
\citet{Wilmoth2021}.  At the time of writing, the HMD covers approximately
40 countries, comprising approximately 50 populations and subpopulations
when constituent parts are counted separately, with temporal coverage
that varies widely:
Sweden's records extend from 1751, France from 1816, and England and Wales
from 1841, while most other countries enter the database in the late
nineteenth or early twentieth century.  Several countries have data through
2024.

The countries represented in the HMD are predominantly European -- spanning
Scandinavia, Western Europe, Southern Europe, Eastern Europe, and the
Baltic states -- together with several high-income countries outside Europe:
the United States, Canada, Australia, New Zealand, Japan, South Korea,
Taiwan, Hong Kong, Chile, and Israel.  The database does not include
countries in Sub-Saharan Africa, South Asia, Southeast Asia, or most of
Latin America.  Consequently, the mortality experience captured by the HMD
reflects the high-income world, and the models developed here should be
understood in that context.

Some countries in the HMD are represented by more than one population.  The
United Kingdom, for example, appears as a national total and separately as
England and Wales, Scotland, and Northern Ireland; Germany is available as
a national total and separately as East and West Germany (covering the
period of division); and New Zealand is split into M\=aori and non-M\=aori
subpopulations.  These subpopulations are treated as distinct entries in
the tensor along the country dimension, bringing the total number of
populations to approximately 50.

The HMD also provides separate civilian-only population series for France
(FRACNP) and England and Wales (GBRCENW).  These civilian series exclude
deaths reported by military authorities and differ from the corresponding
total-population series (FRATNP, GBRTENW) only during wartime --
specifically 1914--1920 and 1940--45 for France, and similar periods for
England and Wales.  Outside those years the two series are identical.  We
exclude the civilian series from the tensor because retaining them would
introduce two problems: during non-war years they are exact duplicates of
the total-population entries, inflating the effective sample size without
adding information; and during war years they omit the military mortality
that constitutes the dominant signal in the exceptional mortality analysis
of \cref{sec:exceptional}, which would dilute the estimated war disruption
profiles.  Using only the total-population series ensures that the tensor
captures the full mortality experience -- civilian and military -- for
these countries.

\subsection{Database architecture}
\label{sec:data:db}

The HMD distributes its data as a collection of text files -- one per
combination of file type, sex, and resolution (e.g.\ \texttt{fltper\_1x1}
for female single-year period life tables).  Each file contains all
populations concatenated with a population-name column.  To make this
data efficiently queryable for the multidimensional analyses in this
paper, we load it into a DuckDB database implementing an
Entity-Attribute-Value (EAV) architecture.

\subsubsection{Schema design}

The database consists of six dimension tables and one central fact table.
The dimension tables define the coordinate system: \emph{geography}
(population codes and labels), \emph{time\_period} (calendar years),
\emph{sex} (female, male, both), \emph{age\_group} (single-year and
five-year intervals with a \texttt{sort\_key} that orders correctly as a
string while remaining human-readable), \emph{resolution}
(age-width~$\times$~period-width, e.g.\ $1 \times 1$, $5 \times 1$),
and \emph{source\_version} (HMD release identifier, with an
\texttt{is\_current} flag for version control).

The fact table \texttt{life\_table\_eav} stores every value as a single
row keyed by the full dimension tuple plus a life-table-column identifier.
The composite primary key
\begin{quote}
    (geography, time\_period, sex, age\_group,
     resolution, lt\_column, source\_version)
\end{quote}
guarantees uniqueness across all HMD resolutions in a single database --
resolving the primary-key collision that would otherwise arise from
mixing $1 \times 1$ and $5 \times 1$ data, where both share age group~0.

The life-table columns stored are the eight standard period life table
quantities ($m_x$, $q_x$, $l_x$, $d_x$, $L_x$, $T_x$, $e_x$, $a_x$)
plus raw death counts, person-years of exposure, and observed (ungraduated)
mortality rates $\hat{m}_x^{\text{obs}} = D_x / E_x$.  The distinction
between the graduated life-table $m_x$ -- which is internally consistent
with $q_x$, $l_x$, etc.\ -- and the raw observed $\hat{m}_x^{\text{obs}}$
is important: this analysis works exclusively with $q_x$ from the
graduated tables, but the raw counts and exposures are retained for
potential future use in pooling and smoothing.

\subsubsection{Query interface}

Eight parameterised SQL table macros reconstruct conventional wide-format
life tables on demand from the normalised EAV store.  The macros range
from single-series lookups (one geography $\times$ period $\times$ sex,
returning the standard age-column life table) to matrix-format outputs
that produce age~$\times$~population arrays suitable for direct input to
SVD or tensor decomposition -- the layout needed by
\cref{sec:preprocessing:tensor}.  All macros accept an optional
\texttt{source\_version\_id} parameter that defaults to the current
version, enabling reproducible queries against any historical HMD release.

\subsubsection{Versioning and reproducibility}

Each ingestion run creates a new \texttt{source\_version} row alongside
existing data rather than overwriting it.  Both old and new versions
remain independently queryable, so the entire analysis pipeline can be
re-run against a specific HMD release without re-downloading.  The
database used for this paper corresponds to HMD release 20260218.

\subsection{The quantity of interest: \texorpdfstring{$\qx$}{1qx}}
\label{sec:data:qx}

The fundamental demographic quantity in this analysis is $\qx$, the
probability that an individual alive at exact age~$x$ will die before
reaching exact age~$x + 1$.  The HMD provides period life tables
containing $\qx$ at single-year ages separately for females and males.

A complete mortality schedule is a vector of $\qx$ values across all ages
for one sex in one country in one year.  These schedules -- one
per sex-country-year observation -- are the elements from which we
construct the four-dimensional tensor~$\M$.  The full set of $\qx$ values determines the life table
entirely: the survival function, life expectancy at every age, and all
summary mortality indices can be computed from $\qx$ alone
\citep{Preston2001}.

Life expectancy at birth, $\ezero$, serves as the single most important
summary of a mortality schedule.  Given a vector of $\qx$ values at ages
$x = 0, 1, \ldots, \omega$, we build a survivorship column
$l_0 = 1$, $l_{x+1} = l_x \cdot (1 - \qx)$, and compute
\begin{equation}
    \label{eq:e0}
    \ezero = \sum_{x=0}^{\omega} L_x\,,
\end{equation}
%
where $L_x = (l_x + l_{x+1})/2$ is the person-years lived in the
interval $[x, x+1)$ under a uniform distribution of deaths assumption.
For the first age interval the code uses a fixed infant separation
factor $a_0 = 0.3$, so that $L_0 = 0.3\,l_0 + 0.7\,l_1$; this
reflects the concentration of neonatal deaths early in the first year
of life and approximates the average of the sex-specific, mortality-dependent
$a_0$ values tabulated in \citet{Preston2001} across the range of modern
HMD populations.

\subsection{The logit transform}
\label{sec:data:logit}

All analysis in this paper operates not on $\qx$ directly but on its logit
transform,
\begin{equation}
    \label{eq:logit}
    y \;=\; \logit(\qx) \;=\; \log \frac{\qx}{1 - \qx}\,.
\end{equation}
%
Applied to every age in every sex-country-year schedule, this defines
the tensor~$\M$ with elements $\M_{s,a,c,t} = y_{s,a,c,t}$.

The logit transform is a deliberate modeling choice with several
important properties:
\begin{enumerate}
    \item \textbf{Domain mapping.}  The raw quantity $\qx$ is a probability
    and therefore bounded to the interval $(0, 1)$.  The logit maps this to
    the entire real line, $y \in (-\infty, +\infty)$, which is necessary
    for the Tucker decomposition (and indeed any linear decomposition) to
    operate without constraint violations.  Any linear combination of logit
    values, when mapped back through the inverse logit, is guaranteed to
    produce a valid probability.

    \item \textbf{Linearization of mortality relationships.}  On the logit
    scale, the Gompertz law of mortality -- exponential increase of death
    rates with age \citep{Gompertz1825} -- becomes approximately linear
    above age~40.  More generally, relationships between mortality
    schedules that are multiplicative on the $\qx$ scale become
    approximately additive on the logit scale, making them amenable to
    the linear operations (summation, projection, averaging) that
    underpin the Tucker decomposition.

    \item \textbf{Variance stabilization.}  The variance of $\qx$ depends
    strongly on its level: low-mortality ages (children and young adults,
    $\qx \sim 10^{-4}$) have far less absolute variation across
    populations than high-mortality ages (infants, the elderly).  The logit
    transform compresses the scale at the extremes, so that a given
    absolute change in $y$ corresponds to a proportional change in the odds
    of dying.  This prevents the decomposition from being dominated by
    variation at old ages where $\qx$ is large.

    \item \textbf{Interpretability of additive effects.}  An additive
    perturbation $\Delta$ on the logit scale corresponds to a
    multiplicative change in the odds of dying.  Suppose a baseline
    mortality rate $\qx$ is perturbed to $\qx^{\ast}$ by adding $\Delta$
    on the logit scale:
    \begin{equation}
        \label{eq:logit_additive}
        \logit(\qx^{\ast}) = \logit(\qx) + \Delta
        \quad\Longleftrightarrow\quad
        \frac{\qx^{\ast}/(1-\qx^{\ast})}{\qx/(1-\qx)} = \exp(\Delta)\,.
    \end{equation}
    %
    The left-hand side of \cref{eq:logit_additive} states the operation in
    logit space: the disrupted logit-mortality is the baseline plus a
    shift~$\Delta$.  The right-hand side gives the equivalent statement on
    the natural scale: the ratio of disrupted to baseline odds of dying
    equals $\exp(\Delta)$.  When $\Delta > 0$, mortality increases (the
    odds ratio exceeds one); when $\Delta < 0$, mortality decreases; and
    $\Delta = 0$ leaves mortality unchanged.  Crucially, the
    multiplicative factor $\exp(\Delta)$ does not depend on the baseline
    level~$\qx$, so the same additive shift in logit space has a
    proportional effect regardless of whether $\qx$ is $10^{-4}$
    (young adults) or $0.3$ (the very old).  This property is exploited
    directly in the exceptional mortality model (\cref{sec:exceptional}),
    where disruptions are modeled as additive shifts in logit space and
    therefore as multiplicative changes to the odds of dying at each age.
\end{enumerate}

The logit transform has a long history in demography, beginning with the
relational model life table system of \citet{Brass1971}, which models the
logit of survival as a linear function of a standard.  The modified logit
system of \citet{MurrayEtAl2003} extended this approach.  In statistics
more broadly, the logit is the canonical link function for binomial data
\citep{McCullagh1989}, and its use here connects the demographic modeling
framework to the broader statistical literature on generalized linear
models.

\subsection{Age range and boundary treatment}
\label{sec:data:age}

The HMD provides $\qx$ at single-year ages from 0 through 110+, where
the final entry is an open interval representing all deaths at age~110
and above.  We truncate to ages $0, 1, 2, \ldots, 109$, dropping the open
interval, so that $A = 110$.  This upper bound is chosen to include
effectively all of the observed mortality experience -- survivorship to
age~110 is negligible even in the lowest-mortality populations
\citep{KannistoEtAl1994} -- while
avoiding the open interval, which has a qualitatively different
interpretation ($\qx = 1$ by construction).

At the lower boundary, $\qx$ values of exactly zero occasionally appear in
the HMD for some age-sex-country-year combinations, typically at
young-adult ages in small populations where no deaths were observed in a
given year.  Because $\logit(0) = -\infty$, these values must be handled
before transformation.  We impose a floor,
\begin{equation}
    \label{eq:qx_floor}
    \tilde{q}_{s,a,c,t} = \max\bigl(q_{s,a,c,t},\; q_{\min}\bigr)\,,
\end{equation}
%
where $q_{\min}$ is a small positive constant (e.g., $10^{-8}$),
well below any empirically observed age-specific mortality rate.  This
floor is applied before the logit transform and ensures that all elements
of~$\M$ are finite.  The choice of $q_{\min}$ is not sensitive: values
anywhere in the range $10^{-8}$ to $10^{-5}$ produce effectively
identical results because the logit of such small values is large and
negative (approximately $-18$ to $-12$), far from the range where the
decomposition operates, and because the affected cells are rare.

Similarly, any $\qx$ values of exactly one (which occur only in the final
open age interval, excluded by truncation) would produce
$\logit(1) = +\infty$ and are avoided by the age-range restriction.

The combination of the floor at $q_{\min}$ and truncation at age~109
guarantees that every element of the tensor~$\M$ is a finite real number,
as required by the decomposition.

\section{Preprocessing}
\label{sec:preprocessing}

The raw HMD life tables require several preprocessing steps before the
tensor~$\M$ can be constructed and decomposed.  These steps fall into
three categories: data curation (\cref{sec:preprocessing:curation}),
tensor construction from ragged data (\cref{sec:preprocessing:tensor}),
and adaptive temporal pooling to address Poisson noise at low-mortality
ages (\cref{sec:preprocessing:smoothing}).

\subsection{Data curation}
\label{sec:preprocessing:curation}

The HMD provides period life tables packaged as a consolidated
statistics archive.  The companion implementation ingests this archive
into a normalized DuckDB database \citep{MuehleisenetAl2019}, which stores
all life-table quantities ($m_x$, $q_x$, $a_x$, $l_x$, $d_x$, $L_x$,
$T_x$, $\ezero$) in a single relational table indexed by population
code, sex, year, and age.  All downstream queries -- including the
extraction of 1$\times$1 period life tables and the retrieval of deaths
and exposures for adaptive temporal pooling -- are expressed as SQL,
ensuring reproducibility and making the pipeline trivially extensible
to future HMD releases.  Several curation steps are applied:
\begin{enumerate}
    \item \textbf{Missing values.}  Any life table containing one or more
    \texttt{NaN} entries is dropped entirely.  These are rare and typically
    result from country-years where the HMD's input data were insufficient
    to compute a complete life table.

    \item \textbf{Flat life tables.}  Some early historical life tables
    exhibit $\qx$ values that are constant across a range of old ages
    (e.g., identical values at ages~105 and~109), indicating that the
    original data lacked the resolution to estimate age-specific mortality
    at extreme ages.  These life tables are detected by comparing $\qx$ at
    two high ages and are removed from the analysis.  When a flat life
    table is detected for one sex of a country-year, the paired sex is
    also removed to maintain the sex-pairing structure described below.

    \item \textbf{Sex pairing.}  The tensor~$\M$ requires that every
    country-year observation include both a female and a male life table.
    Any country-year for which only one sex is available is dropped.  In
    practice, unpaired observations are extremely rare in the HMD.

    \item \textbf{Boundary treatment.}  The $\qx$ floor and age truncation
    described in \cref{sec:data:age} are applied, ensuring that all values
    are finite after the logit transform.
\end{enumerate}

After curation, the logit transform (\cref{sec:data:logit}) is applied to
every element.

\subsection{Identification and exclusion of exceptional years}
\label{sec:preprocessing:exceptional}

Wars and pandemics produce transient mortality shocks that are
qualitatively different from the secular trends the Tucker decomposition
is designed to capture.  If these exceptional country-years are included
in the tensor, their extreme mortality schedules distort the factor
matrices, forcing the decomposition to allocate rank to
short-lived disruptions rather than to the slowly varying structure
of interest.  Exceptional years must therefore be identified and
\emph{excluded} before the decomposition, so that the resulting factor
matrices and core tensor describe the baseline mortality surface free
of transient shocks.

Exceptional country-years are identified using the event dictionaries
described in \cref{sec:exceptional:events}, which classify each
country-year as non-exceptional or as affected by one of three
disruption types: armed conflict, respiratory pandemic, or enteric
pandemic.  The dictionaries are compiled from standard historical
sources and cover the full temporal span of the HMD (see
\cref{app:events} for the complete listing).

Let $\mathcal{E} = \{(c,t) : d_{c,t} \neq 0\}$ denote the set of
exceptional country-years, where $d_{c,t}$ is the disruption label
defined in \cref{sec:exceptional:events}.  These observations are
removed from the tensor before the decomposition.  Specifically:
\begin{enumerate}
    \item The observed mask is updated:
    $O_{c,t} \leftarrow 0$ for all $(c,t) \in \mathcal{E}$.
    \item The corresponding slices of the tensor are treated as missing
    and are imputed along with the other missing entries during tensor
    construction (\cref{sec:preprocessing:tensor}).
\end{enumerate}
The exceptional country-years are retained separately for use in the
disruption model of \cref{sec:exceptional}, where their residuals
relative to the baseline decomposition provide the signal from which
disruption profiles are estimated.  The clustering
(\cref{sec:clustering}) and trajectory reconstruction
(\cref{sec:reconstruction}) are likewise computed from non-exceptional
data only, ensuring that the baseline model is not contaminated by
transient shocks.

\subsection{Tensor construction from ragged data}
\label{sec:preprocessing:tensor}

The curated life tables span different year ranges for different countries.
Sweden's data extend from 1751, while many countries enter the database
only in the late nineteenth or early twentieth century.  To form a complete
four-dimensional tensor $\M \in \R^{S \times A \times C \times T}$, we
must decide which countries and years to include and how to handle the
resulting gaps.

\subsubsection{Country and year selection}

Countries with fewer than a minimum number of observed years (e.g., 5) are
excluded from the tensor, as they provide insufficient data to estimate
even a stable country mean.  The threshold is kept deliberately low so that
countries with shorter but valuable series (e.g., countries that joined the
HMD recently) are retained.  The year dimension of the tensor is defined as
the union of all years in which at least one retained country has data.

\subsubsection{The observed mask}

Not every country has data for every year in the tensor.  An \emph{observed
mask} $\bm{O} \in \{0,1\}^{C \times T}$ records which country-year
combinations are directly observed:
\begin{equation}
    \label{eq:obs_mask}
    O_{c,t} =
    \begin{cases}
        1 & \text{if country $c$ has data in year $t$,} \\
        0 & \text{otherwise.}
    \end{cases}
\end{equation}
%
Because temporal coverage varies by country, the observed mask is ragged:
countries with long series (Sweden, France, England and Wales) contribute
observed entries across almost all years, while countries with shorter
series contribute only to recent decades.

\subsubsection{Imputation of missing entries}
\label{sec:preprocessing:imputation}

The HOSVD requires a complete tensor.  Missing country-years ($O_{c,t} =
0$) must be filled before the decomposition can proceed.  The simplest
strategy -- and the one we adopt initially -- is to impute each missing
entry with the temporal mean of the observed entries for that country:
\begin{equation}
    \label{eq:imputation}
    \M_{s,a,c,t}^{\text{imputed}} =
    \frac{1}{|\{t' : O_{c,t'} = 1\}|}
    \sum_{\{t' : O_{c,t'} = 1\}} \M_{s,a,c,t'}\,,
    \qquad \text{for all } O_{c,t} = 0.
\end{equation}
%
Imputed entries are identifiable via the mask~$\bm{O}$ and are excluded
from downstream analyses (e.g., clustering) where only genuine
observations should contribute.

\subsubsection{Imputation bias: the up-weighting problem}
\label{sec:preprocessing:imputation_bias}

Country-mean imputation has an important limitation that must be
acknowledged.  Consider a country with $T_c$ observed years out of a
total of $T$ years in the tensor.  Imputation fills $T - T_c$ cells with
identical copies of the country's mean schedule.  In the mode unfoldings
that feed the SVD, these repeated entries act as additional
``observations'' of the country mean:
\begin{itemize}
    \item In the \textbf{age unfolding} ($\bm{M}_{(2)}$, with ages as
    rows), the country's mean age schedule is replicated $T - T_c$ extra
    times across columns.  This pulls the age basis functions~$\bA$ toward
    the age pattern of countries with more missing data, giving them
    disproportionate influence on the age basis relative to their actual
    observation count.

    \item In the \textbf{country unfolding} ($\bm{M}_{(3)}$, with
    countries as rows), a country with many imputed years has most of its
    row filled with identical values.  The SVD sees this as a strong, low-rank
    signal, inflating the country's apparent importance while suppressing
    its true temporal variation.

    \item In the \textbf{year unfolding} ($\bm{M}_{(4)}$, with years as
    rows), early years -- where most countries are missing -- are filled with
    country means that carry no temporal information, diluting the genuine
    temporal signal from the few countries that do have data in those
    periods.
\end{itemize}

The severity of this bias depends on the ratio of imputed to observed
entries.  When the tensor spans 1751--2024 ($T = 274$ years) and a
country has only 30 years of data, the imputed entries outnumber the
observed by approximately $8 {:} 1$.  The country's mean is effectively
replicated 242 times, giving it far more influence on the decomposition
than its 30 genuine observations warrant.

Several strategies can mitigate this problem:
\begin{enumerate}
    \item \textbf{Weighted decomposition.}  Weight each entry in the
    mode unfoldings by $O_{c,t}$ (or by $1/n_c$ for imputed entries,
    where $n_c$ is the number of imputed years for country $c$), so that
    imputed values contribute minimally to the SVD.

    \item \textbf{Year-range restriction.}  Restrict the year dimension
    to a range where most countries have data (e.g., 1900--2024), reducing
    the number of imputed cells at the cost of discarding the longest
    historical series.

    \item \textbf{Incomplete tensor decomposition.}  Use an iterative
    algorithm that operates only on observed entries, estimating the factor
    matrices and core tensor by minimizing the reconstruction error over
    the observed mask only \citep{AcarEtAl2011}.  This avoids imputation
    entirely but is computationally more demanding and may require careful
    initialization and convergence monitoring.

    \item \textbf{Multiple imputation or regularization.}  Replace the
    single country-mean impute with draws from a model of temporal
    variation \citep{Rubin1987}, or add a regularization penalty that
    shrinks imputed entries toward zero influence.
\end{enumerate}

The choice among these strategies involves a trade-off between
methodological purity, computational cost, and the practical goal of
retaining as much data as possible.  We discuss the approach taken and
its consequences further in \cref{sec:application:imputation}.

\subsection{Adaptive temporal pooling}
\label{sec:preprocessing:smoothing}

At ages where mortality is very low -- typically from the late childhood
minimum (around age~10) through young adulthood (around age~40) -- the
annual probability of dying in most HMD countries is on the order of
$10^{-4}$ to $10^{-3}$.  In small populations, the number of deaths at
these ages in a single year can be very small, and $\qx$ estimates are
dominated by Poisson sampling noise.  On the logit scale, this noise is
amplified: the difference between $\logit(3 \times 10^{-5})$ and
$\logit(5 \times 10^{-5})$, for example, is approximately~0.5, a
substantial fluctuation relative to the smooth underlying age pattern.

Left unsmoothed, these fluctuations introduce high-frequency noise into
the tensor that the Tucker decomposition would need to capture, either by
allocating additional components (increasing rank) or by introducing
oscillations into the age basis functions~$\bA$.  Neither outcome is
desirable: the fluctuations are stochastic artifacts, not features of the
underlying mortality surface.

Rather than smoothing across ages after the fact, we address the noise at
its source by pooling across adjacent years -- following the approach
developed in \citet{clark2001investigation}.  For each adaptive block
the raw death counts $D_x$ and person-years of exposure $E_x$ are summed
across years and the pooled central death rate is computed as
$m_x^{\text{pooled}} = \sum D_x \big/ \sum E_x$ -- the demographically
correct exposure-weighted rate.  The $a_x$ values (average person-years
lived by those dying in the interval) are averaged arithmetically across
the same years.  The pooled $m_x$ and $a_x$ are then used to recompute a
full, internally consistent life table ($\qx$, $l_x$, $d_x$, $L_x$,
$T_x$, $\ezero$).

The pooling window is \emph{adaptive}: for each population and sex, we
start from the earliest non-exceptional year and accumulate
adjacent non-exceptional years until the pooled $m_x$ has no zeros at
valley ages.  Once zero-free, that block is finalized and we start the
next.  This means large populations (which rarely have zero $m_x$ at any
age) get small or no pools, while small populations such as Iceland get
wider pools.

The procedure is as follows:
\begin{enumerate}
    \item \textbf{Identify populations needing pooling.}  A population
    requires pooling if $m_x = 0$ appears at a \emph{valley age} -- defined
    as an age where the overall median $\qx$ across all countries and years
    falls below a threshold $q_{\text{thresh}}$ (e.g., $0.005$) -- in any
    non-exceptional year.

    \item \textbf{Build adaptive blocks.}  For each flagged population,
    iterate through its non-exceptional years chronologically.  Accumulate
    consecutive years into a block; after each addition, check whether the
    block's mean $m_x$ is nonzero at all valley ages.  If so, finalize the
    block and start the next.  If a population's entire non-exceptional
    history must be pooled into one block to eliminate zeros, that single
    block is used.

    \item \textbf{Pool deaths and exposures.}  Within each block, sum
    $D_x$ and $E_x$ across the block's years, separately for each age
    and sex, and compute $m_x^{\text{pooled}} = \sum D_x / \sum E_x$.
    Average $a_x$ arithmetically across the same years.

    \item \textbf{Recompute the life table.}  From the pooled $m_x$ and
    $a_x$, recompute $\qx$, $l_x$, $d_x$, $L_x$, $T_x$, and $\ezero$
    at every age using the standard life table identities
    \citep{Preston2001}.  This ensures internal consistency: the
    life table columns are jointly coherent rather than being
    smoothed independently.

    \item \textbf{Replace individual years.}  Every year within a pooled
    block receives the single pooled life table in place of its original
    yearly values.  Populations that required no pooling are left
    unchanged.
\end{enumerate}

The effect of this procedure is to replace noisy year-to-year
fluctuations in small populations at low-mortality ages with stable rates
that reflect the underlying risk profile averaged over a slightly wider
temporal window.  On the natural $\qx$ scale, the changes are negligible
in absolute terms because the affected rates are already very small.  On
the logit scale, however, the pooling substantially reduces the rank
required to represent the data faithfully, because the decomposition no
longer needs to allocate components to capture stochastic noise in the
valley region.

\begin{remark}[Temporal pooling as a modeling choice]
\label{rem:valley_smoothing}
Temporal pooling is not merely a data-cleaning step: it reflects a
substantive modeling assumption that the underlying mortality risk in the
low-mortality valley changes slowly enough over time that averaging $m_x$
across a few adjacent years does not suppress genuine temporal dynamics.
This assumption is well-supported: the secular mortality transition
operates on decadal timescales, and year-to-year fluctuations in $m_x$
at valley ages in small populations are dominated by Poisson sampling
noise rather than by genuine shifts in risk.  The adaptive block width
ensures that large populations -- where single-year rates are already
precise -- are pooled minimally or not at all.
\end{remark}

A second, distinct smoothing step is applied \emph{after} the Tucker
decomposition, targeting the age basis functions~$\bA$.  This step uses
variable-bandwidth Gaussian kernel smoothing across ages
(not across time) and is described in \cref{sec:tucker:bsmooth}.



We now describe the HMD data as they appear after the
preprocessing steps developed above.

\subsection{Data landscape}
\label{sec:results:data}

\begin{figure}[!htbp]
\centering
\includegraphics[width=\textwidth]{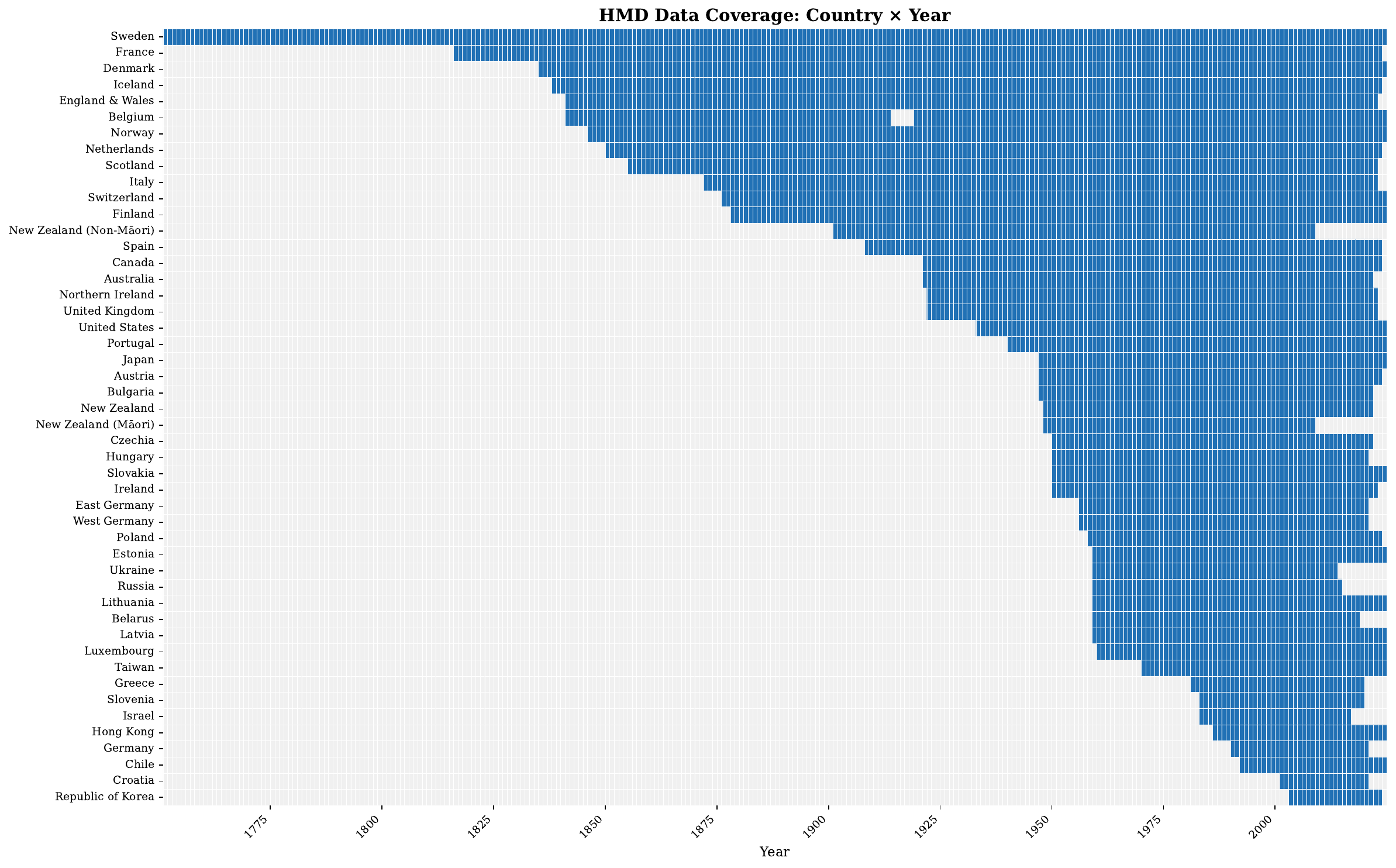}
\caption{HMD temporal coverage by population.  Each colored cell
indicates that a 1$\times$1 period life table is available for that
population-year.  Populations are sorted by the first year of available
data.}
\label{fig:s1_coverage}
\end{figure}

The HMD provides 1$\times$1 period life tables for approximately 50
populations spanning from 1751 (Sweden) through 2024.
\Cref{fig:s1_coverage} displays the temporal coverage map: each colored cell
indicates an available life table for a given population and calendar year.
The ragged structure is immediately apparent -- Sweden, Denmark, and France
provide more than two centuries of continuous data, whereas many Eastern
European and non-European populations enter the database only in the
mid-twentieth century.  \Cref{tab:s1_country_summary} summarizes each
population's year range and the number of usable life tables after
curation (\cref{sec:preprocessing:curation}).

{\setlength\LTleft{0pt}\setlength\LTright{0pt}
\begin{longtable}{@{\extracolsep{\fill}}llcrcc@{}}
\caption{Summary of HMD data by country/population} \label{tab:s1_country_summary} \\
\toprule
Code & Country & Year Range & Years & $e_0$ Range & Median $e_0$ \\
\midrule
\endfirsthead
\caption[]{Summary of HMD data by country/population (continued)} \\
\toprule
Code & Country & Year Range & Years & $e_0$ Range & Median $e_0$ \\
\midrule
\endhead
\midrule
\multicolumn{6}{r}{\textit{Continued on next page}} \\
\endfoot
\bottomrule
\endlastfoot
SWE & Sweden & 1751–2024 & 274 & 17.2–85.4 & 50.400 \\
FRATNP & France & 1816–2023 & 208 & 27.3–85.6 & 51.900 \\
DNK & Denmark & 1835–2024 & 190 & 36.7–83.9 & 62.400 \\
ISL & Iceland & 1838–2023 & 186 & 16.9–84.5 & 61.100 \\
GBRTENW & England \& Wales & 1841–2022 & 182 & 33.4–83.4 & 60.000 \\
BEL & Belgium & 1841–2024 & 179 & 31.7–84.3 & 58.700 \\
NOR & Norway & 1846–2024 & 179 & 43.4–84.9 & 64.600 \\
NLD & Netherlands & 1850–2023 & 174 & 29.9–83.6 & 65.600 \\
GBR\_SCO & Scotland & 1855–2022 & 168 & 38.8–81.3 & 59.800 \\
ITA & Italy & 1872–2022 & 151 & 23.5–85.4 & 61.200 \\
CHE & Switzerland & 1876–2024 & 149 & 38.5–85.9 & 67.700 \\
FIN & Finland & 1878–2024 & 147 & 26.4–84.8 & 64.400 \\
NZL\_NM & New Zealand (Non-Māori) & 1901–2008 & 108 & 50.6–83.1 & 69.200 \\
ESP & Spain & 1908–2023 & 116 & 29.9–86.3 & 70.200 \\
CAN & Canada & 1921–2023 & 103 & 55.9–84.4 & 73.400 \\
AUS & Australia & 1921–2021 & 101 & 59.1–85.7 & 73.000 \\
GBR\_NP & United Kingdom & 1922–2022 & 101 & 55.2–83.2 & 72.800 \\
GBR\_NIR & Northern Ireland & 1922–2022 & 101 & 53.7–82.8 & 71.500 \\
USA & United States & 1933–2024 & 92 & 58.3–81.5 & 73.400 \\
PRT & Portugal & 1940–2024 & 85 & 45.9–85.2 & 71.100 \\
AUT & Austria & 1947–2023 & 77 & 59.0–84.2 & 74.200 \\
JPN & Japan & 1947–2024 & 78 & 49.8–87.8 & 77.000 \\
BGR & Bulgaria & 1947–2021 & 75 & 52.6–78.7 & 69.900 \\
NZL\_NP & New Zealand & 1948–2021 & 74 & 67.0–84.5 & 74.900 \\
NZL\_MA & New Zealand (Māori) & 1948–2008 & 61 & 50.3–76.4 & 65.900 \\
SVK & Slovakia & 1950–2024 & 75 & 59.0–81.6 & 72.600 \\
HUN & Hungary & 1950–2020 & 71 & 59.9–79.7 & 70.200 \\
IRL & Ireland & 1950–2022 & 73 & 63.5–84.0 & 73.900 \\
CZE & Czechia & 1950–2021 & 72 & 62.0–82.1 & 73.400 \\
DEUTW & West Germany & 1956–2020 & 65 & 65.8–83.5 & 75.100 \\
DEUTE & East Germany & 1956–2020 & 65 & 65.8–83.6 & 74.400 \\
POL & Poland & 1958–2023 & 66 & 62.6–82.0 & 72.800 \\
RUS & Russia & 1959–2014 & 56 & 57.4–76.5 & 68.200 \\
BLR & Belarus & 1959–2018 & 60 & 62.2–79.4 & 70.700 \\
LVA & Latvia & 1959–2024 & 66 & 58.7–81.2 & 71.800 \\
EST & Estonia & 1959–2024 & 66 & 60.8–83.4 & 73.600 \\
LTU & Lithuania & 1959–2024 & 66 & 62.5–81.7 & 72.700 \\
UKR & Ukraine & 1959–2013 & 55 & 61.2–76.2 & 70.300 \\
LUX & Luxembourg & 1960–2024 & 65 & 65.4–85.5 & 75.900 \\
TWN & Taiwan & 1970–2024 & 55 & 66.3–84.2 & 75.800 \\
GRC & Greece & 1981–2019 & 39 & 73.3–84.2 & 78.600 \\
ISR & Israel & 1983–2016 & 34 & 73.1–84.2 & 78.700 \\
SVN & Slovenia & 1983–2019 & 37 & 66.8–84.2 & 77.100 \\
HKG & Hong Kong & 1986–2024 & 39 & 74.1–88.7 & 81.200 \\
DEUTNP & Germany & 1990–2020 & 31 & 71.9–83.5 & 78.600 \\
CHL & Chile & 1992–2024 & 33 & 71.4–83.3 & 77.800 \\
HRV & Croatia & 2001–2020 & 20 & 70.9–81.4 & 76.700 \\
KOR & Republic of Korea & 2003–2023 & 21 & 73.8–86.6 & 80.700 \\
\end{longtable}}

\begin{figure}[!htbp]
\centering
\includegraphics[width=\textwidth]{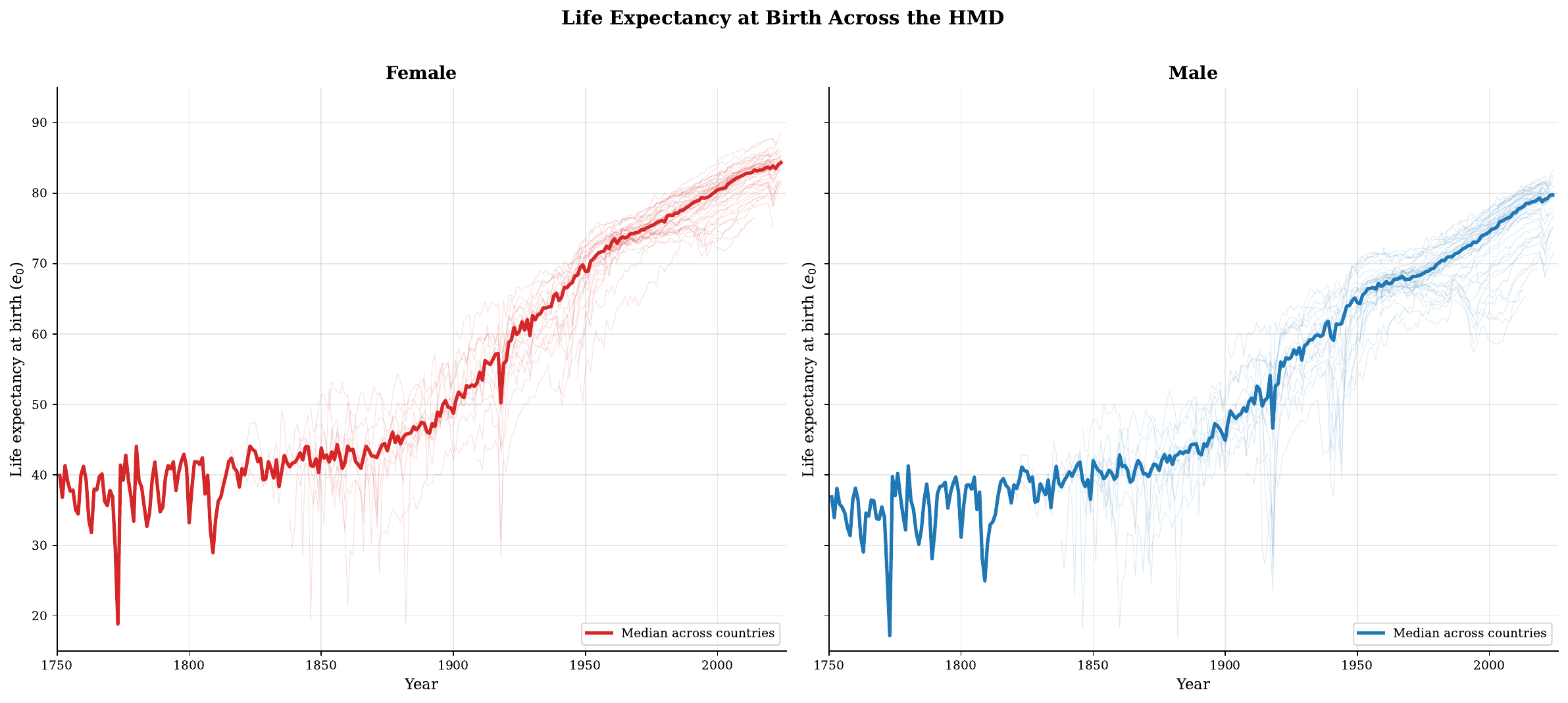}
\caption{Life expectancy at birth ($\ezero$) over time for all HMD
populations, separately by sex.  Thin lines represent individual
countries; the bold line is the cross-country median.  Sharp dips from
wars and pandemics punctuate the secular upward trend.}
\label{fig:s1_e0_trends}
\end{figure}

\Cref{fig:s1_e0_trends} plots life expectancy at birth ($\ezero$) over time
for all HMD populations, separately by sex.  The secular mortality
transition is clearly visible: a sustained upward trend from around
$\ezero = 30$--$40$ in the eighteenth century to $\ezero > 80$ in recent
decades.  Superimposed on this trend are sharp dips corresponding to the
events catalogued in the event dictionaries
(\cref{sec:exceptional:events}): the Napoleonic Wars, the 1918 influenza
pandemic, the two World Wars, and COVID-19.  The cross-country spread
narrows markedly over time, reflecting the convergence of mortality
patterns among high-income countries during the twentieth century.

\begin{figure}[!htbp]
\centering
\includegraphics[width=\textwidth]{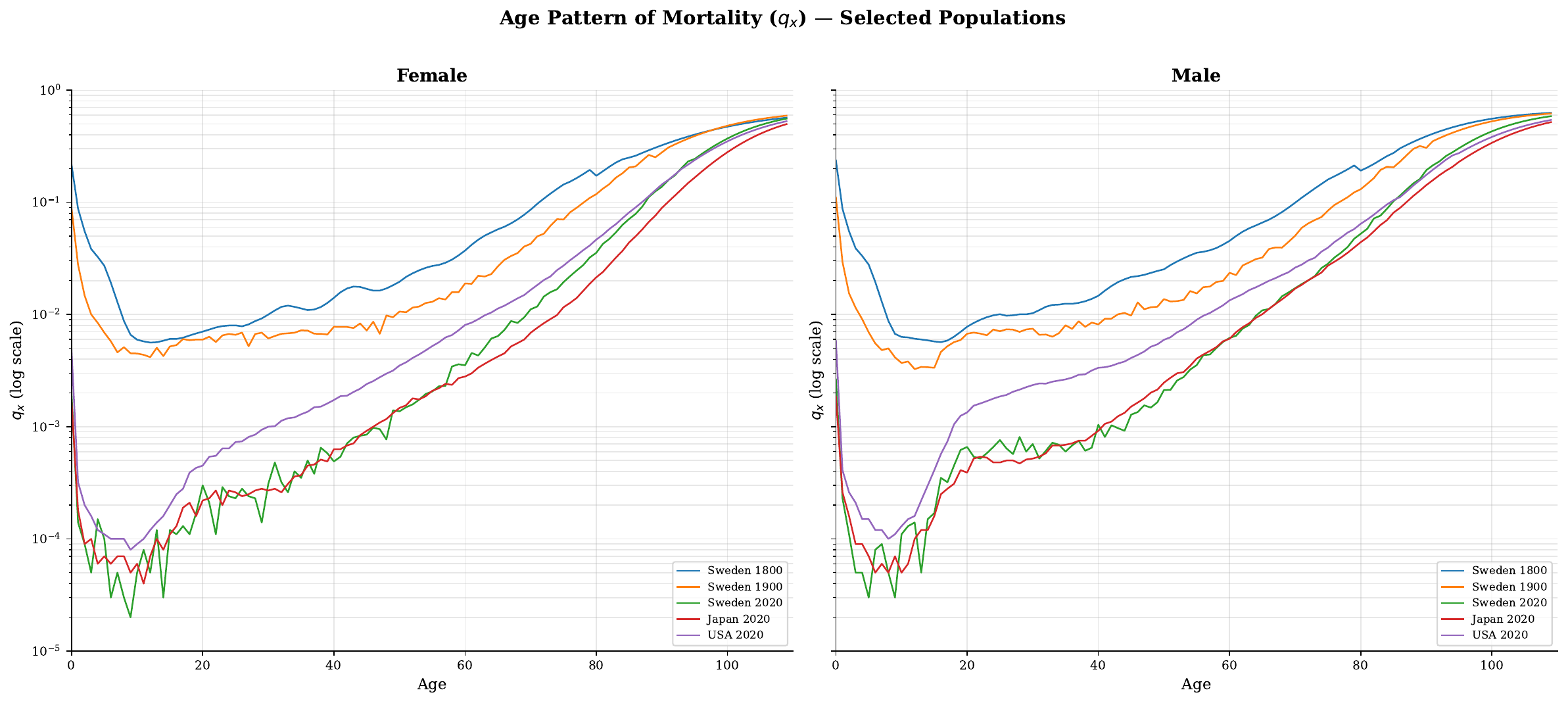}
\caption{Age-specific mortality ($\qx$) on a log scale for selected
countries and years.  The characteristic U-shape of human mortality is
evident: high infant mortality, a childhood minimum, and exponential
increase at older ages.}
\label{fig:s1_qx_profiles}
\end{figure}

\begin{figure}[!htbp]
\centering
\includegraphics[width=\textwidth]{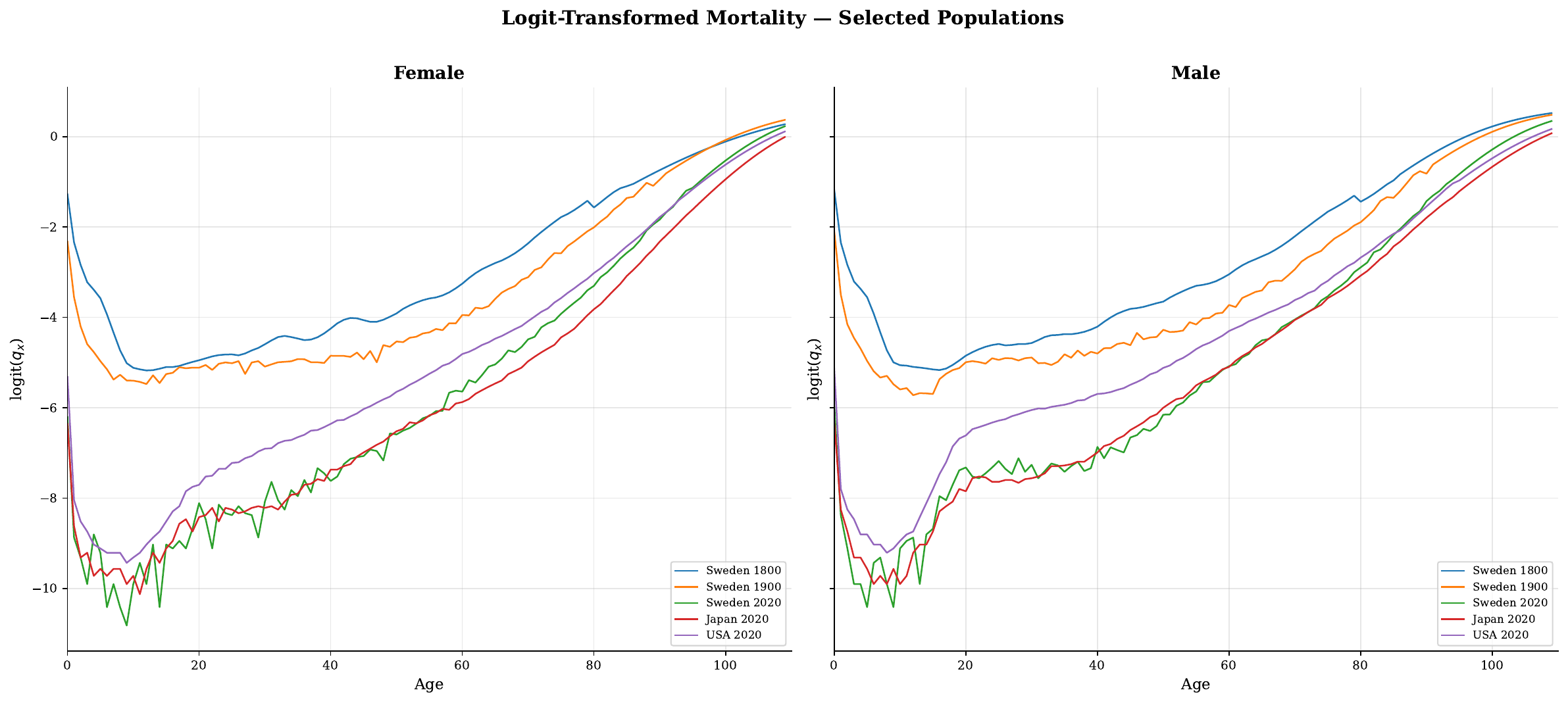}
\caption{Logit-transformed mortality for selected populations.  On this
scale, the Gompertz law appears as an approximately linear increase
above age~40.  The logit transform stabilizes variance and enables
additive modeling of mortality differences.}
\label{fig:s1_logit_qx}
\end{figure}

\Cref{fig:s1_qx_profiles} illustrates the characteristic U-shape of human
mortality on a log scale for selected country-years, while
\cref{fig:s1_logit_qx} shows the same schedules on the logit scale
(\cref{sec:data:logit}).  On the logit scale, the Gompertz law
(\cref{sec:data:logit}) manifests as the approximately linear increase
above age~40, and the variance-stabilizing property of the transform is
evident: the visual spread across populations is roughly uniform across
ages, whereas on the natural scale it is dominated by variation at old
ages.

\begin{figure}[!htbp]
\centering
\includegraphics[width=\textwidth]{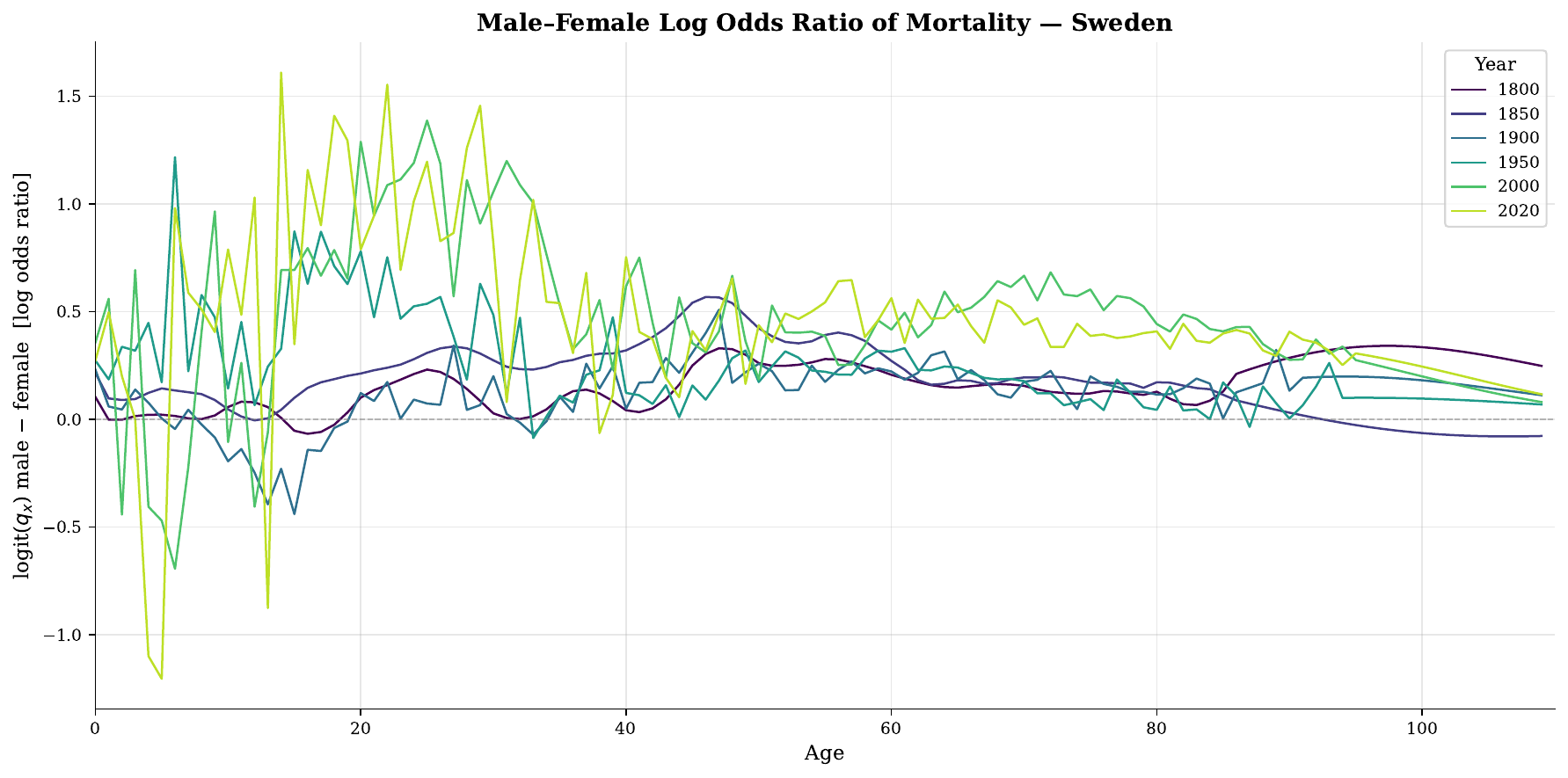}
\caption{Male--female difference in logit mortality by age, for Sweden
across selected years.  Values above zero indicate excess male mortality.
This quantity -- the log odds ratio -- is additive on the logit scale
used throughout.}
\label{fig:s1_sex_diff}
\end{figure}

\Cref{fig:s1_sex_diff} shows the male--female difference in logit mortality
for Sweden across selected years, foreshadowing the sex-differential
structure that the decomposition will separate via the second sex
component (\cref{sec:application:interpretation}).  The excess male
mortality is concentrated at young-adult and old ages, and its magnitude
has evolved substantially over two centuries.

\subsection{Preprocessing and data quality}
\label{sec:results:preprocessing}

\begin{figure}[!htbp]
\centering
\includegraphics[width=\textwidth]{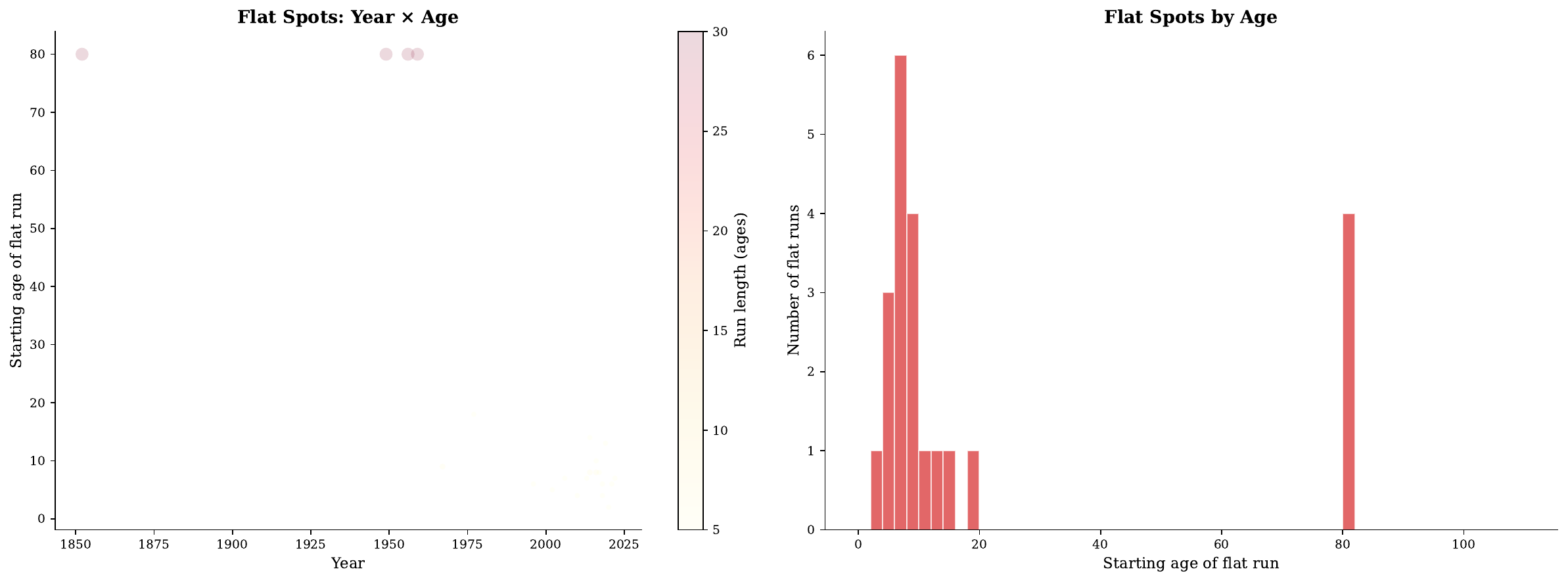}
\caption{Flat spots in $\qx$ by age and year.  Each point marks the
starting age of a run of five or more consecutive ages with identical
nonzero $\qx$.  Flat spots concentrate at old ages in early historical
periods, indicating limited data resolution.}
\label{fig:s1_flat_spots}
\end{figure}

The curation steps described in \cref{sec:preprocessing:curation} removed
a small number of life tables containing missing values, flat $\qx$
schedules at extreme ages, or unpaired sexes.  \Cref{fig:s1_flat_spots}
maps the flat spots detected in the raw data: they concentrate at old ages
in early historical periods, confirming that the original data lacked the
resolution to estimate age-specific mortality beyond approximately age~100
in many early life tables.

\subsubsection{Exceptional year identification}

\begin{figure}[!htbp]
\centering
\includegraphics[width=\textwidth]{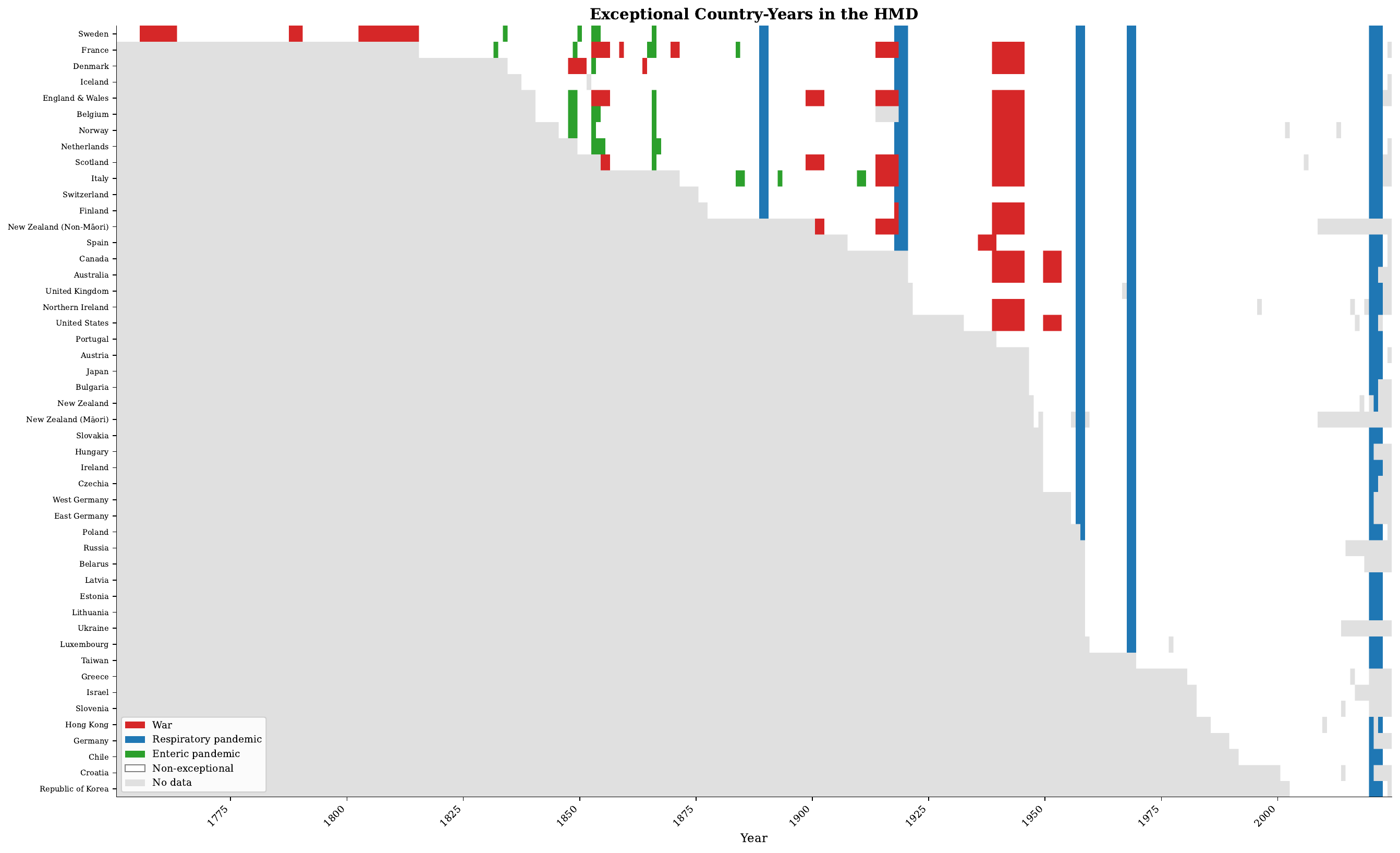}
\caption{Exceptional country-years in the HMD.  Red = war, blue =
respiratory pandemic, green = enteric pandemic.  The dominance of wars
in the first half of the twentieth century and respiratory pandemics
in recent years is clearly visible.}
\label{fig:s2_events}
\end{figure}

\begin{figure}[!htbp]
\centering
\includegraphics[width=\textwidth]{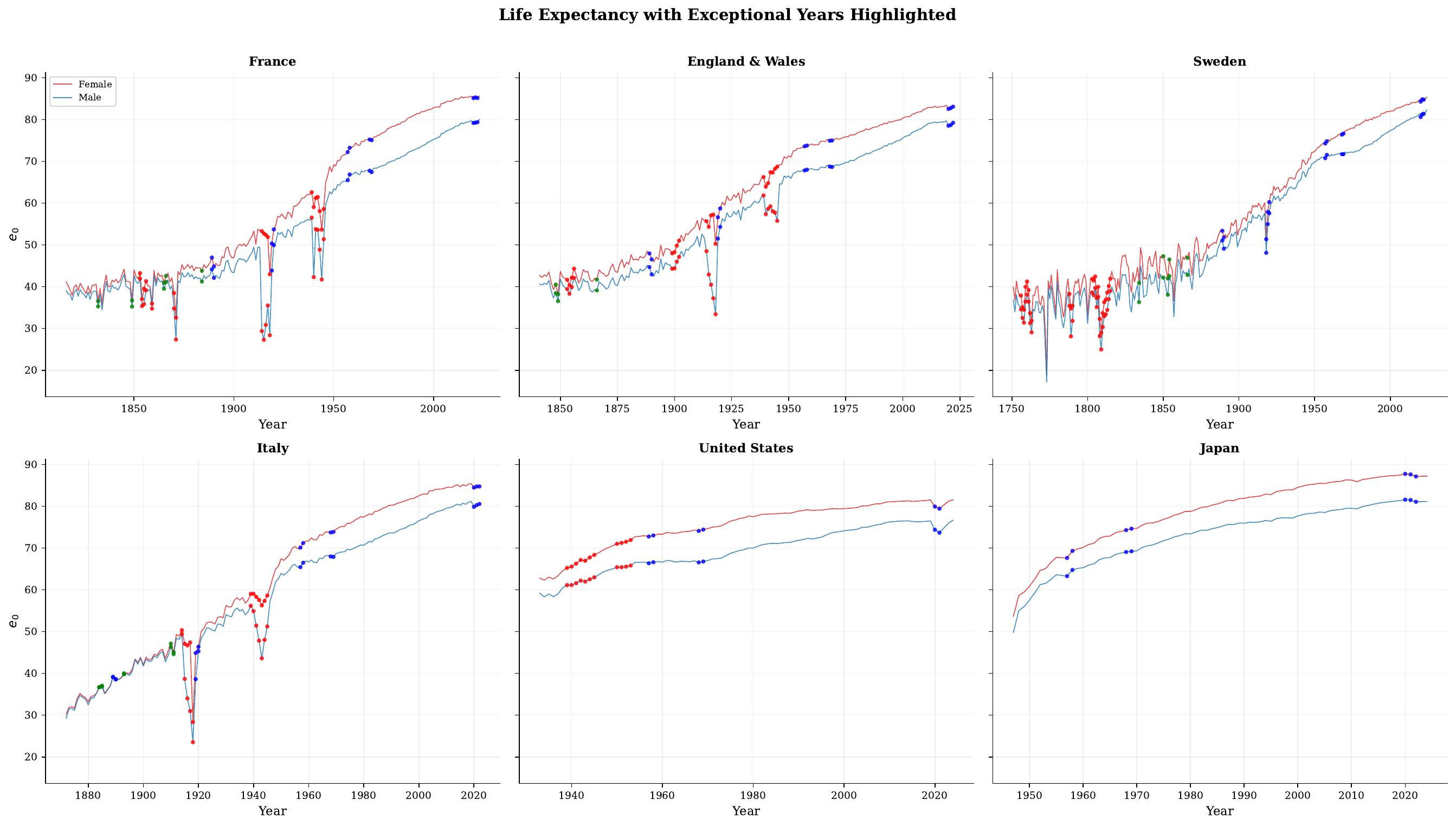}
\caption{Life expectancy at birth for selected countries, with
exceptional years highlighted (red = war, blue = respiratory pandemic).
The flagged years align with the visible sharp dips in $\ezero$.}
\label{fig:s2_events_e0}
\end{figure}

The event dictionaries (\cref{sec:exceptional:events};
\cref{app:events}) flag country-years affected by armed conflicts,
respiratory pandemics, and enteric pandemics.  \Cref{fig:s2_events} presents
the full exceptional-year map: each colored cell marks a flagged
country-year, with red for wars, blue for respiratory pandemics, and green
for enteric pandemics.  The dominance of wars in the first half of the
twentieth century and of respiratory pandemics (especially COVID-19) in
recent years is clearly visible.  \Cref{fig:s2_events_e0} confirms that the
flagged years align with the sharp dips visible in the $\ezero$ time series
of \cref{fig:s1_e0_trends}.

\subsubsection{Adaptive temporal pooling}

\begin{figure}[!htbp]
\centering
\includegraphics[width=\textwidth]{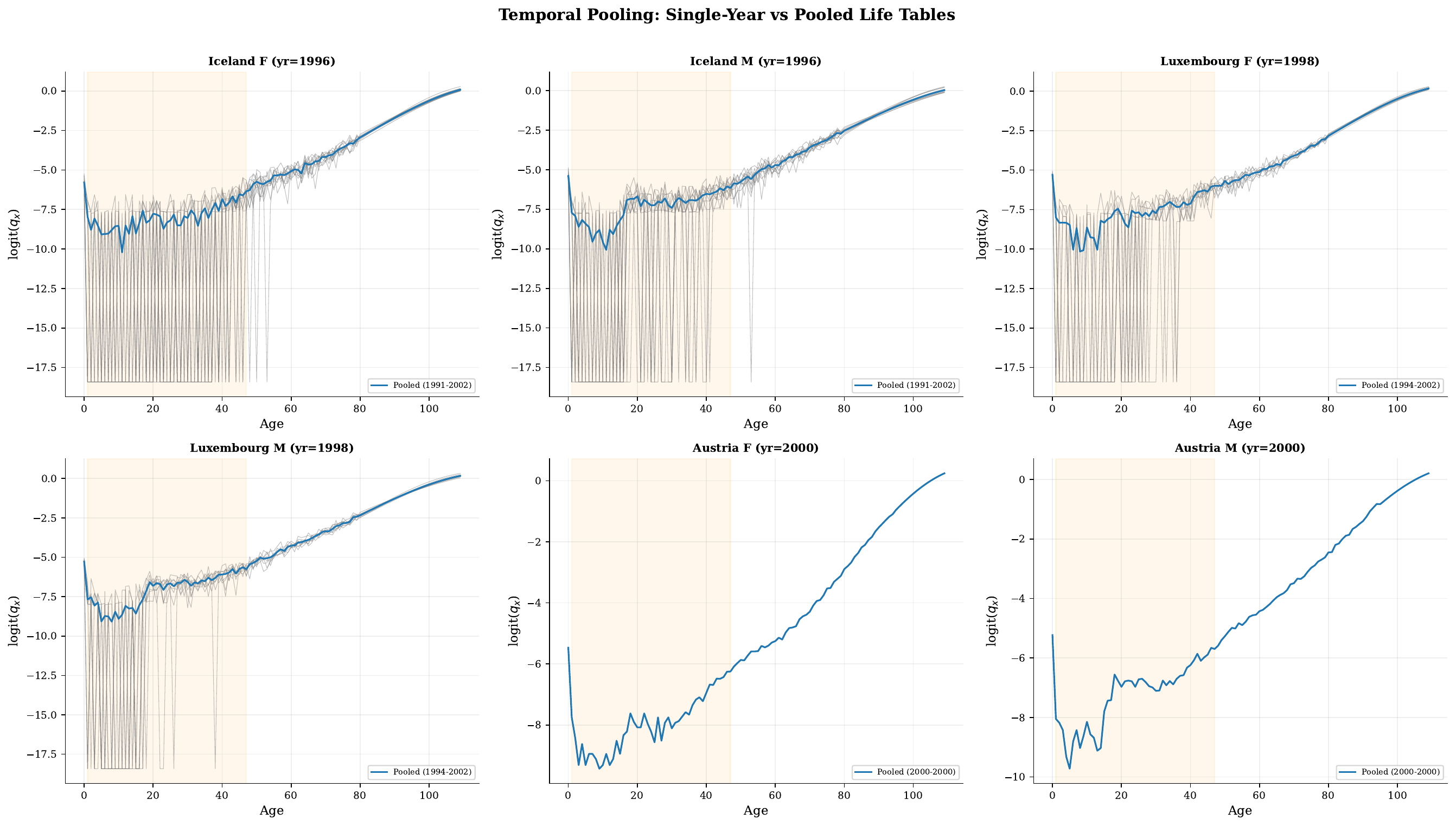}
\caption{Single-year vs.\ temporally pooled life tables for selected
small populations.  Thin gray lines show the original single-year
logit($\qx$) curves that were averaged into each pooled life table.
Heavy blue lines show the pooled result.  Orange shading marks valley
ages.}
\label{fig:s2_smoothing}
\end{figure}

\begin{figure}[!htbp]
\centering
\includegraphics[width=\textwidth]{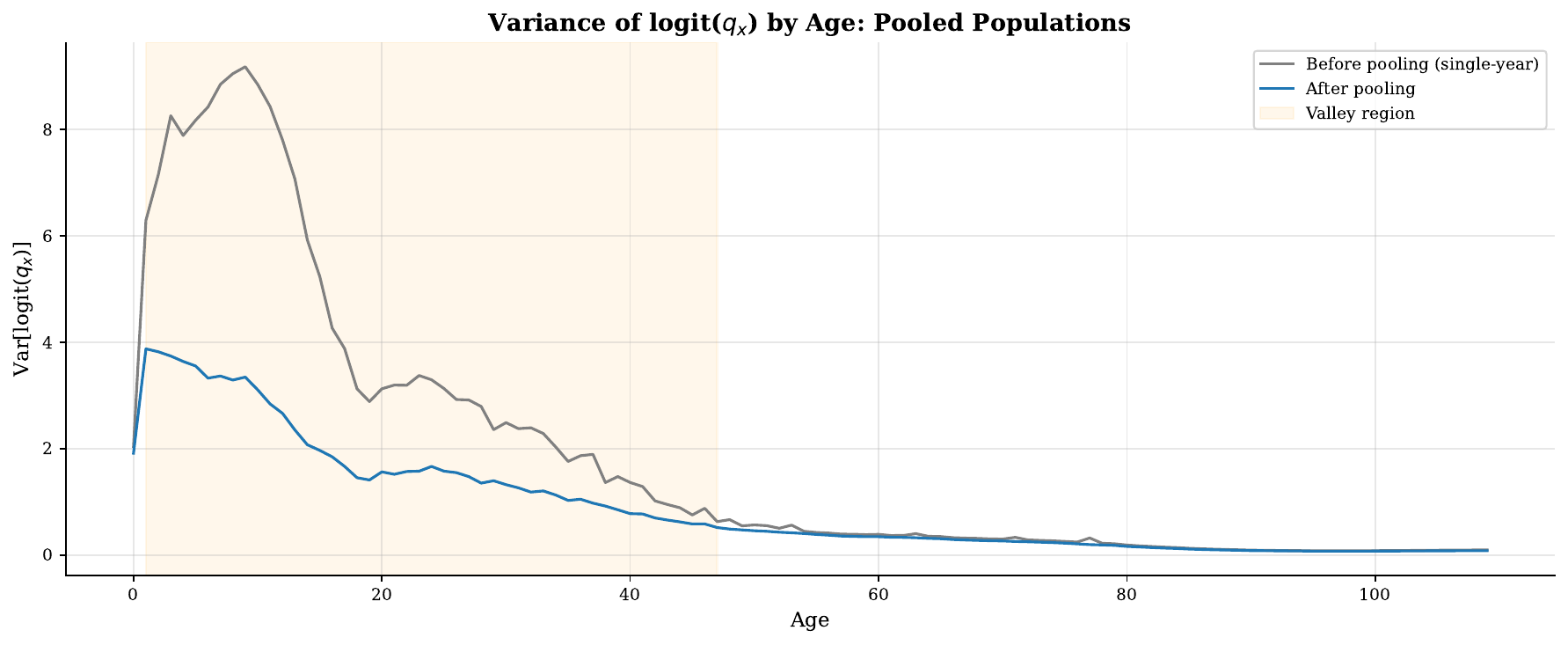}
\caption{Variance of logit($\qx$) by age for pooled populations,
before (gray) and after (blue) temporal pooling.  The pooling reduces
variance at all ages, with the largest effect in the low-mortality
valley where Poisson noise was dominant.}
\label{fig:s2_smoothing_by_age}
\end{figure}

We applied the adaptive temporal pooling of \cref{sec:preprocessing:smoothing}
to populations in which Poisson noise produced zero $m_x$ values
at valley ages.  As expected, small populations -- Iceland, Luxembourg,
and several early historical series -- required pooling across multiple
years, while large populations such as England and Wales, France, and the
United States required no pooling.  \Cref{fig:s2_smoothing} compares the
original single-year logit($\qx$) curves with the pooled result for
selected small populations: the thin gray lines show the noisy
originals, and the heavy blue line shows the smoothed product.  The
orange shading marks the valley ages where zeros were problematic.
\Cref{fig:s2_smoothing_by_age} confirms that the pooling reduces variance
at all ages, with the largest effect at valley ages where Poisson noise
was dominant, consistent with \cref{rem:valley_smoothing}.

\subsection{Tensor construction}
\label{sec:results:tensor}

After preprocessing, we constructed the mortality tensor
$\M \in \R^{S \times A \times C \times T}$ with the
dimensions shown in \cref{tab:s4_tensor_dims}.  Every
element $\M_{s,a,c,t}$ is the logit-transformed probability of dying
$\logit(\qx)$.

{\setlength\LTleft{0pt}\setlength\LTright{0pt}
\begin{longtable}{@{\extracolsep{\fill}}lcl@{}}
\caption{Dimensions of the HMD mortality tensor. Total elements: 2,893,440; observed fraction: 0.277.} \label{tab:s4_tensor_dims} \\
\toprule
Dimension & Size & Description \\
\midrule
\endfirsthead
\caption[]{Dimensions of the HMD mortality tensor. Total elements: 2,893,440; observed fraction: 0.277. (continued)} \\
\toprule
Dimension & Size & Description \\
\midrule
\endhead
\midrule
\multicolumn{3}{r}{\textit{Continued on next page}} \\
\endfoot
\bottomrule
\endlastfoot
Sex ($S$) & 2 & Female, male \\
Age ($A$) & 110 & Single-year ages 0--109 \\
Country ($C$) & 48 & HMD populations \\
Year ($T$) & 274 & 1751--2024 \\
\end{longtable}}

\begin{figure}[!htbp]
\centering
\includegraphics[width=\textwidth]{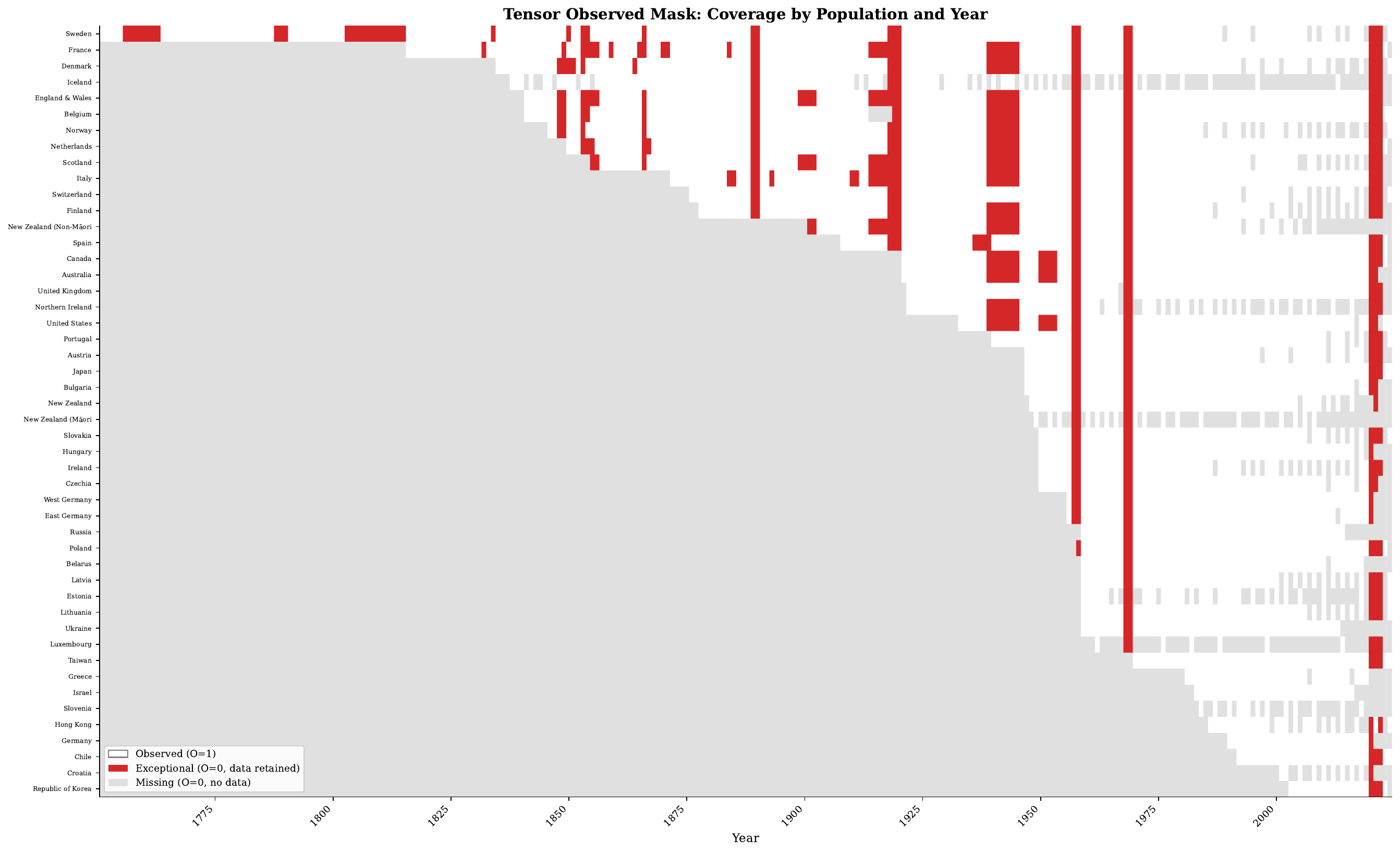}
\caption{Observed mask for the mortality tensor.  White = observed
(non-exceptional), red = exceptional (data exists but masked for
decomposition), gray = missing (no data).  The ragged structure reflects
the varying temporal coverage across HMD populations.}
\label{fig:s3_obs_mask}
\end{figure}

\Cref{fig:s3_obs_mask} displays the observed mask $\bm{O}$
(\cref{eq:obs_mask}): white cells are observed non-exceptional
country-years (used in the decomposition), red cells are exceptional
(data exists but is masked), and gray cells are missing (no data).

We imputed missing entries with the country temporal mean as described
in \cref{sec:preprocessing:imputation}.  \Cref{tab:missing_year_audit}
classifies the $C \times T$ country-year cells that are \emph{not}
observed (i.e., $O_{c,t} = 0$) into five mutually exclusive categories.
The vast majority of missing cells simply fall outside a population's
data range in the HMD -- the tensor spans from 1751, but most
populations enter the database in the mid-twentieth century.  A smaller
fraction are exceptional years masked by the event dictionaries, years
dropped during quality-control curation, or years whose individual life
tables were replaced by pooled tables during adaptive temporal pooling
(\cref{sec:preprocessing:smoothing}).  Only five cells represent true
gaps within a population's nominal range (Belgium 1914--1918).

{\setlength\LTleft{0pt}\setlength\LTright{0pt}
\begin{longtable}{@{\extracolsep{\fill}}lrr@{}}
\caption{Classification of missing country-years in the mortality tensor. The tensor spans all HMD populations over the full year range (1751--most recent HMD year); a country-year cell is ``missing'' if the corresponding population was not observed in that year after all preprocessing steps.  Each missing cell is assigned to exactly one of the five categories below.} \label{tab:missing_year_audit} \\
\toprule
Reason & Count & \% \\
\midrule
\endfirsthead
\caption[]{Classification of missing country-years in the mortality tensor. The tensor spans all HMD populations over the full year range (1751--most recent HMD year); a country-year cell is ``missing'' if the corresponding population was not observed in that year after all preprocessing steps.  Each missing cell is assigned to exactly one of the five categories below. (continued)} \\
\toprule
Reason & Count & \% \\
\midrule
\endhead
\midrule
\multicolumn{3}{r}{\textit{Continued on next page}} \\
\endfoot
\bottomrule
\noalign{\vskip 2pt}
\multicolumn{3}{@{}p{\textwidth}@{}}{\footnotesize \emph{Not in HMD}: year falls before the population's first or after its last HMD life table. \emph{Exceptional}: data exist but the year is flagged by the event dictionaries and excluded from the baseline decomposition. \emph{Dropped in curation}: a life table existed but was removed during quality checks (missing $q_x$ values, flat spots at old ages, or unpaired sexes). \emph{Replaced by pooling}: individual yearly life tables were replaced by a single pooled life table during adaptive temporal pooling (\cref{sec:preprocessing:smoothing}); the pooled table is assigned to the block's midpoint year, so component years no longer appear. \emph{Gap in HMD}: year falls within the population's nominal range but no life table was provided (e.g., Belgium 1914--1918).} \\
\endlastfoot
Not in HMD (outside data range) & 8,558 & 90.0 \\
Exceptional (war/pandemic, masked) & 529 & 5.6 \\
Dropped in curation (quality checks) & 15 & 0.2 \\
Replaced by pooling (valley smoothing) & 397 & 4.2 \\
Gap in HMD (within range, no data) & 5 & 0.1 \\
\midrule
\textbf{Total missing} & \textbf{9,504} & \textbf{100.0} \\
\end{longtable}}

\begin{figure}[!htbp]
\centering
\includegraphics[width=\textwidth]{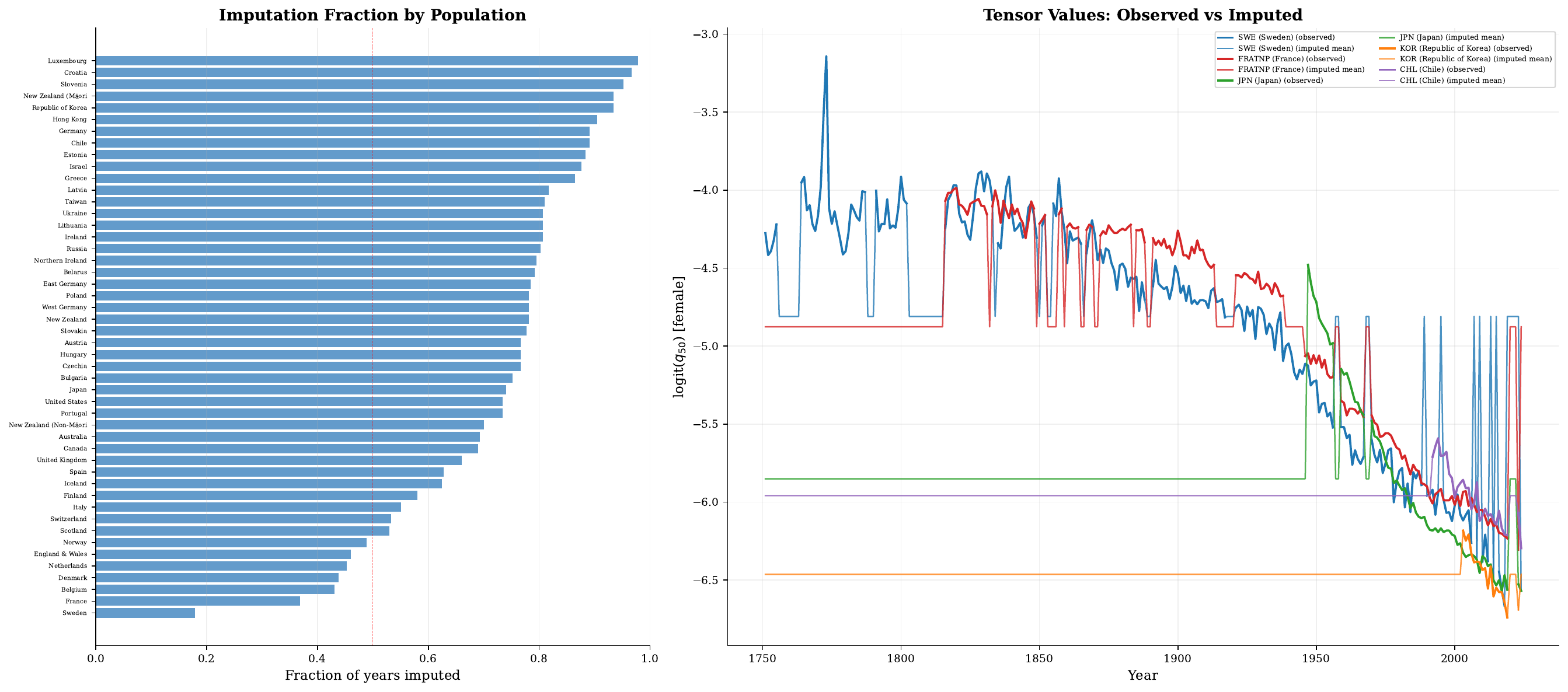}
\caption{Imputation fraction by population (left) and a spot-check of
imputed vs.\ observed logit($q_{50}$) trajectories for selected
countries (right).  Solid lines throughout; observed data are drawn
thicker, while the thinner segments show the imputed country temporal
mean, which is constant across all missing years and therefore appears
as a horizontal line in the pre-data region.}
\label{fig:s3_imputation}
\end{figure}

\Cref{fig:s3_imputation}
shows the imputation fraction by population (left panel) and a
spot-check of imputed versus observed trajectories for selected countries
(right panel).  Countries entering the HMD late -- such as South Korea,
Taiwan, and several Eastern European nations -- have high imputation
fractions, motivating the weighted HOSVD of \cref{rem:weighted_hosvd}
to prevent these replicated means from dominating the decomposition.

\section{Tucker Decomposition}
\label{sec:tucker}

We develop the Tucker tensor decomposition from first
principles, beginning with a review of the singular value decomposition
(SVD) for matrices (\cref{sec:tucker:svd}), generalizing to higher-order
tensors (\cref{sec:tucker:definition}), and developing the geometric
interpretation (\cref{sec:tucker:geometry}).  We then explain how
projections onto subspaces work within the Tucker framework
(\cref{sec:tucker:projections}), how the core tensor encodes joint
structure across all dimensions (\cref{sec:tucker:core}), and how to
reconstruct individual mortality schedules from the decomposition
(\cref{sec:tucker:reconstruction}).  The section concludes with the
practical matters of rank selection (\cref{sec:tucker:rank}) and
post-decomposition smoothing of the age basis
(\cref{sec:tucker:bsmooth}).

\subsection{From the SVD to higher-order decompositions}
\label{sec:tucker:svd}

The singular value decomposition (SVD) is the foundation for
understanding the Tucker decomposition.  Given a matrix $\bm{X} \in
\R^{I \times J}$, the SVD factors it as
\begin{equation}
    \label{eq:svd}
    \bm{X} = \bm{U} \, \bm{\Sigma} \, \bm{V}^\top,
\end{equation}
%
where $\bm{U} \in \R^{I \times I}$ and $\bm{V} \in \R^{J \times J}$ are
orthogonal matrices and $\bm{\Sigma}$ is a diagonal matrix of singular
values $\sigma_1 \geq \sigma_2 \geq \cdots \geq 0$
\citep{Eckart1936,GolubVanLoan2013}.  The columns of~$\bm{U}$ are
orthonormal basis vectors for the column space, and the columns
of~$\bm{V}$ are orthonormal basis vectors for the row space.  Each singular value
$\sigma_k$ measures the importance of the $k$-th component: the rank-$r$
truncation $\bm{X}_r = \sum_{k=1}^{r} \sigma_k \, \bm{u}_k \bm{v}_k^\top$
is the best rank-$r$ approximation of~$\bm{X}$ in the Frobenius norm.

In the context of mortality, if $\bm{X}$ is a matrix of logit-transformed
$\qx$ values with ages as rows and populations as columns, then the SVD
identifies: a set of \emph{age basis functions} (columns of~$\bm{U}$) that
capture the dominant shapes of the mortality curve, and a set of
\emph{population loadings} (columns of~$\bm{V}$, scaled by~$\sigma_k$)
that specify how much of each shape is present in each population.  This
is the mathematical structure underlying the Lee--Carter model
\citep{LeeCarter1992} and its extensions.

The SVD, however, operates on matrices -- two-dimensional objects.  When the
data have more than two natural dimensions, the SVD must be applied after
collapsing some dimensions together.  The Tucker decomposition generalizes
the SVD to tensors of arbitrary order, decomposing along all dimensions
simultaneously \citep{Tucker1966}.

\subsection{The Tucker decomposition}
\label{sec:tucker:definition}

Let $\M \in \R^{I_1 \times I_2 \times \cdots \times I_N}$ be an
$N$-th order tensor.  The Tucker decomposition expresses $\M$ as
\begin{equation}
    \label{eq:tucker_general}
    \M = \G \nmode{1} \bm{U}^{(1)} \nmode{2} \bm{U}^{(2)} \cdots
         \nmode{N} \bm{U}^{(N)},
\end{equation}
%
where $\G \in \R^{r_1 \times r_2 \times \cdots \times r_N}$ is the
\emph{core tensor} and each $\bm{U}^{(n)} \in \R^{I_n \times r_n}$ is a
matrix whose columns form an orthonormal basis for the $n$-th mode
\citep{Tucker1966,DeLathauwer2000,KoldaBader2009}.  The operation
$\nmode{n}$ denotes the \emph{$n$-mode product}, which multiplies the
tensor by a matrix along its $n$-th mode:
\begin{equation}
    \label{eq:nmode_product}
    \bigl(\M \nmode{n} \bm{U}\bigr)_{i_1, \ldots, i_{n-1},\, j,\,
    i_{n+1}, \ldots, i_N}
    = \sum_{i_n=1}^{I_n} \M_{i_1, \ldots, i_N} \; U_{j, i_n}\,.
\end{equation}

For the four-dimensional mortality tensor $\M \in \R^{S \times A \times C
\times T}$, the decomposition becomes
\begin{equation}
    \label{eq:tucker_mortality}
    \M = \G \nmode{1} \bS \nmode{2} \bA \nmode{3} \bC \nmode{4} \bT,
\end{equation}
%
where $\bS \in \R^{S \times r_1}$, $\bA \in \R^{A \times r_2}$, $\bC \in
\R^{C \times r_3}$, and $\bT \in \R^{T \times r_4}$ are the factor
matrices, and $\G \in \R^{r_1 \times r_2 \times r_3 \times r_4}$ is the
core tensor.  Crucially, the tensor that enters the decomposition
excludes exceptional country-years
(\cref{sec:preprocessing:exceptional}), so the factor matrices and core
tensor describe the \emph{baseline} mortality surface, free of transient
shocks from wars and pandemics.  Written element-wise, the reconstructed value at sex~$s$,
age~$a$, country~$c$, year~$t$ is
\begin{equation}
    \label{eq:tucker_elementwise}
    \hat{\M}_{s,a,c,t}
    = \sum_{\fs=1}^{r_1} \sum_{\fa=1}^{r_2} \sum_{\fc=1}^{r_3} \sum_{\ft=1}^{r_4}
      \G_{\fs,\fa,\fc,\ft} \; S_{s,\fs} \; A_{a,\fa} \; C_{c,\fc} \; T_{t,\ft}\,.
\end{equation}

We compute the Tucker decomposition using the \emph{higher-order SVD}
(HOSVD) of \citet{DeLathauwer2000}.  The HOSVD proceeds by unfolding the
tensor along each mode, computing the SVD of the resulting matrix, and
retaining the leading left singular vectors as the factor matrix for that
mode.  Specifically, the \emph{mode-$n$ unfolding} of~$\M$, denoted
$\bm{M}_{(n)} \in \R^{I_n \times \prod_{m \neq n} I_m}$, rearranges the
tensor into a matrix with the $n$-th dimension as rows and all other
dimensions flattened into columns.  The SVD of each unfolding yields
\begin{equation}
    \label{eq:mode_svd}
    \bm{M}_{(n)} = \bm{U}^{(n)} \, \bm{\Sigma}^{(n)} \,
    (\bm{V}^{(n)})^\top,
\end{equation}
%
and the factor matrix is taken as the leading $r_n$ columns of
$\bm{U}^{(n)}$.  The core tensor is then computed as the projection of the
original tensor onto the truncated factor matrices:
\begin{equation}
    \label{eq:core_projection}
    \G = \M \nmode{1} \bS^\top \nmode{2} \bA^\top
           \nmode{3} \bC^\top \nmode{4} \bT^\top.
\end{equation}

\begin{remark}[Weighted HOSVD]
\label{rem:weighted_hosvd}
The standard HOSVD treats all entries of the mode unfoldings equally.
When the tensor contains imputed entries
(\cref{sec:preprocessing:imputation_bias}), this gives disproportionate
influence to countries with many missing years.  A \emph{weighted} variant
replaces the ordinary SVD of $\bm{M}_{(n)}$ with a weighted SVD in which
columns corresponding to imputed country-year cells receive reduced
weight, so that the factor matrices are determined primarily by observed
data.  The core projection~\cref{eq:core_projection} proceeds as before.
\end{remark}

\subsection{Geometric interpretation: basis functions and loadings}
\label{sec:tucker:geometry}

The Tucker decomposition has a clear geometric interpretation.  Each
factor matrix $\bm{U}^{(n)}$ defines an orthonormal basis for a
low-dimensional subspace of dimension~$r_n$ within the full
$I_n$-dimensional space of mode~$n$.  Any point in the original space can
be projected onto this subspace, yielding a set of \emph{coordinates}
(loadings) in the new basis.

In the mortality context:
\begin{itemize}
    \item $\bS \in \R^{S \times r_1}$ defines a basis for \emph{sex
    space}.  With $S = 2$ and $r_1 = 2$ (full rank), this is a rotation
    of the original female--male coordinates into an orthonormal basis that
    separates the sex-average mortality level from the sex differential.

    \item $\bA \in \R^{A \times r_2}$ defines a set of \emph{age basis
    functions}.  Each column of~$\bA$ is a function of age that captures a
    particular pattern of variation in the mortality curve: the first
    column captures the dominant shape (approximately the mean mortality
    curve), and subsequent columns capture increasingly fine-scale
    deviations.  These are the higher-dimensional analogue of the
    Lee--Carter $b_x$ vector, but here there are $r_2$ such functions
    rather than one.

    \item $\bC \in \R^{C \times r_3}$ contains \emph{country loadings}.
    Each row locates a country in an $r_3$-dimensional space that
    summarizes its mortality structure.  Countries with similar age-pattern
    shapes appear close together in this space.

    \item $\bT \in \R^{T \times r_4}$ contains \emph{year loadings}.
    Each row locates a calendar year in an $r_4$-dimensional space that
    traces the secular evolution of mortality.  The trajectory of year
    loadings over time captures the mortality transition.
\end{itemize}

The key distinction between the SVD and Tucker lies in the core
tensor~$\G$.  In the SVD, the ``core'' is diagonal ($\bm{\Sigma}$): the
$k$-th row basis vector interacts only with the $k$-th column basis vector,
scaled by $\sigma_k$.  In the Tucker decomposition, $\G$ is a full
(generally non-diagonal) tensor: every combination of basis vectors across
all modes can interact, and the strength and sign of each interaction is
encoded in the corresponding element of~$\G$.  This is what gives the
Tucker decomposition its expressive power -- and why it is essential for
modeling mortality data where age patterns, sex differentials, country
effects, and temporal trends are deeply intertwined.

\subsection{Projections and subspaces}
\label{sec:tucker:projections}

One of the most useful operations within the Tucker framework is
\emph{projection}: fixing the coordinate(s) along one or more dimensions
and examining the resulting lower-dimensional object.  This is equivalent
to slicing the tensor along the fixed dimensions and expressing the
result in terms of the basis functions of the remaining dimensions.

Consider projecting onto a specific sex~$s$.  From
\cref{eq:tucker_elementwise}, fixing $s$ gives
\begin{equation}
    \label{eq:project_sex}
    \hat{\M}_{s,a,c,t}
    = \sum_{\fa=1}^{r_2} \sum_{\fc=1}^{r_3} \sum_{\ft=1}^{r_4}
      \underbrace{\left(\sum_{\fs=1}^{r_1} \G_{\fs,\fa,\fc,\ft} \; S_{s,\fs}\right)}
      _{\displaystyle \G^{(s)}_{\fa,\fc,\ft}}
      A_{a,\fa} \; C_{c,\fc} \; T_{t,\ft}\,,
\end{equation}
%
where $\G^{(s)} \in \R^{r_2 \times r_3 \times r_4}$ is the
\emph{sex-projected core tensor} obtained by contracting~$\G$ with the
$s$-th row of~$\bS$.  The resulting three-dimensional tensor
$\G^{(s)}_{\fa,\fc,\ft}$ encodes the age--country--year structure specific to
sex~$s$.

This operation is a projection in the linear-algebraic sense: the row
$\bm{s}_s = (S_{s,1}, \ldots, S_{s,r_1})$ specifies a point in the
$r_1$-dimensional sex space, and the contraction $\sum_\fs \G_{\fs,\fa,\fc,\ft}
S_{s,\fs}$ evaluates the core tensor at that point.  Different sexes produce
different projected core tensors, and therefore different age patterns,
country effects, and temporal trends -- even though all of these emerge from
the \emph{same} underlying decomposition.

The same logic applies to projecting onto a specific country~$c$ (using
row~$c$ of~$\bC$), a specific year~$t$ (using row~$t$ of~$\bT$), or any
combination of these.

\subsection{Joint structure and the core tensor}
\label{sec:tucker:core}

The core tensor~$\G$ is the most important -- and most easily
misunderstood -- element of the Tucker decomposition.  Its role is to encode
the \emph{interactions} between the basis functions of different
dimensions.  An element $\G_{\fs,\fa,\fc,\ft}$ specifies the strength with which
sex component~$\fs$, age component~$\fa$, country component~$\fc$, and time
component~$\ft$ jointly contribute to the mortality tensor.

This joint encoding is what distinguishes the Tucker decomposition from
applying the SVD separately to each dimension.  Separate SVDs would
produce, for example, an age basis~$\bA$ and a time basis~$\bT$, but
would provide no information about how age patterns change over time or how
that change differs between countries.  The core tensor provides exactly
this information.

\begin{remark}[Sex differentials as joint structure]
\label{rem:sex_diff}
The sex differential in mortality is perhaps the clearest illustration of
why the core tensor matters.  Males die at higher rates than females at
nearly every age, but the magnitude of the differential varies
systematically with age (large in young adulthood, smaller at the extremes)
and has changed over time (widening through most of the twentieth century,
then narrowing; \citealp{PrestonWang2006}).  In the Tucker decomposition, this sex--age--time
interaction is captured by those elements of~$\G$ that couple different sex
components ($\fs$) with different age components ($\fa$) and time
components~($\ft$).  Because the sex differential is represented through
$\G$, it is not modeled as a separate quantity but emerges from the same
low-rank structure that governs the overall mortality surface.  Any
reconstruction at a given sex, country, and year automatically produces a
sex differential that is consistent with the age pattern and historical
context.
\end{remark}

\subsection{Reconstruction of mortality schedules}
\label{sec:tucker:reconstruction}

The Tucker decomposition provides a direct formula for reconstructing a
complete mortality schedule at any observed (or imputed) combination of
sex, country, and year.  From \cref{eq:tucker_elementwise}, the
reconstructed logit($\qx$) for sex~$s$ ($s = 1$ female, $s = 2$ male)
in country~$c$ and year~$t$ is
\begin{equation}
    \label{eq:reconstruct_logit}
    \hat{y}_{s,a,c,t}
    = \sum_{\fs,\fa,\fc,\ft} \G_{\fs,\fa,\fc,\ft} \; S_{s,\fs} \; A_{a,\fa} \; C_{c,\fc} \;
    T_{t,\ft}\,,
    \qquad a = 1, 2, \ldots, A.
\end{equation}
%
This produces a vector of $A$ logit values for each sex.  Applying the
inverse transform yields the $\qx$ schedule:
\begin{equation}
    \label{eq:reconstruct_qx}
    \widehat{\qx}_{s,a,c,t} = \expit\bigl(\hat{y}_{s,a,c,t}\bigr)
    = \frac{1}{1 + \exp(-\hat{y}_{s,a,c,t})}\,,
    \qquad x = a = 1, 2, \ldots, A,
\end{equation}
%
from which the full life table (survival function, $\ezero$, etc.) can be
computed for each sex.

\subsubsection{Reconstruction at reduced rank}

Only a fraction of the Tucker components are needed for an accurate
reconstruction.  Because the HOSVD selects basis vectors in order of
decreasing singular value, the leading components capture the most
important structure.  A reconstruction using ranks $(r_1, r_2', r_3',
r_4')$ with $r_2' \leq r_2$, $r_3' \leq r_3$, $r_4' \leq r_4$
simply truncates the sums in \cref{eq:tucker_elementwise}:
\begin{equation}
    \label{eq:reconstruct_reduced}
    \hat{\M}^{(r_2', r_3', r_4')}_{s,a,c,t}
    = \sum_{\fs=1}^{r_1} \sum_{\fa=1}^{r_2'} \sum_{\fc=1}^{r_3'}
      \sum_{\ft=1}^{r_4'}
      \G_{\fs,\fa,\fc,\ft} \; S_{s,\fs} \; A_{a,\fa} \; C_{c,\fc} \; T_{t,\ft}\,.
\end{equation}
%
This property is essential for the reconstruction model of
\cref{sec:reconstruction}: it means that mortality schedules can be
described with fewer parameters while retaining the joint structure encoded
in~$\G$.

\subsection{Rank selection}
\label{sec:tucker:rank}

The ranks $r_1, r_2, r_3, r_4$ control the trade-off between compression
and fidelity.  Each rank determines how many basis vectors are retained
along the corresponding mode.

\subsubsection{Sex: \texorpdfstring{$r_1 = S = 2$}{r1 = S = 2}}

The sex dimension has only two levels (female and male), so its
mode-unfolding $\bm{M}_{(1)} \in \R^{2 \times (A \cdot C \cdot T)}$ is a
rank-2 matrix.  We retain both singular vectors ($r_1 = 2$), preserving
the sex dimension at full rank.  No information about sex is lost, and the
two components can be interpreted as the sex-average mortality structure
(corresponding to the larger singular value) and the sex differential
(corresponding to the smaller singular value).

\subsubsection{Age, country, and year: variance thresholds}

For the remaining modes, the rank is selected by retaining enough
components to capture a specified fraction of the variance.  The
\emph{mode-$n$ variance} attributable to the $k$-th component is
proportional to the square of its singular value:
\begin{equation}
    \label{eq:variance_fraction}
    v^{(n)}_k = \frac{(\sigma^{(n)}_k)^2}
    {\sum_{j=1}^{I_n} (\sigma^{(n)}_j)^2}\,,
    \qquad
    \text{cumulative: } V^{(n)}_r = \sum_{k=1}^{r} v^{(n)}_k\,.
\end{equation}
%
The rank~$r_n$ is set to the smallest value such that $V^{(n)}_{r_n}$ exceeds a
threshold, subject to minimum and maximum bounds:
\begin{equation}
    \label{eq:rank_selection}
    r_n = \min\bigl\{ r : V^{(n)}_r \geq \tau \bigr\},
    \qquad \text{bounded by } r_n^{\min} \leq r_n \leq r_n^{\max}.
\end{equation}

We use a variance threshold of $\tau = 0.9999$ (99.99\%) for all three
non-sex modes.  This high threshold ensures that the decomposition
captures essentially all the structure in the data, including the
fine-grained variation needed to reconstruct individual life tables
accurately.  Post-decomposition smoothing of the age basis functions
(\cref{sec:tucker:bsmooth}) removes high-frequency noise from the
higher-order age components without discarding them.

\subsubsection{Singular value decay}

The singular values of each mode unfolding decay at different rates,
reflecting the different degrees of complexity along each dimension.  The
age mode has a small number of dominant components (reflecting
the limited number of qualitatively distinct age patterns in human
mortality) followed by a rapid decay.  The country and year modes have
more gradually decaying spectra, because the variation across
countries and over time is more complex.  Despite this, moderate ranks
suffice at the 99.99\% threshold, representing a fraction of the
full dimensions.

\subsection{Post-decomposition smoothing of the age basis}
\label{sec:tucker:bsmooth}

The HOSVD computes the age basis functions~$\bA$ as the leading left
singular vectors of the mode-2 (age) unfolding.  While these are
mathematically optimal in the least-squares sense, they may contain
high-frequency oscillations -- particularly in the higher-order components
that capture fine-scale variation.  At ages where mortality is low (the
``valley'' discussed in \cref{sec:preprocessing:smoothing}), these
oscillations reflect residual stochastic noise that the pre-decomposition
temporal pooling did not fully remove.

To enforce the biologically motivated assumption that age-specific
mortality varies smoothly with age, we apply a Gaussian kernel smoother
with \emph{age-varying bandwidth} to each column of~$\bA$ (except the
first) after the decomposition.  At each age~$x$, the smoothed value is a
weighted average of the original basis vector with Gaussian weights
centered at~$x$; the standard deviation $\sigma(x)$ of the kernel controls
the degree of smoothing and varies with age:
\begin{equation}
    \label{eq:varbw_sigma}
    \sigma(x) = \sigma_{\max} \,\Bigl[\,
        s_{\min} + (1 - s_{\min})\,\min\!\bigl(x / x_{\mathrm{ramp}},\; 1\bigr)
    \,\Bigr],
\end{equation}
where $\sigma_{\max}$ is the maximum kernel width (reached at
age~$x_{\mathrm{ramp}}$ and beyond), $s_{\min} \in (0,1)$ scales~$\sigma$
down at age~$0$, and $x_{\mathrm{ramp}}$ is the age at which the ramp
reaches full width.  This design preserves genuine high-frequency structure
at young ages (the infant peak, childhood valley, and accident hump) while
aggressively smoothing the low-frequency, noisy tail at older ages.  The
procedure is:
\begin{enumerate}
    \item \textbf{Skip the first component.}  The leading column
    of~$\bA$ captures the mean age structure of mortality and is already
    smooth by virtue of averaging over all countries and years.  It is left
    unchanged.

    \item \textbf{Per-component parameterization.}  Each of the first few
    higher-order components (2--5) receives its own triple
    $(x_{\mathrm{ramp}},\, s_{\min},\, \sigma_{\max})$, tuned to its
    specific frequency content.  All remaining components share a single
    parameter set with a wider $\sigma_{\max}$.

    \item \textbf{Boundary preservation.}  Ages~0 and~1 are preserved
    exactly (set to their raw HOSVD values after smoothing), because the
    infant-to-childhood transition involves a genuine discontinuity that
    the smoother should not attenuate.

    \item \textbf{Re-orthonormalization.}  Smoothing may slightly
    perturb the orthonormality of~$\bA$.  If the orthonormality error
    (measured as $\max |\bA^\top \bA - \bm{I}|$) exceeds a tolerance,
    the smoothed~$\bA$ is re-orthonormalized via QR decomposition
    \citep[Ch.~5]{GolubVanLoan2013}.  The
    core tensor is then recomputed from \cref{eq:core_projection} using
    the smoothed, orthonormal~$\bA$.
\end{enumerate}

The result is a set of age basis vectors that are smooth, orthonormal,
and capture the biologically meaningful structure of the mortality curve
without the high-frequency noise artifacts.

\section{Application to the HMD Mortality Tensor}
\label{sec:application}

We now apply the Tucker decomposition to the HMD mortality tensor.  This
section describes the concrete dimensions of the tensor, the resulting
decomposition, and the demographic interpretation of each component.

\subsection{Tensor dimensions}
\label{sec:application:dimensions}

After the preprocessing steps of \cref{sec:preprocessing}, the mortality
tensor $\M \in \R^{S \times A \times C \times T}$ has four dimensions:
sex ($S = 2$: female, male), single-year age ($A$ ages from 0 through
the truncation boundary), country ($C$ HMD populations meeting the
minimum-coverage threshold), and calendar year ($T$ spanning the full
HMD range, with missing country-years imputed as described in
\cref{sec:preprocessing:tensor}).  The concrete values are shown in
\cref{tab:s4_tensor_dims}.

Every element $\M_{s,a,c,t}$ is the logit-transformed probability of dying
between exact ages $a$ and $a+1$, for sex~$s$, in country~$c$, during
year~$t$.

\subsection{Decomposition and selected ranks}
\label{sec:application:ranks}

The HOSVD of~$\M$ produces factor matrices and a core tensor with ranks
determined by the $\tau = 0.9999$ (99.99\%) variance threshold
(\cref{eq:rank_selection}), with $r_1 = S = 2$ fixed for the sex mode.
The total number of decomposition
parameters is $S \cdot r_1 + A \cdot r_2 + C \cdot r_3 + T \cdot r_4
+ r_1 \cdot r_2 \cdot r_3 \cdot r_4$, which is a small fraction of the
$S \cdot A \cdot C \cdot T$ elements in the original tensor.  The
concrete ranks, compression ratio, and reconstruction accuracy are
reported in \cref{tab:s4_tucker_ranks,tab:s4_compression}.

\subsection{Demographic interpretation of the factor matrices}
\label{sec:application:interpretation}

\subsubsection{Sex components (\texorpdfstring{$\bS$}{S})}

With $r_1 = S = 2$, the sex factor matrix $\bS \in \R^{2 \times 2}$ is an
orthogonal rotation.  Its two columns define two sex ``components'':
\begin{itemize}
    \item The first component ($\fs = 1$) corresponds approximately to the
    sex-averaged mortality level, with loadings of similar magnitude and
    the same sign for both females and males.

    \item The second component ($\fs = 2$) corresponds to the sex
    differential, with loadings of opposite sign for the two sexes.
\end{itemize}
Through the core tensor, the second sex component interacts with specific
age and time components to produce the age- and time-varying sex
differential described in \cref{rem:sex_diff}.

\subsubsection{Age basis functions (\texorpdfstring{$\bA$}{A})}

The columns of $\bA \in \R^{A \times r_2}$ are the age basis vectors
$\bm{a}_1, \ldots, \bm{a}_{r_2}$, each capturing a distinct feature
of the mortality curve.  (Following demographic convention, we refer to
these vectors as ``basis functions'' when interpreting them as
discretized functions of age.)  Crucially,
because these are orthogonal components of the logit($\qx$) surface,
each component beyond the first must cross zero: it cannot shift all
ages in the same direction (that would correlate with the first
component) but instead \emph{rebalances} mortality across age groups.
\begin{itemize}
    \item The first basis function ($\fa = 1$) is approximately the mean
    logit($\qx$) schedule across all countries and years: high at birth,
    declining through childhood, reaching a minimum around age~10, then
    rising approximately linearly (on the logit scale) through adulthood
    and old age.  This component captures overall mortality level.

    \item The second basis function ($\fa = 2$) captures the dominant
    \emph{age-rebalancing} mode.  It crosses zero once, near age~18,
    and is opposite in sign on either side: positive for childhood ages
    and negative for adult ages (or vice versa, depending on sign
    convention).  A positive weight on this component shifts relative
    mortality toward childhood and away from adulthood, independently
    of the overall level set by the first component.  This single
    zero-crossing makes it the coarsest possible rebalancing --
    children versus adults.

    \item Higher-order basis functions rebalance mortality at
    progressively finer scales.  Each successive component crosses zero
    more frequently, partitioning the age axis into increasingly narrow
    bands of positive and negative deviation.  The third component
    typically isolates the infant--childhood transition; the fourth and
    fifth capture the ``accident hump'' in young-adult mortality and
    deviations from Gompertz linearity at old ages; and still higher
    components encode fine-grained age-specific corrections.
\end{itemize}
Because the basis functions have been smoothed
(\cref{sec:tucker:bsmooth}), they represent smooth functions of age,
consistent with the biological expectation that underlying mortality risk
does not change discontinuously from one age to the next (with the
exception of the infant-to-childhood transition).

\subsubsection{Country loadings (\texorpdfstring{$\bC$}{C})}

Each row of $\bC \in \R^{C \times r_3}$ locates a country in an
$r_3$-dimensional space.  The distances between countries in this space
reflect the similarity of their mortality structures.  Countries with
similar mortality patterns -- similar age shapes, similar levels, similar
sex differentials -- are neighbors.  The country loadings thus provide a
natural basis for the clustering analysis of \cref{sec:clustering}.

\subsubsection{Year loadings (\texorpdfstring{$\bT$}{T})}

Each row of $\bT \in \R^{T \times r_4}$ locates a calendar year in an
$r_4$-dimensional space.  Over time, the year loadings trace a trajectory
through this space that reflects the secular mortality transition: from
high-mortality regimes of the eighteenth and nineteenth centuries through
the rapid gains of the twentieth century to the low-mortality plateau of
the present.  Because exceptional years (wars and pandemics) have been
excluded from the tensor (\cref{sec:preprocessing:exceptional}), the year
loadings capture the secular trend and shorter-term non-exceptional
fluctuations only; the exceptional patterns are modeled separately in
\cref{sec:exceptional}.

Because the year loadings are shared across all countries, the temporal
basis functions represent \emph{HMD-wide} patterns of mortality
change.  Country-specific deviations from these HMD-wide patterns are encoded
in the interaction between the country and year components through the core
tensor~$\G$.

\subsection{The role of imputed entries}
\label{sec:application:imputation}

As discussed in \cref{sec:preprocessing:imputation_bias}, country-mean
imputation introduces a systematic up-weighting of countries with short
time series.  When the tensor spans the full HMD year range
(1751--2024), countries entering the database in the late twentieth
century have the vast majority of their year slices filled with identical
copies of their mean schedule.  These replicated entries inflate the
country's influence on all mode unfoldings, pulling the age basis
functions, country loadings, and year loadings toward the mean patterns of
short-series countries at the expense of the long-series countries that
provide the genuine historical signal.

Of the mitigation strategies outlined in
\cref{sec:preprocessing:imputation_bias}, we favor \textbf{weighted
decomposition} as the most promising approach.  The idea is to weight
each entry in the mode unfoldings by the observed mask, so that imputed
values contribute minimally -- or not at all -- to the SVD.
Concretely, when forming the mode-$n$ unfolding $\bm{M}_{(n)}$ for the
SVD step, columns corresponding to imputed country-year cells receive
weight~$w$ while observed entries receive weight~$1$.  Setting $w = 0$
recovers a decomposition that depends only on observed data, while
intermediate values $0 < w < 1$ allow imputed entries to regularize the
solution without dominating it.  This preserves the complete-tensor
structure that the HOSVD requires while ensuring that the factor
matrices reflect the actual data rather than artifacts of the
imputation.

Year-range restriction is a useful complement: trimming the early
centuries (e.g., restricting to 1900--2024) dramatically reduces the
fraction of imputed cells at the cost of discarding the longest
historical series.  Incomplete tensor decomposition avoids imputation
entirely but is computationally more demanding.  We intend to compare all
three strategies empirically, evaluating their effect on reconstruction
error, rank selection, and the demographic interpretability of the factor
matrices.

Regardless of the imputation strategy chosen for the decomposition itself,
downstream analyses such as clustering (\cref{sec:clustering}) use only
truly observed country-years, so the imputation does not affect the
substantive results of those analyses.



We now report the results of applying the weighted HOSVD to the
HMD mortality tensor.

\subsection{Tucker decomposition}
\label{sec:results:decomposition}

The weighted HOSVD (\cref{sec:tucker:definition}; \cref{rem:weighted_hosvd})
was applied to the mortality tensor with observation-based weighting,
down-weighting imputed entries so that the factor matrices are driven
primarily by observed data.  Ranks were selected by the
$\tau = 0.9999$ (99.99\%) cumulative variance threshold
(\cref{eq:rank_selection}), with $r_1 = 2$ fixed for the sex dimension.

\subsubsection{Singular value decay and rank selection}

\begin{figure}[!htbp]
\centering
\includegraphics[width=\textwidth]{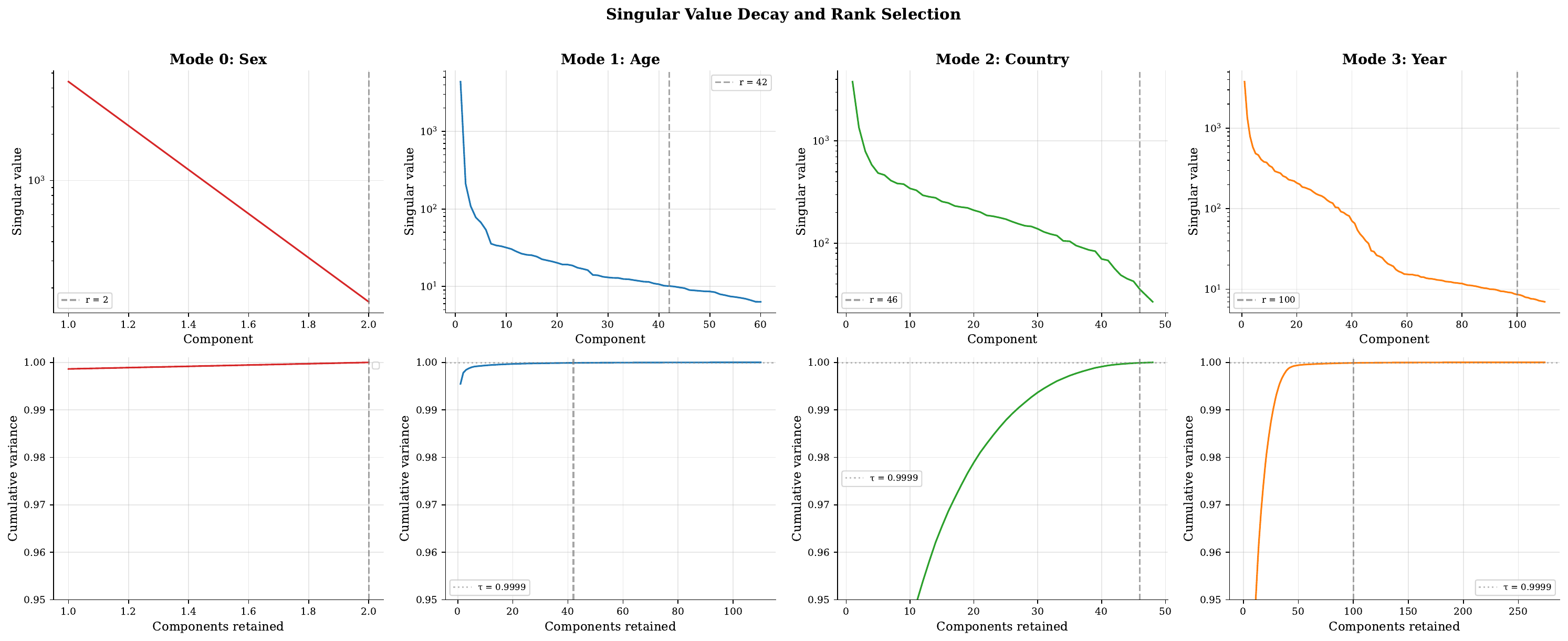}
\caption{Singular value decay and cumulative variance for each tensor
mode.  Vertical dashed lines mark the selected ranks at the
$\tau = 0.9999$ (99.99\%) threshold.  The age mode has rapid decay
(few dominant age patterns), while country and year modes require more
components.}
\label{fig:s4_sv_decay}
\end{figure}

\Cref{fig:s4_sv_decay} shows the singular value decay and cumulative
variance explained for each tensor mode.  The patterns confirm the
expectations laid out in \cref{sec:tucker:rank}: the age mode exhibits
the most rapid decay, with a small number of dominant components
capturing the limited repertoire of qualitatively distinct age patterns
in human mortality, followed by a sharp drop.  The country and year modes
decay more gradually, reflecting the greater complexity of cross-country
and temporal variation.  At the respective thresholds (vertical dashed
lines in \cref{fig:s4_sv_decay}), the selected ranks are $r_2$ (age),
$r_3$ (country), and $r_4$ (year), representing a dramatic compression
of the original tensor with negligible loss of information
(\cref{tab:s4_compression}).

\Cref{tab:s4_tucker_ranks} records the selected rank for each mode
alongside the cumulative variance captured, and
\cref{tab:s4_compression} summarizes the resulting compression and
reconstruction accuracy.

{\setlength\LTleft{0pt}\setlength\LTright{0pt}
\begin{longtable}{@{\extracolsep{\fill}}lcrc@{}}
\caption{Tucker decomposition: selected ranks by mode ($\tau = 0.9999$)} \label{tab:s4_tucker_ranks} \\
\toprule
Mode & Dimension & Rank & Variance captured \\
\midrule
\endfirsthead
\caption[]{Tucker decomposition: selected ranks by mode ($\tau = 0.9999$) (continued)} \\
\toprule
Mode & Dimension & Rank & Variance captured \\
\midrule
\endhead
\midrule
\multicolumn{4}{r}{\textit{Continued on next page}} \\
\endfoot
\bottomrule
\endlastfoot
Sex & 2 & 2 & 1.000000 \\
Age & 110 & 42 & 0.999901 \\
Country & 48 & 46 & 0.999913 \\
Year & 274 & 100 & 0.999898 \\
\end{longtable}}

{\setlength\LTleft{0pt}\setlength\LTright{0pt}
\begin{longtable}{@{\extracolsep{\fill}}lr@{}}
\caption{Tucker decomposition: compression and reconstruction accuracy.  The decomposition captures the full tensor in a fraction of the original parameters with negligible loss.} \label{tab:s4_compression} \\
\toprule
Quantity & Value \\
\midrule
\endfirsthead
\caption[]{Tucker decomposition: compression and reconstruction accuracy.  The decomposition captures the full tensor in a fraction of the original parameters with negligible loss. (continued)} \\
\toprule
Quantity & Value \\
\midrule
\endhead
\midrule
\multicolumn{2}{r}{\textit{Continued on next page}} \\
\endfoot
\bottomrule
\endlastfoot
Original tensor elements & 2,893,440 \\
Factor matrix parameters & 34,232 \\
Core tensor parameters & 386,400 \\
Total decomposition parameters & 420,632 \\
Compression ratio & 6.9x \\
Relative Frobenius error (all entries) & 0.376316 \\
Relative Frobenius error (observed only) & 0.061078 \\
R-squared (observed entries) & 0.986222 \\
\end{longtable}}

\subsubsection{Age basis functions}

\begin{figure}[!htbp]
\centering
\includegraphics[width=\textwidth]{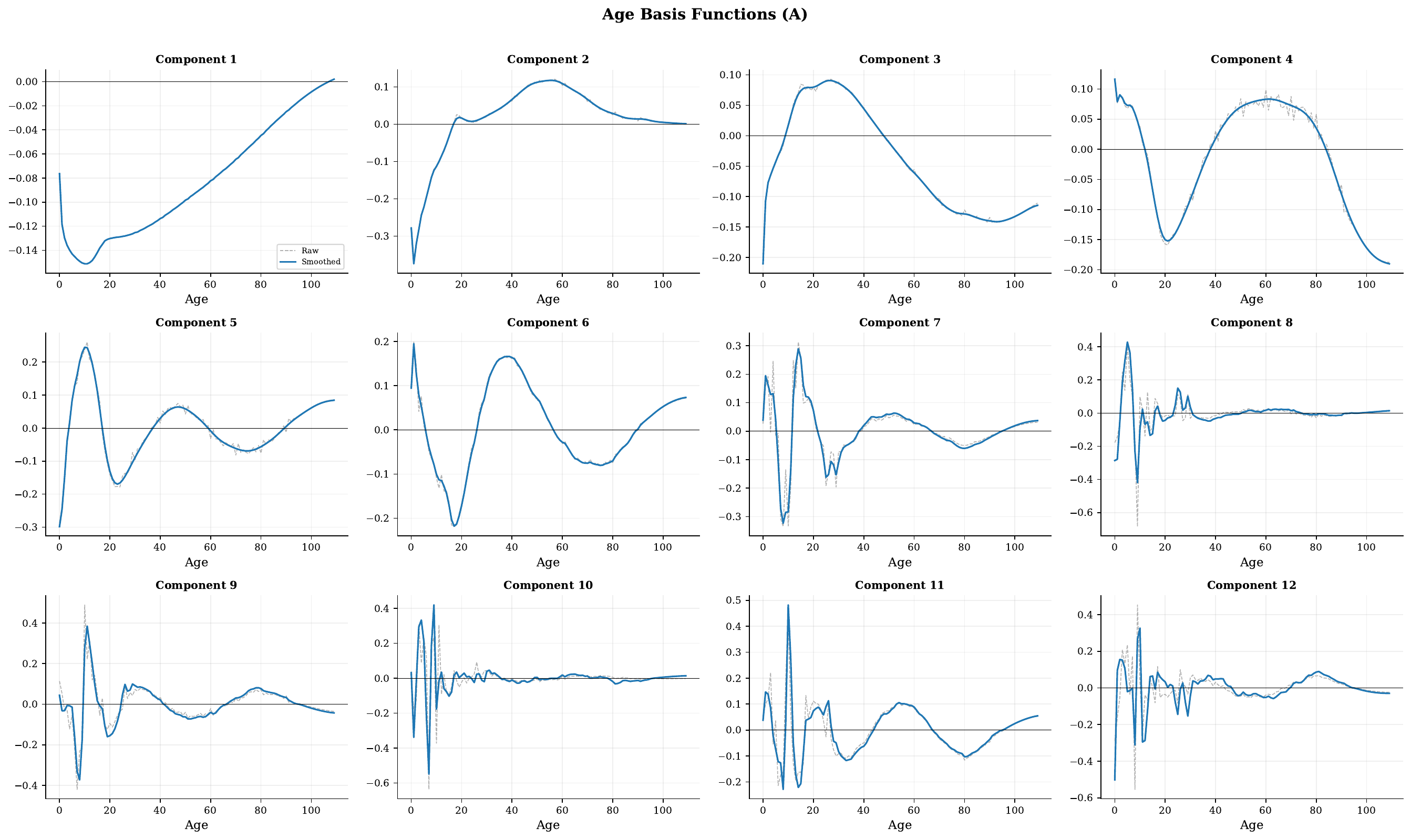}
\caption{Age basis functions from the Tucker decomposition.  Dashed
gray = raw HOSVD; solid colored = after variable-bandwidth Gaussian kernel smoothing.
The first component captures the mean logit($\qx$) schedule; the second
crosses zero once (near age~18), rebalancing childhood vs.\ adulthood
mortality; higher components cross zero with increasing frequency,
correcting progressively finer age-specific features.}
\label{fig:s4_age_basis}
\end{figure}

\Cref{fig:s4_age_basis} displays the age basis functions (columns of
$\bA$), comparing the raw HOSVD output (dashed gray) with the smoothed
versions after variable-bandwidth Gaussian kernel smoothing
(\cref{sec:tucker:bsmooth}).  The first component captures the mean
logit($\qx$) schedule: monotonically decreasing from birth through
childhood, reaching a minimum around age~10, then rising through
adulthood and old age.  The second component crosses zero once, near
age~18, rebalancing childhood mortality against adult mortality -- the
coarsest possible age redistribution.  Each successive higher-order
component crosses zero more frequently, partitioning the age range into
increasingly narrow bands of positive and negative deviation.  The
third component isolates the infant--childhood transition; the fourth
and fifth capture the young-adult ``accident hump'' and old-age
curvature; and components beyond the fifth encode fine-grained
age-specific corrections that become progressively noisier in the raw
HOSVD output.  The smoothing removes these high-frequency oscillations
without materially altering the first few components, which are already
smooth by virtue of averaging.

\subsubsection{Country and year loadings}

\begin{figure}[!htbp]
\centering
\includegraphics[width=\textwidth]{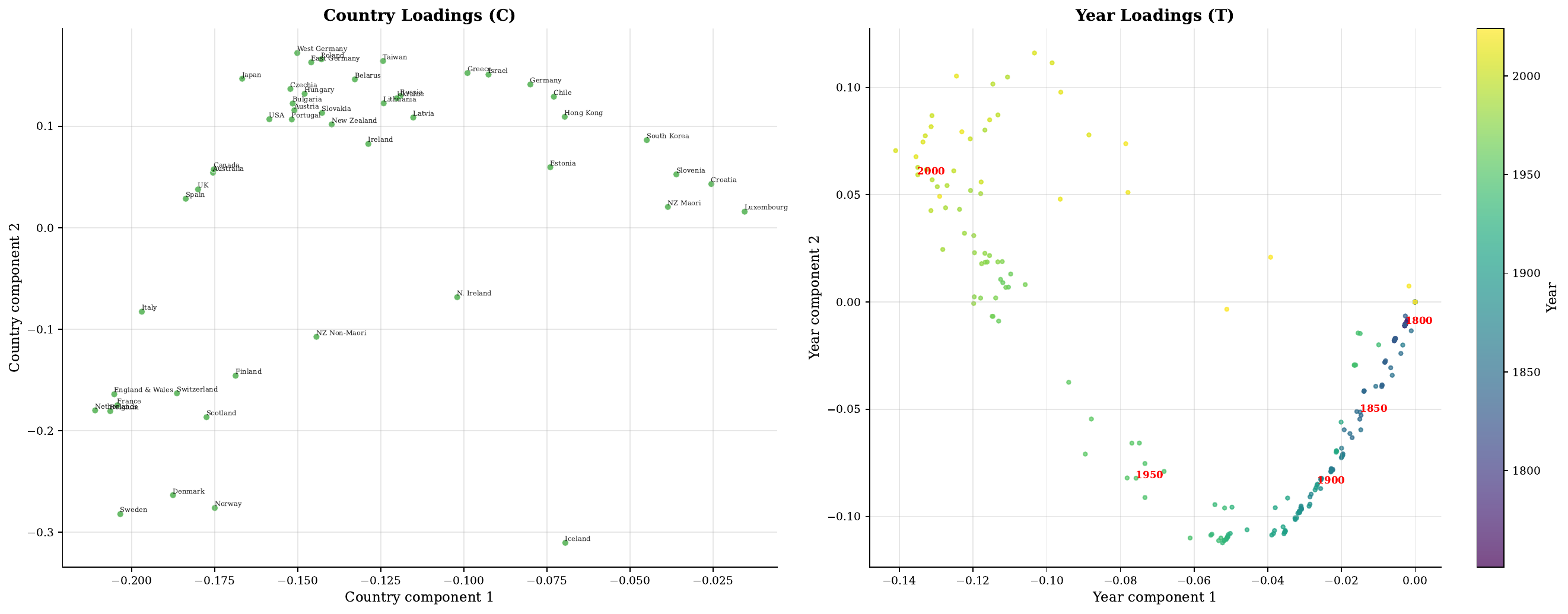}
\caption{Leading country and year loadings.  Left: countries plotted in
the space of their first two Tucker components -- nearby countries have
similar mortality structures.  Right: year loadings trace the secular
mortality transition over time.}
\label{fig:s4_loadings}
\end{figure}

\Cref{fig:s4_loadings} plots the leading country and year loadings.  In
the left panel, countries are positioned in the space of their first two
Tucker components: nearby countries share similar mortality structures.
Countries with similar underlying age-pattern shapes appear close
together, confirming that the country loadings capture systematic
differences in the age structure of mortality.
In the right panel, the year loadings trace the secular mortality
transition over time (\cref{sec:application:interpretation}): the
dominant temporal component shows a smooth, monotonic trend reflecting
the long-run decline in mortality levels.

\subsubsection{Reconstruction accuracy}

\begin{figure}[!htbp]
\centering
\includegraphics[width=\textwidth]{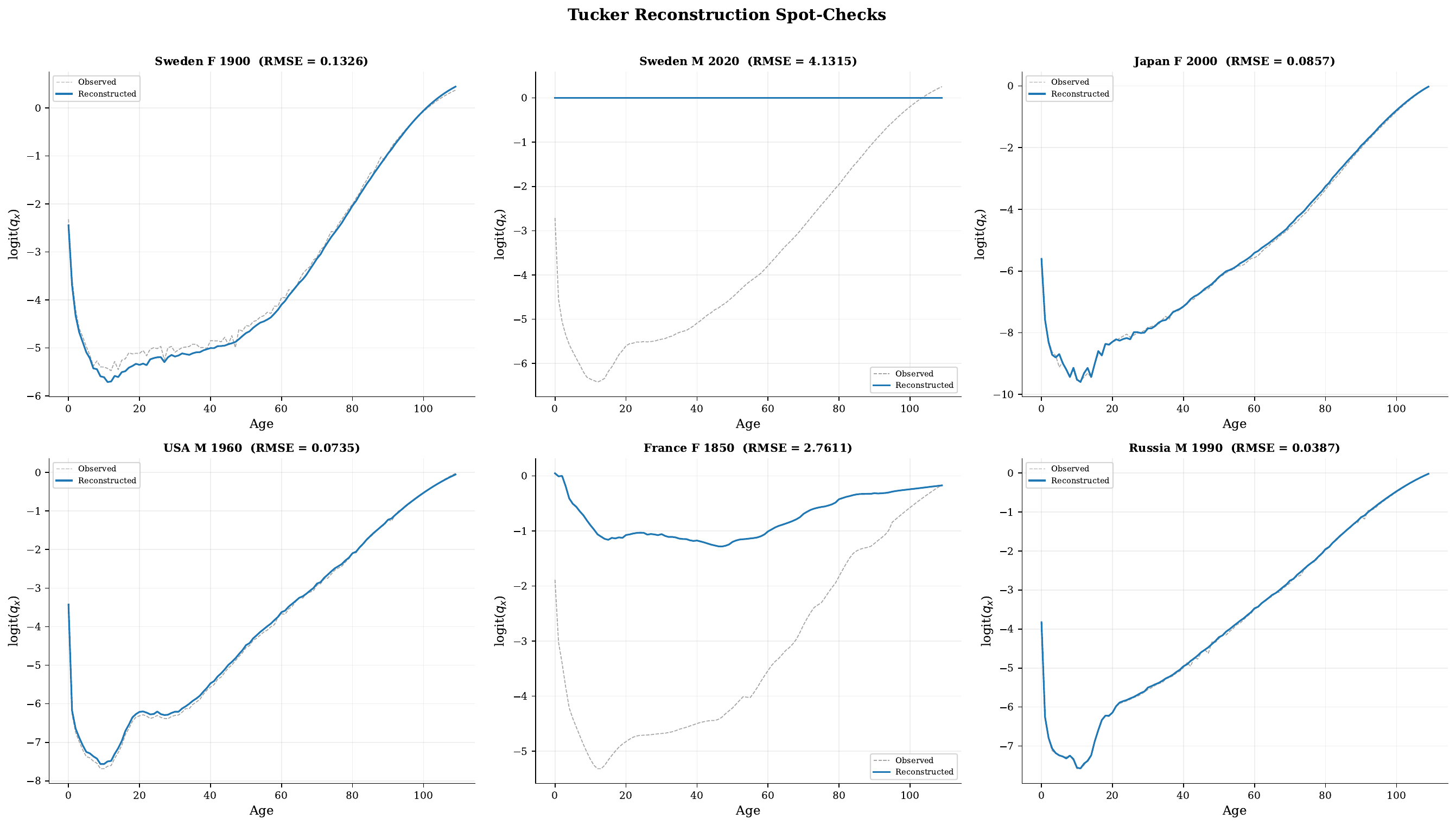}
\caption{Reconstruction spot-checks: observed logit($\qx$) (dashed
gray) vs.\ Tucker reconstruction (solid blue) for selected
country-years.  Four panels show genuinely observed schedules with
excellent reconstruction (RMSE~$<$~0.08).  Sweden~M~2020 (COVID,
excluded as exceptional): the reconstruction is \emph{identically zero}
because the null year-loading vector makes the Tucker product vanish --
exceptional mortality is handled by the disruption framework
(\cref{sec:exceptional}).  France~F~1850 (data quality exclusion):
the reconstruction is non-zero because other countries provide
observations for~1850, giving a non-null year loading; the large RMSE
reflects the gap between the model's interpolation and the excluded
low-quality data.}
\label{fig:s4_reconstruction}
\end{figure}

\Cref{fig:s4_reconstruction} provides spot-checks of reconstruction quality:
observed logit($\qx$) schedules (dashed gray) are overlaid with the
Tucker reconstruction (solid blue) for selected country-years spanning a
wide range of mortality levels and time periods.  Four of the six panels
show genuinely observed country-years, where the decomposition captures
both the overall shape and fine-scale features of individual mortality
schedules, including the infant--childhood transition, the valley
minimum, the accident hump, and the old-age increase.  The remaining
two panels -- Sweden M~2020 and France F~1850 -- illustrate what happens
when the tensor entry is \emph{not genuinely observed}, each via a different
mechanism.

Sweden~2020 is a COVID year excluded as exceptional.  Because 2020 is
exceptional for most HMD countries, very few non-exceptional observations
exist for that year.  With zero imputation weight
(\cref{sec:preprocessing:imputation_bias}), the weighted HOSVD assigns a null
year-loading vector $\bm{t}_{2020} = \bm{0}$, so the Tucker product
$\mathcal{G} \times_1 \bm{s} \times_2 \bm{a} \times_3 \bm{c}
\times_4 \bm{0} = \bm{0}$ regardless of sex, age, or country --
the reconstruction is \emph{identically zero}.  Exceptional mortality
is handled by the disruption modelling framework
(\cref{sec:exceptional}), not by the base reconstruction.

France~F~1850 was excluded for data quality reasons, but year~1850 has
genuine observations from other countries (e.g.\ Sweden), so
$\bm{t}_{1850} \neq \bm{0}$.  The reconstruction therefore produces a
non-zero age-varying schedule -- the model's interpolation of what
France's mortality would look like in~1850 based on France's country
loading and the~1850 year loading estimated from other countries.  The
large RMSE reflects the gap between this model-based interpolation and
the excluded low-quality data shown in gray.

The $R^2$ of the reconstruction on genuinely observed
entries confirms that the decomposition captures effectively all of the
systematic variation in the data, as anticipated in
\cref{sec:application:ranks}.

\section{Clustering Mortality Regimes}
\label{sec:clustering}

The Tucker decomposition provides a compact representation of every
observed mortality schedule, but it does not by itself identify the
discrete regimes that structure the mortality landscape.  We now develop
a clustering procedure that \emph{controls for overall mortality
level} and groups country-year observations by the \emph{age-specific
residual} -- the part of the mortality schedule that describes its shape
independently of how high or low mortality is overall.

\subsection{The unit of clustering: country-year observations}
\label{sec:clustering:unit}

The objects being clustered are not countries and not years, but
\emph{country-year observations}: each observed pair $(c, t)$ with
$O_{c,t} = 1$ constitutes a single data point -- a deliberate
choice, because a given country may occupy one mortality regime in 1900 and a
very different one in 2000, and assigning a single cluster label to the
country as a whole would obscure this historical transition.  Conversely,
different countries may reach the same mortality regime at different
calendar times.  Clustering at the country-year level respects both of
these realities.

Only truly observed country-years are clustered.  Imputed entries
(\cref{sec:preprocessing:imputation}) are excluded, so the clustering
results are determined entirely by genuine data.

\subsection{Feature space: level-controlled age-structure features}
\label{sec:clustering:features}

A na\"{\i}ve clustering approach would operate on the full reconstructed
sex-age logit($\qx$) schedules $\bm{z}_{c,t} \in \R^{2A}$, which
concatenate the female and male age vectors:
\begin{equation}
    \label{eq:cluster_feature}
    \bm{z}_{c,t} =
    \bigl(\hat{y}_{1,1,c,t}, \ldots, \hat{y}_{1,A,c,t},\;
           \hat{y}_{2,1,c,t}, \ldots, \hat{y}_{2,A,c,t}\bigr)^\top
    \in \R^{2A}.
\end{equation}
%
However, because the first Tucker age component captures the dominant axis
of variation -- the mortality-level trajectory of the epidemiological
transition -- clustering on $\bm{z}_{c,t}$ primarily stratifies
observations by $\ezero$ rather than by differences in the
\emph{shape} of the age schedule.  Two country-years with the same
life expectancy but qualitatively different age patterns would be
grouped together, while a single country observed 30 years apart might
be split across clusters solely because its $\ezero$ has changed.

The Tucker decomposition provides a natural solution.  The reconstructed
logit($\qx$) schedule at any country-year $(c,t)$ is a linear
combination of rank-one terms, each formed by the outer product of a sex
component, an age basis function, and the corresponding element of the
core tensor contracted along the country and year dimensions.  The first
of these terms -- involving the leading age basis function $\bm{a}_1$,
which is approximately proportional to the mean logit($\qx$) schedule
across all countries and years -- captures the overall \emph{level} of
mortality.  The remaining terms capture \emph{departures} from that mean
age pattern: the tilt, the accident hump, the old-age curvature, the sex
differential structure, and all other age-specific features that
distinguish one mortality schedule from another at a given level.

We formalize this decomposition by working with the \emph{effective core
matrix} for each country-year and separating it into a level component
and an age-specific residual.

\subsubsection{The effective core matrix}
\label{sec:clustering:Gct}

For each observed country-year $(c, t)$, contract the core tensor~$\G$
along the country and year dimensions using the corresponding rows of
$\bC$ and~$\bT$:
\begin{equation}
    \label{eq:Gct}
    G_{c,t}[\fs, \fa]
        = \sum_{\fc=1}^{r_3} \sum_{\ft=1}^{r_4}
          \G_{\fs,\fa,\fc,\ft} \; C_{c,\fc} \; T_{t,\ft}\,,
    \qquad \fs = 1, \ldots, r_1, \quad \fa = 1, \ldots, r_2.
\end{equation}
%
The result is an $(r_1 \times r_2)$ matrix that encodes how much of each
sex-component $\times$ age-component combination is active for this
particular country-year.  The full Tucker reconstruction at $(c,t)$ can
be written as a sum over the columns of $G_{c,t}$:
\begin{equation}
    \label{eq:level_residual_decomposition}
    \hat{\M}_{s,a,c,t}
    = \underbrace{\sum_{\fs=1}^{r_1}
      G_{c,t}[\fs, 1] \; S_{s,\fs} \; A_{a,1}}
      _{\text{level component}}
    + \underbrace{\sum_{\fs=1}^{r_1} \sum_{\fa=2}^{r_2}
      G_{c,t}[\fs, \fa] \; S_{s,\fs} \; A_{a,\fa}}
      _{\text{age-specific residual}}\,.
\end{equation}
%
The first term is the \emph{level component}: it is the schedule's
projection onto the leading age basis function $\bm{a}_1$ (column~1
of~$\bA$), weighted by the sex--level interaction coefficients
$G_{c,t}[\fs, 1]$, $\fs = 1, \ldots, r_1$.  Because
$\bm{a}_1$ approximates the mean mortality curve, this term captures the
overall mortality level -- its magnitude is strongly correlated with
$\ezero$.  The second term is the \emph{age-specific residual}: it
captures everything that the leading age basis function cannot represent
-- the shape features that distinguish one mortality schedule from another
at a given level.

\subsubsection{The age-structure feature: clustering on the residual}
\label{sec:clustering:age_structure}

The clustering strategy follows directly from this decomposition: we
\textbf{control for level by removing the level component and cluster on
the age-specific residual}.  Concretely, the \emph{age-structure feature}
is obtained by dropping the first age component (column) from $G_{c,t}$:
\begin{equation}
    \label{eq:age_structure_feature}
    \bm{f}_{c,t}
        = \operatorname{vec}\!\bigl(G_{c,t}[\,\cdot,\, 2{:}r_2]\bigr)
        \in \R^{r_1(r_2 - 1)}.
\end{equation}
%
This is the vectorized set of weights on all age basis functions
\emph{except} the first -- precisely the coefficients that generate the
age-specific residual in
\cref{eq:level_residual_decomposition}.  Two country-years with the same
$\bm{f}_{c,t}$ produce identical age-pattern shapes (identical
departures from the mean curve) regardless of how different their overall
mortality levels are.  Conversely, two country-years with different
$\bm{f}_{c,t}$ differ in the shape of their age schedules even if they
happen to share the same $\ezero$.

This approach is the HOSVD analogue of a standard technique in matrix
factor analysis: dropping the first right singular vector to remove the
dominant source of variation (here, level) and clustering on the
remaining components (here, age-pattern shape).  The key advantage of
doing this within the Tucker framework, rather than through ad hoc
mean-centering or residualization of the reconstructed schedules, is that
the decomposition into level and residual is \emph{exact} -- it follows
from the orthogonality of the age basis functions -- and the resulting
features are low-dimensional, interpretable, and free of any approximation
beyond the Tucker truncation itself.

This level-controlled feature space has several advantages over clustering
on the full $\bm{z}_{c,t}$:
\begin{enumerate}
    \item \textbf{Exact level control.}  The decomposition into level
    component and age-specific residual
    (\cref{eq:level_residual_decomposition}) is exact, given the Tucker
    truncation.  The level contribution is removed completely -- not
    approximately, as would be the case with ad hoc mean-centering or
    regression adjustment -- because the age basis functions are
    orthogonal.  The clustering is therefore driven entirely by
    age-pattern shape.

    \item \textbf{Low dimensionality.}  The feature vector
    $\bm{f}_{c,t} \in \R^{r_1(r_2 - 1)}$ is compact: with $r_1 = 2$
    and $r_2$ typically around 10--20, the dimensionality is on the
    order of 20--40 rather than $2A = 220$.  This makes clustering more
    efficient and reduces the risk of overfitting.

    \item \textbf{Interpretability.}  Each element of $\bm{f}_{c,t}$
    has a direct interpretation: it is the weight on a specific
    sex--age interaction in the Tucker decomposition -- the coefficient
    that generates one term of the age-specific residual.  Large values
    in a particular element indicate that the corresponding age basis
    function (accident hump, old-age curvature, etc.)\ is unusually
    active for this country-year.
\end{enumerate}

\begin{remark}[The rotation of mortality and residual level correlation]
\label{rem:rotation_feature}
Although the level component has been removed exactly, the age-specific
residual is not entirely uncorrelated with $\ezero$.  The second age
component $\bm{a}_2$ (column~2 of~$\bA$) typically captures the tilt of
the age schedule: it is positive at young ages and negative at old ages
(or vice versa), encoding the rotation of the age pattern of mortality
decline as $\ezero$ rises
(\cref{sec:reconstruction:trajectories}).  Because this rotation co-moves
with the mortality transition, the feature $G_{c,t}[\cdot, 2]$ -- which
is part of the age-specific residual -- retains a modest correlation with
$\ezero$.  This is not a defect -- it is the signature of the
substantively important phenomenon that the \emph{shape} of mortality
change depends on where a population stands in the transition.  Removing
this component would discard genuine age-structure information; retaining
it allows the clustering to distinguish, for example, early-transition
populations (where infant mortality dominates the residual) from
late-transition populations (where old-age mortality dominates) even at
similar $\ezero$ values.
\end{remark}

The feature vectors $\bm{f}_{c,t}$ are further reduced by applying
principal component analysis \citep[PCA;][]{Jolliffe2002} to the matrix
$\bm{F}$ whose rows are $\bm{f}_{c,t}^\top$ for all observed
country-years, retaining the leading $d$ components (chosen to explain
99.9\% of the variance).  This PCA step removes residual collinearity
and produces a moderate-dimensional space in which standard clustering
algorithms perform well.

\subsection{Gaussian mixture model with BIC selection}
\label{sec:clustering:gmm}

We fit a Gaussian mixture model \citep[GMM;][]{McLachlanPeel2000} to the PCA-reduced
age-structure feature matrix $\tilde{\bm{F}}$.  The GMM assumes that the
$n_{\mathrm{obs}}$ feature vectors are drawn from a mixture of $k$
multivariate Gaussian distributions, each with its own mean vector and
covariance matrix.  The model is fit by expectation-maximization
\citep[EM;][]{DempsterEtAl1977} for a range of candidate
values of~$k$, and the number of clusters is selected by the Bayesian
information criterion \citep[BIC;][]{Schwarz1978}:
\begin{equation}
    \label{eq:bic}
    \mathrm{BIC}(k) = -2 \, \ell(\hat{\theta}_k) + p_k \, \log(n_{\mathrm{obs}}),
\end{equation}
%
where $\ell(\hat{\theta}_k)$ is the maximized log-likelihood under the
$k$-component model and $p_k$ is the number of free parameters.  The
BIC penalizes model complexity, favoring parsimonious solutions.  The
selected $k$ is the value that minimizes $\mathrm{BIC}(k)$ over the
search range $k = 2, \ldots, 15$.

Two levels of granularity are informative.  A coarse clustering (small
$k$) identifies broad age-pattern families: for example, a single cluster
might unite all country-years sharing a distinctive age-pattern shape,
regardless of when or where they were observed.  A finer
clustering (larger $k$) subdivides these broad families into more specific
age-pattern variants -- distinguishing, for instance, countries with large
young-adult male accident humps from those with smaller humps, or
populations with steep versus gradual old-age mortality increase.  Both
levels of clustering are computed and reported; the specific values of $k$
are determined empirically by the BIC.

\begin{remark}[Validation by hierarchical clustering]
\label{rem:ward}
The GMM results are cross-validated against a second, independent
clustering method: Ward's minimum-variance hierarchical clustering
\citep{Ward1963} applied to the same PCA-reduced features.  The hierarchical
dendrogram provides a visual check on the structure identified by the
GMM and on the plausibility of the selected~$k$.  In practice, the two
methods produce broadly concordant groupings, with differences primarily
at the boundaries between clusters where observations are genuinely
intermediate.
\end{remark}

\subsection{Derived clusterings: countries and time periods}
\label{sec:clustering:derived}

The observation-level cluster assignments $\{g_{c,t}\}_{O_{c,t}=1}$,
where $g_{c,t} \in \{1, \ldots, k\}$, are the fundamental output.  From
these, two derived clusterings are constructed by aggregation.

\subsubsection{Country clusters}

A country-level cluster label is assigned by \emph{majority vote}:
country~$c$ receives the label of the cluster to which the plurality of
its observed country-years belong:
\begin{equation}
    \label{eq:country_cluster}
    g_c = \arg\max_{j \in \{1,\ldots,k\}}
    \bigl|\{t : O_{c,t} = 1 \text{ and } g_{c,t} = j\}\bigr|.
\end{equation}
%
This label reflects the \emph{modal} age-pattern regime of the country
over its observed history.  Because the clustering is level-controlled,
countries with persistent age-pattern signatures tend to maintain a single
regime label across their entire time series, even as $\ezero$ changes
substantially.  Countries that do transition between regimes over time
reflect genuine shifts in the shape of their mortality schedule -- for
example, a country whose age profile shifts toward a different canonical
pattern as its epidemiological context evolves -- rather than mere
changes in overall mortality level.

\subsubsection{Time-period clusters}

A year-level cluster label is assigned analogously:
\begin{equation}
    \label{eq:year_cluster}
    g_t = \arg\max_{j \in \{1,\ldots,k\}}
    \bigl|\{c : O_{c,t} = 1 \text{ and } g_{c,t} = j\}\bigr|.
\end{equation}
%
This label identifies the dominant age-pattern regime across countries in
each calendar year.  The sequence $\{g_t\}_t$ traces the evolution of the
observed mix of age-pattern profiles over time: in early periods when
the HMD contains fewer populations, one set of regimes dominates; as
additional populations enter the database in the twentieth century,
the composition shifts to reflect the broader diversity of age-pattern
families.

\subsection{Demographic interpretation of clusters}
\label{sec:clustering:interpretation}

Each cluster defines a canonical mortality regime characterized by a
distinctive age-specific residual -- a particular pattern of departures
from the mean age schedule -- independent of overall mortality level.
Its demographic character is summarized by the \emph{cluster centroid}:
the mean of the reconstructed sex-age mortality schedules $\bm{z}_{c,t}$
across all observations assigned to the cluster.  The centroid can be
back-transformed from logit to the $\qx$ scale to produce a
representative life table and a corresponding $\ezero$ for each sex.

Because the clustering operates on the age-specific residual
$\bm{f}_{c,t}$ (\cref{sec:clustering:age_structure}) -- the Tucker
coefficients that remain after the level component has been removed --
the resulting regimes primarily reflect \textbf{canonical age-pattern
families}: groups of country-years that share distinctive age patterns,
such as the magnitude of the accident hump, the pace of old-age
mortality increase, the size of the sex differential, and the shape of
infant-to-childhood mortality decline, irrespective of their position in
the mortality transition.  A given cluster may draw members from diverse
geographies and time periods, because the clustering criterion is
similarity of the age-pattern shape, not proximity in space or time
\citep{Mesle2004,VallinMesle2004}.

The historical mortality transition -- the secular decline from high to
low mortality -- is captured not by the cluster structure but by the
\emph{trajectories within each cluster}
(\cref{sec:reconstruction:trajectories}).  As a country's $\ezero$ rises,
its cluster membership may remain stable (reflecting a persistent age-pattern
signature) while its position \emph{along the cluster's trajectory}
advances.  The combination of cluster identity (age-pattern shape) and
trajectory position (mortality level) fully specifies the mortality
schedule.

The cluster centroids and their associated life tables serve as the basis
for the reconstruction model of \cref{sec:reconstruction}, which
interpolates within and between clusters to produce mortality schedules at
arbitrary target levels of $\ezero$.

\subsection{Level trajectory and epoch classification}
\label{sec:clustering:epochs}

The age-structure clustering of
\crefrange{sec:clustering:features}{sec:clustering:interpretation}
operates on the age-specific residual $\bm{f}_{c,t}$ and deliberately
discards the level component $G_{c,t}[\cdot, 1]$.  The level information
is not wasted: it is analyzed separately to classify \emph{how the
overall mortality level is changing over time} within each country.

The grand-mean Tucker loading $G_{c,t}[1, 1]$, which is strongly
correlated with $\ezero$ (\cref{sec:clustering:Gct}), encodes the
mortality level of country~$c$ in year~$t$ within the HOSVD basis.  For
each country, the time series $\{G_{c,t}[1, 1]\}_t$ traces the secular
trajectory of mortality -- the arc of the demographic transition.  We
classify the local behavior of this trajectory into five
\emph{epoch categories} using a two-stage threshold approach:
\begin{enumerate}
    \item \textbf{Smooth the level series.}  For each country, apply
    LOWESS \citep{Cleveland1979} to both the $G_{c,t}[1,1]$ series
    (the classification signal) and the corresponding $\ezero$ series
    (for interpretable visualization), using non-exceptional years only.

    \item \textbf{Extract rolling-window slopes.}  Slide a window of
    width $W$ years (default $W = 15$) across the smoothed $G_{c,t}[1,1]$
    series.  In each window, fit an OLS linear model of level against year
    and extract the slope.  A negative slope indicates declining mortality
    (rising $\ezero$); a positive slope indicates worsening mortality;
    a slope near zero indicates stagnation.

    \item \textbf{Classify by fixed thresholds.}  Each window slope is
    assigned to one of five categories using two threshold parameters,
    $\delta$ (the dead-zone half-width) and $\delta_{\text{rapid}}$
    (the rapid-change boundary):
    \begin{itemize}
        \item \textbf{Rapid improvement}: slope $< -\delta_{\text{rapid}}$
        \item \textbf{Slow improvement}: $-\delta_{\text{rapid}} \le$
              slope $< -\delta$
        \item \textbf{Stagnation}: $|\text{slope}| \le \delta$
        \item \textbf{Slow worsening}: $\delta <$ slope
              $\le \delta_{\text{rapid}}$
        \item \textbf{Rapid worsening}: slope $> \delta_{\text{rapid}}$
    \end{itemize}
    This guarantees that improvement and worsening are always separated
    -- unlike a purely data-driven approach (e.g., GMM on slopes) that
    may lump worsening with stagnation if worsening is rare.

    \item \textbf{Assign years to categories.}  For each calendar year
    in each country's observed range, collect all overlapping windows
    that contain that year and assign the modal (most frequent)
    category.

    \item \textbf{Identify epochs.}  Contiguous stretches of years
    assigned to the same trajectory category within a country define
    \emph{epochs}: named periods during which the country's mortality
    level was, for example, declining rapidly (the core of the mortality
    transition), stagnating (the post-transition period in many
    high-income countries), or worsening (the Soviet-era mortality crisis
    in Eastern Europe).
\end{enumerate}

The epoch classification provides a second, complementary axis of
description that pairs naturally with the age-structure clusters.
Together, the two components yield a two-layer characterization of each
country-year: the age-structure cluster identifies \emph{what kind of
mortality schedule} the population has (its shape), while the level epoch
identifies \emph{where the population stands in the mortality transition
and how fast it is changing} (its trajectory phase).

The epoch classification also serves a practical role in the
neural trajectory model of
\cref{sec:reconstruction:neural_trajectory}: the epoch categories
can parameterize transition timing, helping the network learn that
the relationship between $\ezero$ and age-pattern shape differs
depending on whether a population is actively transitioning, has
stagnated, or is experiencing a reversal.



We now apply the clustering procedure to the HMD mortality tensor.

\subsection{Clustering of mortality regimes}
\label{sec:results:clustering}

The level-controlled clustering procedure of \cref{sec:clustering} was
applied to the age-structure features $\bm{f}_{c,t}$
(\cref{eq:age_structure_feature}), obtained by contracting the core
tensor along the country and year dimensions
(\cref{eq:Gct}) and dropping the first age component.  The resulting
feature vectors have dimension $r_1(r_2 - 1)$, substantially smaller
than the $2A = 220$-dimensional reconstructed schedules.

\subsubsection{Validation of level removal}

\begin{figure}[!htbp]
\centering
\includegraphics[width=\textwidth]{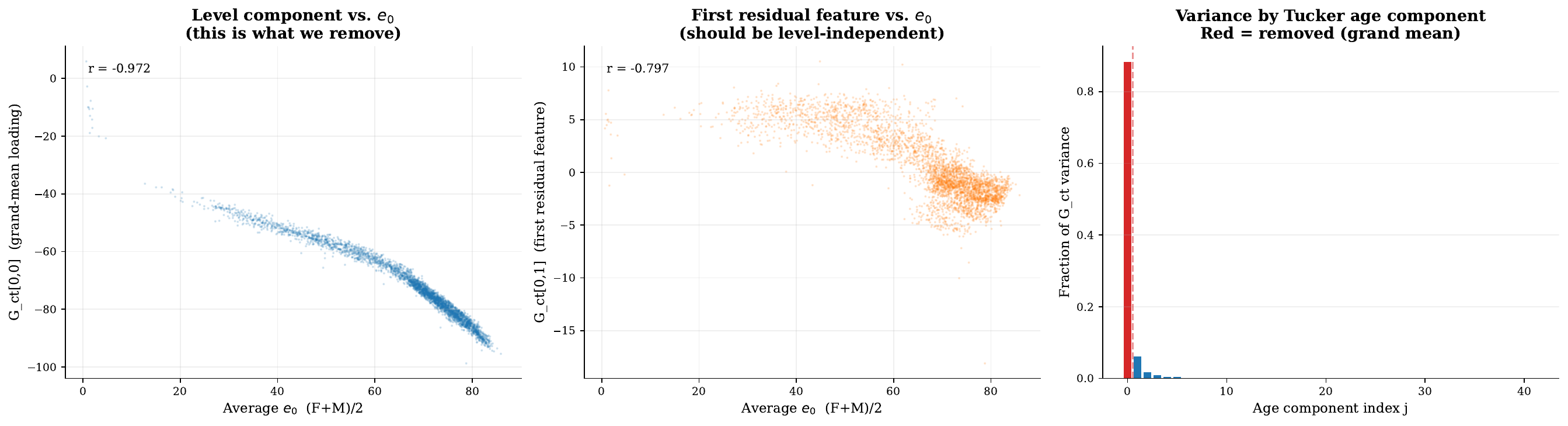}
\caption{Validation of the level-removal strategy.  Left: grand-mean
Tucker loading $G_{c,t}[1,1]$ vs.\ average $\ezero$, showing the
strong monotone relationship that confirms this component encodes
mortality level.  Center: first residual feature $G_{c,t}[1,2]$ vs.\
$\ezero$, showing reduced but nonzero correlation (the rotation
signature).  Right: fraction of $G_{c,t}$ variance by Tucker age
component -- the removed component (red) dominates.}
\label{fig:s5_grand_mean}
\end{figure}

\Cref{fig:s5_grand_mean} confirms the validity of the level-removal
strategy: the grand-mean Tucker loading $G_{c,t}[1,1]$ is strongly
correlated with $\ezero$, confirming that this component encodes overall
mortality level and is the right quantity to remove before age-structure
clustering.  The first remaining feature dimension $G_{c,t}[\cdot, 2]$
retains a modest residual correlation with $\ezero$, reflecting the
rotation of the age pattern of mortality decline as $\ezero$ rises
(\cref{rem:rotation_feature}).  The right panel shows the fraction of
$G_{c,t}$ variance attributable to each Tucker age component: the
removed component dominates, while the retained components collectively
capture the age-structure signal.

\subsubsection{BIC selection and cluster structure}

\begin{figure}[!htbp]
\centering
\includegraphics[width=\textwidth]{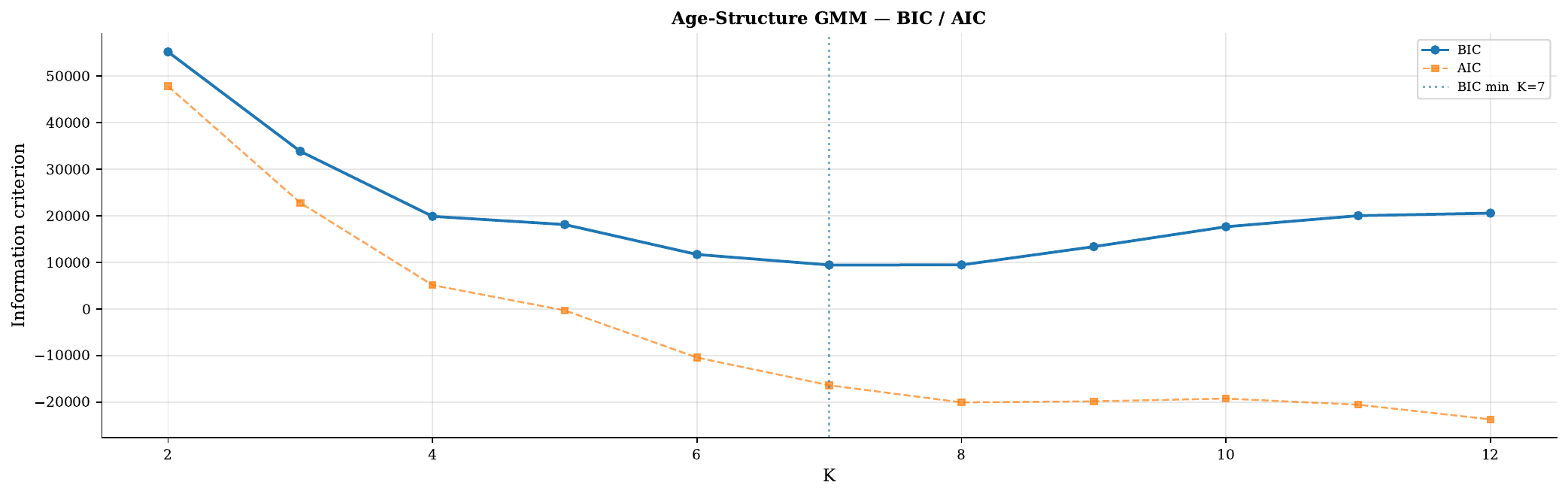}
\caption{BIC and AIC scores for Gaussian mixture models fit to the
PCA-reduced age-structure Tucker features $\tilde{\bm{F}}$.  The BIC
minimum (vertical dashed line) identifies the optimal $K$.}
\label{fig:s5_bic}
\end{figure}

\begin{figure}[!htbp]
\centering
\includegraphics[width=\textwidth]{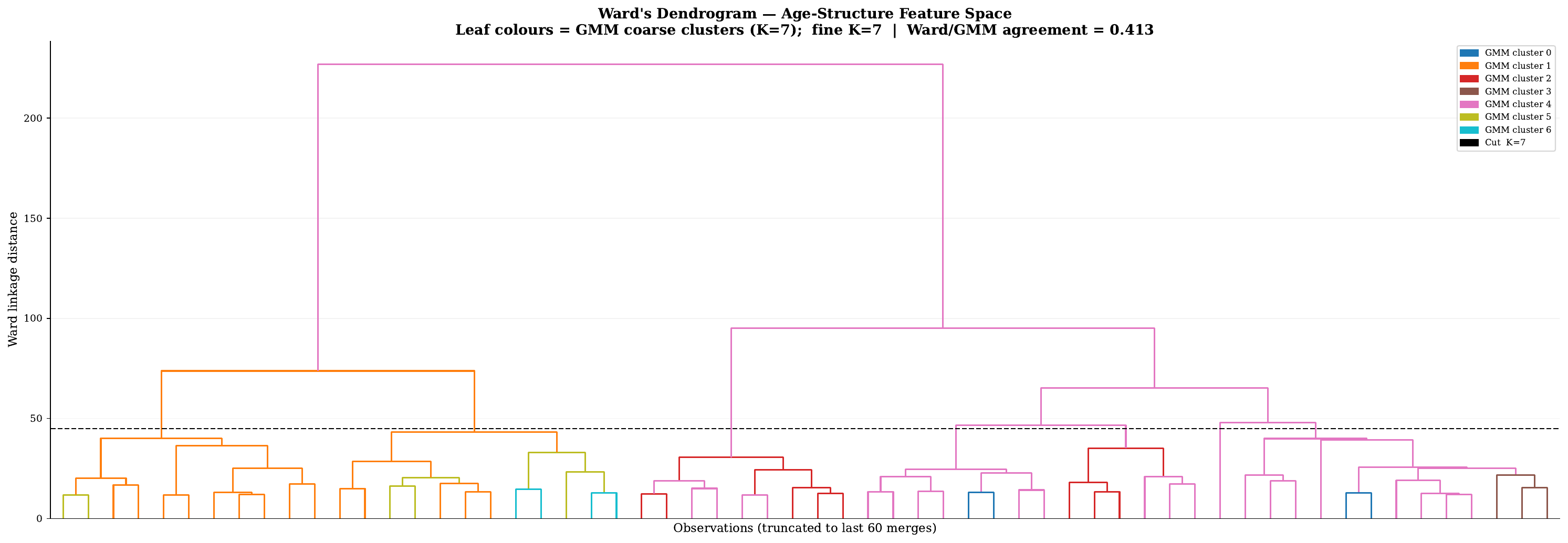}
\caption{Truncated Ward's dendrogram (last 60 merges) for the
age-structure feature space.  Leaves are colored by GMM coarse-cluster
assignment.  The dashed horizontal line marks the cut height for
$K$ clusters.  Tight monochromatic subtrees indicate strong agreement
between Ward's and GMM; mixed subtrees indicate ambiguity at cluster
boundaries.}
\label{fig:s5_dendrogram}
\end{figure}

PCA reduced the age-structure feature space to $d$ components capturing
99.9\% of the variance (\cref{sec:clustering:features}).  Gaussian
mixture models were then fit for a range of component counts $K$
(\cref{sec:clustering:gmm}).
\Cref{fig:s5_bic} shows the BIC and AIC scores (\cref{eq:bic}) for $K = 2$
through $12$.  The BIC-optimal $K$ defines the clustering.  Ward's
hierarchical clustering (\cref{rem:ward}) was applied independently to
the same PCA features as a cross-validation check.
\Cref{fig:s5_dendrogram} shows the truncated Ward's dendrogram with leaves
colored by GMM cluster assignment: tight monochromatic subtrees indicate
strong agreement between the two methods; mixed subtrees mark cluster
boundaries where observations are genuinely intermediate.

\subsubsection{Cluster age-pattern profiles}

\begin{figure}[!htbp]
\centering
\includegraphics[width=\textwidth]{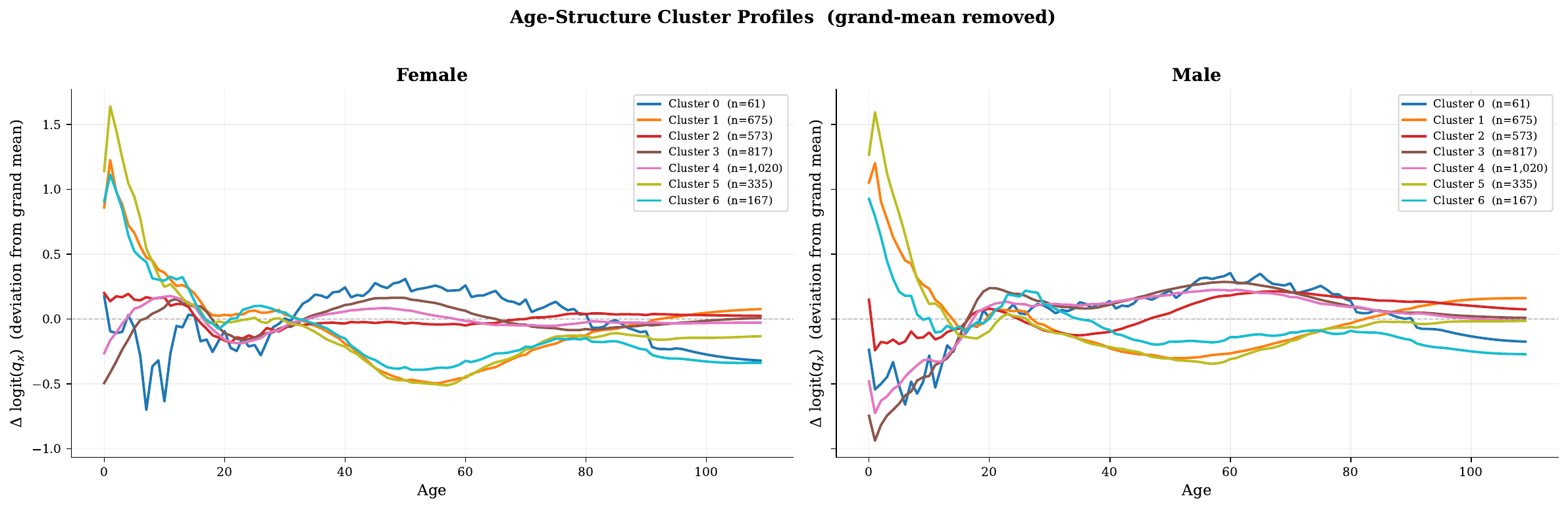}
\caption{Age-structure cluster profiles: the deviation in logit($\qx$)
from the grand mean associated with each cluster, separately for females
and males.  These show the pure age-pattern signal that defines each
cluster, with the overall mortality level removed.}
\label{fig:s5_profiles}
\end{figure}

\Cref{fig:s5_profiles} presents the age-pattern profiles that define each
cluster.  These are computed by reconstructing each cluster centroid back
to logit($\qx$) space after zeroing the grand-mean (first age) component,
so that only the \emph{departures} from the mean age schedule are shown.
The profiles reveal the distinctive age-structure features that drive
the clustering: some clusters are characterized by a pronounced
young-adult accident hump in males, others by a steep or shallow old-age
Gompertz slope, and others by the relative magnitude of infant versus
adult mortality or by the size of the sex differential.

\begin{figure}[!htbp]
\centering
\includegraphics[width=\textwidth]{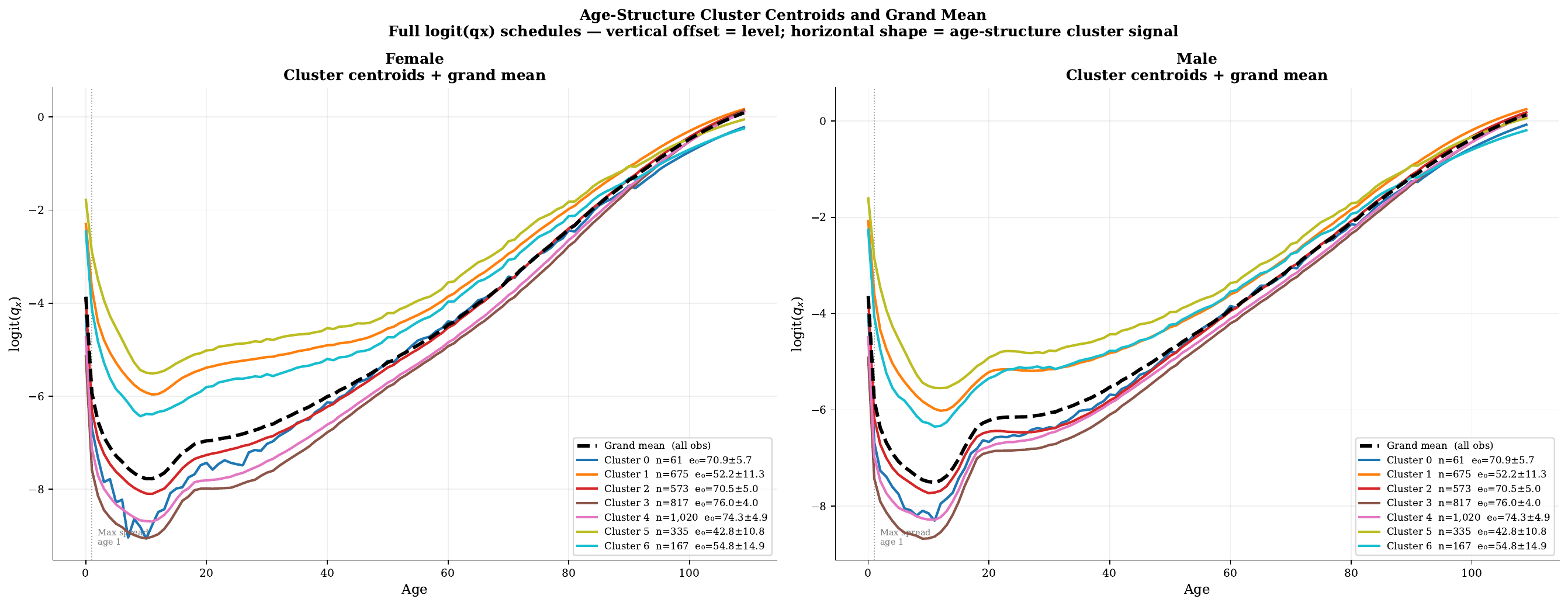}
\caption{Cluster centroids (colored lines) and grand mean (dashed
black) in logit($\qx$) space, for females (left) and males (right).
Vertical offsets reflect level differences; shape differences
(departures from the grand mean) reflect the age-structure clustering
signal.}
\label{fig:s5_centroid_overlay}
\end{figure}

\Cref{fig:s5_centroid_overlay} shows the full cluster centroids in
logit($\qx$) space (including both the level and shape components),
overlaid with the grand mean across all observations.  The vertical
offset between centroids reflects level differences within the
cluster membership (because each cluster spans a wide $\ezero$ range,
the centroid $\ezero$ depends on the cluster's temporal composition),
while the shape differences -- visible as departures from the grand-mean
curve -- are the age-structure features that the clustering identifies.

\subsubsection{Composition and validation}

\begin{figure}[!htbp]
\centering
\includegraphics[width=\textwidth]{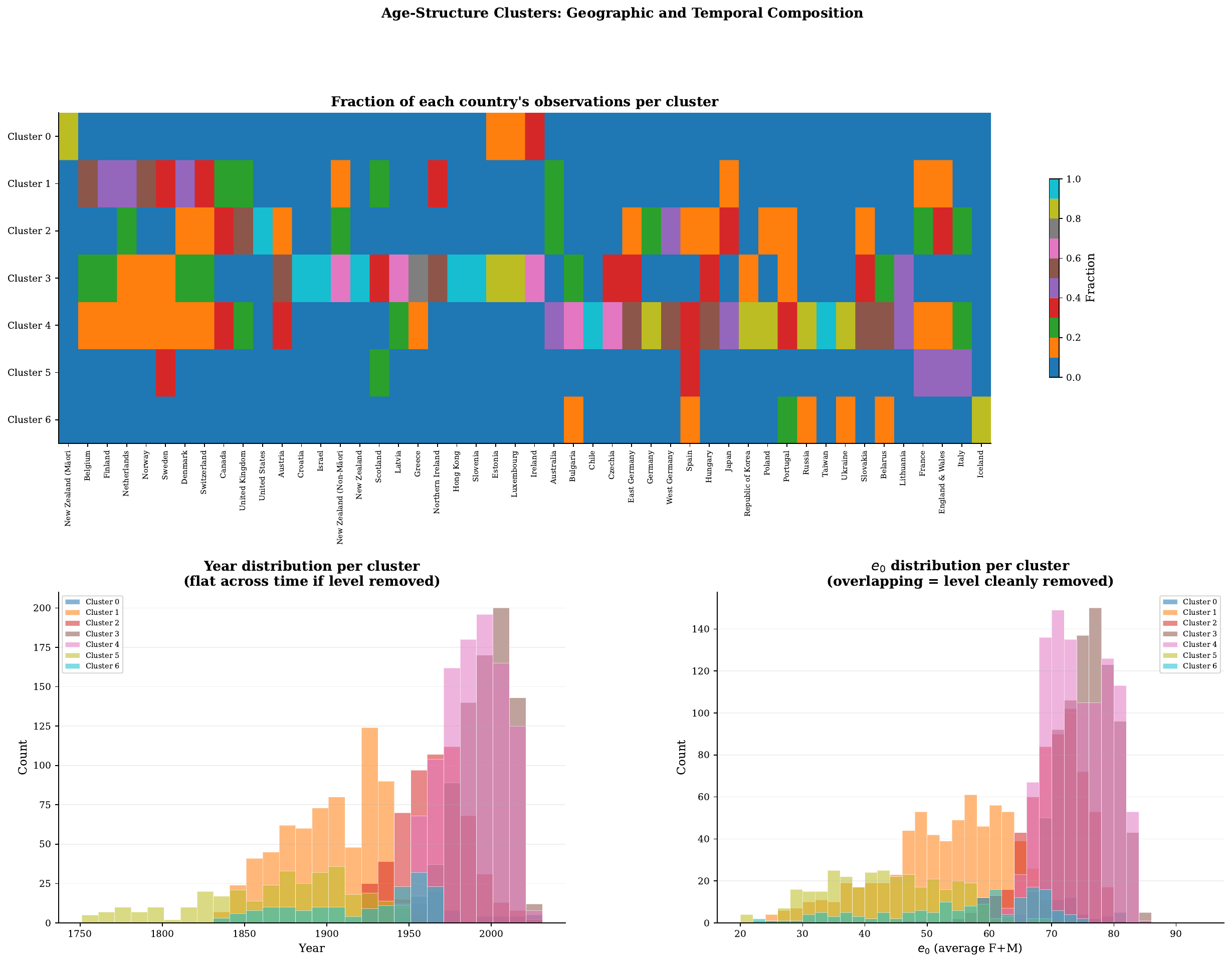}
\caption{Geographic structure of age-structure clusters.  Top: fraction
of each country's country-years assigned to each cluster (countries
sorted by dominant cluster).  Bottom left: year distribution per cluster
-- age-structure clusters should not be strongly year-stratified.
Bottom right: $\ezero$ distribution per cluster -- broad, overlapping
distributions confirm that level has been cleanly separated.}
\label{fig:s5_geography}
\end{figure}

\begin{figure}[!htbp]
\centering
\includegraphics[width=\textwidth]{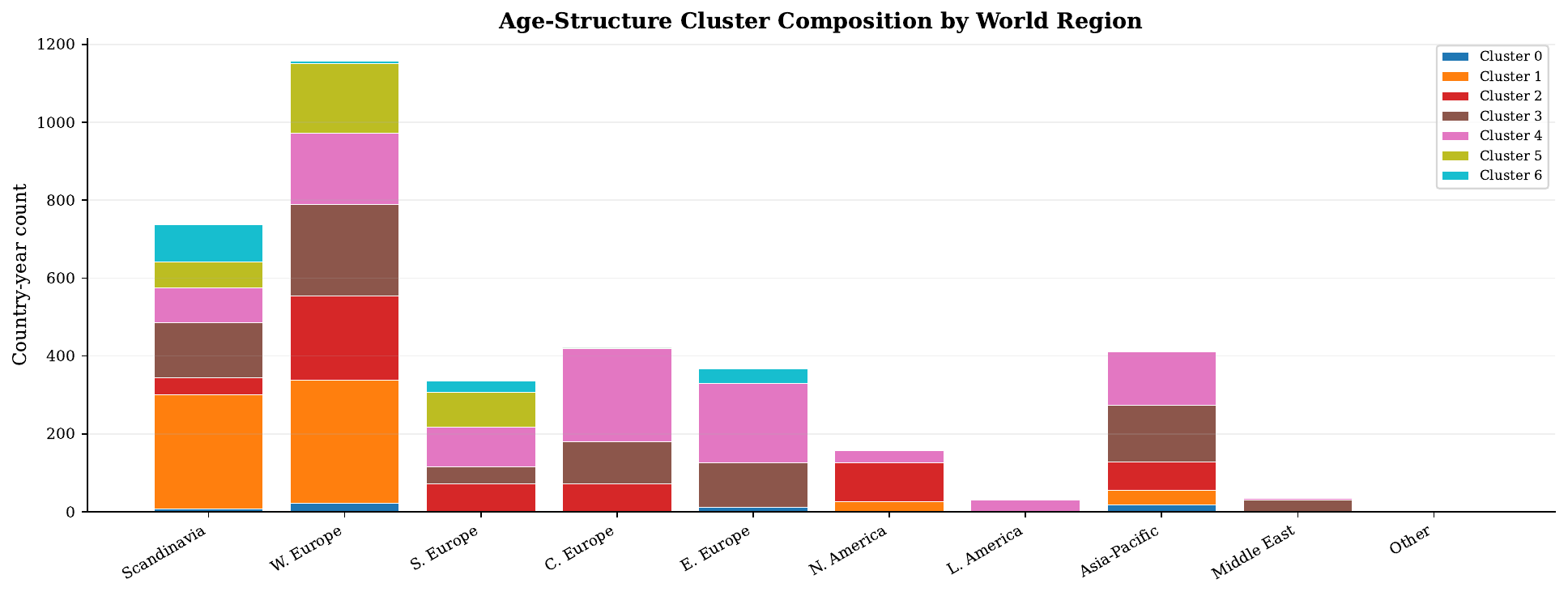}
\caption{Region-level cluster composition.  Each bar shows the total
country-year count per region, colored by cluster assignment.
Nearly monochromatic regions map tightly onto one age-structure
pattern; mixed regions contain countries with diverse age patterns.}
\label{fig:s5_region_summary}
\end{figure}

\Cref{fig:s5_geography} shows the composition of each cluster:
the fraction of each country's observed country-years assigned to each
age-structure cluster, alongside the year and $\ezero$ distributions
per cluster.  Each cluster draws members from multiple countries and
time periods, confirming that the clusters represent canonical
age-pattern families rather than geographic or temporal strata.
The $\ezero$ distributions per cluster are broad and
overlapping, confirming that each cluster spans the full range of
mortality levels rather than being confined to a particular $\ezero$
stratum.  \Cref{fig:s5_region_summary} shows the cluster composition
aggregated by world region; some regions are dominated by a single
age-pattern family while others contain a mix, reflecting the diversity
of epidemiological profiles within those regions.

\subsubsection{Temporal stability and within-cluster structure}

\begin{figure}[!htbp]
\centering
\includegraphics[width=\textwidth]{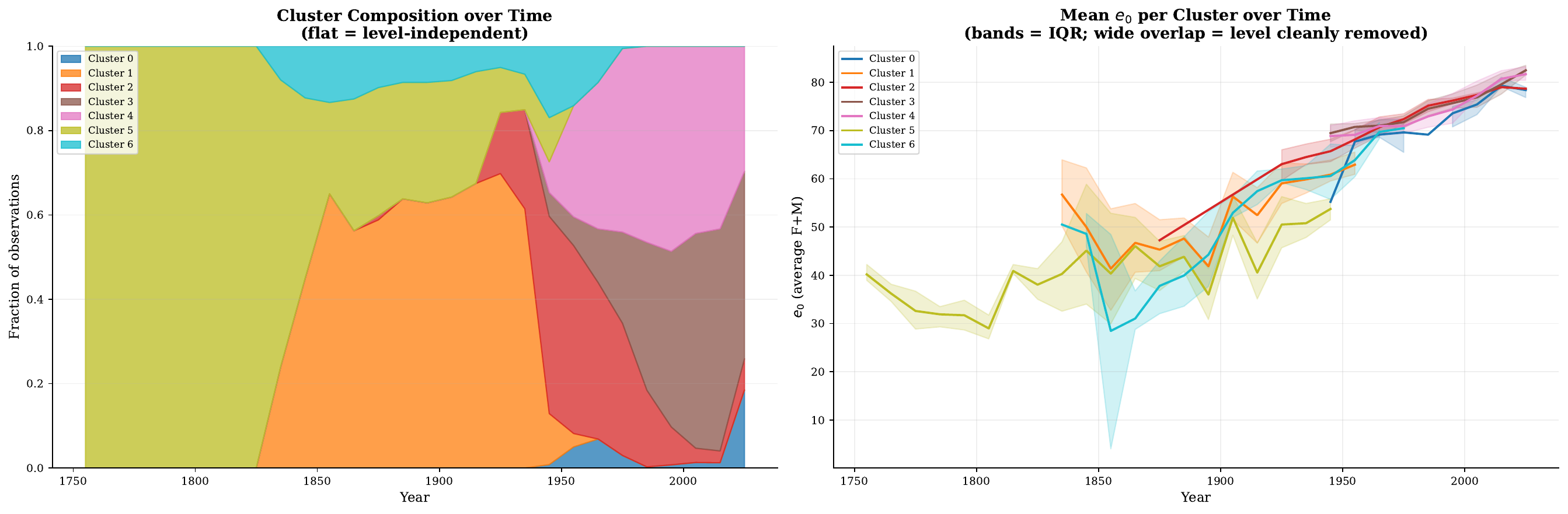}
\caption{Left: stacked area chart of cluster composition across
calendar time (10-year bins).  A flat profile confirms temporal
stability; a trend would indicate residual level contamination.
Right: mean $\ezero$ per cluster over time -- wide overlapping bands
confirm that each cluster spans the full $\ezero$ range.}
\label{fig:s5_time_structure}
\end{figure}

A key prediction of the level-controlled approach is that cluster
membership should be temporally stable: because the clusters capture
persistent age-pattern signatures rather than transient mortality levels,
a country should remain in the same cluster as its $\ezero$ changes.
\Cref{fig:s5_time_structure} tests this prediction.  The left panel shows
a stacked area chart of cluster composition over calendar time
(10-year bins): a flat profile indicates temporal stability, while a
trend would indicate residual temporal structure.  The right panel shows
the mean $\ezero$ per cluster over time -- wide, overlapping bands
confirm that each cluster spans the full $\ezero$ range.

\begin{figure}[!htbp]
\centering
\includegraphics[width=\textwidth]{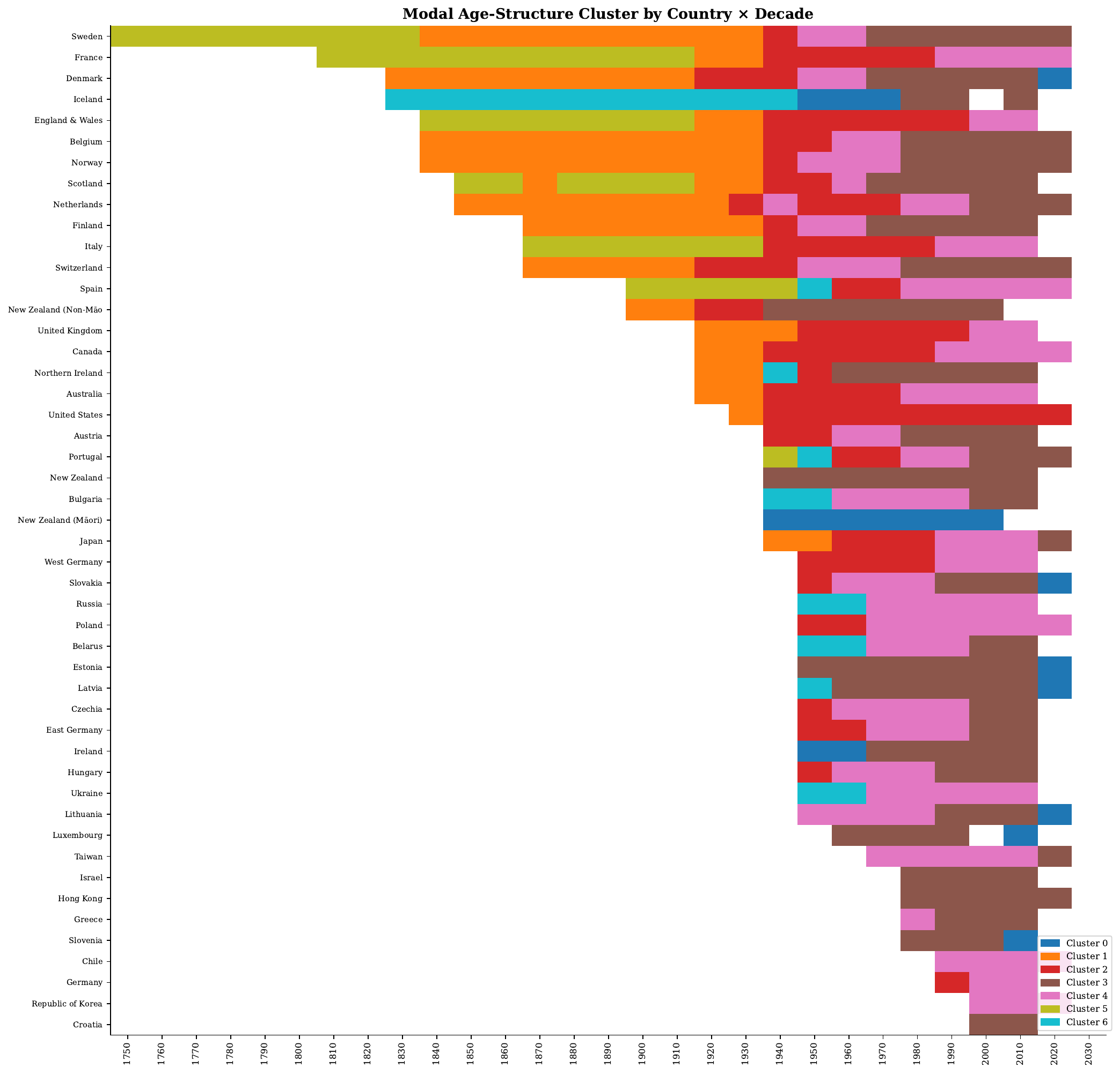}
\caption{Country $\times$ decade cluster heatmap.  Each cell shows
the modal age-structure cluster for that country in that decade.
Countries sorted by first data decade.  Stable horizontal bands
confirm persistent age-pattern signatures.}
\label{fig:s5_time_heatmap}
\end{figure}

\Cref{fig:s5_time_heatmap} provides a comprehensive country-by-decade
cluster membership heatmap.  Synchronized color switches across
countries indicate HMD-wide regime changes (which are rare, confirming
level independence); asynchronous switches indicate country-specific
transitions in age-pattern structure.

\begin{figure}[!htbp]
\centering
\includegraphics[width=\textwidth]{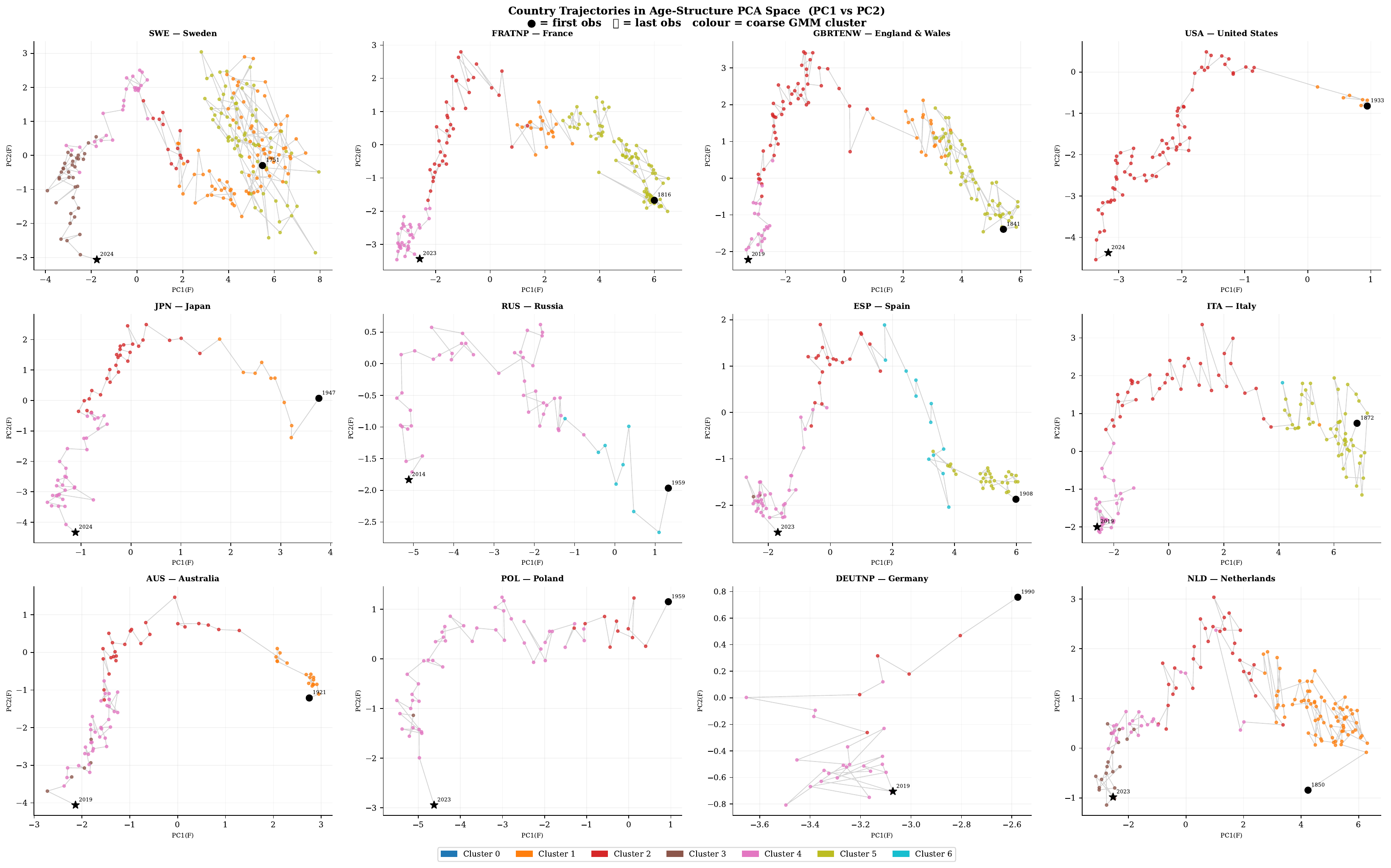}
\caption{Country trajectories in the PCA-reduced age-structure feature
space (PC1 vs.\ PC2), colored by coarse cluster assignment.  Each panel
shows one country; filled circle = first observation, star = last.
Smooth single-color trajectories indicate stable age-structure
membership.}
\label{fig:s5_trajectories}
\end{figure}

\Cref{fig:s5_trajectories} shows selected country trajectories in the
PCA-reduced age-structure feature space, colored by cluster assignment.
Countries with stable age-pattern signatures trace smooth, single-color
paths through the feature space; countries that undergo genuine
age-structure transitions (visible as color changes along the trajectory)
are the exception rather than the rule.

\begin{figure}[!htbp]
\centering
\includegraphics[width=\textwidth]{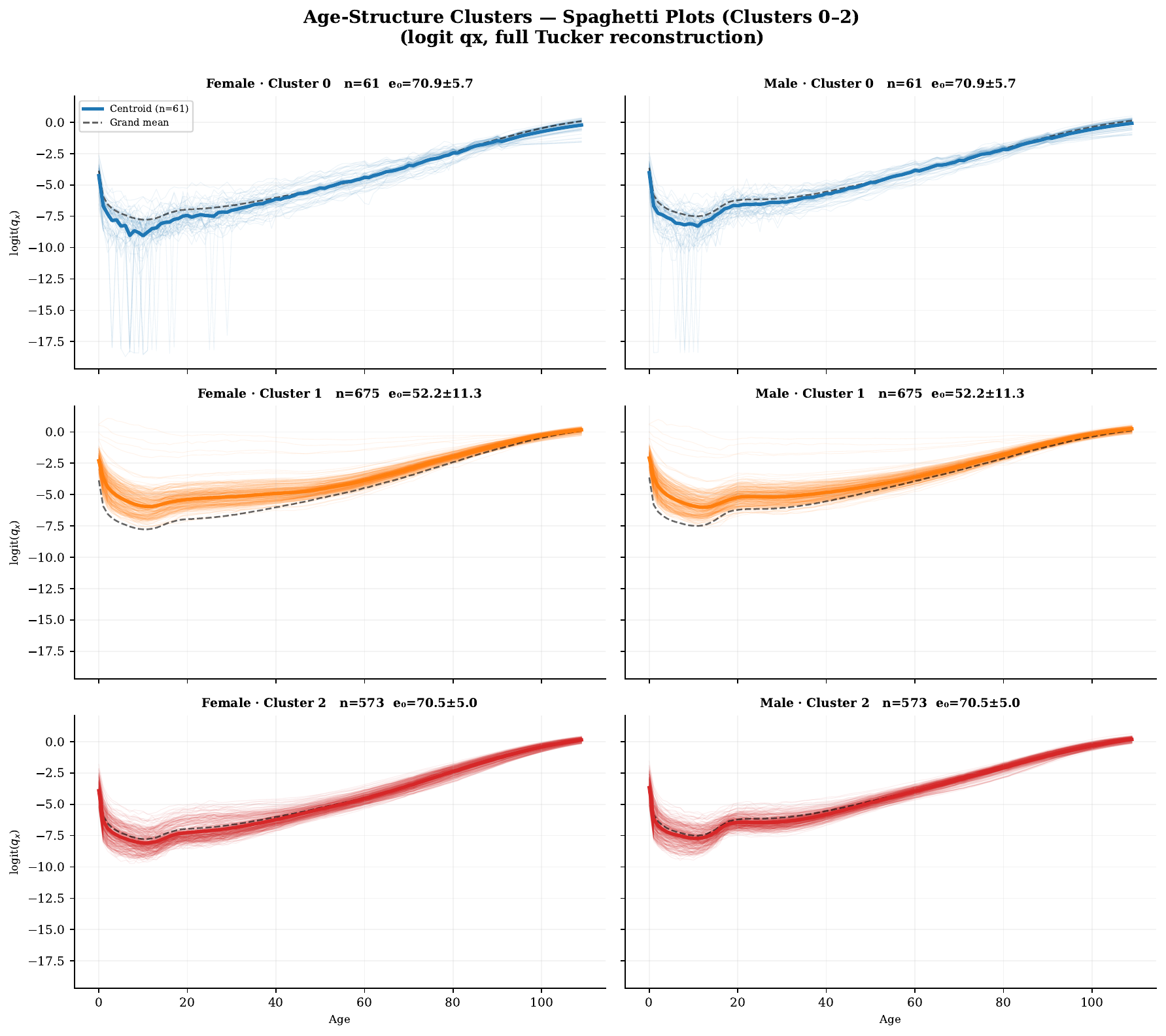}
\caption{Spaghetti plots for age-structure clusters 0--2.  Each faint
line is the full Tucker reconstruction for one country-year member; the
thick line is the cluster centroid; dashed black is the grand mean.
Female (left) and male (right).}
\label{fig:s5_spaghetti_a}
\end{figure}

\begin{figure}[!htbp]
\centering
\includegraphics[width=\textwidth]{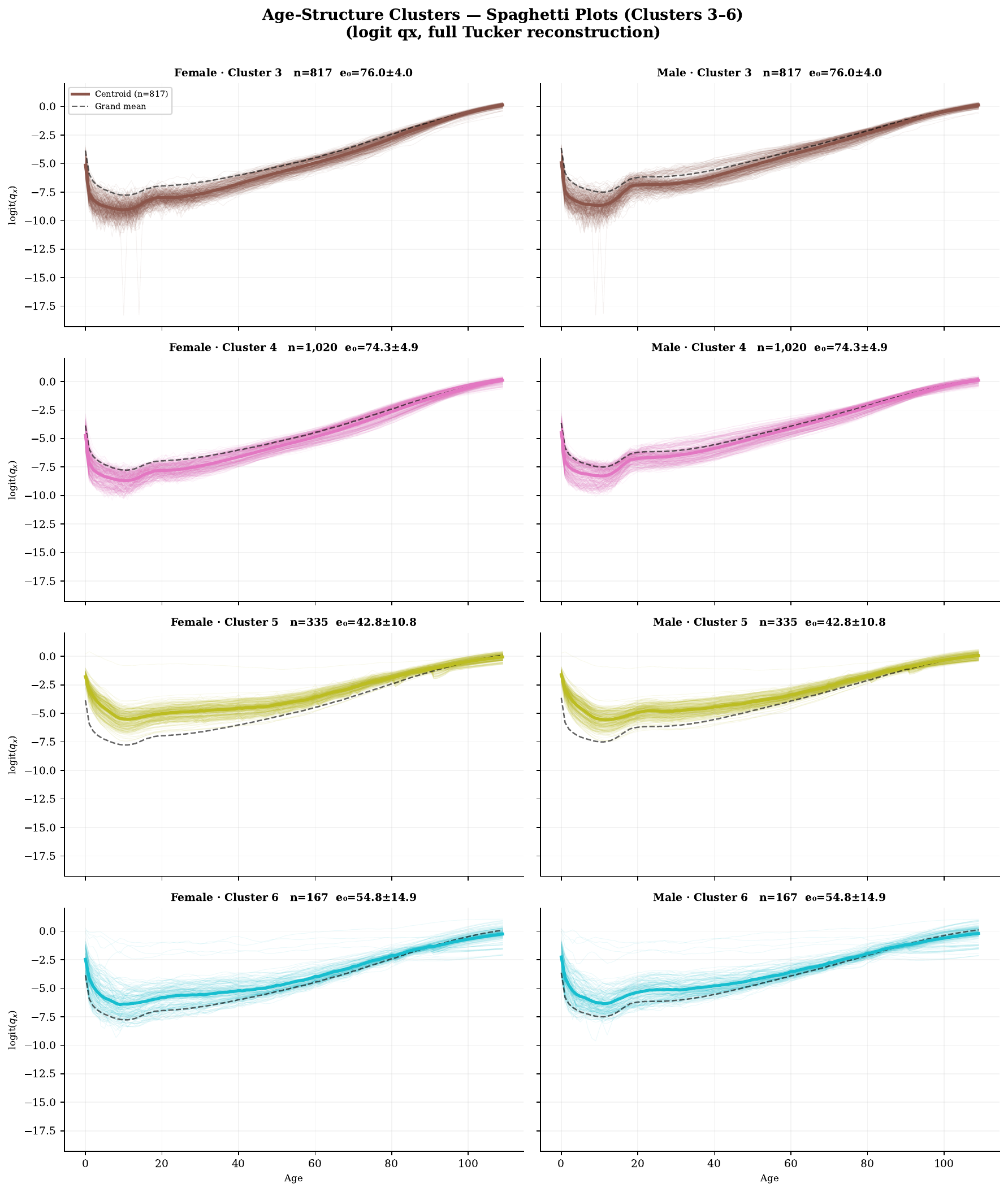}
\caption{Spaghetti plots for age-structure clusters 3--6 (continued).
Same conventions as \cref{fig:s5_spaghetti_a}.}
\label{fig:s5_spaghetti_b}
\end{figure}

\Cref{fig:s5_spaghetti_a,fig:s5_spaghetti_b} display
spaghetti plots of the full logit($\qx$) schedules within each coarse
cluster.  Because the clusters span a wide range of $\ezero$ values, the
within-cluster dispersion in \emph{level} (vertical spread) is large;
however, the \emph{age-pattern shapes} (relative curvature, hump
magnitude, sex-differential pattern) within each cluster are notably
homogeneous, confirming that the level-controlled features successfully
group country-years by schedule shape.

\subsubsection{Level trajectory epochs}

The complementary epoch analysis of \cref{sec:clustering:epochs} was
applied to the grand-mean Tucker loading $G_{c,t}[1,1]$ for all
countries with sufficient temporal coverage.  Rolling-window slopes were
computed across each country's smoothed level trajectory and classified
by the two-stage threshold scheme with $\delta = 0.05$ and
$\delta_{\text{rapid}} = 0.20$ (in $G_{c,t}$ units per year).

\begin{figure}[!htbp]
\centering
\includegraphics[width=\textwidth]{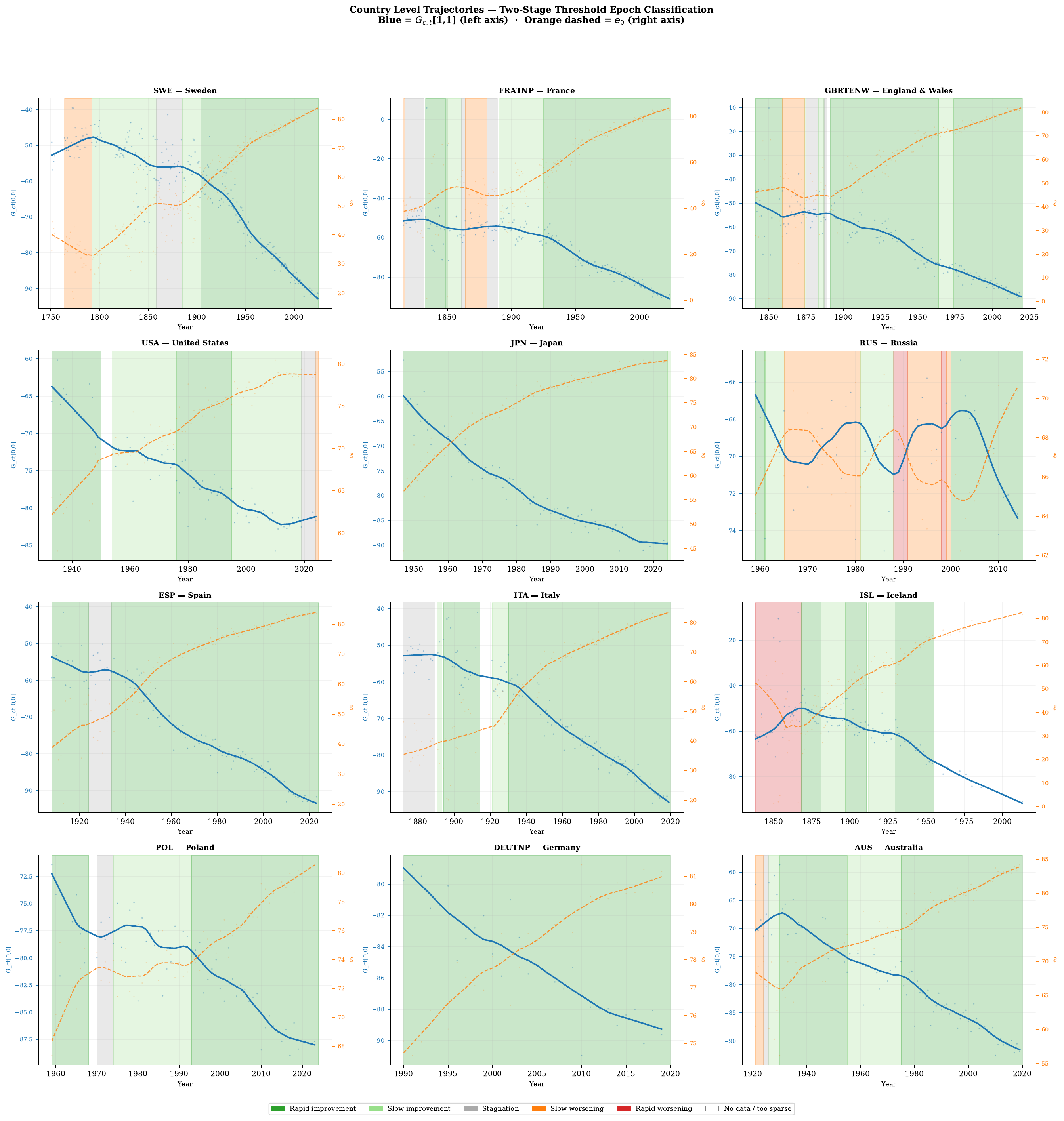}
\caption{$G_{c,t}[1,1]$ (blue, left axis) and $\ezero$ (orange dashed,
right axis) over time for selected countries.  Background shading encodes
epoch category: dark green = rapid improvement; light green = slow
improvement; gray = stagnation; orange = slow worsening; red = rapid
worsening.  White (unshaded) regions indicate periods where observations
are too sparse for the rolling-window classifier -- e.g., Iceland after
the mid-1950s, where multi-year gaps between observations prevent any
15-year window from accumulating enough data points.  The LOWESS curves
extend through these gaps by interpolation, but no epoch is assigned.}
\label{fig:s5_epoch_trajectories}
\end{figure}

\Cref{fig:s5_epoch_trajectories} displays the level trajectory
(blue, left axis) and the corresponding $\ezero$ (orange dashed, right
axis) for selected countries, with background shading indicating the
epoch category at each point in time.  The epochs are historically
recognizable: sustained dark green (rapid improvement) corresponds to
the core of the mortality transition; light green (slow improvement)
marks the post-transition deceleration; gray (stagnation) captures
periods of no net change; and orange or red (slow or rapid worsening)
corresponds to documented mortality crises such as the Soviet-era
reversals in Eastern Europe and the recent US mortality slowdown.

\begin{figure}[!htbp]
\centering
\includegraphics[width=\textwidth]{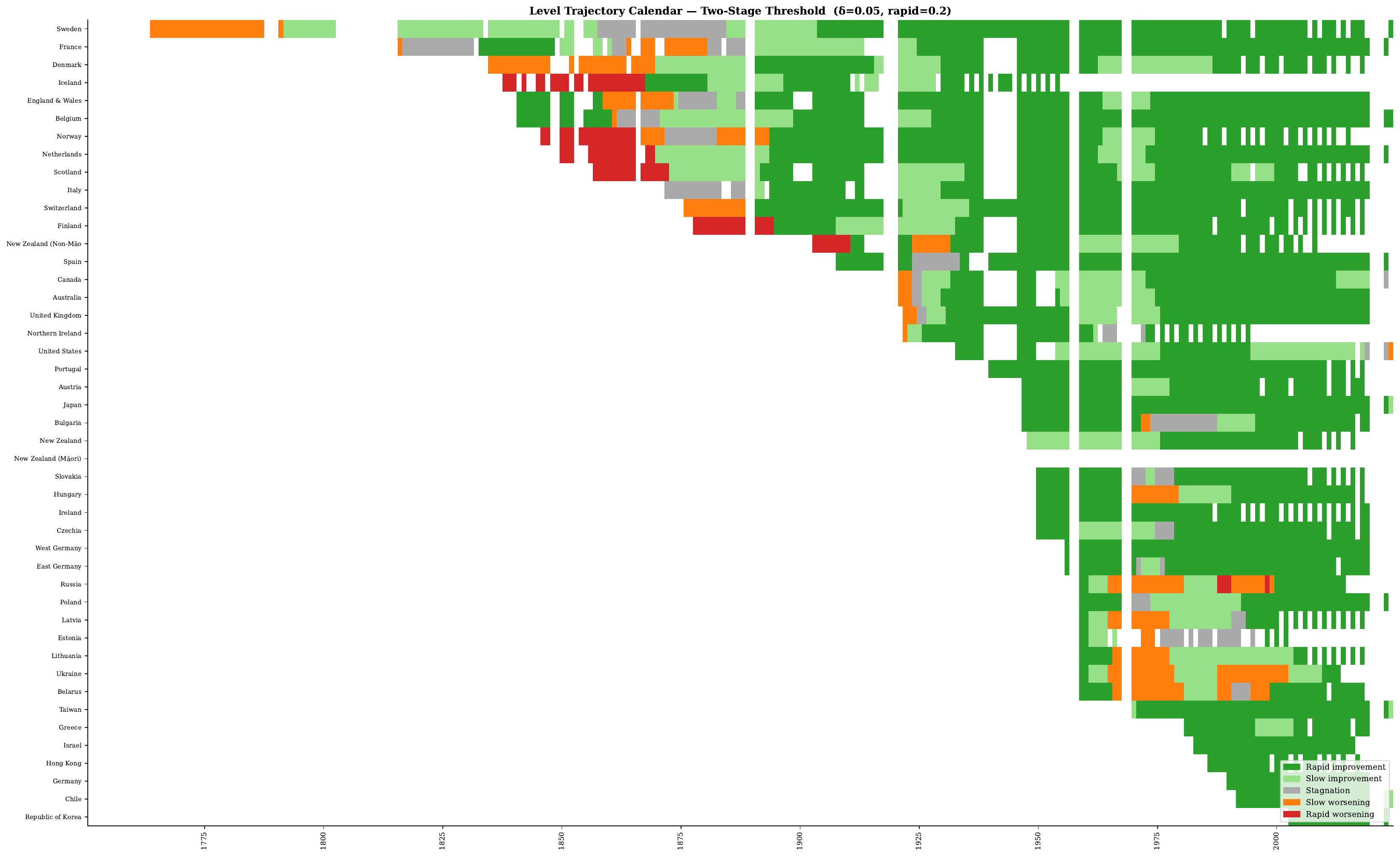}
\caption{Epoch calendar: each cell shows the trajectory category for a
country-year.  Green tones = improving mortality; gray = stagnation;
warm tones = worsening.  Vertical green bands correspond to HMD-wide waves
of mortality improvement; isolated orange/red patches indicate
country-specific crises.}
\label{fig:s5_epoch_calendar}
\end{figure}

\Cref{fig:s5_epoch_calendar} presents the epoch assignments as a
country-by-year calendar, providing a comprehensive visualization of when
each country was in each phase of the mortality transition.

\section{Reconstruction at Arbitrary Mortality Levels}
\label{sec:reconstruction}

The Tucker decomposition and clustering together provide a compact
description of every observed mortality schedule in the HMD.  But the
practical goal of a model life table system is to produce mortality
schedules at \emph{arbitrary} mortality levels -- including levels not
directly observed in any country-year.  We now develop a
reconstruction model that, given a cluster label and a target life
expectancy~${\ezero}^{\ast}$, returns a complete sex-specific age schedule
of mortality.

\subsection{The forward model}
\label{sec:reconstruction:forward}

The path from the Tucker decomposition to a summary measure of mortality
is well-defined and smooth.  For any observed country-year~$(c,t)$, the
decomposition produces a reconstructed logit($\qx$) vector for each sex
(\cref{sec:tucker:reconstruction}).  Concatenating the two sex-specific
vectors yields the feature vector $\bm{z}_{c,t} \in \R^{2A}$ defined in
\cref{eq:cluster_feature}.  Applying the expit transform and the
life-table machinery of \cref{eq:e0} to each sex-specific half of
$\bm{z}_{c,t}$ yields a pair of life expectancies,
${\ezero}^{\mathrm{F}}_{c,t}$ and ${\ezero}^{\mathrm{M}}_{c,t}$.  Denote by
${\ezero}_{c,t}$ a scalar summary -- either the female value, the male
value, or a weighted average of the two.  The \emph{forward model} is the
composite map
\begin{equation}
    \label{eq:forward}
    \bm{z} \;\longmapsto\; \qx = \expit\!\bigl(z_{s,a}\bigr)
    \;\longmapsto\; \ezero(\bm{z}),
    \qquad x = a,
\end{equation}
%
which is continuous and differentiable almost everywhere (the only
non-smoothness arises from the piecewise structure of the life-table
calculation at the final age).  The forward model is straightforward to
evaluate: given any point in the $2A$-dimensional feature space, it
returns a life expectancy.

\subsection{The inverse problem}
\label{sec:reconstruction:inverse}

The reconstruction task is the \emph{inverse} of the forward model:
given a target ${\ezero}^{\ast}$, find a feature vector
$\bm{z}^{\ast} \in \R^{2A}$ such that $\ezero(\bm{z}^{\ast}) =
{\ezero}^{\ast}$.  This is a fundamentally underdetermined problem.  The
forward model maps $\R^{2A}$ to $\R^{1}$ (or $\R^{2}$ if female and
male $\ezero$ are specified separately), and the pre-image of any
target value is a $(2A - 1)$-dimensional manifold.  Infinitely many
mortality schedules produce the same life expectancy: one could achieve
$\ezero = 70$ through low infant mortality and high adult mortality, or
through moderate mortality at all ages, or through many other
configurations.

Any useful reconstruction model must therefore impose additional
structure to select a \emph{single} schedule from the pre-image.  The
clustering of \cref{sec:clustering} provides exactly this structure.

\subsection{The rotation of mortality and the trajectory idea}
\label{sec:reconstruction:trajectories}

The trajectory approach to reconstruction is motivated by a well-documented
empirical regularity: as life expectancy rises, the age pattern of
mortality decline \emph{rotates}.  In the early stages of the mortality
transition, gains are concentrated at young ages -- infant and child
mortality fall rapidly while old-age mortality changes little.  As the
transition progresses, gains at young ages decelerate (because mortality
there is already low) and gains at older ages accelerate
\citep{KannistoEtAl1994, HoriuchiWilmoth1998}.  \citet{LiLeeGerland2013}
formalized this phenomenon in the Lee--Carter framework, showing that the
age pattern of mortality decline ($b_x$ in Lee--Carter notation) is not
fixed but rotates over time, with implications for long-term projections.
\citet{Vekas2019} documented the same rotation across EU countries,
finding that it has been more prevalent among women and in Western Europe.

The rotation means that a mortality schedule at $\ezero = 50$ differs
from one at $\ezero = 75$ not merely in level but in \emph{shape}: the
relative contributions of different age groups to total mortality shift
systematically as overall mortality declines.  The sex differential
evolves in parallel, widening through much of the twentieth century and
narrowing more recently.  These correlated changes in age pattern and
sex differential define a \emph{trajectory} through the
$2A$-dimensional sex-age feature space, parameterized by $\ezero$.

\subsubsection{The HMD-wide trajectory}

The simplest trajectory treats the entire dataset as a single group.
All observed country-year schedules -- regardless of cluster
membership -- contribute to a \emph{HMD-wide trajectory} $\bm{\mu}_0$
that traces the average evolution of the sex-age mortality schedule with
$\ezero$ across the full HMD.  This HMD-wide trajectory captures the
dominant rotation: the shift from infant-dominated to old-age-dominated
mortality decline, the broad evolution of the sex differential, and the
gradual changes in the shape of the Gompertz slope at older ages.  It
provides a useful baseline and a reconstruction tool when no cluster
label is available or when a universal (non-regime-specific) schedule is
desired.

\subsubsection{Within-cluster trajectories}

The HMD-wide trajectory averages over the diversity of age-pattern
families revealed by the clustering (\cref{sec:clustering}).  Within each
cluster~$k$, the observed schedules span a range of $\ezero$ values, and
the way the sex-age pattern evolves with $\ezero$ may differ from the
HMD-wide average.  For example, the rotation may proceed at a different
pace in one cluster than in another; the sex differential may narrow
earlier or later; the accident hump may persist or disappear at different
mortality levels.

Each cluster therefore has its own trajectory $\bm{\mu}_k$: the mean
sex-age schedule at each mortality level within the cluster.  We
formalize this as a smooth function
\begin{equation}
    \label{eq:trajectory}
    \bm{\mu}_k : \R \longrightarrow \R^{2A}, \qquad
    {\ezero}^{\ast} \longmapsto \bm{\mu}_k({\ezero}^{\ast}),
\end{equation}
%
that maps a target life expectancy to a complete sex-age schedule.  The
index $k = 0$ denotes the HMD-wide trajectory (all data); $k = 1, \ldots,
K$ denote the cluster-specific trajectories.  The trajectory
$\bm{\mu}_k$ averages over the country-specific and year-specific
idiosyncrasies within the cluster, retaining only the systematic
dependence of the age-sex pattern on the overall mortality level -- the
rotation as it manifests within that particular mortality regime.

\subsubsection{Estimation}

The trajectory $\bm{\mu}_k$ is estimated from the observed schedules
assigned to cluster~$k$ (or from all observations for $k = 0$).  Let
$\{(\bm{z}_{c,t},\, {\ezero}_{c,t})\}$ be the set of feature vectors
and corresponding life expectancies for the relevant observations.
The trajectory is estimated by smoothing each of the $2A$ components of
$\bm{z}$ as a function of $\ezero$:
\begin{equation}
    \label{eq:trajectory_smooth}
    \hat{\mu}_{k,j}({\ezero}^{\ast})
    = \hat{f}_{k,j}({\ezero}^{\ast}),
    \qquad j = 1, \ldots, 2A,
\end{equation}
%
where $\hat{f}_{k,j}$ is a smooth estimate of the conditional mean of
the $j$-th component of $\bm{z}$ given $\ezero$.  Suitable smoothers
include local polynomial regression
\citep[loess/lowess;][]{Cleveland1979}, smoothing splines
\citep{Wahba1990,GreenSilverman1994}, or penalized B-splines
\citep{EilersMarx1996}.  The choice of smoother is not critical provided
that it captures the curvature of the relationship without overfitting.

Assembling the $2A$ smoothed components yields the estimated trajectory
$\hat{\bm{\mu}}_k({\ezero}^{\ast})$, a smooth curve through the feature
space that traces how the sex-age mortality schedule evolves with
$\ezero$ within cluster~$k$.

\subsubsection{Neural trajectory model}
\label{sec:reconstruction:neural_trajectory}

The LOWESS-based trajectory estimation has three limitations.  First,
\emph{hard cluster boundaries}: a country transitioning between clusters
experiences an abrupt change in the predicted schedule, because each
cluster's trajectory is estimated independently.  Second, \emph{no
extrapolation}: querying an $\ezero$ value outside a cluster's observed
range produces edge artifacts, because the smoother has no data to anchor
the prediction.  Third, \emph{no cross-cluster information sharing}: a
cluster with sparse observations at low $\ezero$ cannot borrow strength
from other clusters that have dense coverage at similar mortality levels.

A neural trajectory model addresses all three limitations by replacing
the per-cluster LOWESS machinery with a single learned function that maps
a cluster identity and a target life expectancy jointly to a complete
sex-age mortality schedule.  The remainder of this section describes the
construction in detail, beginning with how the inputs are encoded and
proceeding through the network architecture, the training procedure, and
the properties of the trained model.

\paragraph{Encoding the inputs.}
A neural network operates on vectors of real numbers.  To supply it with
a cluster label (a categorical variable) and a life expectancy value (a
scalar), we must first convert these into numerical vectors -- a process
called \emph{encoding} or \emph{featurization}.  The design of these
encodings has a substantial effect on what the network can learn
efficiently.

\paragraph{Cluster embedding.}
Each cluster~$k$ has a centroid -- the mean reconstructed sex-age
schedule across all observations assigned to that cluster.  These
centroids are $2A$-dimensional vectors that encode the distinctive age
pattern of each cluster.  We apply principal component analysis (PCA) to
the $K$ cluster centroids, retaining the first $d_k = \min(8, K)$
components.  The $k$th cluster is then represented by its scores on
these components:
\begin{equation}
    \bm{e}_k \;=\; \operatorname{PCA}(\text{centroid}_k) \;\in\; \R^{d_k}.
    \label{eq:cluster_embedding}
\end{equation}
This \emph{cluster embedding} is a compact numerical vector that
summarizes the distinctive age-pattern shape of cluster~$k$.  Crucially,
because PCA preserves distances, clusters with similar age patterns
receive similar embedding vectors.  This means the network can generalize
across clusters -- unlike a one-hot encoding (a vector of length~$K$ with
a single~1 and the rest~0s), which treats every cluster as equally
different from every other.

\paragraph{Life expectancy features.}
The target life expectancy $\ezero^{\ast}$ is a single scalar.  Rather
than pass it directly to the network, we expand it into a 7-dimensional
feature vector $\bm{\psi}(\ezero^{\ast})$ that includes polynomial and
trigonometric terms:
\begin{equation}
    \label{eq:e0_features}
    \bm{\psi}({\ezero}^{\ast}) = \bigl[
        \tilde{e},\; \tilde{e}^2,\; \tilde{e}^3,\;
        \sin(\pi\tilde{e}),\; \cos(\pi\tilde{e}),\;
        \sin(2\pi\tilde{e}),\; \cos(2\pi\tilde{e})
    \bigr]^\top \in \R^{7},
\end{equation}
%
where $\tilde{e} = ({\ezero}^{\ast} - {\ezero}_{\min}) /
({\ezero}_{\max} - {\ezero}_{\min})$ normalizes the target life
expectancy to $[0,1]$ using the observed range across all clusters.
A polynomial basis ($\tilde{e}, \tilde{e}^2, \tilde{e}^3$) lets the
network represent smooth trends, curvature, and inflection points.  The
trigonometric terms provide periodic basis functions that help capture
the nonlinear rotation of the age pattern over the $\ezero$ range
-- the shift from infant-dominated to old-age-dominated mortality decline.
Together, these seven features give the network a head start in
representing the complex relationship between life expectancy and age
pattern, so it can focus its learning capacity on the residual details.
This is analogous to using polynomial regression with a rich set of basis
functions, except that the neural network learns a flexible nonlinear
transformation of them rather than a fixed linear combination.

The full input to the network is the concatenation of the cluster
embedding and the $\ezero$ features:
$\bm{x} = [\bm{e}_k,\; \bm{\psi}(\ezero^{\ast})] \in \R^{d_k + 7}$.
With $d_k = 8$ and 7 life expectancy features, this is a 15-dimensional
input vector.

\paragraph{Network architecture.}
The prediction function is a \emph{multi-layer perceptron} (MLP) -- the
simplest widely used neural network architecture -- consisting of a
sequence of \emph{layers}, each of which performs a linear transformation
followed by a nonlinear activation function.  Given an input
$\bm{x} \in \R^{d_{\text{in}}}$, the first layer computes
\begin{equation}
    \bm{h}_1 = \phi\!\bigl(W_1 \bm{x} + \bm{b}_1\bigr),
    \label{eq:nn_layer}
\end{equation}
where $W_1 \in \R^{d_1 \times d_{\text{in}}}$ is a matrix of
\emph{weights}, $\bm{b}_1 \in \R^{d_1}$ is a vector of \emph{biases},
and $\phi(\cdot)$ is a nonlinear function applied element-wise.  The
weights and biases are the parameters that the network learns from data.
The output $\bm{h}_1 \in \R^{d_1}$ is called the \emph{hidden
representation} or \emph{activation} of the first layer.

The process is repeated: the second layer takes $\bm{h}_1$ as input
and produces $\bm{h}_2 = \phi(W_2 \bm{h}_1 + \bm{b}_2)$, and so on.
The final (output) layer omits the nonlinearity and produces:
\begin{equation}
    \label{eq:neural_trajectory}
    \hat{\bm{z}}_{\mathrm{NN}}({\ezero}^{\ast}, k)
    = W_L \bm{h}_{L-1} + \bm{b}_L \;\in\; \R^{2A},
\end{equation}
where $L$ is the number of layers.

Each layer can be understood as two operations: (1)~a linear
transformation that mixes and rescales the features from the previous
layer, and (2)~a nonlinear activation that introduces the capacity to
represent nonlinear relationships.  Without the nonlinearity, stacking
multiple layers would reduce to a single linear transformation (because a
product of matrices is still a matrix), so the nonlinearity is essential.

The activation function used here is the \emph{rectified linear unit}
(ReLU): $\phi(x) = \max(0, x)$.  This simply sets all negative values to
zero and leaves positive values unchanged.  Despite its simplicity, ReLU
has proven highly effective in practice: it is computationally cheap and
avoids the vanishing-gradient problems that affect other activation
functions like the sigmoid or hyperbolic tangent
\citep{GlorotBengioAISTATS2011}.

The specific architecture is: input (15) $\to$ hidden layer~1 (256 units,
ReLU) $\to$ hidden layer~2 (128 units, ReLU) $\to$ output (220 units,
linear).  The total number of learnable parameters is
$15 \times 256 + 256 + 256 \times 128 + 128 + 128 \times 220 + 220
= 65{,}372$.
With ${\sim}4{,}000$ training observations, each of dimension~220, the
effective number of scalar training targets is
$4{,}000 \times 220 \approx 880{,}000$, which comfortably exceeds the
parameter count -- though the targets are not independent (mortality
schedules are smooth functions of age), so regularization is still
important.

\paragraph{Training.}
The weights and biases are learned by minimizing a loss function that
measures how well the predictions match the observed data:
\begin{equation}
    \label{eq:neural_traj_loss}
    \mathcal{L} = \frac{1}{N}
    \sum_{(c,t) \in \mathcal{O}}
    \lVert \hat{\bm{z}}_{\mathrm{NN}}({\ezero}_{c,t},\, k_{c,t})
           - \bm{z}_{c,t} \rVert^2
    + \lambda_{\mathrm{reg}} \sum_{\ell=1}^{L}
    \lVert W_\ell \rVert_F^2\,.
\end{equation}
%
The first term is the mean squared error between predicted and observed
$\logit(\qx)$ schedules across all $N$ training observations.  The
second term is \emph{weight decay} ($L_2$ regularization): it penalizes
large weight values, discouraging overfitting by keeping the learned
function smooth.  The Frobenius norm
$\lVert W_\ell \rVert_F^2 = \sum_{ij} (W_\ell)_{ij}^2$ is the sum of
squared entries of each weight matrix, and
$\lambda_{\mathrm{reg}} = 10^{-4}$ controls the penalty strength.

Minimization proceeds by \emph{gradient descent}: starting from random
initial weights, the algorithm repeatedly computes the gradient
$\partial \mathcal{L} / \partial \theta$ of the loss with respect to all
parameters $\theta = \{W_1, \bm{b}_1, \ldots, W_L, \bm{b}_L\}$ and
updates each parameter in the direction that reduces the loss:
$\theta \leftarrow \theta - \eta \,
\partial \mathcal{L} / \partial \theta$,
where $\eta$ is the learning rate (step size).
The gradient is computed by \emph{backpropagation}
\citep{RumelhartHintonWilliams1986} -- an efficient application of the
chain rule that works backward through the network, composing the local
derivatives at each layer.  The cost is approximately twice that of a
forward pass.

In practice, the gradient is estimated from a random subset (a
\emph{mini-batch}) of the training data rather than the full dataset.
This stochastic gradient descent introduces noise into the gradient
estimate but converges much faster and provides an implicit
regularization effect.  The specific optimizer is Adam
\citep{KingmaBa2015}, which maintains per-parameter adaptive learning
rates based on running estimates of the first and second moments of the
gradient.  Weights are initialized using the He scheme
\citep{HeZhangRenSun2015}, which scales initial random values by
$\sqrt{2/d_{\text{in}}}$ to prevent the variance of activations from
exploding or vanishing as signals propagate through the network.
Training uses a learning rate of $10^{-3}$, mini-batch size~512, and runs
for 500~epochs (complete passes through the training data).

\paragraph{Properties of the trained network.}
Because the cluster identity enters the network through a continuous
embedding vector $\bm{e}_k$ rather than a discrete label, the trained
network supports smooth \emph{inter-cluster interpolation}: querying a
point between two cluster embeddings produces a blended schedule that
transitions smoothly between the two regimes.
A country that is gradually shifting from one mortality regime to another
receives a schedule that interpolates between the relevant cluster
trajectories, weighted by proximity in embedding space, rather than
jumping abruptly from one to the other.

The network also supports modest \emph{extrapolation} beyond each
cluster's observed $\ezero$ range.  If cluster~$k$ has no observations at
$\ezero = 90$ but cluster~$j$ does, the network can borrow the
relationship between $\ezero$ and schedule shape learned from cluster~$j$,
adapted through the cluster embedding to reflect cluster~$k$'s
distinctive age pattern.  The extrapolated schedules degrade gracefully --
maintaining demographic plausibility (monotonic $\qx$ at older ages,
reasonable sex differentials) rather than producing artifacts.

Finally, because the network maps $\ezero^{\ast}$ directly to
$\hat{\bm{z}}$, applying the forward model to the network output and
checking whether it returns the target $\ezero$ measures the
self-consistency of the learned mapping.  In practice the discrepancy is
small, and the Brent's method refinement of
\cref{sec:reconstruction:trajectory_eval} becomes optional.

The neural trajectory does not replace the LOWESS trajectories but
complements them: the LOWESS approach is interpretable and requires no
hyperparameter tuning, while the neural model offers smoothness across
cluster boundaries and safe extrapolation at the cost of an opaque
mapping.  Both are computed and compared below
(\cref{sec:results:reconstruction,sec:results:neural_trajectory}).

An obvious question is whether the neural trajectory could serve as the
\emph{sole} trajectory engine, eliminating the LOWESS machinery entirely.
We investigated this by training a single MLP to map cluster identity and
$\ezero$ directly to complete schedules, with the LOWESS grids replaced
by NN-evaluated grids in the same cache format so that all downstream
code (the fitter, the forecasting module, and the visualization pipeline)
would work without modification.  The results were mixed: with a
PCA-based cluster embedding, the network smoothed away genuine
between-cluster differences because similar embeddings forced similar
outputs; with a one-hot cluster encoding and an aggressive architecture,
the network overfit to individual country-year scatter and produced noisy
age-rotation profiles.  Intermediate configurations -- training-time
noise injection on the $\ezero$ features to enforce smoothness, stronger
weight decay -- improved matters but introduced a systematic bias in the
HMD-wide trajectory that proved difficult to eliminate without extensive
hyperparameter search.  We conclude that the unified NN trajectory is a
promising direction for future work but requires careful calibration that
is beyond the scope of this paper.  The dual LOWESS + neural architecture
is therefore retained: LOWESS provides the primary reconstruction
pathway on which all cross-validation results are based, and the neural
model provides the complementary capabilities described above.

\subsubsection{Reconstruction as trajectory evaluation}
\label{sec:reconstruction:trajectory_eval}

Given a cluster label $k$ and a target ${\ezero}^{\ast}$, the
reconstructed mortality schedule is simply the trajectory evaluated at
the target:
\begin{equation}
    \label{eq:reconstruct_trajectory}
    \bm{z}^{\ast} = \hat{\bm{\mu}}_k({\ezero}^{\ast}).
\end{equation}
%
The logit($\qx$) schedule for each sex is read off from the appropriate
half of $\bm{z}^{\ast}$, and the $\qx$ schedule is obtained by applying
the expit transform.  This produces a complete life table for each sex.

The reconstructed $\ezero$ will not in general equal the target
${\ezero}^{\ast}$ exactly, because the trajectory is estimated in
logit($\qx$) space (where the smoothing is performed) rather than in
$\ezero$ space directly.  In practice, the discrepancy is small because
$\ezero$ is a smooth function of the logit($\qx$) schedule.  If exact
agreement is required, a one-dimensional root-finding step (e.g.,
bisection or Brent's method; \citealp{Brent1973}) can adjust the evaluation point along the
trajectory until the forward model (\cref{eq:forward}) returns the
desired $\ezero$.

\subsection{Proactively defined clusters for exceptional years}
\label{sec:reconstruction:exceptional_clusters}

The data-driven clusters of \cref{sec:clustering} partition the
non-exceptional country-years into groups that share a common mortality
regime.  The exceptional country-years excluded from the decomposition
(\cref{sec:preprocessing:exceptional}) are not assigned to these clusters
by the GMM.  Instead, they are grouped into three
\emph{proactively defined clusters} based on their disruption type:
armed conflict, respiratory pandemic, and enteric pandemic.

These clusters differ fundamentally from the data-driven clusters.  The
data-driven clusters group country-years that are similar in the
\emph{shape and level} of their baseline mortality; they emerge from
the structure of the data.  The proactively defined clusters group
country-years that share a common \emph{type of exceptional exposure},
regardless of their baseline mortality.  A war cluster, for instance,
may include observations from countries with very different baseline
mortality levels -- eighteenth-century Sweden and twentieth-century
France -- united only by the fact that both experienced armed conflict
in the year in question.

The proactively defined clusters serve a different purpose from the
data-driven clusters.  They are not used for trajectory-based
reconstruction of baseline schedules; rather, they provide the
observations from which the disruption profiles of
\cref{sec:exceptional} are estimated.  Each exceptional country-year
contributes a residual (the difference between its observed mortality
and its projected baseline), and these residuals are averaged within each
proactively defined cluster to estimate the canonical disruption profile
for each type.

\subsection{Properties of the trajectory-based reconstruction}
\label{sec:reconstruction:properties}

The trajectory approach has several desirable properties.

\subsubsection{Sex differential preservation}

Because the trajectory is estimated jointly over all $2A$ components of
the feature vector -- which includes both the female and male age
schedules -- the sex differential at any point along the trajectory is the
empirically observed differential at that mortality level within the
cluster.  The differential is not modeled separately or imposed
externally; it emerges from the data as part of the trajectory.  As the
trajectory moves from high to low $\ezero$, the sex differential
evolves in whatever way the data within the cluster dictate: widening,
narrowing, or remaining stable.

\subsubsection{Age-pattern evolution}

Similarly, the shape of the age schedule is not held fixed as $\ezero$
changes.  The trajectory captures the subtle shifts in age structure
that accompany the mortality transition within each regime: the relative
pace of infant vs.\ adult mortality decline, the emergence or
disappearance of the accident hump, changes in the curvature of old-age
mortality.  These features are encoded in the trajectory automatically,
without requiring parametric assumptions about how they depend on
$\ezero$.

\subsubsection{Cluster-specific structure}

Different clusters may have different trajectories even at the same
$\ezero$.  Two clusters, both at $\ezero = 70$, may differ in the shape
of the age curve, the magnitude of the sex differential, and the rate at
which mortality declines with further improvements.  The trajectory
approach preserves these regime-specific features because each cluster
has its own estimated $\hat{\bm{\mu}}_k$.

\subsection{Interpolation, extrapolation, and limitations}
\label{sec:reconstruction:limits}

The trajectory $\hat{\bm{\mu}}_k$ is well-supported over the range of
$\ezero$ values observed within cluster~$k$.  Within this range,
reconstruction amounts to interpolation along a well-estimated curve,
and the results are expected to be reliable.

Extrapolation beyond the observed range is possible but should be treated
with caution.  At the low-$\ezero$ end, the cluster may not contain
enough observations at very high mortality levels to constrain the
trajectory.  At the high-$\ezero$ end, extrapolation projects into
mortality regimes not yet observed -- a task that requires assumptions
about how current trends will continue.  We recommend reporting the
range of $\ezero$ values supported by each cluster and flagging
reconstructions that fall outside this range.

That said, modest extensions of the trajectory beyond the observed
$\ezero$ range are often desirable in practice, as the boundary of
observed data in a cluster is an artifact of sample coverage rather than
a hard demographic limit.  We will explore methods for extending each
cluster's mean trajectory slightly beyond its observed range under
suitably conservative assumptions -- for instance, requiring that the
extrapolated trajectory preserve the direction and curvature of the
trajectory near the boundary, that the age-pattern changes implied by
the extension remain monotone in $\ezero$, and that the resulting
schedules satisfy basic demographic constraints (non-negative $\qx$,
plausible sex differentials).  The goal is to widen the usable domain of
each cluster modestly without venturing into speculative territory.

Between-cluster reconstruction -- producing a schedule at an $\ezero$
that falls outside all clusters, or at a level where two clusters
overlap but with different age patterns -- is not addressed by the
single-trajectory model.  Possible extensions include weighted averaging
of trajectories from adjacent clusters, or fitting a single HMD-wide
trajectory with cluster-specific offsets.  These extensions are left to
future work.



We now report the results of estimating the trajectory-based
reconstruction model on the HMD.

\subsection{Reconstruction at arbitrary mortality levels}
\label{sec:results:reconstruction}

Following \cref{sec:reconstruction}, the trajectory-based reconstruction
model was estimated for both the HMD-wide dataset and within each coarse
cluster.  The LOWESS smoother (\cref{eq:trajectory_smooth}) was applied
independently to each of the $2A = 220$ components of the feature
vector $\bm{z}$ as a function of $\ezero$, producing smooth trajectories
$\hat{\bm{\mu}}_k(\ezero^{\ast})$ through the sex-age feature space.

\subsubsection{Trajectories and reconstruction accuracy}

\begin{figure}[!htbp]
\centering
\includegraphics[width=\textwidth]{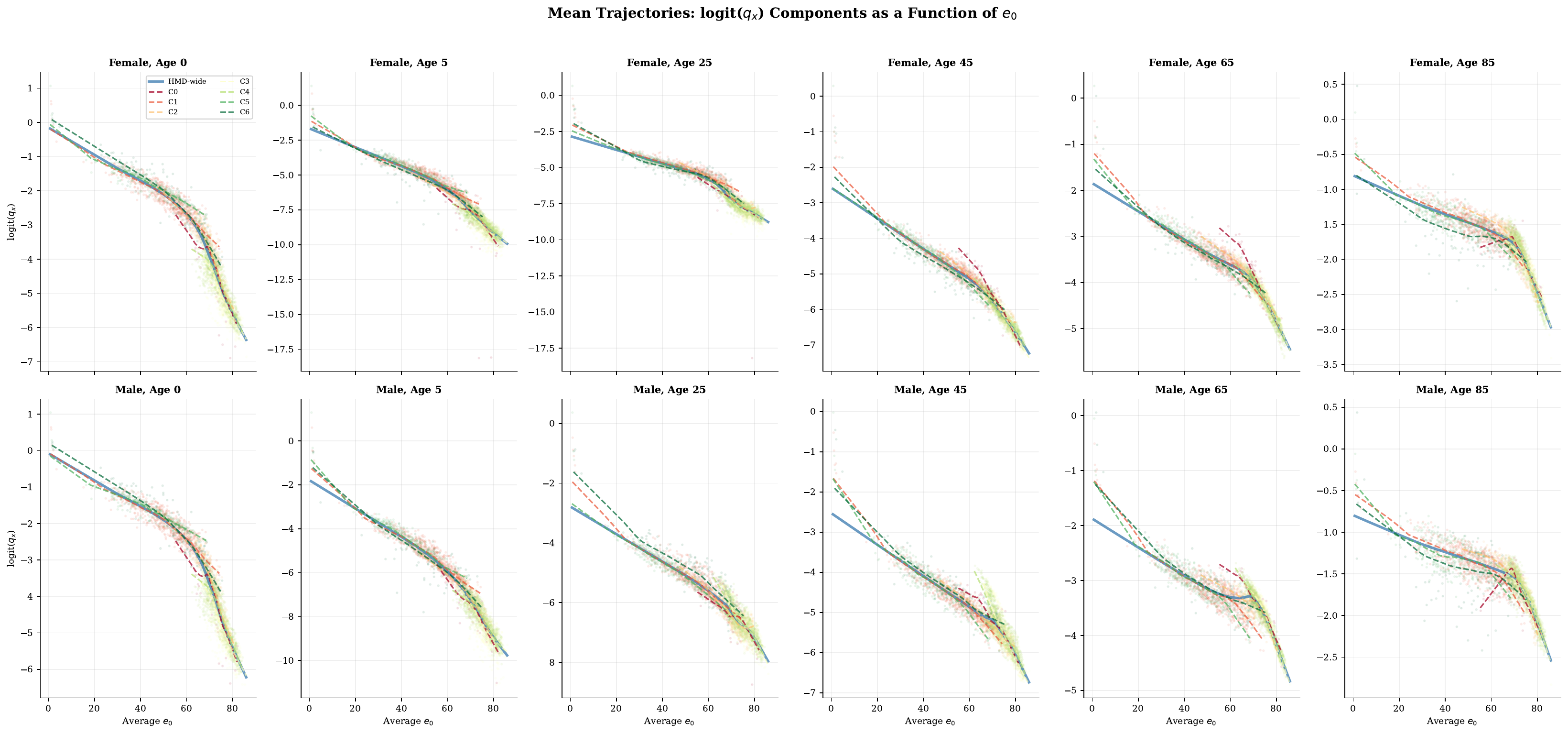}
\caption{Smoothed mean trajectories overlaid on the scatter of observed
country-year data.  Blue = HMD-wide trajectory; colored lines = coarse
cluster trajectories.  The trajectories capture the systematic
dependence of the age-sex pattern on overall mortality level.}
\label{fig:s6_trajectory_scatter}
\end{figure}

\Cref{fig:s6_trajectory_scatter} shows the smoothed mean trajectories
overlaid on the scatter of observed country-year data.  Each panel
displays one component of the feature vector against average $\ezero$:
the blue line is the HMD-wide trajectory, and the colored lines are the
coarse cluster trajectories.  The trajectories capture the systematic
dependence of the age-sex pattern on overall mortality level -- the
rotation of mortality described in
\cref{sec:reconstruction:trajectories} -- with regime-specific
deviations visible where the cluster trajectories diverge from the HMD-wide
trend.

\begin{figure}[!htbp]
\centering
\includegraphics[width=\textwidth]{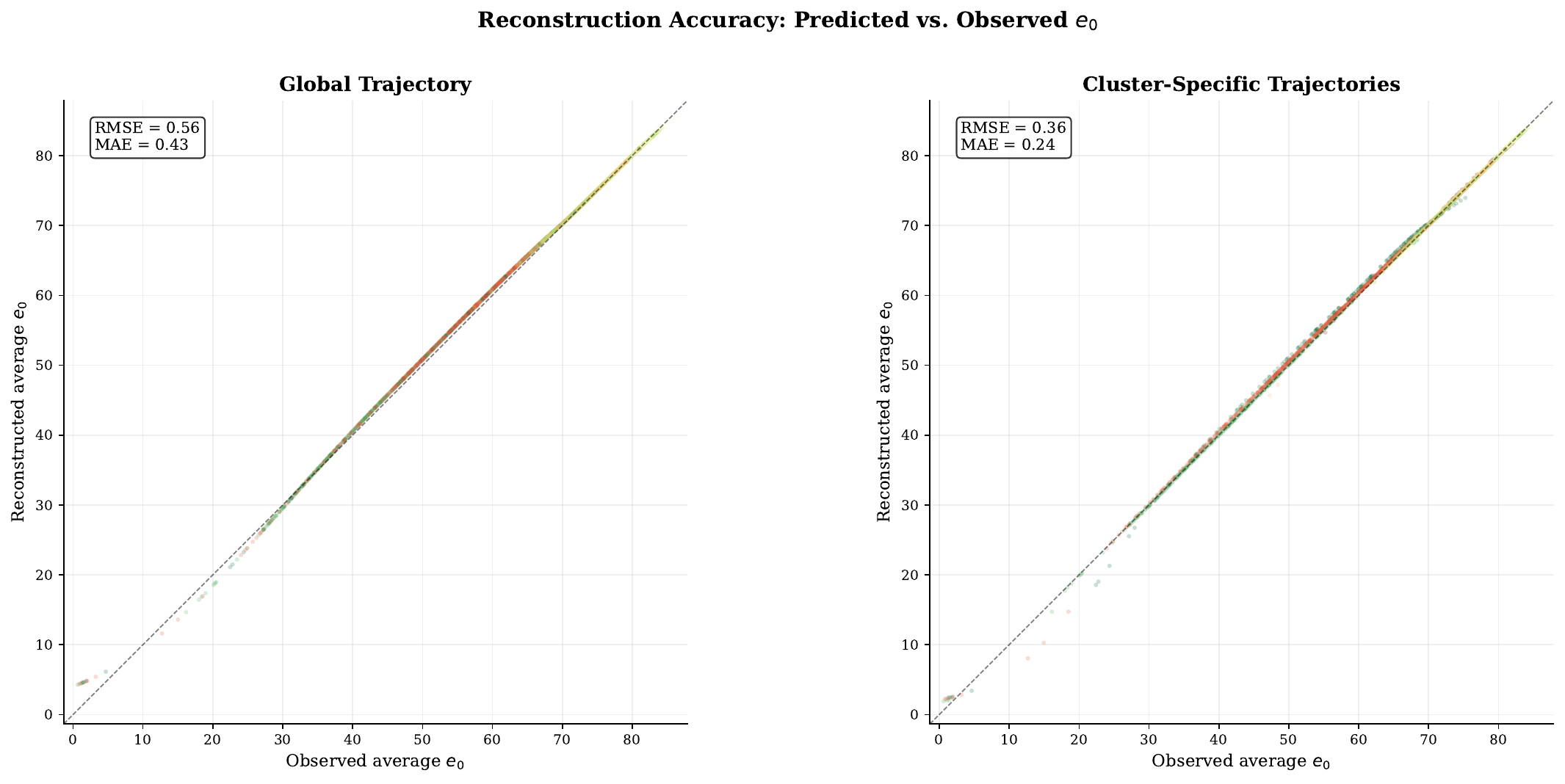}
\caption{Reconstruction accuracy: forward-model $\ezero$ of the
reconstructed schedule vs.\ observed $\ezero$.  Left: HMD-wide
trajectory.  Right: within-cluster (coarse) trajectories.  The
diagonal line is perfect agreement.  Points are colored by coarse
cluster.}
\label{fig:s6_recon_accuracy}
\end{figure}

\Cref{fig:s6_recon_accuracy} evaluates reconstruction accuracy by comparing
the forward-model $\ezero$ of the reconstructed schedule
(\cref{eq:forward}) against the observed $\ezero$, for both the HMD-wide
and within-cluster trajectories.  Points lie close to the diagonal,
indicating good self-consistency between the trajectory evaluation and
the target $\ezero$.  The within-cluster trajectories produce tighter
agreement than the HMD-wide trajectory, particularly at the extremes of
the $\ezero$ range, confirming the value of regime-specific
reconstruction.

\Cref{tab:s6_recon_verification} confirms that the reconstruction
produces valid $\ezero$ values at each coarse cluster centroid, and
\cref{tab:s6_recon_accuracy} quantifies the per-cluster reconstruction
error (bias, MAE, RMSE) between the forward-model $\ezero$ and the
observed $\ezero$.

{\setlength\LTleft{0pt}\setlength\LTright{0pt}
\begin{longtable}{@{\extracolsep{\fill}}crrrrc@{}}
\caption{Reconstruction at coarse cluster centroid $e_0$ values} \label{tab:s6_recon_verification} \\
\toprule
Cluster & Target $e_0$ & Recon $e_0$(F) & Recon $e_0$(M) & Avg & $\Delta$ \\
\midrule
\endfirsthead
\caption[]{Reconstruction at coarse cluster centroid $e_0$ values (continued)} \\
\toprule
Cluster & Target $e_0$ & Recon $e_0$(F) & Recon $e_0$(M) & Avg & $\Delta$ \\
\midrule
\endhead
\midrule
\multicolumn{6}{r}{\textit{Continued on next page}} \\
\endfoot
\bottomrule
\endlastfoot
0 & 72.110 & 74.820 & 69.400 & 72.110 & -0.000 \\
1 & 53.620 & 55.140 & 52.110 & 53.620 & -0.000 \\
2 & 71.360 & 74.330 & 68.380 & 71.360 & -0.000 \\
3 & 76.420 & 79.500 & 73.330 & 76.420 & 0.000 \\
4 & 74.890 & 78.000 & 71.790 & 74.890 & -0.000 \\
5 & 43.360 & 44.610 & 42.100 & 43.360 & -0.000 \\
6 & 57.640 & 60.260 & 55.020 & 57.640 & -0.000 \\
\end{longtable}}

{\setlength\LTleft{0pt}\setlength\LTright{0pt}
\begin{longtable}{@{\extracolsep{\fill}}lrrrrr@{}}
\caption{Reconstruction accuracy: $e_0$ error (observed $-$ reconstructed)} \label{tab:s6_recon_accuracy} \\
\toprule
Method & Mean & SD & MAE & Max $|\Delta|$ & RMSE \\
\midrule
\endfirsthead
\caption[]{Reconstruction accuracy: $e_0$ error (observed $-$ reconstructed) (continued)} \\
\toprule
Method & Mean & SD & MAE & Max $|\Delta|$ & RMSE \\
\midrule
\endhead
\midrule
\multicolumn{6}{r}{\textit{Continued on next page}} \\
\endfoot
\bottomrule
\endlastfoot
Global & 0.400 & 0.399 & 0.431 & 3.697 & 0.565 \\
Cluster-specific & 0.214 & 0.292 & 0.241 & 4.643 & 0.362 \\
\end{longtable}}

\subsubsection{Example reconstructed schedules}

\begin{figure}[!htbp]
\centering
\includegraphics[width=\textwidth]{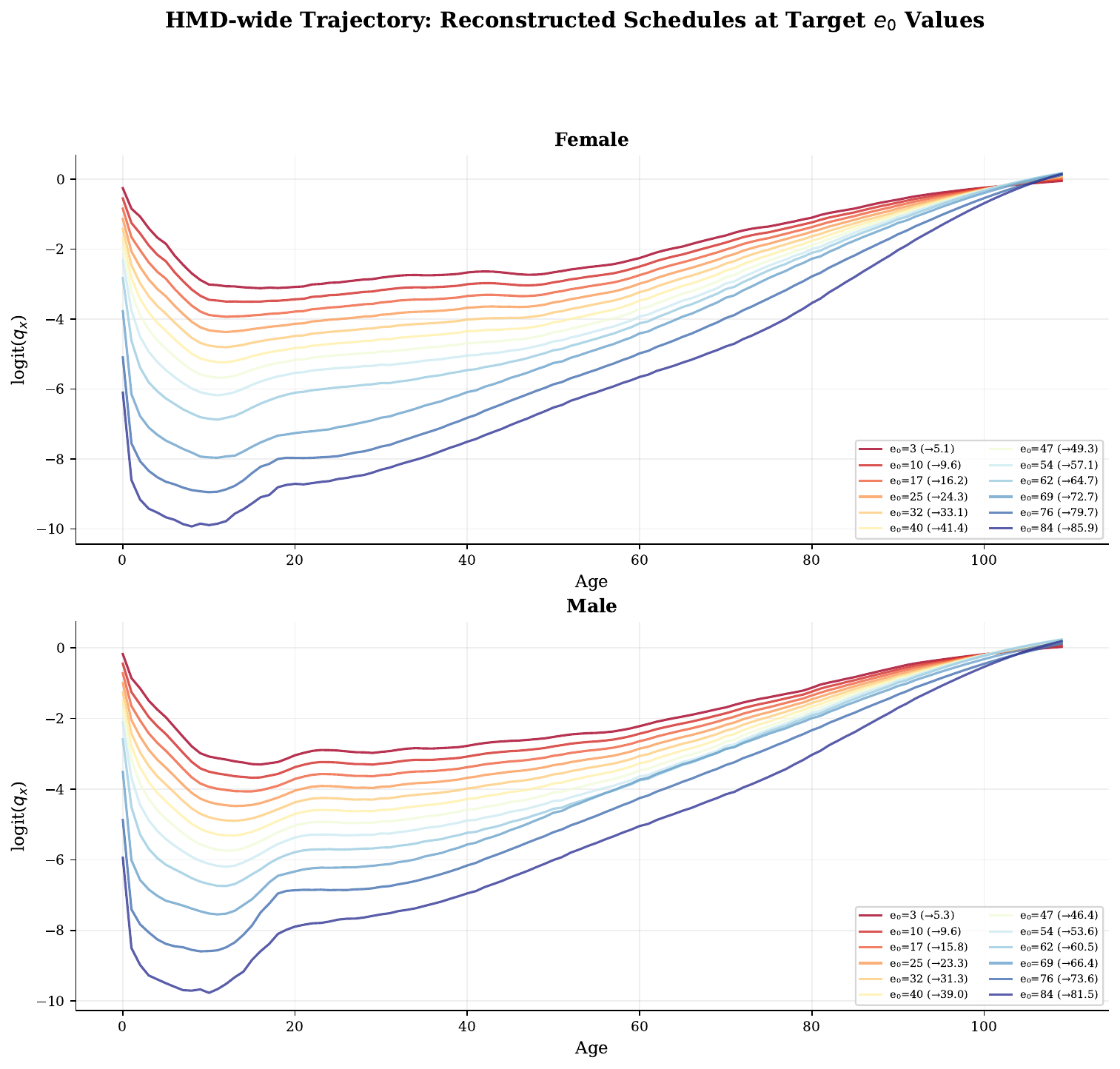}
\caption{Reconstructed mortality schedules at selected target $\ezero$
values from the HMD-wide trajectory on the logit($\qx$) scale.  Colors
progress from warm (low $\ezero$) to cool (high $\ezero$).  The rotation
of the age pattern is clearly visible.}
\label{fig:s6_example_schedules}
\end{figure}

\Cref{fig:s6_example_schedules} displays reconstructed mortality schedules
at selected target $\ezero$ values from the HMD-wide trajectory on the
logit($\qx$) scale.  Colors progress from warm (low $\ezero$) to cool
(high $\ezero$).  The rotation of the age pattern
(\cref{sec:reconstruction:trajectories}) is clearly visible: at low
$\ezero$, infant and child mortality are high and dominate the overall
mortality level; at high $\ezero$, mortality gains are concentrated at
older ages, and the infant--childhood transition is compressed near zero.

\subsubsection{Sex differential and age-pattern evolution}

\begin{figure}[!htbp]
\centering
\includegraphics[width=\textwidth]{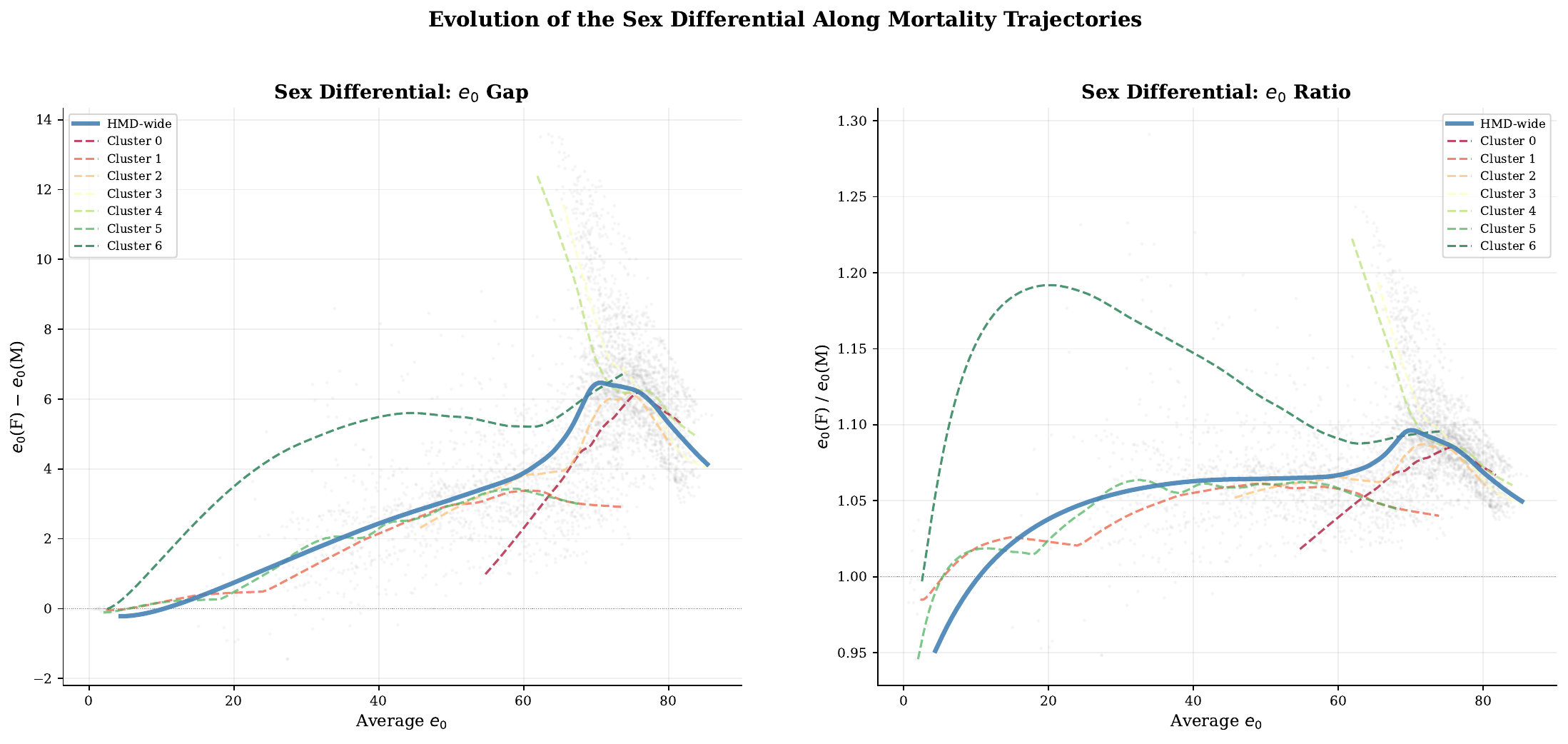}
\caption{Evolution of the sex differential along the HMD-wide and
cluster-specific trajectories.  Left: $\ezero$ gap (female minus male).
Right: ratio of female to male $\ezero$.}
\label{fig:s1_sex_diff_trajectory}
\end{figure}

\begin{figure}[!htbp]
\centering
\includegraphics[width=\textwidth]{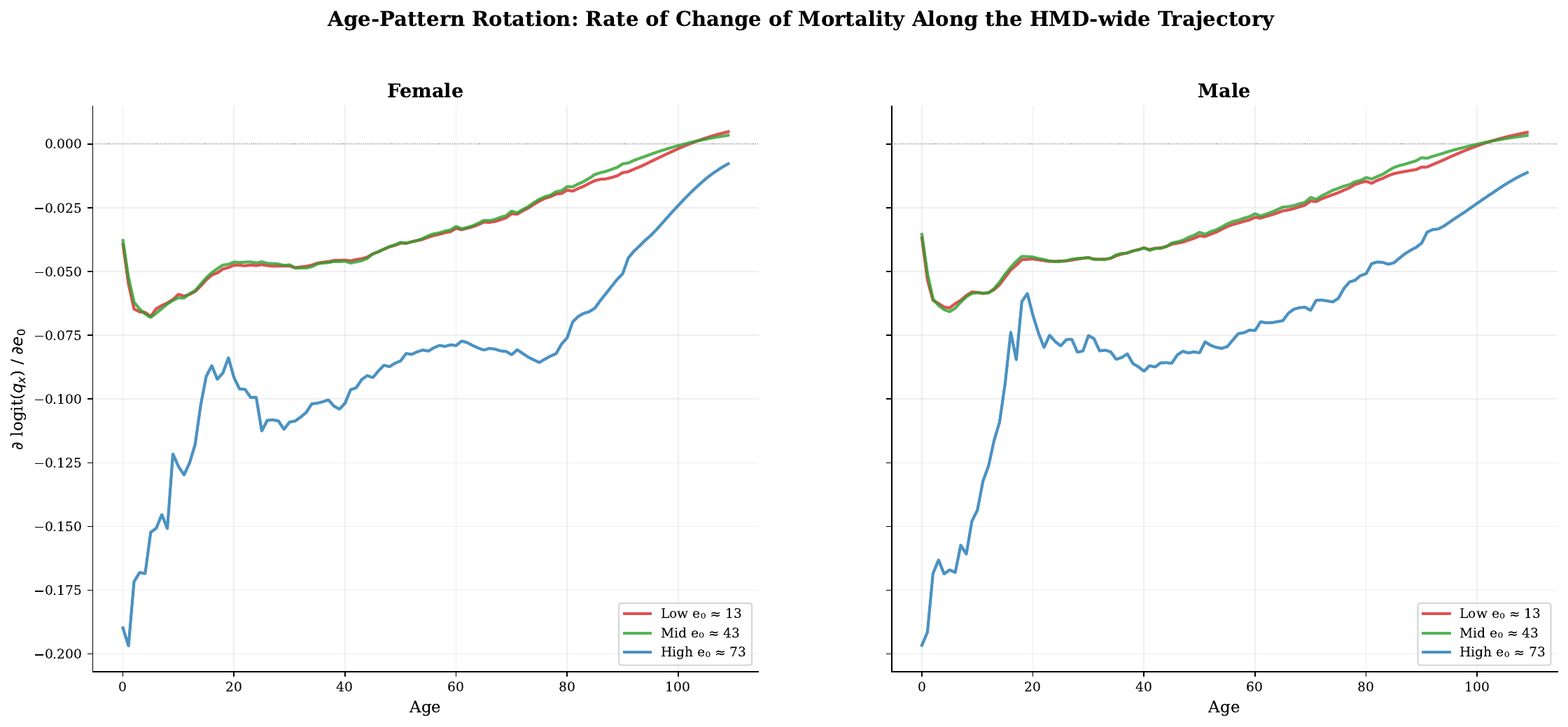}
\caption{Age-pattern rotation: the age-specific rate of decline of
logit($\qx$) per unit increase in $\ezero$, evaluated at three points
along the HMD-wide trajectory (low, medium, high $\ezero$).}
\label{fig:s6_age_rotation}
\end{figure}

\Cref{fig:s1_sex_diff_trajectory} tracks the evolution of the sex
differential along the HMD-wide and cluster-specific trajectories: the
left panel shows the $\ezero$ gap (female minus male), and the right
panel shows the ratio of female to male $\ezero$.  The differential
widening during the mortality transition and the tentative narrowing at
the lowest mortality levels -- the pattern well-documented in the
demographic literature -- emerges naturally from the trajectory model
without being imposed.  \Cref{fig:s6_age_rotation} quantifies the
age-pattern rotation by computing the age-specific rate of decline of
logit($\qx$) per unit increase in $\ezero$ at three points along the
HMD-wide trajectory (low, medium, high $\ezero$), confirming the shift
from young-age-dominated to old-age-dominated mortality decline predicted
in \cref{sec:reconstruction:trajectories}.

\subsubsection{Cluster-specific vs.\ HMD-wide reconstruction}

\begin{figure}[!htbp]
\centering
\includegraphics[width=\textwidth]{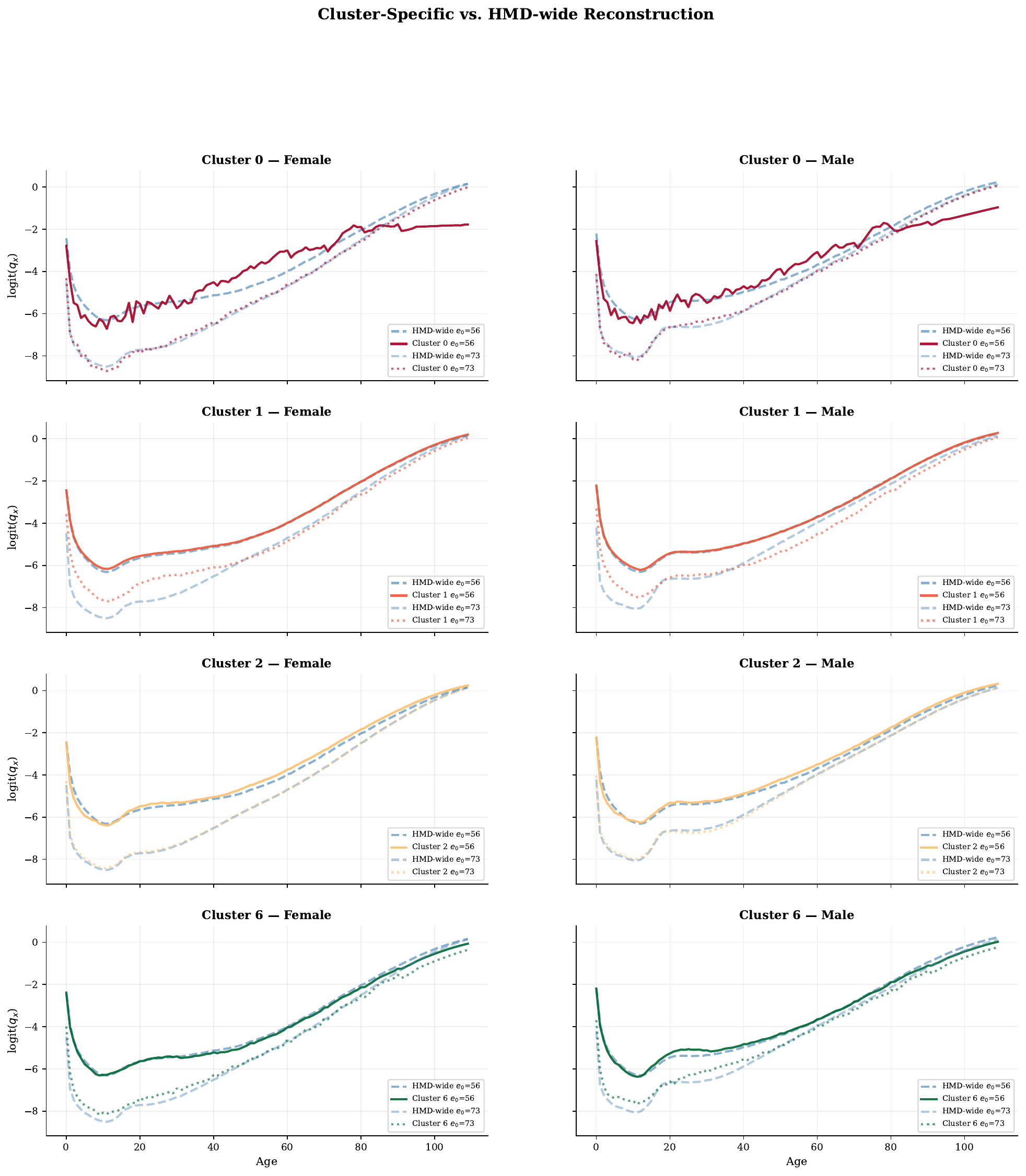}
\caption{Cluster-specific vs.\ HMD-wide reconstruction at two target
$\ezero$ levels.  Solid lines = cluster-specific; dashed blue = HMD-wide.
Differences reveal regime-specific mortality patterns.}
\label{fig:s6_cluster_vs_global}
\end{figure}

\Cref{fig:s6_cluster_vs_global} compares cluster-specific with HMD-wide
reconstructions at matched $\ezero$ levels.  Solid lines show the
cluster-specific schedules; dashed blue lines show the HMD-wide trajectory
at the same target.  The differences reveal regime-specific mortality
patterns that the HMD-wide trajectory averages away: the pace of the
Gompertz slope at old ages, the magnitude of the accident hump at
young-adult ages, and the size and timing of the sex differential all
vary across regimes, illustrating why the within-cluster trajectories
of \cref{sec:reconstruction:trajectories} provide a richer description
than a single HMD-wide model.

\subsubsection{Neural trajectory validation}
\label{sec:results:neural_trajectory}

\begin{figure}[!htbp]
\centering
\includegraphics[width=\textwidth]{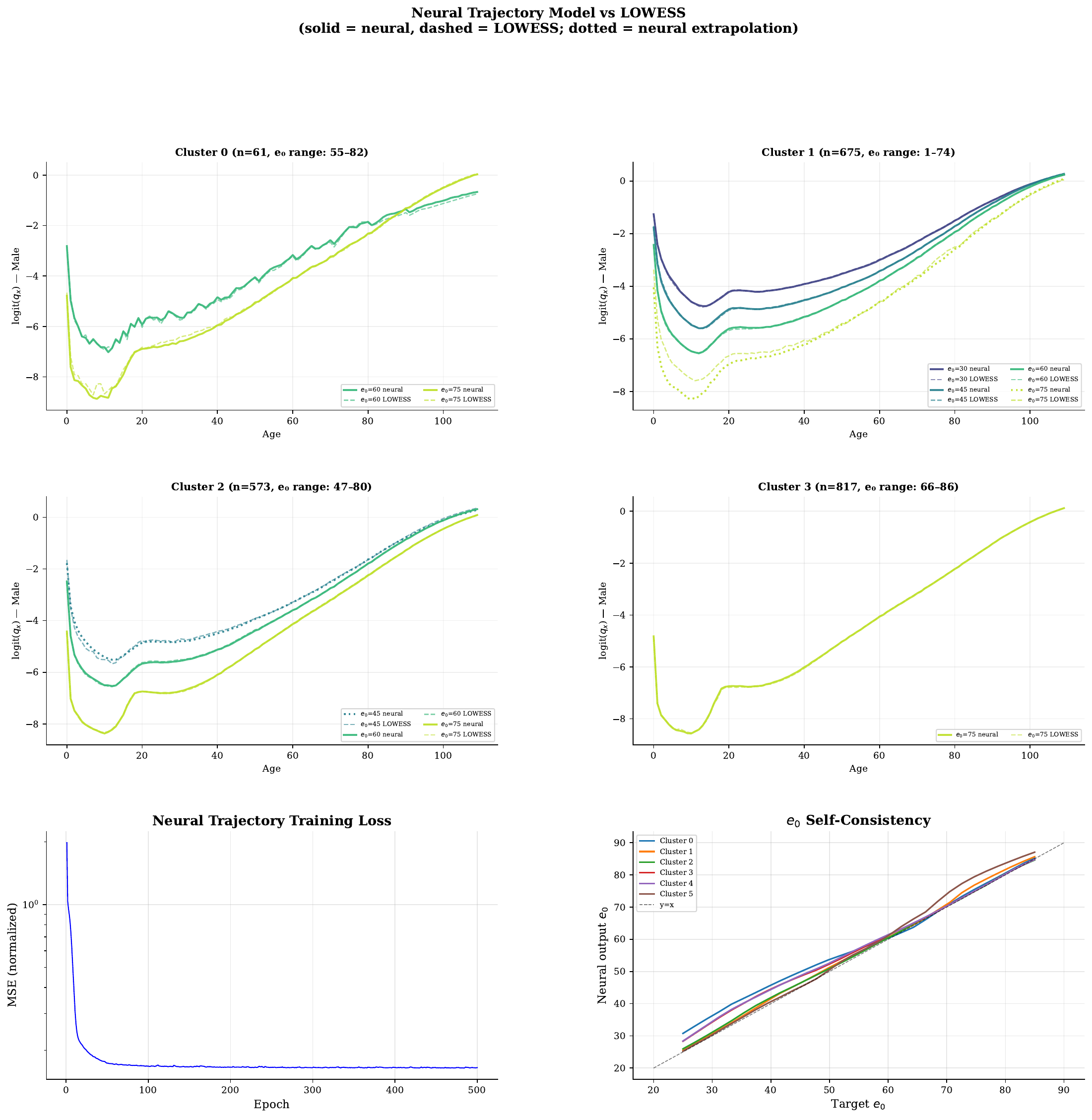}
\caption{Neural trajectory model vs.\ LOWESS.  Top: logit($\qx$)
profiles at selected $\ezero$ values for two clusters, comparing neural
(solid) with LOWESS (dashed).  Bottom left: training loss.  Bottom
right: $\ezero$ self-consistency.}
\label{fig:s7_neural_trajectory}
\end{figure}

The neural trajectory MLP (\cref{eq:neural_trajectory}) was trained on
all observed non-exceptional country-years with their cluster labels and
life expectancies.  \Cref{fig:s7_neural_trajectory} compares the neural
trajectory with the LOWESS-based trajectories: the top row shows
age-specific logit($\qx$) profiles at selected $\ezero$ values for two
clusters, the bottom left shows the training loss curve, and the bottom
right evaluates $\ezero$ self-consistency -- how well the forward model
applied to the neural output returns the target $\ezero$.  The neural
trajectory produces smooth schedules that closely match the LOWESS
estimates within each cluster's observed $\ezero$ range while providing
the smooth inter-cluster interpolation and safe extrapolation advantages
described in \cref{sec:reconstruction:neural_trajectory}.

\Cref{tab:s7_neural_traj_errors} quantifies the within-sample agreement
between the neural trajectory and the LOWESS trajectories it replaces.

{\setlength\LTleft{0pt}\setlength\LTright{0pt}
\begin{longtable}{@{\extracolsep{\fill}}lrr@{}}
\caption{Within-sample trajectory reconstruction error: LOWESS vs.\ neural trajectory ($\|\hat{z} - z\|_2$)} \label{tab:s7_neural_traj_errors} \\
\toprule
Method & Mean & Median \\
\midrule
\endfirsthead
\caption[]{Within-sample trajectory reconstruction error: LOWESS vs.\ neural trajectory ($\|\hat{z} - z\|_2$) (continued)} \\
\toprule
Method & Mean & Median \\
\midrule
\endhead
\midrule
\multicolumn{3}{r}{\textit{Continued on next page}} \\
\endfoot
\bottomrule
\endlastfoot
LOWESS & 3.089 & 2.828 \\
Neural trajectory & 3.097 & 2.889 \\
\end{longtable}}

\begin{figure}[!htbp]
\centering
\includegraphics[width=\textwidth]{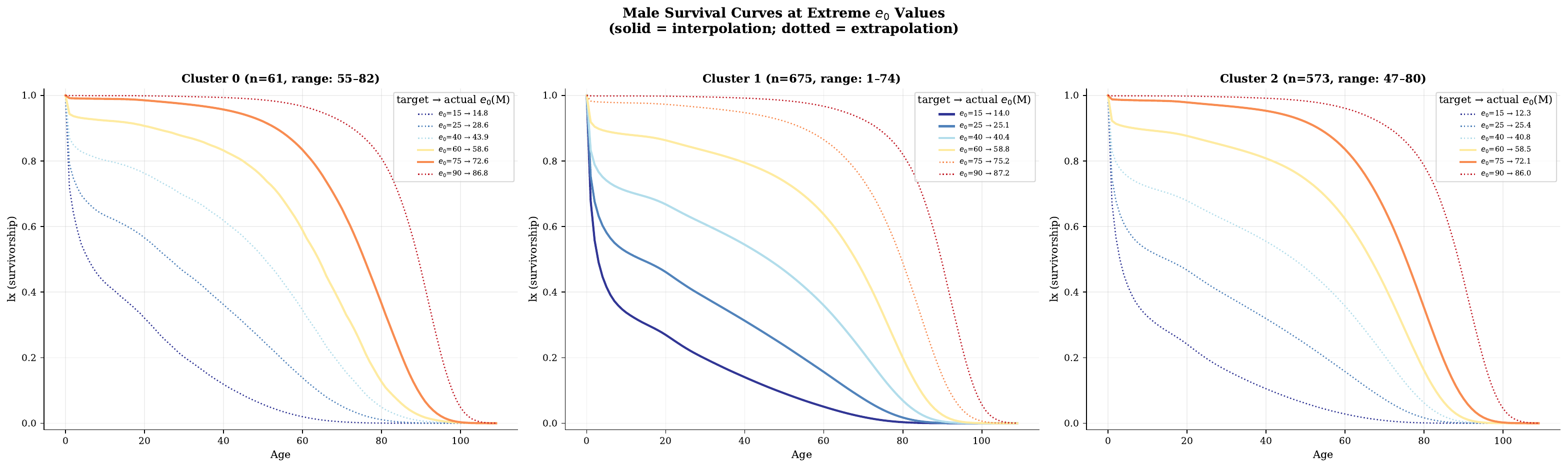}
\caption{Extreme extrapolation behavior of the neural trajectory model.
Top: logit($\qx$) profiles at extreme $\ezero$ values.  Middle: same
on the $\qx$ scale.  Bottom: sex-specific $\ezero$ and sex differential
as a function of target $\ezero$.}
\label{fig:s7_neural_extrapolation}
\end{figure}

\Cref{fig:s7_neural_extrapolation} evaluates the extreme extrapolation
behavior of the neural trajectory model.  The top row shows logit($\qx$)
age profiles at extreme $\ezero$ values, the middle row shows the same
on the natural $\qx$ scale, and the bottom row tracks the sex-specific
$\ezero$ and sex differential as a function of target $\ezero$.  The
model degrades gracefully at boundaries, maintaining demographic
plausibility -- monotonic $\qx$ at older ages, reasonable sex
differentials -- rather than producing artifacts, confirming the expected
behavior described in \cref{sec:reconstruction:neural_trajectory}.

\Cref{tab:s7_extrapolation} reports the reconstructed sex-specific
$\ezero$ values at extreme target values, confirming graceful
degradation outside the training range.

{\setlength\LTleft{0pt}\setlength\LTright{0pt}
\begin{longtable}{@{\extracolsep{\fill}}clcccc@{}}
\caption{Neural trajectory extrapolation: reconstructed $e_0$(F)/$e_0$(M) at extreme target values} \label{tab:s7_extrapolation} \\
\toprule
Cluster & Obs range & $e_0=15$ & $e_0=20$ & $e_0=90$ & $e_0=95$ \\
\midrule
\endfirsthead
\caption[]{Neural trajectory extrapolation: reconstructed $e_0$(F)/$e_0$(M) at extreme target values (continued)} \\
\toprule
Cluster & Obs range & $e_0=15$ & $e_0=20$ & $e_0=90$ & $e_0=95$ \\
\midrule
\endhead
\midrule
\multicolumn{6}{r}{\textit{Continued on next page}} \\
\endfoot
\bottomrule
\endlastfoot
0 & 55--82 & 17.2/14.8 & 25.7/22.1 & 90.7/86.8 & 93.0/89.6 \\
1 & 1--74 & 13.7/14.0 & 20.4/20.1 & 90.4/87.2 & 92.7/89.8 \\
2 & 47--80 & 12.5/12.3 & 20.8/19.9 & 90.0/86.0 & 92.4/88.8 \\
\end{longtable}}

\section{Exceptional Mortality}
\label{sec:exceptional}

The Tucker decomposition and within-cluster trajectory together describe
the \emph{secular} evolution of mortality: the smooth, long-run decline
from high to low mortality that characterizes the demographic transition.
But the historical record is punctuated by acute departures from
this baseline -- wars, pandemics, and famines that elevate mortality
sharply for one or a few years before receding.  These exceptional
events are poorly described by the low-rank baseline, which is designed
to capture slowly varying structure, not transient shocks.  We now
develop a separable model that isolates exceptional mortality from the
baseline and represents it as a product of a type-specific
\emph{disruption profile} and a scalar \emph{intensity}.

\subsection{Baseline reconstruction for exceptional years}
\label{sec:exceptional:residuals}

Exceptional country-years were excluded from the Tucker decomposition
(\cref{sec:preprocessing:exceptional}), so the factor matrices $\bS$,
$\bA$, $\bC$, $\bT$ and core tensor $\G$ describe only the baseline
mortality surface.  To compute disruption residuals, we need a baseline
estimate for each exceptional year -- the mortality schedule that
\emph{would have been observed} had the disruption not occurred.

Because the exceptional years were not used to estimate the factor
matrices, their observed schedules are not directly represented in the
decomposition.  However, the baseline factor matrices span a low-rank
subspace that captures the full range of secular mortality variation.
We consider four approaches to estimating the counterfactual baseline,
each with different strengths and limitations: na\"{\i}ve projection onto
the Tucker subspace, temporal interpolation from neighboring
non-exceptional years, penalized projection toward the temporal
interpolation, and a neural core that predicts the Tucker core tensor
slices from country loadings and year features.  We present each in turn and then explain why the neural core is the
best of the four.

\subsubsection{Na\"{\i}ve projection and its bias}
\label{sec:exceptional:naive}

For an exceptional country-year $(c,t) \in \mathcal{E}$, let
$\bm{y}_{c,t} \in \R^{2A}$ be the observed logit($\qx$) vector
(concatenating both sexes as in \cref{eq:cluster_feature}).  The
na\"{\i}ve baseline reconstruction $\hat{\bm{y}}_{c,t}$ is the point in
the range of the Tucker model that best fits the observed schedule:
\begin{equation}
    \label{eq:baseline_projection}
    \hat{\bm{y}}_{c,t}^{\text{proj}} = \arg\min_{\bm{z} \in \mathcal{R}}
    \lVert \bm{y}_{c,t} - \bm{z} \rVert^2\,,
\end{equation}
%
where $\mathcal{R} = \{\bm{z} \in \R^{2A} :
\bm{z} \text{ is representable by } \G, \bS, \bA, \bC, \bT
\text{ at some country and year loadings}\}$ is the set of schedules
that can be produced by the baseline decomposition.  In practice, this
projection amounts to finding the country and year loading vectors
(rows of $\bC$ and $\bT$, or their extensions into the loading spaces)
that produce the closest reconstruction.

This approach has a systematic flaw.  The projection minimizes squared
error across \emph{all} $2A$ sex-age cells simultaneously.  When the
observed schedule contains a disruption that elevates mortality at
certain ages, the optimizer compromises: it shifts the entire baseline
estimate upward to partially accommodate the excess, because the smooth
basis functions in~$\bA$ cannot represent the sharp disruption shape
without also shifting adjacent ages.  The result is a baseline estimate
that is biased in a characteristic way:
\begin{itemize}
    \item At \textbf{affected ages} (where the disruption elevates
    mortality), the projected baseline sits above the true baseline but
    below the observed schedule.  The disruption signal is attenuated.

    \item At \textbf{unaffected ages} (where the observed schedule
    equals the true baseline), the projected baseline sits above the
    true baseline.  Spurious negative residuals appear at ages that
    experienced no excess mortality.
\end{itemize}
The net effect is that the estimated disruption profile inherits an
artifact: a genuine excess at affected ages paired with a spurious
deficit at unaffected ages.  The disruption intensity is
underestimated, and the profile shape is distorted.  The severity of
this bias depends on the magnitude of the disruption relative to the
baseline and on how many ages are affected.  A war that concentrates
severe excess mortality in a narrow age band of one sex produces a
worse bias than a broadly distributed pandemic, because the narrow
spike is harder for the smooth basis functions to accommodate without
distorting the rest of the schedule.

\subsubsection{Temporal interpolation of the baseline}
\label{sec:exceptional:temporal}

The key insight is that baseline mortality evolves slowly -- the
secular trend changes little from one year to the next.  For a country
$c$ in exceptional year $t$, the non-exceptional years near $t$ have
well-estimated baseline schedules from the Tucker reconstruction
(\cref{sec:tucker:reconstruction}).  The baseline for the exceptional
year can therefore be interpolated from these neighbors without
consulting the contaminated observed schedule.

Let $\hat{\bm{z}}_{c,t'} \in \R^{2A}$ denote the reconstructed
baseline feature vector (from the Tucker decomposition) for country $c$
in non-exceptional year~$t'$.  The temporally interpolated baseline for
the exceptional year~$t$ is
\begin{equation}
    \label{eq:baseline_temporal}
    \hat{\bm{y}}_{c,t}^{\text{interp}}
    = \hat{f}_{c}(t)\,,
\end{equation}
%
where $\hat{f}_{c} : \R \to \R^{2A}$ is a smooth function estimated
from the non-exceptional reconstructions
$\{(t',\, \hat{\bm{z}}_{c,t'})\}_{t' \notin \mathcal{E}_c}$, with
$\mathcal{E}_c = \{t : (c,t) \in \mathcal{E}\}$ the set of
exceptional years for country~$c$.  The smoother operates independently
on each of the $2A$ components as a function of calendar year.
Suitable smoothers include local polynomial regression
\citep{Cleveland1979}, smoothing splines \citep{GreenSilverman1994},
or even linear interpolation from the nearest non-exceptional years on
either side.

This approach has two important strengths.  First, the contaminated
observed schedule $\bm{y}_{c,t}$ is never consulted, so there is no
mechanism for the disruption to bias the baseline estimate.  Second,
the method rests on a well-justified assumption: baseline mortality
evolves smoothly within the secular trend, and the baseline in an
exceptional year is well-predicted by the trajectory of non-exceptional
years surrounding it.

The practical challenge arises when exceptional years are consecutive.
For belligerent countries during World War I, the exceptional spell
1914--1918 spans five years, and linear interpolation from 1913 to 1919
must bridge a six-year gap.  A spline through all non-exceptional years
for the country handles this more gracefully, as it borrows information
from the entire non-exceptional trajectory rather than relying solely on
the nearest neighbors.  Nevertheless, the interpolation uncertainty
grows with the length of the exceptional spell, and baselines for years
deep within a prolonged disruption should be treated with appropriate
caution.

\subsubsection{Penalized projection toward temporal interpolation}
\label{sec:exceptional:penalized}

A hybrid strategy retains the projection framework of
\cref{eq:baseline_projection} but regularizes it toward the temporal
interpolation of \cref{eq:baseline_temporal}.  The penalized baseline
estimate is
\begin{equation}
    \label{eq:baseline_penalized}
    \hat{\bm{y}}_{c,t}^{\text{pen}}
    = \arg\min_{\bm{z} \in \mathcal{R}}
    \Bigl\{
        \lVert \bm{y}_{c,t} - \bm{z} \rVert^2
        + \alpha \, \lVert \bm{z} - \hat{\bm{y}}_{c,t}^{\text{interp}} \rVert^2
    \Bigr\}\,,
\end{equation}
%
where $\hat{\bm{y}}_{c,t}^{\text{interp}}$ is the temporally
interpolated baseline from \cref{eq:baseline_temporal} and $\alpha \geq
0$ is a penalty parameter controlling the strength of the prior.  At
$\alpha = 0$, the estimator reduces to the na\"{\i}ve projection
\cref{eq:baseline_projection}; as $\alpha \to \infty$, it converges to
the temporal interpolation.  Intermediate values allow the observed data
to inform the baseline at sex-age cells where the disruption is small,
while preventing the projection from being pulled into the disruption
at affected ages.

The penalty has a Bayesian interpretation: $\alpha$ is the ratio of the
observational variance to the prior variance, and the temporal
interpolation serves as a prior mean for the baseline.  Selection
of~$\alpha$ can be guided by cross-validation on non-exceptional years:
artificially designate a held-out non-exceptional year as exceptional,
estimate its baseline under both methods, and compare the reconstruction
error against the known baseline.

\subsubsection{Neural core approach}
\label{sec:exceptional:neural_core}

The three approaches above all operate in the reconstructed feature
space, \emph{after} the Tucker decomposition.  A fourth approach
addresses the baseline contamination problem at its source: \emph{within}
the decomposition itself.

Recall that the Tucker reconstruction of any country-year $(c,t)$ depends
on the core tensor~$\G$ through the country and year factor matrices
$\bC$ and~$\bT$.  For observed non-exceptional country-years, the core
tensor faithfully encodes the interactions between country and temporal
loadings.  For exceptional country-years that were excluded during tensor
construction (\cref{sec:preprocessing:imputation_bias}), the
zero observation weight causes the corresponding year loadings
$\bm{t}_t$ to be null vectors, so the core slices
$\G_{:,:,c,t}$ are identically zero rather than meaningful estimates.

The neural core approach replaces these missing core slices with the
output of a multi-layer perceptron trained on non-exceptional data --
learning the mapping from country identity and calendar year to the core
tensor slice, and then evaluating that mapping at the exceptional
country-years to produce counterfactual baselines.  The contaminated
observed schedule is never consulted; instead, the baseline is predicted
entirely from the country's time-invariant loading vector and the year's
temporal features.

\subsubsection{Comparison and preferred method}
\label{sec:exceptional:baseline_discussion}

Of the four baseline estimation strategies, we adopt the neural core
in the companion implementation
(\cref{sec:results:exceptional}).  It has three advantages over the
alternatives.  First, it solves the contamination problem \emph{by
construction}: because the network is trained only on non-exceptional
country-years, the baseline predictions cannot be biased by disruption
signals -- unlike the na\"{\i}ve projection
(\cref{sec:exceptional:naive}), which absorbs the disruption into the
baseline estimate.  Second, it \emph{pools information across countries}:
the shared network weights learn how core slices vary with country
loadings and year features from the entire non-exceptional dataset, not
just from a single country's temporal neighbors.  This is particularly
valuable for countries with sparse non-exceptional data near event
boundaries, where the temporal interpolation
(\cref{sec:exceptional:temporal}) may lack support on one or both sides
of the disruption spell.  Third, it handles \emph{consecutive exceptional
years} naturally: the MLP predicts the baseline for any year from the
country's time-invariant loading and the year's feature encoding, without
needing temporal neighbors on both sides -- a practical advantage over
the temporal interpolation for prolonged events like the World Wars.

The temporal interpolation remains a useful cross-check on the neural
core baselines and is preferred when the non-exceptional temporal
neighbors provide adequate support -- that is, when non-exceptional years
are available on both sides of the disruption spell, and the gap is not
excessively long.  The penalized projection
(\cref{sec:exceptional:penalized}) provides an additional robustness
check: if the methods produce similar baselines (as they should for
well-identified events with good temporal coverage), this increases
confidence in all of them.

The implementation uses the neural core as the primary method and reports
the temporal interpolation and penalized projection for comparison
(\cref{sec:results:exceptional}).

\subsubsection{Neural core implementation}
\label{sec:exceptional:neural_core_impl}

The same Poisson-noise problem that motivates the pre-decomposition
adaptive temporal pooling (\cref{sec:preprocessing:smoothing}) also affects the
exceptional country-years retained for the disruption model.  Small
populations can have $m_x = 0$ at valley ages even during wars and
pandemics, producing extreme logit values that distort the residual
analysis.  Before computing residuals, the adaptive temporal pooling
procedure is therefore applied to the exceptional data as well,
pooling $m_x$ and $a_x$ across the years of each exceptional spell
(e.g., France 1914--1918 for World War~I).  If an isolated exceptional
year has zeros at valley ages, it is pooled with the nearest one or two
non-exceptional years on each side, borrowing the exposure base without
materially contaminating the disruption signal.

With the exceptional data smoothed, the neural core MLP is defined as:
\begin{equation}
    \label{eq:neural_core}
    \hat{G}_{c,t} = \operatorname{MLP}\!\bigl(
        \bm{u}_c,\; \bm{\phi}(t)
    \bigr)\,,
\end{equation}
%
where $\bm{u}_c \in \R^{r_3}$ is the country loading vector (row~$c$ of
$\bC$), $\bm{\phi}(t) \in \R^{9}$ is a feature encoding of the calendar
year (including normalized year, polynomial terms, and sinusoidal basis
functions at multiple frequencies), and the output is the $r_1 \times r_2$
core slice.  Crucially, the year features are raw temporal basis functions,
\emph{not} the HOSVD year loadings~$\bT$, which inherit contamination from
the imputed tensor.  The function $\bm{\phi}$ uses polynomial and Fourier
features:
\begin{equation}
    \label{eq:year_features}
    \bm{\phi}(t) = \bigl[
        \tilde{t},\; \tilde{t}^2,\; \tilde{t}^3,\;
        \sin(2\pi\tilde{t}),\; \cos(2\pi\tilde{t}),\;
        \sin(4\pi\tilde{t}),\; \cos(4\pi\tilde{t}),\;
        \sin(10\pi\tilde{t}),\; \cos(10\pi\tilde{t})
    \bigr]^\top \in \R^{9},
\end{equation}
%
where $\tilde{t} = (t - t_{\min}) / (t_{\max} - t_{\min})$ normalizes
the year to $[0, 1]$.

The MLP is trained on all observed non-exceptional $(c,t)$ pairs.
Rather than targeting the HOSVD core contraction (which would introduce
round-trip approximation error from the Tucker truncation), the training
target is the \emph{direct projection} of the observed data onto the
sex and age factor matrices:
\begin{equation}
    \label{eq:neural_core_target}
    \tilde{G}_{c,t} = \bS^\top \, \M_{:,:,c,t} \, \bA
    \;\in\; \R^{r_1 \times r_2}\,,
\end{equation}
%
where $\M_{:,:,c,t}$ is the observed $(S \times A)$ logit-mortality
slice at country~$c$ and year~$t$.  This target is more principled than
the HOSVD core slice $\G_{:,:,c,t}^{\text{HOSVD}}$ because it retains
all of the observed information projected onto the sex--age subspace,
including any structure that the Tucker truncation along the country and
year modes would discard.  The loss function is
\begin{equation}
    \label{eq:neural_core_loss}
    \mathcal{L} = \frac{1}{|\mathcal{O}|}
    \sum_{(c,t) \in \mathcal{O}}
    \lVert \hat{G}_{c,t} - \tilde{G}_{c,t} \rVert_F^2
    + \lambda_{\mathrm{reg}} \sum_{\ell} \lVert W_\ell \rVert_F^2\,,
\end{equation}
%
where $\mathcal{O}$ is the set of observed non-exceptional country-years
and the second term is $L_2$ weight regularization to
prevent overfitting.  Training uses the Adam optimizer
\citep{KingmaBa2015}.

The reconstructed baseline schedule at an exceptional $(c,t)$ is then
obtained by feeding the MLP-predicted core slice through the standard
Tucker reconstruction formula (\cref{sec:tucker:reconstruction}):
\begin{equation}
    \label{eq:neural_baseline}
    \hat{\bm{y}}_{c,t}^{\mathrm{NN}}
    = \hat{G}_{c,t} \nmode{1} \bS \nmode{2} \bA\,.
\end{equation}

\subsubsection{Residuals}
\label{sec:exceptional:residual_def}

Regardless of which baseline estimation method is used, the
\emph{residual} is defined as the difference between the observed and
estimated baseline on the logit scale:
\begin{equation}
    \label{eq:residual}
    r_{s,a,c,t} = y_{s,a,c,t} - \hat{y}_{s,a,c,t}\,,
\end{equation}
%
where $y_{s,a,c,t} = \logit(\qx)$ is the observed logit-mortality and
$\hat{y}_{s,a,c,t}$ is the baseline estimate (from the neural core
\cref{eq:neural_baseline}, the temporal interpolation
\cref{eq:baseline_temporal}, or the penalized projection
\cref{eq:baseline_penalized}).  By the
additive property of \cref{eq:logit_additive}, $\exp(r_{s,a,c,t})$ is
the factor by which the observed odds of dying exceed (or fall below)
the baseline odds at sex~$s$ and age~$a$.  A positive residual
indicates excess mortality; a negative residual indicates a mortality
deficit relative to baseline.

At ages unaffected by the disruption, the residuals should be small --
close to the reconstruction noise of the Tucker model.  If the baseline
has been estimated well, the residuals at unaffected ages will scatter
symmetrically around zero, without the systematic negative bias that
afflicts the na\"{\i}ve projection.  At affected ages, the residuals
exhibit a distinctive, large-amplitude pattern characteristic of the
disruption type: concentrated in young-adult males for wars, broadly
distributed for respiratory pandemics, and broadly elevated from young
through adult ages for enteric pandemics.

\subsection{Event dictionaries}
\label{sec:exceptional:events}

To identify which country-years are exceptional and what type of
disruption they reflect, we rely on the external historical knowledge
encoded in the \emph{event dictionaries} introduced in
\cref{sec:preprocessing:exceptional}.  These same dictionaries were used
to exclude exceptional years from the decomposition; here we describe
their structure in more detail.  Each dictionary records, for a specific
class of disruption, the countries and years affected.  Three classes are
defined:
\begin{enumerate}
    \item \textbf{Wars.}  Armed conflicts that caused substantial
    mortality in at least one HMD country, from the Seven Years' War
    (1756--1763) through the wars of the late twentieth century.

    \item \textbf{Respiratory pandemics.}  Epidemic events driven by
    airborne pathogens, notably the influenza pandemics of 1889--1890,
    1918--1920, and 1957--1958, as well as the COVID-19 pandemic of
    2020--2022 \citep{KarlinskyKobak2021}.

    \item \textbf{Enteric pandemics.}  Epidemic events driven by
    waterborne or fecal--oral pathogens, principally the cholera pandemics
    of the nineteenth century.
\end{enumerate}
The complete event dictionaries are provided in \cref{app:events}.  Each
entry specifies the countries affected, the year(s) of impact, and the
type classification.  These dictionaries are compiled from standard
historical sources; the classification into types is based on the
dominant transmission pathway and is not intended to capture every
epidemiological nuance.

Using the event dictionaries, each observed country-year $(c,t)$ is
assigned a disruption label $d_{c,t} \in \{0, 1, 2, 3\}$, where
$0$~denotes a non-exceptional year and $1$, $2$, $3$ denote war,
respiratory pandemic, and enteric pandemic, respectively.  Country-years
affected by multiple simultaneous events (e.g., war and pandemic in
1918) receive the war label, on the grounds that combat mortality
dominates the overall mortality signal in belligerent populations.

\subsection{Disruption profiles}
\label{sec:exceptional:profiles}

Each type of disruption produces a characteristic pattern of excess
mortality across sex and age.  We call this pattern the \emph{disruption
profile}.  The profile is a vector $\bm{\delta}_d \in \R^{2A}$
(concatenating female and male age schedules, as in
\cref{eq:cluster_feature}) that describes the \emph{shape} of the excess
mortality -- which sexes and ages are most affected -- independently of its
overall magnitude.

The profiles differ qualitatively across types:
\begin{itemize}
    \item \textbf{War} profiles concentrate excess mortality in
    young-adult males (roughly ages 15--45), reflecting combat deaths and
    associated trauma.  Female excess mortality is typically much smaller
    and may be concentrated at different ages (e.g., civilian casualties
    affecting older ages and children).

    \item \textbf{Respiratory pandemic} profiles are characterized by
    elevated mortality across a broad age range, often with a distinctive
    pattern that depends on the pathogen.  The 1918 influenza pandemic, for
    example, produced a W-shaped age profile with peaks in infants, young
    adults, and the elderly \citep{TaubenbergerMorens2006,MurrayEtAl2006}.

    \item \textbf{Enteric pandemic} profiles would, on epidemiological
    grounds, be expected to concentrate excess mortality in infants and
    young children, reflecting the vulnerability of the very young to
    dehydration from diarrheal disease.  As discussed below
    (\cref{sec:results:profile_results}), the estimated profiles show a
    broader pattern than this expectation.
\end{itemize}
These qualitative differences are what makes the type classification
useful: knowing the type of disruption substantially constrains the
expected shape of excess mortality.

\subsection{The separable disruption model}
\label{sec:exceptional:model}

The disruption model represents the logit-mortality in an exceptional
country-year as the baseline plus a scaled disruption profile:
\begin{equation}
    \label{eq:disruption_model}
    y_{s,a,c,t} = \hat{y}_{s,a,c,t}
    + \lambda_{c,t} \; \delta_{d_{c,t},\, s,a}\,,
\end{equation}
%
where $\hat{y}_{s,a,c,t}$ is the baseline estimate (from the neural core
\cref{eq:neural_baseline}, the temporal interpolation
\cref{eq:baseline_temporal}, or the penalized projection
\cref{eq:baseline_penalized}),
$\bm{\delta}_{d_{c,t}}$ is the smoothed disruption profile
(\cref{eq:profile_smooth}) for the event type
affecting $(c,t)$, and $\lambda_{c,t} \geq 0$ is a scalar
\emph{intensity} that scales the profile to match the severity of the
event in that particular country-year.

The model is \emph{separable} in the sense that it factors the
disruption into a type-specific shape ($\bm{\delta}_d$, shared across
all country-years of the same type) and an event-specific magnitude
($\lambda_{c,t}$, varying by country-year).  By the additive property
of the logit (\cref{eq:logit_additive}), the intensity $\lambda_{c,t}$
acts as a uniform scaling of the log-odds ratio across all ages: an
intensity of $\lambda = 1$ produces the canonical disruption, while
$\lambda = 2$ doubles the log-odds ratio at every sex and age.

\subsection{Estimation}
\label{sec:exceptional:estimation}

\subsubsection{Profile estimation}

The disruption profile $\bm{\delta}_d$ for type~$d$ is estimated from
the residuals of all country-years labeled as type~$d$.  Let
$\mathcal{E}_d = \{(c,t) : d_{c,t} = d\}$ be the set of country-years
affected by type-$d$ disruptions.  The raw mean residual for type~$d$ is
\begin{equation}
    \label{eq:mean_residual}
    \bar{\bm{r}}_d = \frac{1}{|\mathcal{E}_d|}
    \sum_{(c,t) \in \mathcal{E}_d} \bm{r}_{c,t}\,,
\end{equation}
%
where $\bm{r}_{c,t} = (r_{1,1,c,t}, \ldots, r_{1,A,c,t},\;
r_{2,1,c,t}, \ldots, r_{2,A,c,t})^\top \in \R^{2A}$ is the residual
vector for country-year $(c,t)$.  The profile is the normalized mean
residual:
\begin{equation}
    \label{eq:profile_norm}
    \hat{\bm{\delta}}_d^{\text{raw}}
    = \frac{\bar{\bm{r}}_d}{\lVert \bar{\bm{r}}_d \rVert}\,,
\end{equation}
%
so that $\hat{\bm{\delta}}_d^{\text{raw}}$ is a unit vector specifying
the direction of excess mortality in the $2A$-dimensional space.  The
normalization ensures that the profile captures shape only, with all
magnitude information absorbed into the intensity scalar.

\subsubsection{Low-pass smoothing of disruption profiles}
\label{sec:exceptional:profile_smooth}

The raw profile $\hat{\bm{\delta}}_d^{\text{raw}}$ is estimated from a
finite number of exceptional country-years and inherits high-frequency
age-to-age noise from both the baseline estimation and the underlying
mortality data.  This noise is particularly pronounced at ages where the
number of contributing events is small or where baseline estimation is
imprecise.  Left unsmoothed, the raw profile would transmit this noise
into every reconstructed disruption schedule.

To suppress this noise while preserving the broad age structure of the
disruption, we apply a Savitzky--Golay low-pass filter
\citep{SavitzkyGolay1964} to each sex half of the profile independently.
The filter is a local polynomial smoother applied across adjacent ages,
with parameters chosen to balance noise removal against preservation of
genuine age-structure features:
\begin{enumerate}
    \item The filter is applied to a window of width $w$ (e.g., $w = 11$
    ages) using a cubic polynomial ($p = 3$), which preserves the location
    and shape of peaks and troughs while smoothing age-to-age fluctuations.

    \item \textbf{Boundary preservation.}  The first $\lfloor w/2 \rfloor$
    and last $\lfloor w/2 \rfloor$ ages within each sex half are left at
    their original (raw) values, avoiding the ringing artifacts that
    Savitzky--Golay filters can produce at boundaries.

    \item After smoothing, the profile is re-normalized to unit length:
    \begin{equation}
        \label{eq:profile_smooth}
        \hat{\bm{\delta}}_d = \frac{\operatorname{SG}
        \bigl(\hat{\bm{\delta}}_d^{\text{raw}}\bigr)}
        {\bigl\lVert \operatorname{SG}
        \bigl(\hat{\bm{\delta}}_d^{\text{raw}}\bigr) \bigr\rVert}\,,
    \end{equation}
    %
    where $\operatorname{SG}(\cdot)$ denotes the per-sex-half
    Savitzky--Golay filter with boundary preservation.
\end{enumerate}
The smoothed profile $\hat{\bm{\delta}}_d$ is the default for all
downstream use (intensity estimation, model reconstruction, and
disruption-impact calculations).  Both raw and smoothed profiles are
retained for comparison: the cosine similarity between the two is
typically very high (above 0.99), confirming that the smoothing removes
noise without materially altering the disruption shape.

\subsubsection{Intensity estimation}

Given the estimated profile $\hat{\bm{\delta}}_d$, the intensity for a
specific country-year $(c,t) \in \mathcal{E}_d$ is estimated by
projecting its residual vector onto the profile:
\begin{equation}
    \label{eq:intensity}
    \hat{\lambda}_{c,t}
    = \bm{r}_{c,t}^\top \hat{\bm{\delta}}_{d_{c,t}}\,.
\end{equation}
%
Because $\hat{\bm{\delta}}_d$ is a unit vector, this is simply the
component of the residual in the direction of the type's canonical
disruption.  A large $\hat{\lambda}_{c,t}$ indicates a severe event; a
value near zero indicates that the country-year, despite being flagged in
the event dictionary, experienced little excess mortality of the expected
type (e.g., a country on the periphery of a conflict, or a pandemic wave
that was mild locally).

The residual component \emph{orthogonal} to the profile,
$\bm{r}_{c,t} - \hat{\lambda}_{c,t} \hat{\bm{\delta}}_{d_{c,t}}$,
captures mortality variation in the exceptional year that is not
explained by the type's canonical pattern.  This orthogonal component is
expected to be small if the type classification is appropriate and the
event is well-characterized by a single profile.

\subsubsection{Sub-clustering of disruption profiles}
\label{sec:exceptional:subclustering}

The single-profile model captures the \emph{average} pattern within each
disruption type, but events within a type can differ substantially.
Among wars, for example, WWI trench warfare concentrates casualties
in young adult males aged 18--35, whereas WWII total war produces
broader civilian casualties across ages and both sexes.  Among
respiratory pandemics, the 1918 influenza produced a distinctive
W-shaped mortality curve with a young-adult peak not seen in other
pandemics, while COVID-19 concentrated excess mortality in the elderly.
Among enteric pandemics, the estimated profiles show a broader
age pattern than the child-concentrated excess that might be expected on
epidemiological grounds -- a point discussed further in
\cref{sec:results:profile_results}.

To capture this within-type heterogeneity, the residual vectors
$\{\bm{r}_{c,t}\}_{(c,t) \in \mathcal{E}_d}$ are sub-clustered
\emph{within each disruption type}.  Unlike the baseline clustering
(\cref{sec:clustering}), which operates on level-controlled Tucker core
features, the sub-clustering operates directly on the $2A$-dimensional
residual vectors, because the disruption residuals have already had the
baseline (including its level component) subtracted.  The procedure is as
follows:

\begin{enumerate}
    \item \textbf{Dimensionality reduction.}  Apply PCA to the $2A$-dimensional
    residual vectors within type~$d$, retaining the leading 10--20 components
    (enough to explain a substantial fraction of the variance).

    \item \textbf{Sub-cluster identification.}  Apply Ward's minimum-variance
    hierarchical clustering \citep{Ward1963} to the PCA scores, testing
    sub-cluster counts $K_d$ over the range $2$ through~$7$ and selecting the
    $K_d$ that maximizes the silhouette score, subject to a minimum cluster size
    constraint (e.g., at least 5 observations per sub-cluster).  The
    optimal $K_d$ is typically small (2--4), reflecting the limited
    number of qualitatively distinct events within each type.

    \item \textbf{Sub-profile estimation.}  For each sub-cluster $k$
    within type~$d$, compute the mean residual, normalize to a unit
    vector, and apply the same Savitzky--Golay low-pass smoothing
    described in \cref{sec:exceptional:profile_smooth}, yielding the
    smoothed sub-profile $\hat{\bm{\delta}}_{d,k}$.
    A sign convention ensures that positive intensity corresponds to
    excess mortality.
\end{enumerate}

The extended disruption model replaces the single profile with a
sub-cluster-specific profile:
\begin{equation}
    \label{eq:disruption_subcluster}
    y_{s,a,c,t} = \hat{y}_{s,a,c,t}
    + \lambda_{c,t} \; \hat{\delta}_{d_{c,t},\, k_{c,t},\, s,a}\,,
\end{equation}
%
where $k_{c,t}$ is the sub-cluster assignment of the exceptional
country-year $(c,t)$ within its disruption type.

The sub-clustering serves two purposes.  First, it improves the
explained fraction of the residual variance: matching each event to its
nearest sub-profile produces a tighter fit than projecting onto a single
average profile.  The improvement in~$R^2$ (the fraction of residual
variance explained by the profile) is a direct measure of within-type
heterogeneity.  Second, the sub-clusters are often \emph{historically
interpretable}: they may separate pre-modern from modern conflicts,
or distinguish pandemic strains with different age selectivity.  The
PCA scatter of residuals, annotated with named historical events,
provides a visual check of whether the sub-clusters correspond to
recognized historical groupings.

A small neural network is trained \emph{within each disruption type}
to map sub-cluster embeddings to
profile vectors, providing smooth interpolation between sub-profiles and
enabling the model to predict profiles for events that do not fit neatly
into a single sub-cluster.  For each type~$d$ with $K_d$ sub-clusters, the
architecture is a multi-layer perceptron:
\begin{equation}
    \label{eq:neural_subprofile}
    \hat{\bm{\delta}}_{d}(\bm{e}_{d,k})
    = \operatorname{MLP}_d\!\bigl(\bm{e}_{d,k}\bigr)\,,
\end{equation}
%
where $\bm{e}_{d,k} \in \R^{d'}$ is a continuous embedding of sub-cluster~$k$
(obtained by PCA of the sub-cluster centroids, with $d' = \min(K_d, 4)$),
and the output is the full $2A$-dimensional profile vector.  A separate
network is trained for each disruption type.

\subsection{The complete model}
\label{sec:exceptional:complete}

Combining the baseline (\cref{sec:reconstruction}) and the disruption
model, the full reconstructed logit($\qx$) at sex~$s$ and age~$a$ for
any country-year is
\begin{equation}
    \label{eq:full_model}
    y_{s,a,c,t} = \underbrace{\hat{\mu}_{k,\,s,a}\!\bigl({\ezero}^{\ast}\bigr)}_{\text{baseline from cluster trajectory}}
    + \underbrace{\lambda \; \hat{\delta}_{d,\,s,a}}_{\text{disruption (if any)}}\,,
\end{equation}
%
where $k$ is the cluster label (from the non-exceptional data-driven
clustering), ${\ezero}^{\ast}$ is the target life expectancy,
$d$ is the disruption type (with $\lambda = 0$ for
non-exceptional conditions), and $\lambda \geq 0$ is the disruption
intensity.  Because the cluster trajectories are estimated from
non-exceptional data only (\cref{sec:preprocessing:exceptional}), the
baseline term is free of contamination from past disruptions.
This equation is the central product of the framework: it
generates a complete sex-specific mortality schedule from three inputs
(cluster, target $\ezero$, and optional disruption specification).

With the extensions developed above, \cref{eq:full_model} can be
enriched in two ways.  The baseline term $\hat{\mu}_k(\ezero^{\ast})$
can be computed either from the LOWESS trajectory
(\cref{eq:trajectory_smooth}) or from the neural trajectory
(\cref{eq:neural_trajectory}), with the latter providing smooth
inter-cluster interpolation and extrapolation.  The disruption profile
$\hat{\delta}_d$ can be replaced by the sub-cluster-specific profile
$\hat{\delta}_{d,k'}$ (\cref{eq:disruption_subcluster}) to capture
within-type heterogeneity.  The most general form of the model is
therefore
\begin{equation}
    \label{eq:full_model_extended}
    y_{s,a,c,t}
    = \hat{\bm{z}}_{\mathrm{NN}}\!\bigl({\ezero}^{\ast},\, k\bigr)_{s,a}
    + \lambda \; \hat{\delta}_{d,\,k',\,s,a}\,,
\end{equation}
%
where $k'$ indexes the sub-cluster within disruption type~$d$.

Applying the expit to \cref{eq:full_model} and computing the life table
yields the disrupted life expectancy.  The reduction in $\ezero$ caused
by a disruption of type~$d$ at intensity~$\lambda$ can be computed by
evaluating the forward model (\cref{eq:forward}) with and without the
disruption term.

\subsection{Properties and limitations}
\label{sec:exceptional:properties}

\subsubsection{Baseline independence}

Because the disruption acts additively on the logit scale, the same
profile and intensity can be applied to \emph{any} baseline schedule.  A
war disruption calibrated from early-twentieth-century European data can
be applied to a contemporary low-mortality baseline or to a
high-mortality historical baseline.  The multiplicative interpretation on
the odds scale (\cref{eq:logit_additive}) ensures that the disruption
produces a proportional increase in mortality risk regardless of the
baseline level.

\subsubsection{Composability}

If two disruptions of different types occur simultaneously (e.g., war
and respiratory pandemic, as in 1918), the model can in principle
represent the joint effect as the sum of two disruption terms:
\begin{equation}
    \label{eq:composite}
    y_{s,a,c,t} = \hat{y}_{s,a,c,t}
    + \lambda^{(1)}_{c,t} \; \hat{\delta}_{d_1,\,s,a}
    + \lambda^{(2)}_{c,t} \; \hat{\delta}_{d_2,\,s,a}\,.
\end{equation}
%
This assumes that the two disruptions act independently on the log-odds
scale, which is plausible when the affected age groups are largely
disjoint (war in young-adult males, pandemic across all ages) but may
overestimate joint effects if there are interactions.

\subsubsection{Limitations}

The separable model assumes that all events of a given type share a
single canonical profile.  In reality, disruption profiles may vary:
the age pattern of combat mortality in the Napoleonic Wars may differ from
that of the Second World War, and the 1918 influenza pandemic had a
distinctive young-adult excess not seen in other influenza pandemics.
The sub-clustering extension (\cref{sec:exceptional:subclustering})
partially addresses this by identifying distinct sub-profiles within
each type, recovering historically recognizable groupings such as
trench warfare versus total war, or infant-concentrated versus
elderly-concentrated pandemics.  However, sub-clustering requires
sufficient data within each type, and disruption types with few events
may not support meaningful subdivision.

The model also assumes that the event dictionaries correctly identify
exceptional country-years.  Misclassification -- labeling a non-exceptional
year as exceptional, or missing a genuine disruption -- will bias the
profile estimates.  The event dictionaries are based on well-established
historical scholarship, but edge cases (minor conflicts, localized
epidemics) inevitably involve judgment.



We now apply the exceptional mortality model to the HMD.

\subsection{Exceptional mortality}
\label{sec:results:exceptional}

The exceptional mortality model of \cref{sec:exceptional} was applied to
all country-years flagged in the event dictionaries.  After the
exceptional-year valley smoothing (the same adaptive temporal pooling of
\cref{sec:preprocessing:smoothing} applied to exceptional spells), the
neural core, disruption profiles, and intensities were estimated as
described below.

\subsubsection{Neural core baseline estimation}

\begin{figure}[!htbp]
\centering
\includegraphics[width=\textwidth]{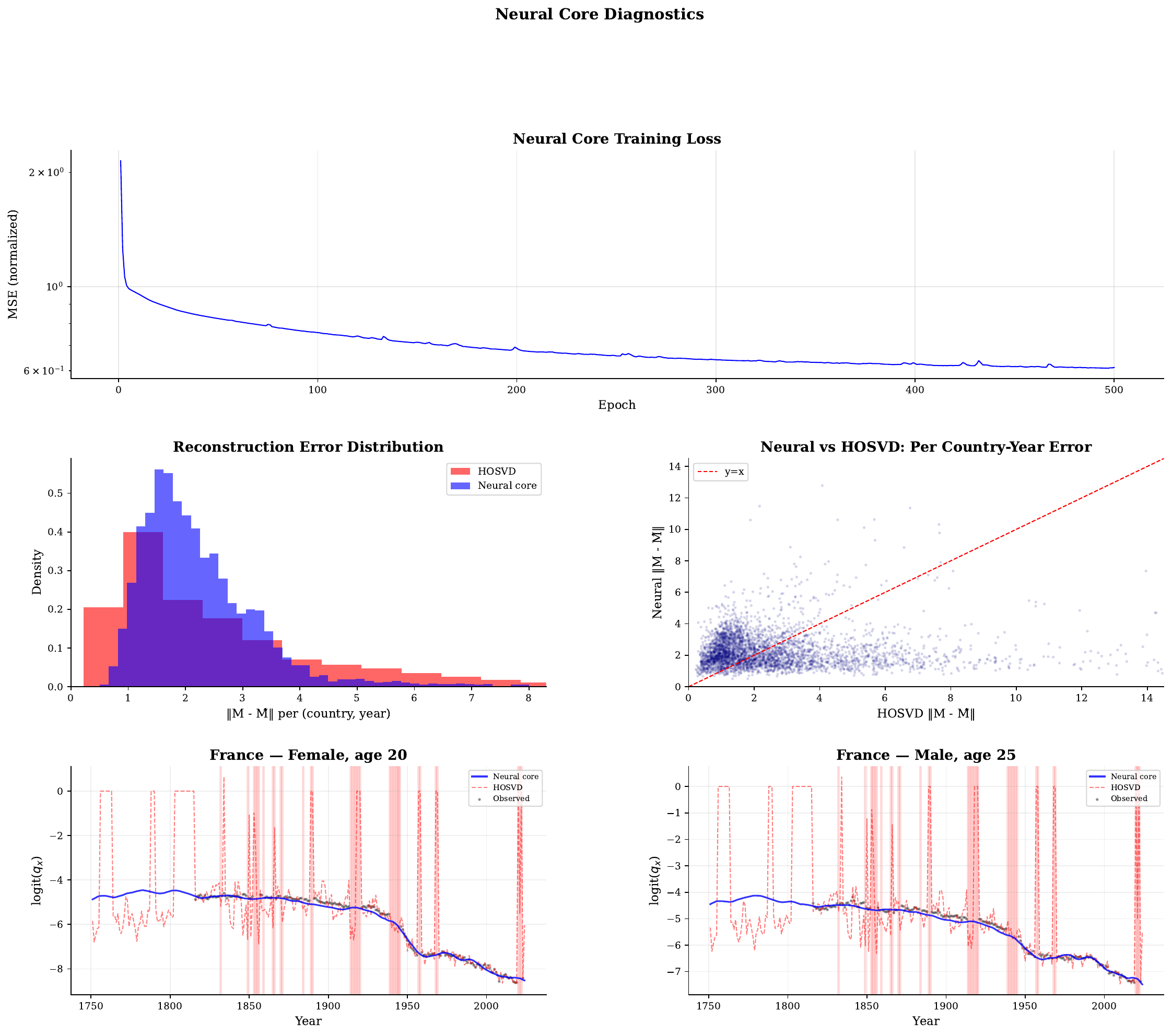}
\caption{Neural core training diagnostics.  Top: training loss curve.
Bottom left: reconstruction error distribution (neural vs.\ HOSVD).
Bottom right: neural core reconstruction at selected countries showing
smooth temporal interpolation through exceptional-year gaps.}
\label{fig:s7_neural_core_diag}
\end{figure}

The neural core MLP (\cref{eq:neural_core}) was trained on all observed
non-exceptional country-years, mapping country loadings $\bm{u}_c$ and
year features $\bm{\phi}(t)$ (\cref{eq:year_features}) to core tensor
slices.  \Cref{fig:s7_neural_core_diag} shows the training diagnostics: the
loss curve, the distribution of reconstruction errors (neural vs.\
HOSVD), and the neural core reconstruction at selected countries
illustrating smooth temporal interpolation through exceptional-year gaps.
The neural core achieves reconstruction accuracy comparable to the HOSVD
at observed non-exceptional years, confirming that it has learned the
underlying baseline structure, while providing smooth predictions at
exceptional years where the HOSVD core slices are contaminated by
imputed values.

\Cref{tab:s7_neural_core_errors} quantifies the per-cluster
reconstruction error (RMSE over logit~$\qx$) for the HOSVD and neural
core, confirming that the neural core matches or improves upon the
HOSVD baseline at every cluster.

{\setlength\LTleft{0pt}\setlength\LTright{0pt}
\begin{longtable}{@{\extracolsep{\fill}}lrrr@{}}
\caption{Core tensor reconstruction error: HOSVD vs.\ neural core (RMSE over logit $q_x$)} \label{tab:s7_neural_core_errors} \\
\toprule
Method & Mean & Median & P95 \\
\midrule
\endfirsthead
\caption[]{Core tensor reconstruction error: HOSVD vs.\ neural core (RMSE over logit $q_x$) (continued)} \\
\toprule
Method & Mean & Median & P95 \\
\midrule
\endhead
\midrule
\multicolumn{4}{r}{\textit{Continued on next page}} \\
\endfoot
\bottomrule
\endlastfoot
HOSVD & 2.853 & 1.914 & 7.419 \\
Neural core & 2.268 & 2.010 & 4.105 \\
\end{longtable}}

\begin{figure}[!htbp]
\centering
\includegraphics[width=\textwidth]{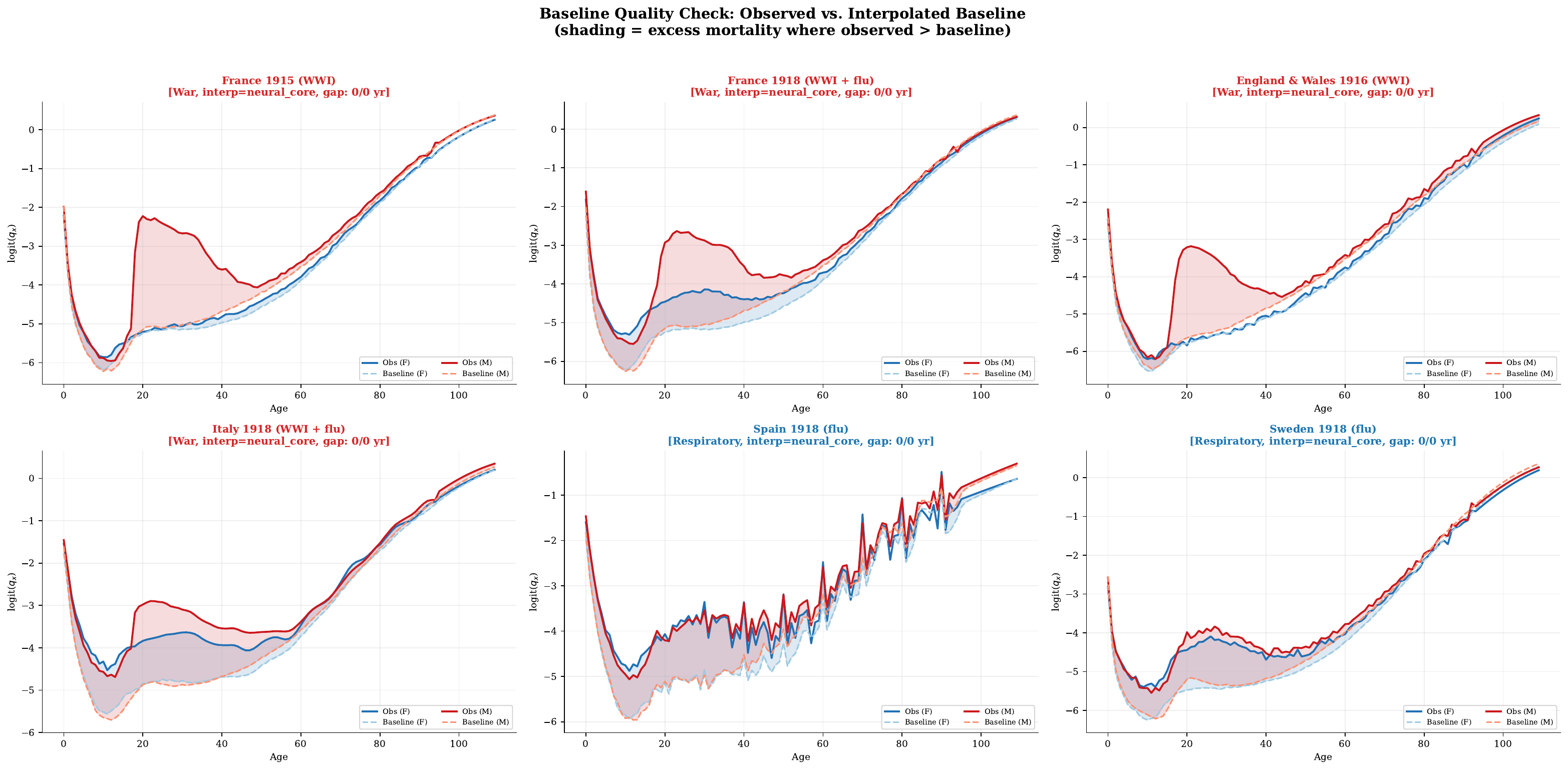}
\caption{Observed (solid) vs.\ neural core baseline (dashed) for
selected exceptional country-years.  Female (blue) and male (red)
logit($\qx$) schedules are shown.  The shaded area highlights the
residual.}
\label{fig:s7_baseline_selected}
\end{figure}

\begin{figure}[!htbp]
\centering
\includegraphics[width=\textwidth]{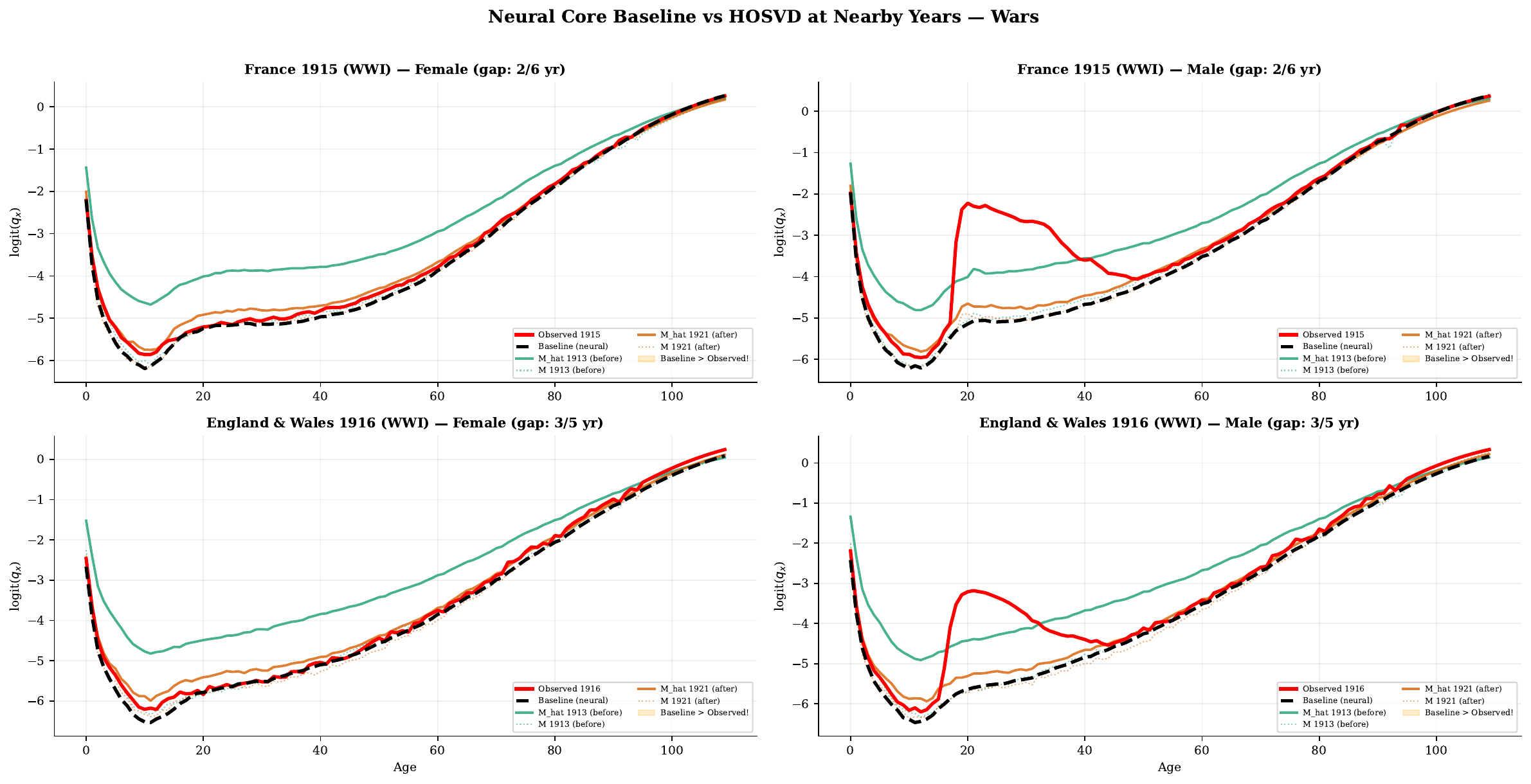}
\caption{Neural core baseline vs.\ HOSVD reconstruction at nearby
years -- war events.  Red solid = observed exceptional schedule.  Black
dashed = neural core baseline.  Green = HOSVD $\hat{\M}$ at nearest
non-exceptional year before; orange = after.  Dotted lines = actual
$\M$ at those years.}
\label{fig:s7_anchor_profiles_wars}
\end{figure}

\begin{figure}[!htbp]
\centering
\includegraphics[width=\textwidth]{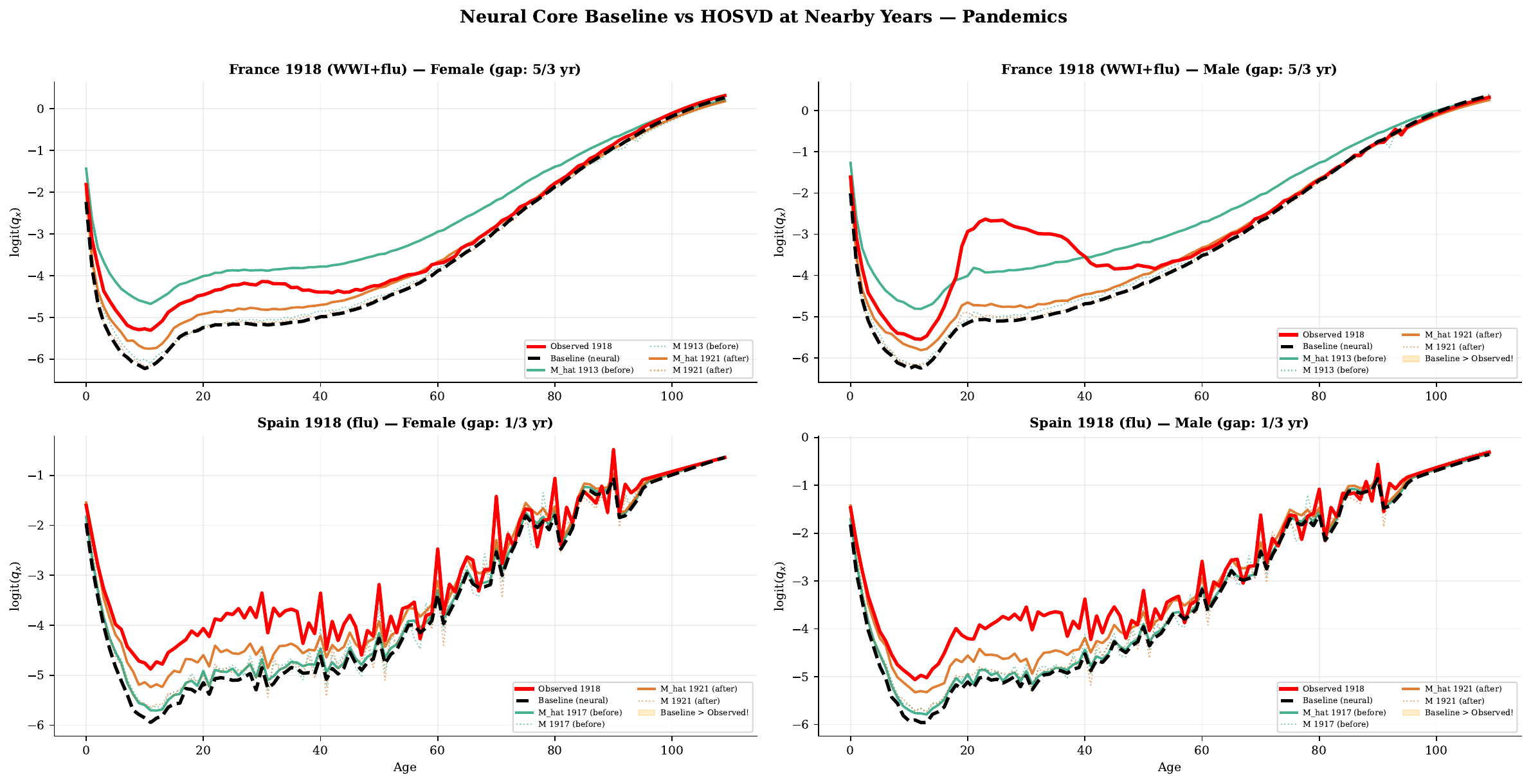}
\caption{Neural core baseline vs.\ HOSVD reconstruction at nearby
years -- pandemic events.  Layout and legend as in
\cref{fig:s7_anchor_profiles_wars}.}
\label{fig:s7_anchor_profiles_pandemics}
\end{figure}

\begin{figure}[!htbp]
\centering
\includegraphics[width=\textwidth]{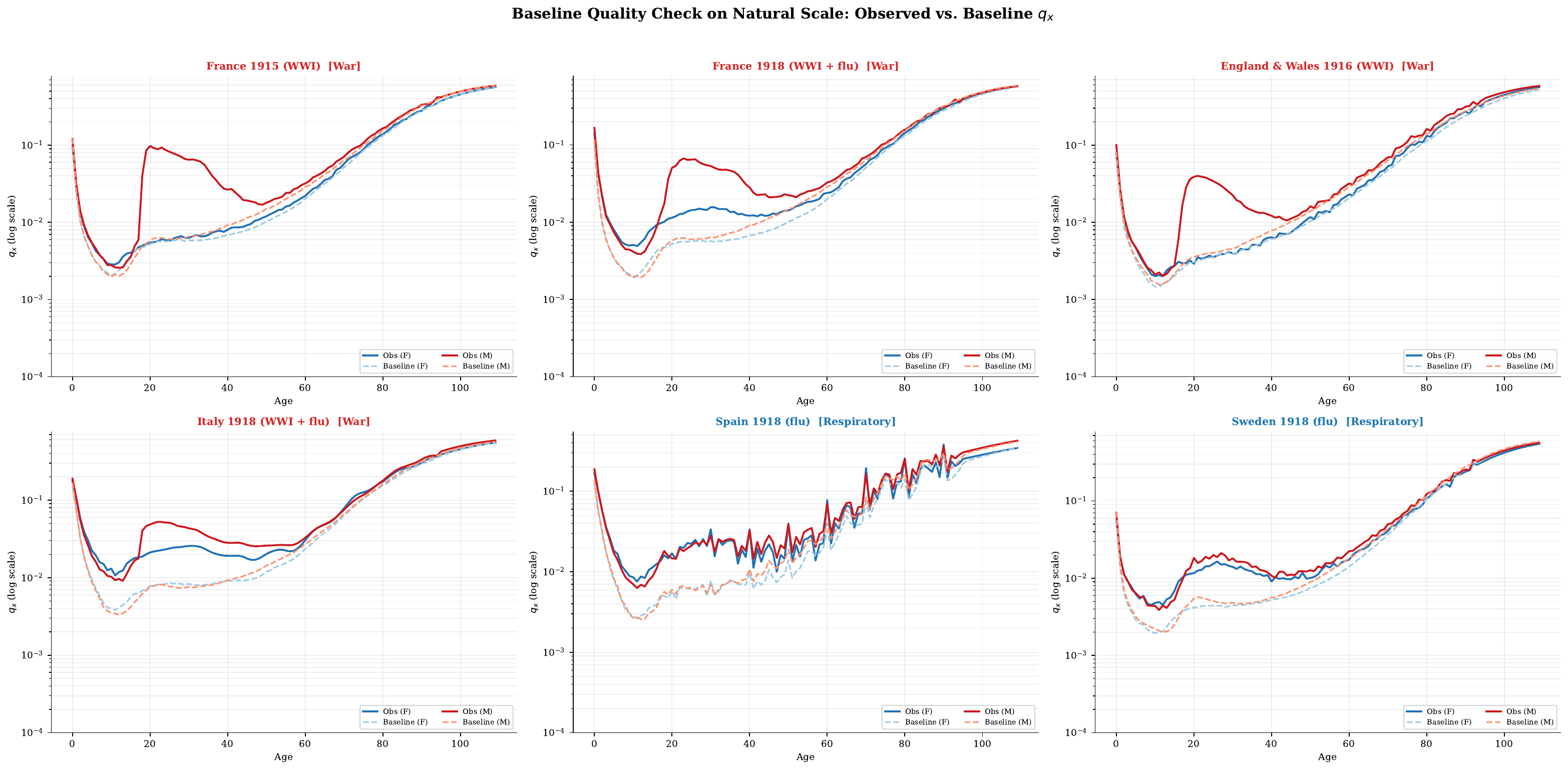}
\caption{Same comparison as \cref{fig:s7_anchor_profiles_wars} but on the
natural $\qx$ scale (log axis), revealing the absolute magnitude of
excess mortality at young-adult ages during wars.}
\label{fig:s7_baseline_natural}
\end{figure}

\Cref{fig:s7_baseline_selected} compares observed and neural core baseline
schedules for selected exceptional country-years: each panel shows one
event, with female and male logit($\qx$) schedules and the residual
(observed minus baseline) highlighted.  Good baselines track the observed
schedule at unaffected ages while sitting below it at ages where excess
mortality occurred.  \Cref{fig:s7_anchor_profiles_wars,fig:s7_anchor_profiles_pandemics}
provide a further validation by comparing the neural core baseline at
exceptional years with the HOSVD reconstruction at nearby
non-exceptional years, confirming temporal continuity and the absence of
the contamination bias described in \cref{sec:exceptional:naive}.
\Cref{fig:s7_baseline_natural} repeats this comparison on the natural
$\qx$ scale, where the absolute magnitude of excess mortality --
especially the dramatic spikes at young-adult ages during wars -- is most
clearly visible.

\begin{figure}[!htbp]
\centering
\includegraphics[width=\textwidth]{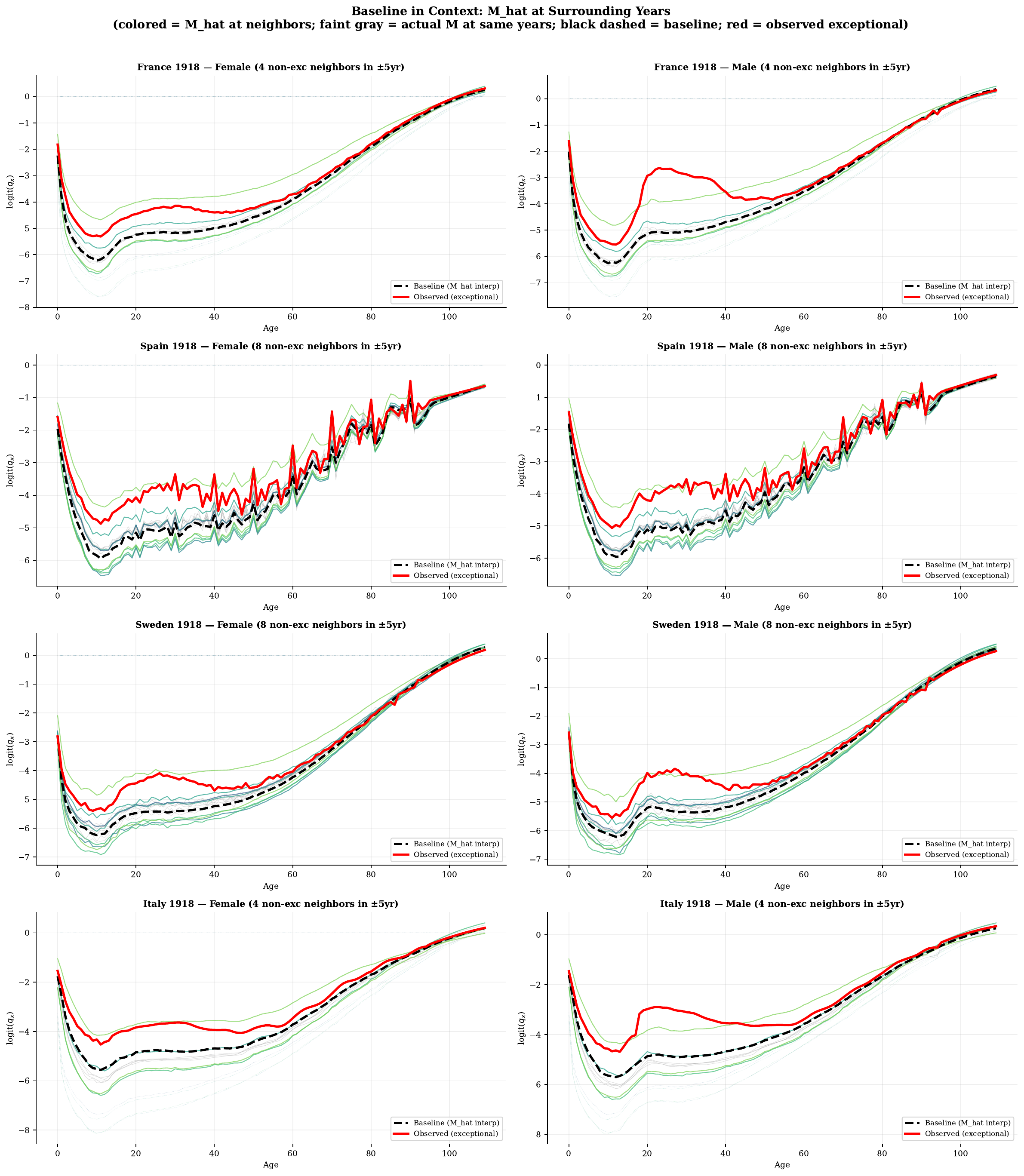}
\caption{Temporal context for baseline estimation.  Thin colored lines =
HOSVD $\hat{\M}$ at surrounding non-exceptional years; black dashed =
neural core baseline; red solid = observed exceptional schedule.}
\label{fig:s7_baseline_context}
\end{figure}

\Cref{fig:s7_baseline_context} provides temporal context for the baseline
estimation: for each event, thin colored lines show the HOSVD
reconstruction at surrounding non-exceptional years, the black dashed
line shows the neural core baseline, and the red solid line shows the
observed exceptional schedule.  The neural core baselines sit within the
envelope of neighboring non-exceptional reconstructions, as expected
from the temporal interpolation argument of
\cref{sec:exceptional:temporal}.

\subsubsection{Disruption profiles and intensities}
\label{sec:results:profile_results}

\begin{figure}[!htbp]
\centering
\includegraphics[width=\textwidth]{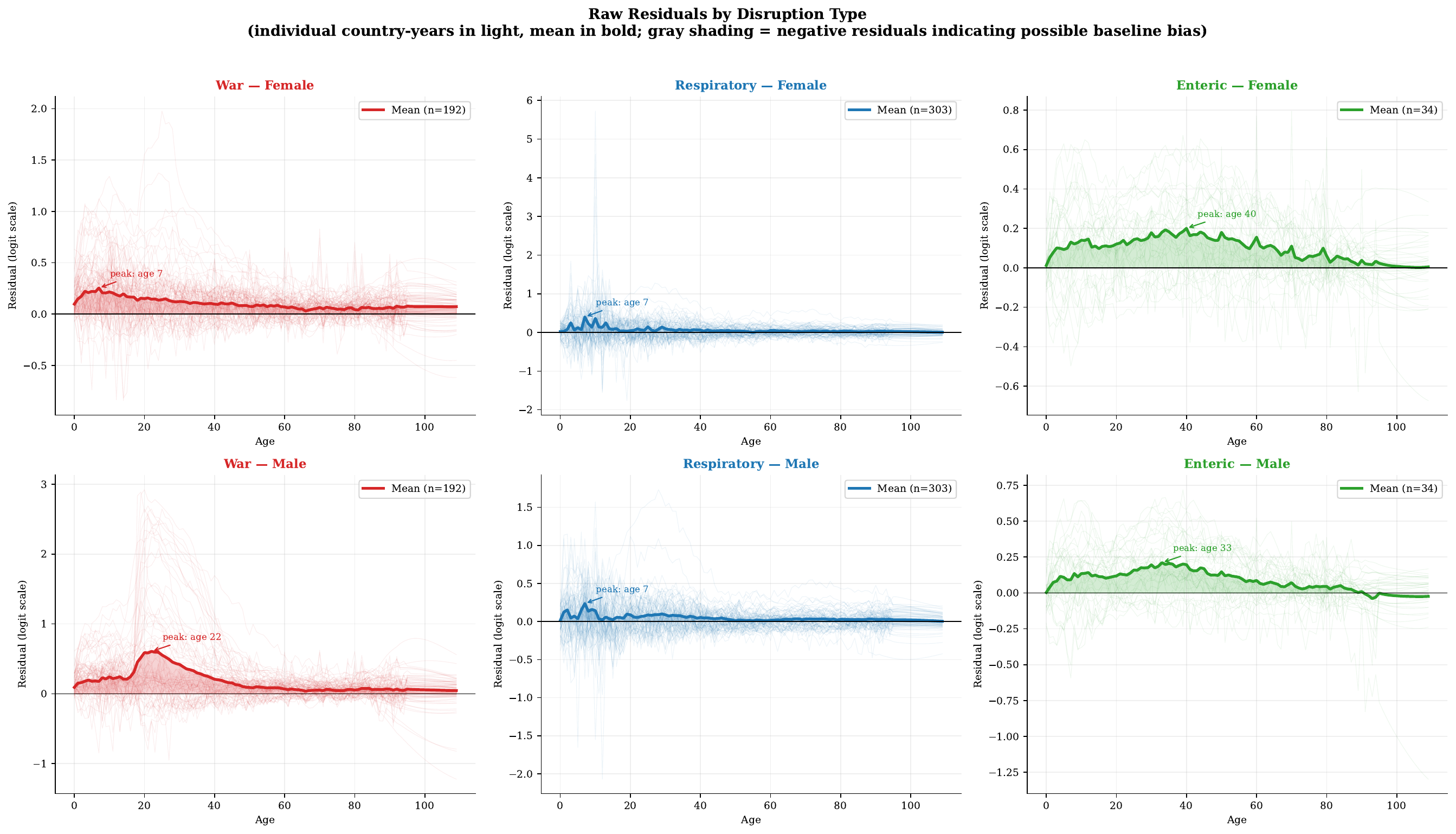}
\caption{Mean residual (observed $-$ baseline) on the logit scale by
disruption type.  Each heatmap shows the average excess across all
country-years of that type.  War residuals concentrate in young-adult
males; respiratory residuals are broadly distributed.}
\label{fig:s7_residual_heatmaps}
\end{figure}

\begin{figure}[!htbp]
\centering
\includegraphics[width=\textwidth]{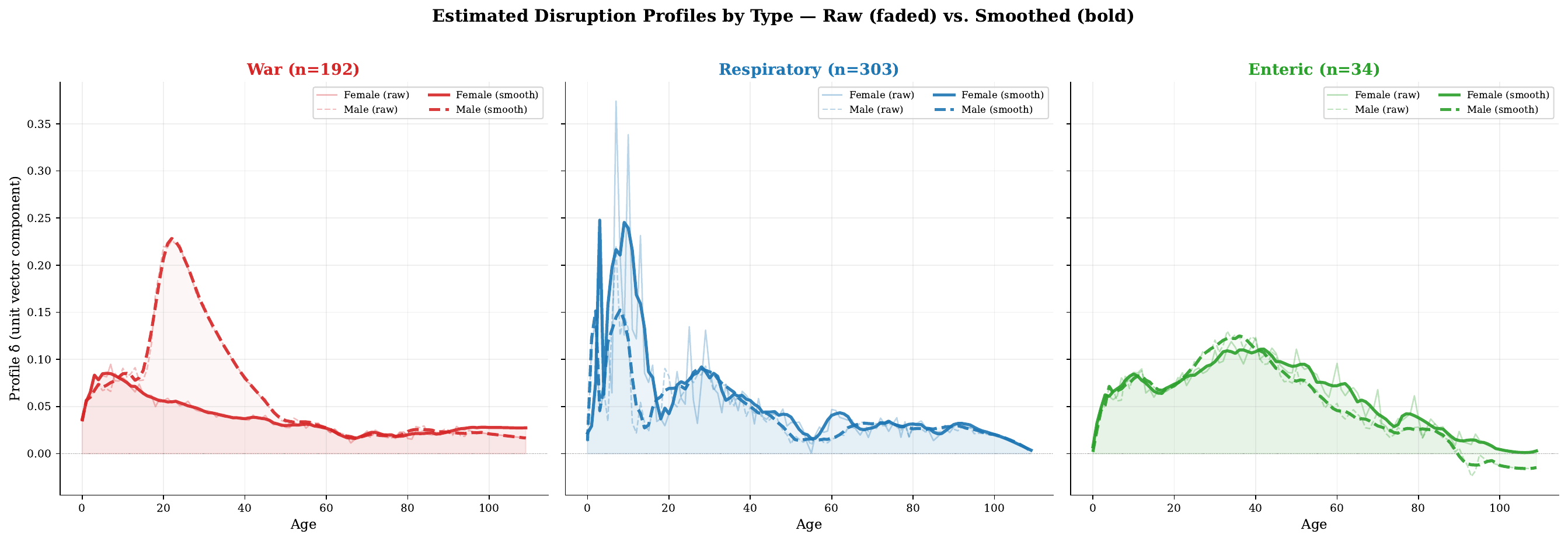}
\caption{Estimated disruption profiles by type.  Each panel shows the
smoothed profile $\hat{\bm{\delta}}_d$ (unit vector) for one disruption
type, split by sex, with the raw (unsmoothed) profile overlaid for
comparison.  The war profile concentrates in young-adult males; the
respiratory profile is broad; the enteric profile shows a
broader-than-expected elevation across young and adult ages.}
\label{fig:s7_disruption_profiles}
\end{figure}

\Cref{fig:s7_residual_heatmaps} displays mean residual heatmaps by
disruption type (\cref{sec:exceptional:residual_def}), with age on the
horizontal axis and country-years stacked vertically.  The type-specific
patterns anticipated in \cref{sec:exceptional:profiles} emerge clearly
from the data.  \Cref{fig:s7_disruption_profiles} presents the estimated
disruption profiles $\hat{\bm{\delta}}_d$
(\cref{eq:profile_smooth}) for each type, split by sex.  Both the raw
profiles (\cref{eq:profile_norm}) and the Savitzky--Golay-smoothed
profiles (\cref{sec:exceptional:profile_smooth}) are shown; the two
are nearly indistinguishable visually (cosine similarity above 0.99),
confirming that the smoothing removes only high-frequency noise without
altering the substantive disruption pattern.  The war profile
concentrates sharply in young-adult males (ages 15--45), as expected
from combat mortality.  The respiratory pandemic profile is broadly
distributed across ages in both sexes, reflecting the nonspecific nature
of airborne pathogens.

The enteric profile is more surprising.  Rather than the
infant-and-child-concentrated pattern expected on epidemiological grounds
-- dehydration from diarrheal disease disproportionately kills the very
young -- the estimated profile shows a broad elevation spanning young
through adult ages.  This broader-than-expected pattern may reflect
heterogeneity in the events classified as ``enteric'' in the event
dictionary: the label encompasses nineteenth-century cholera pandemics,
typhus outbreaks, and other waterborne disease episodes that may have
had different age patterns of excess mortality, and some of these events
may have co-occurred with other causes of excess mortality (famine,
social disruption) that affect adults.  The finding warrants further
investigation of the event classification and the underlying
cause-of-death structure, but for the purposes of the disruption model
the estimated profile captures the empirical pattern in the data
regardless of its epidemiological interpretation.

\begin{figure}[!htbp]
\centering
\includegraphics[width=\textwidth]{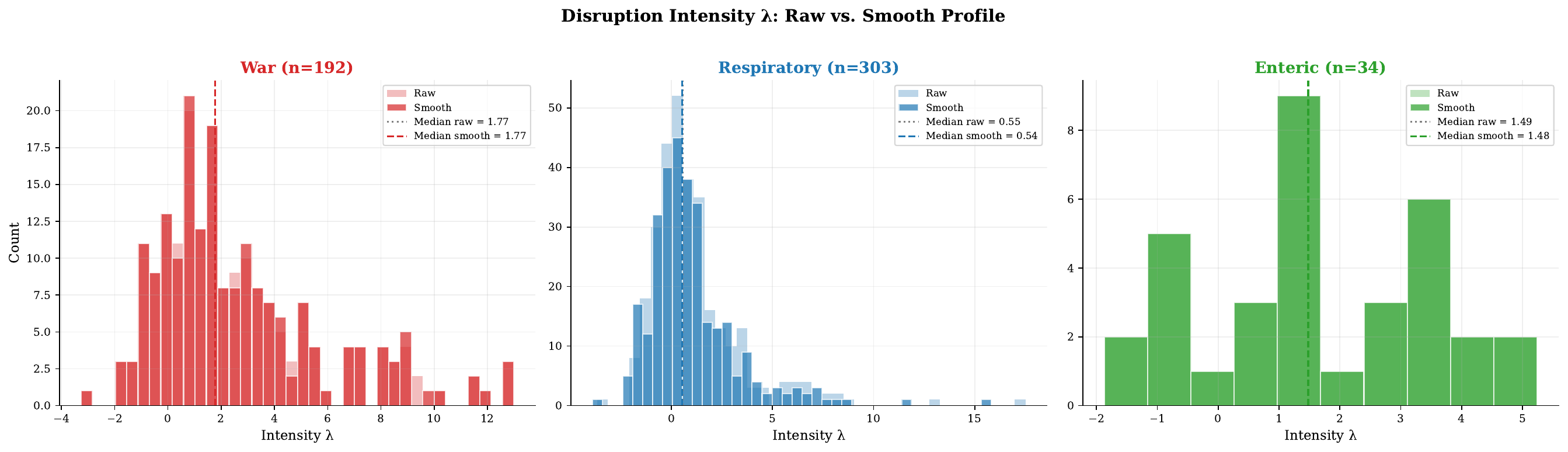}
\caption{Distribution of estimated disruption intensities
$\hat{\lambda}_{c,t}$ by type.  Higher intensity means more severe
disruption.  Values near zero indicate mild or marginal events.}
\label{fig:s7_intensity_dist}
\end{figure}

\Cref{fig:s7_intensity_dist} shows the distribution of estimated disruption
intensities $\hat{\lambda}_{c,t}$ (\cref{eq:intensity}) by type.  Higher
intensity indicates a more severe mortality disruption.  The war
distribution shows the greatest spread, reflecting the enormous range of
conflict severity -- from peripheral involvement to total war.  Values
near zero indicate country-years that were flagged in the event
dictionaries but experienced only mild excess mortality of the expected
type.

\subsubsection{Disruption sub-clustering}

\begin{figure}[!htbp]
\centering
\includegraphics[width=\textwidth]{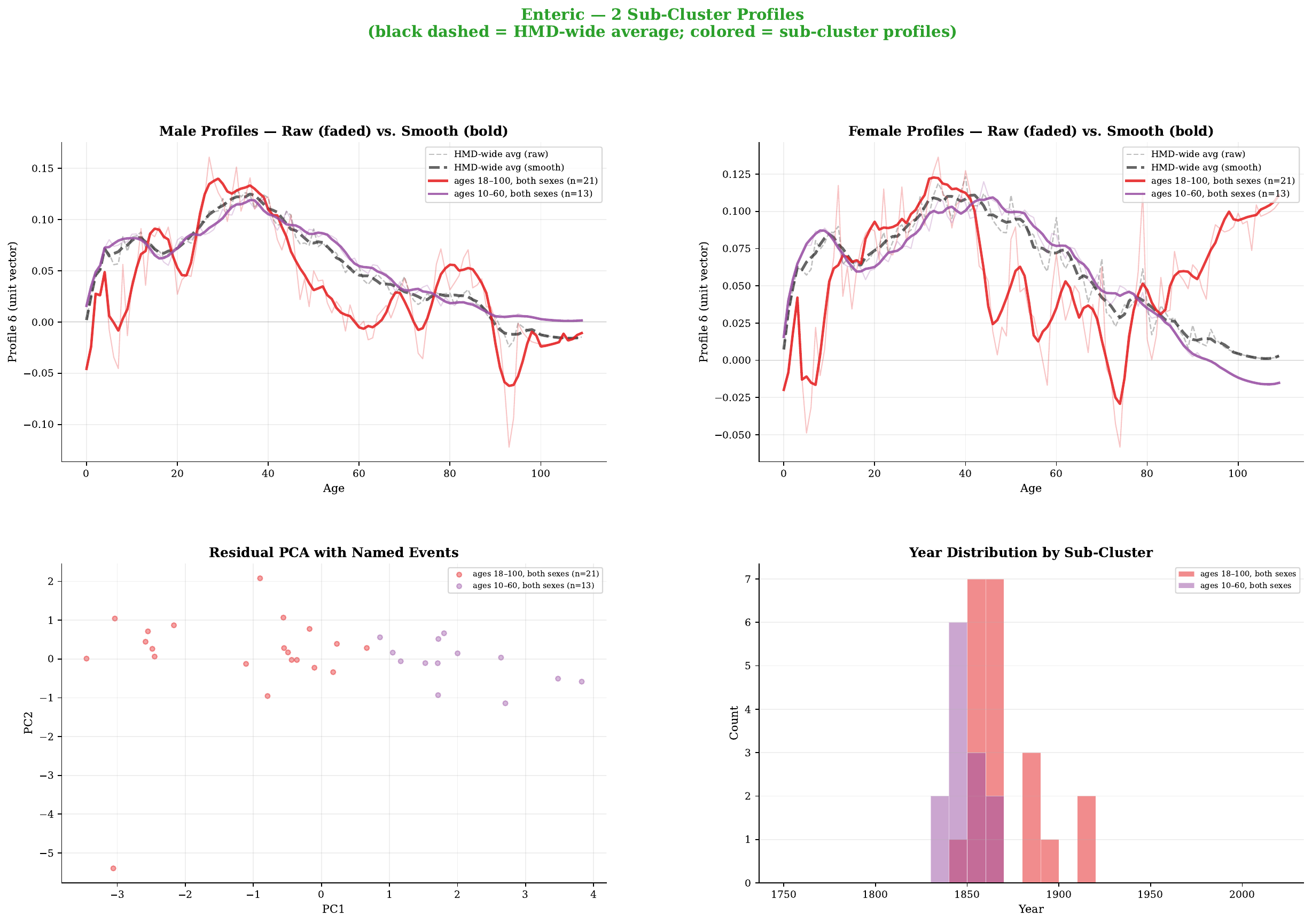}
\caption{Disruption sub-cluster profiles.  For each type: top-left =
male profiles (solid = sub-clusters, dashed = HMD-wide profile);
top-right = female profiles; bottom-left = PCA scatter with labeled
events; bottom-right = year distribution per sub-cluster.}
\label{fig:s7_disruption_subclusters}
\end{figure}

The sub-clustering procedure of \cref{sec:exceptional:subclustering} was
applied within each disruption type.  \Cref{fig:s7_disruption_subclusters}
presents the results: for each type, the figure shows the male and female
sub-cluster profiles alongside the single HMD-wide profile, a PCA scatter
of residuals annotated with named historical events, and the year
distribution per sub-cluster.  The sub-clusters are historically
interpretable.  Among wars, distinct sub-profiles emerge that correspond
to qualitatively different conflict types -- separating pre-modern from
modern warfare and distinguishing conflicts with primarily
military casualties from those with significant civilian mortality.
Among respiratory pandemics, the sub-clustering distinguishes events with
different age selectivity, consistent with the known epidemiological
differences between pathogen strains.  The improvement in explained
variance relative to the single-profile model confirms the within-type
heterogeneity anticipated in \cref{sec:exceptional:subclustering}.

\Cref{tab:s7_subclustering} reports the variance explained ($R^2$)
by the single HMD-wide disruption profile versus the neural sub-cluster
profiles for each disruption type, quantifying the improvement from
sub-clustering.

{\setlength\LTleft{0pt}\setlength\LTright{0pt}
\begin{longtable}{@{\extracolsep{\fill}}lcrrr@{}}
\caption{Variance explained ($R^2$): single disruption profile vs.\ neural sub-cluster profiles} \label{tab:s7_subclustering} \\
\toprule
Disruption type & $K_{\mathrm{sub}}$ & Single $R^2$ & Sub-cluster $R^2$ & $\Delta$ (\%) \\
\midrule
\endfirsthead
\caption[]{Variance explained ($R^2$): single disruption profile vs.\ neural sub-cluster profiles (continued)} \\
\toprule
Disruption type & $K_{\mathrm{sub}}$ & Single $R^2$ & Sub-cluster $R^2$ & $\Delta$ (\%) \\
\midrule
\endhead
\midrule
\multicolumn{5}{r}{\textit{Continued on next page}} \\
\endfoot
\bottomrule
\endlastfoot
war & 2 & 0.457 & 0.475 & +4.0 \\
respiratory & 2 & 0.230 & 0.250 & +8.9 \\
enteric & 2 & 0.575 & 0.527 & -8.3 \\
\end{longtable}}

\subsubsection{Model fit and residual quality}

\begin{figure}[!htbp]
\centering
\includegraphics[width=\textwidth]{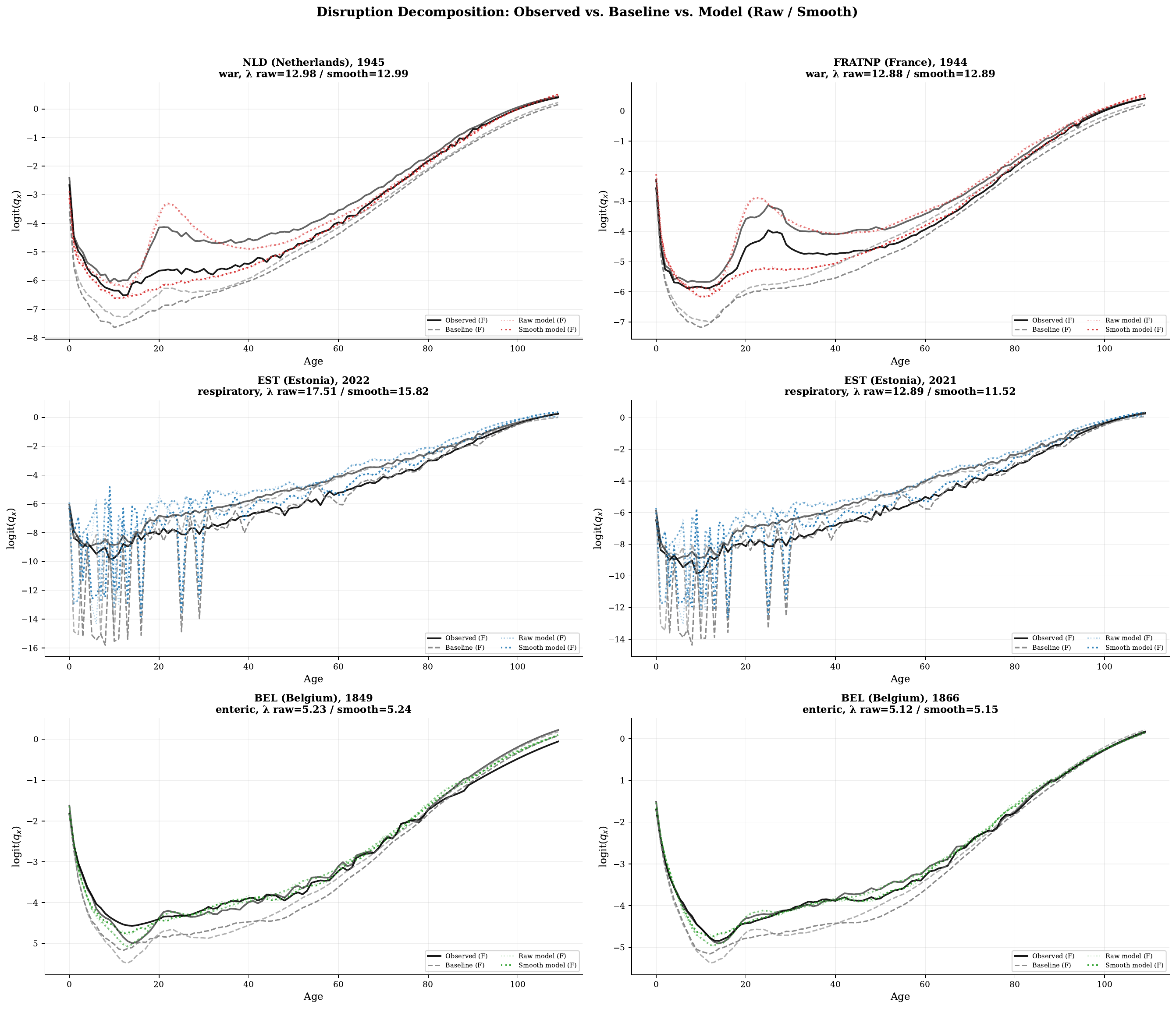}
\caption{Example disruption decompositions for selected exceptional
country-years.  Solid = observed; dashed = neural core baseline;
dotted = baseline + $\hat{\lambda}\hat{\bm{\delta}}_d$ (model).}
\label{fig:s7_residual_examples}
\end{figure}

\begin{figure}[!htbp]
\centering
\includegraphics[width=\textwidth]{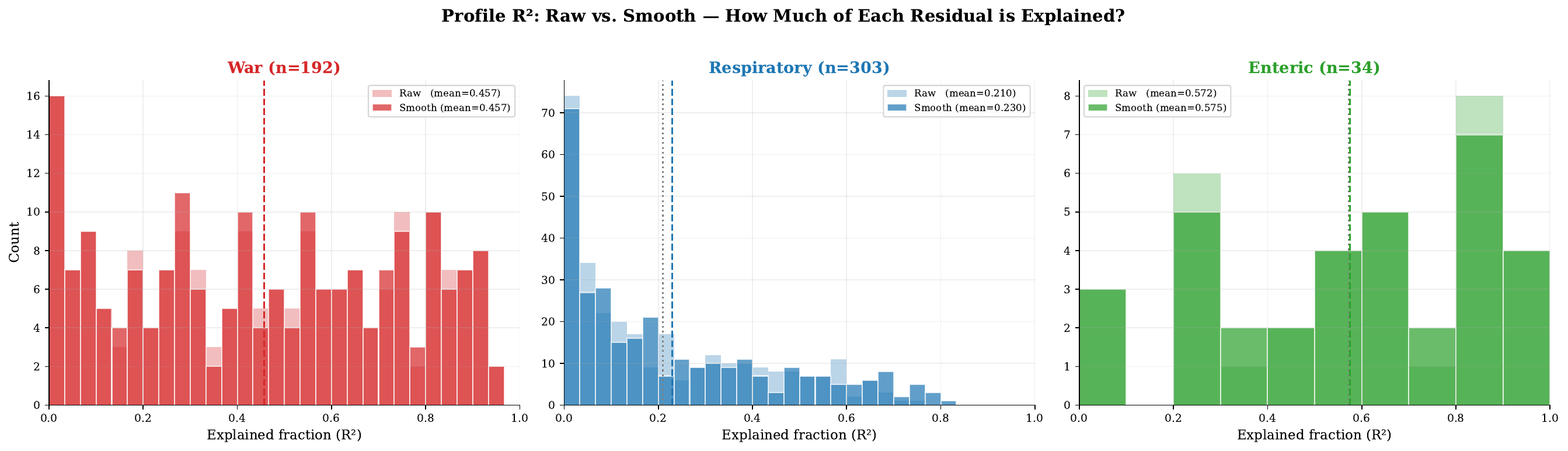}
\caption{Fraction of residual variance explained by the canonical
disruption profile, by type.  Higher values indicate better fit of the
single-profile model.}
\label{fig:s7_explained_fraction}
\end{figure}

\Cref{fig:s7_residual_examples} shows example disruption decompositions for
selected exceptional country-years: the observed schedule (solid), the
neural core baseline (dashed), and the model-predicted schedule
(baseline + $\hat{\lambda}\hat{\bm{\delta}}_d$, dotted).  The separable
model (\cref{eq:disruption_model}) captures the characteristic shape of
each disruption type.  \Cref{fig:s7_explained_fraction} reports the fraction
of residual variance explained by the canonical disruption profile
(\cref{eq:profile_norm}) for each type -- a direct measure of how well
the single-profile assumption of \cref{sec:exceptional:model} holds.

\begin{figure}[!htbp]
\centering
\includegraphics[width=\textwidth]{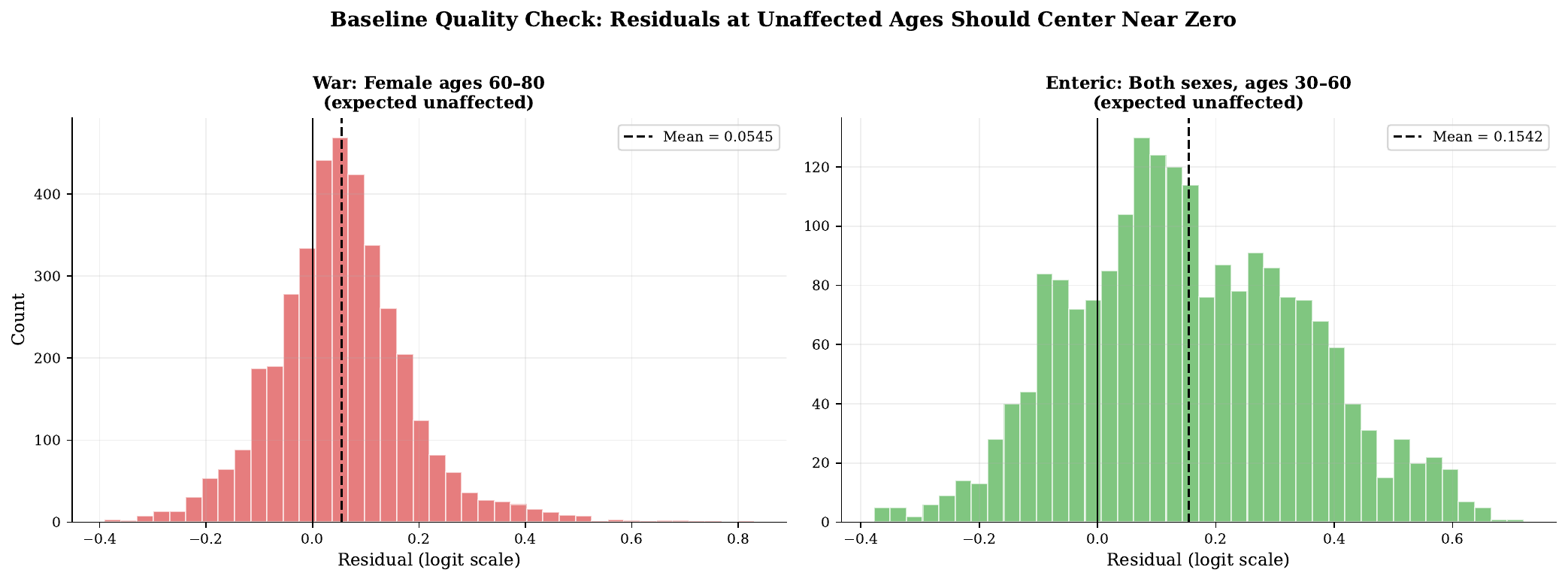}
\caption{Residual distributions at ages expected to be unaffected by
the disruption.  Left: war residuals at ages 60--80 (females only).
Right: enteric residuals at ages 70+.  The distributions center near
zero, confirming unbiased baseline estimation.}
\label{fig:s7_residual_quality}
\end{figure}

The quality of the baseline estimation is assessed in
\cref{fig:s7_residual_quality}, which examines residual distributions at
ages expected to be \emph{unaffected} by the disruption.  For wars, this
means ages 60--80 among females (where no combat deaths are expected);
for enteric pandemics, this means ages 70 and above (beyond the range of
the observed broad elevation).  If the baseline estimation is unbiased,
these residuals should center near zero -- in contrast with the
systematic negative bias expected from the na\"{\i}ve projection approach
described in \cref{sec:exceptional:naive}.  The residual distributions
confirm that the neural core baselines avoid this bias, centering near
zero at unaffected ages.

\subsubsection{Impact of disruptions on life expectancy}

\begin{figure}[!htbp]
\centering
\includegraphics[width=\textwidth]{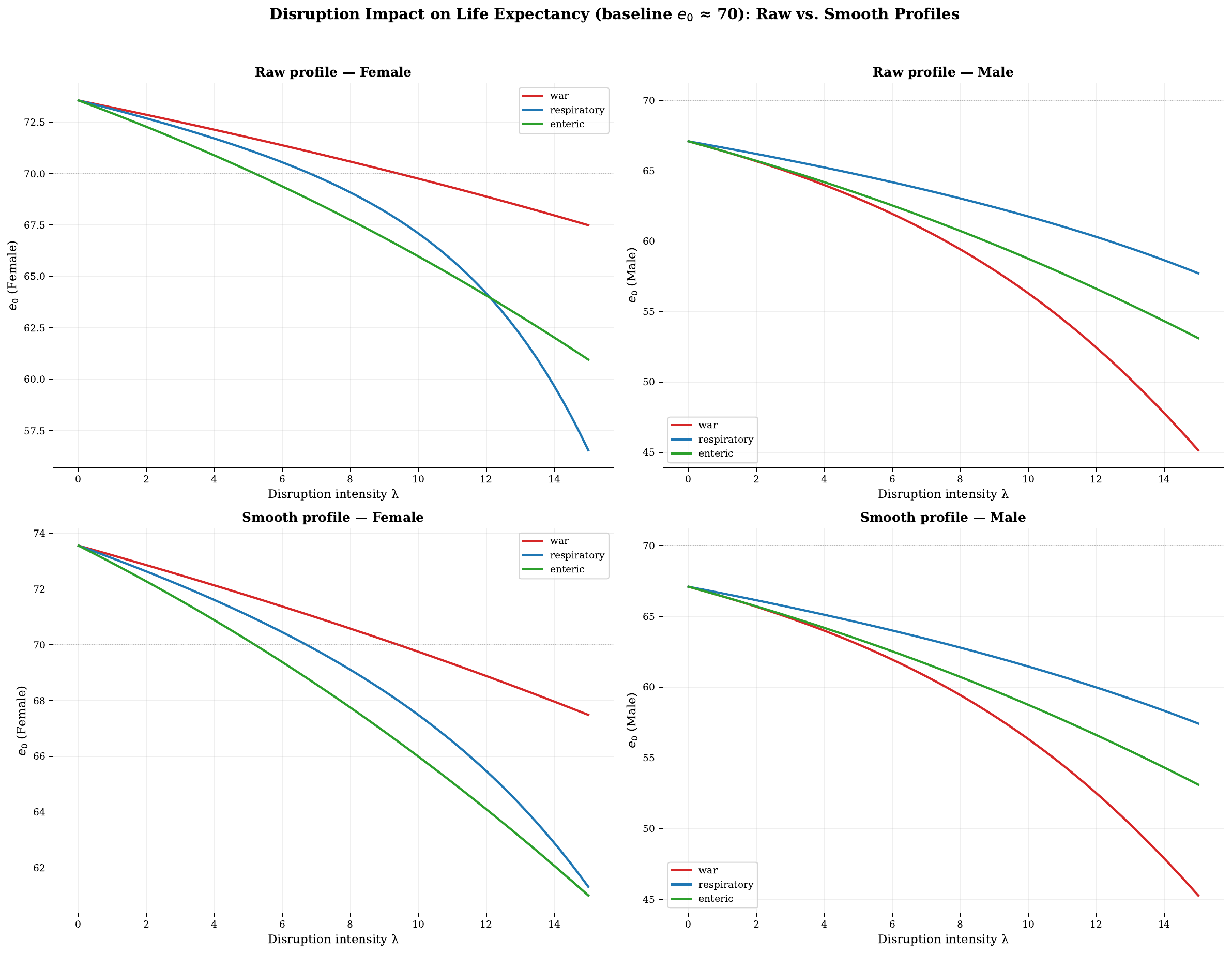}
\caption{Impact of disruption on life expectancy by type and intensity.
Each curve shows how female (solid) and male (dashed) $\ezero$ decline
as intensity increases, applied to a baseline of $\ezero \approx 70$.
War produces the sharpest male-specific decline.}
\label{fig:s7_disruption_impact}
\end{figure}

\Cref{fig:s7_disruption_impact} illustrates the impact of each disruption
type on life expectancy as a function of intensity, applied to a baseline
of approximately $\ezero = 70$.  Each curve shows how female and male
$\ezero$ decline as intensity $\lambda$ increases
(\cref{sec:exceptional:complete}).  War produces the sharpest decline in
male $\ezero$ with a much smaller female effect, consistent with the
combat-concentrated war profile.  Respiratory pandemics reduce $\ezero$
for both sexes roughly equally.  Enteric pandemics reduce $\ezero$
through a broad elevation of mortality across young and adult ages,
producing a moderate decline in both sexes.

\section{Life Table Fitting}
\label{sec:fitting}

The preceding sections have developed the MDMx framework as a
\emph{forward model}: given a cluster label~$k$, a target life
expectancy~$\estar$, and an optional disruption specification
$(d, \lambda)$, the system produces a complete sex-specific mortality
schedule via the generative equation
(\cref{eq:full_model}).  We now address the \emph{inverse problem}:
given only an observed schedule~$\yobs$, estimate the parameters
$(k, \estar, d, \lambda)$.  We call this the \emph{life table fitting
problem} because the schedule arrives without country, year, or
temporal context -- the primary use case for applications where a user
provides a single mortality schedule and requests the best-fit
decomposition from the model.

\subsection{The three-stage algorithm}
\label{sec:fitting:algorithm}

The fitter decomposes the life table fitting problem into three stages
that address progressively deeper questions about the observed schedule.
The first question is structural: \emph{which mortality regime does this
schedule belong to, and at what mortality level?}  This is answered by a
coarse grid search across all cluster--$\ezero$ combinations, using a
linearized model that is cheap enough to evaluate exhaustively.  The
second question is whether a disruption is present, and if so, which
type: the same grid search simultaneously tests each disruption type
against the null hypothesis of no disruption, selecting the best
$(k, d, \ezero)$ combination by BIC.  The third question is precise
calibration: \emph{exactly how much disruption is there, and does the
evidence justify including it at all?}  This is answered by
Gauss--Newton refinement of $\ezero$ followed by an exact (non-linearized)
evaluation with Bayes-factor model comparison.

The three stages thus perform a nested model selection.  The fitter does
not assume that a disruption is present -- it discovers whether one is
needed, identifies which type best explains the residual pattern, and
estimates its intensity, all from a single observed schedule with no
temporal or geographic context.  This is a substantially harder problem
than the forward model (which is given the disruption specification as
input), and the estimation procedure must be both fast enough for
interactive use and statistically principled enough to avoid false
positives.

\subsubsection{Stage~1: Linearized grid search}
\label{sec:fitting:stage1}

For fixed $(k, \eref)$, we linearize the trajectory around $\eref$:
\begin{equation}
  \label{eq:fitting:linearized}
  \yobs \approx \zk(\eref) + \De \cdot \tk(\eref) + \lambda \cdot \dd
       + \boldsymbol{\varepsilon},
\end{equation}
where $\tk(\eref) = \partial \zk / \partial \ezero \big|_{\eref}$ is the
trajectory tangent.  This is a linear regression with design matrix
$\mathbf{X} = [\tk \mid \dd]$ and response
$\mathbf{r} = \yobs - \zk(\eref)$.  The ordinary least squares (OLS)
solution gives $(\De, \lhat)$ in closed form via a $2 \times 2$ matrix
inverse.

By the Frisch--Waugh--Lovell theorem \citep{FrischWaugh1933,Lovell1963},
$\lhat$ reflects only the component of the disruption signal orthogonal
to the trajectory tangent.  This provides automatic protection against
the \emph{absorption problem}: when a disruption profile is partially
collinear with $\tk$, a shift in $\estar$ can absorb part of the
disruption signal.  The regression partitions variance correctly without
explicit orthogonalization.

For efficiency, the computation is vectorized.  For each cluster~$k$,
$\zk$ and~$\tk$ are evaluated at all $N_{\ezero} = 150$ grid points
simultaneously using array interpolation, and the regressions for
$d = 0$ and $d = 1,2,3$ are solved via batch linear algebra.  The total
cost is $K \times 4$ batch operations, each processing 150 regressions
in parallel.  Model selection at this stage uses BIC
\citep{Schwarz1978} to identify the most promising $(k, d, \eref)$
combinations for refinement.

\subsubsection{Stage~2: Gauss--Newton refinement}
\label{sec:fitting:stage2}

The linearization~\eqref{eq:fitting:linearized} is accurate for
$|\De| \lesssim 5$ years.  For larger shifts, the procedure iterates: at
the best $(k, d)$ from Stage~1, the reference point is updated
$\eref \leftarrow \eref + \De$, and the regression is re-solved at the
new reference.  Three iterations suffice for convergence to
$|\De| < 0.3$~years in all cases tested.

\subsubsection{Stage~3: Exact evaluation with Bayes factor}
\label{sec:fitting:stage3}

After refinement, the trajectory is evaluated \emph{exactly} at the
converged $\estar$ -- no linearization.  The residual is
\begin{equation}
  \label{eq:fitting:exact-resid}
  \mathbf{r} = \yobs - \zk(\estar),
\end{equation}
and the disruption intensity is estimated by projection:
\begin{equation}
  \label{eq:fitting:lam-proj}
  \lhat = \frac{\dd^\top \mathbf{r}}{\dd^\top \dd},
  \qquad \text{constrained to } \lhat \geq 0.
\end{equation}
Crucially, the trajectory tangent $\tk$ does not appear in the Stage~3
model.  This eliminates the collinearity that inflated false positives
in the linearized model.

A \emph{multi-disruption projection} is also computed by regressing
$\mathbf{r}$ onto $[\dd[1] \mid \dd[2] \mid \dd[3]]$ simultaneously.
This prevents one disruption type from absorbing variance that belongs
to another and reveals compound events (e.g., World War~I combined with
the 1918 influenza pandemic).

\paragraph{Model comparison.}
For each disruption hypothesis $d > 0$ versus $d = 0$, a
Laplace-approximated log Bayes factor \citep{KassRaftery1995,
TierneyKadane1986} is computed:
\begin{equation}
  \label{eq:fitting:log-bf}
  \log \mathrm{BF}(d \text{ vs.\ } 0)
    \approx \frac{p}{2} \log \frac{\mathrm{RSS}_0}{\mathrm{RSS}_d}
    - \frac{\lhat^2}{2\sigma_\lambda^2}
    - \frac{1}{2} \log I_{\lambda\lambda}
    + \frac{1}{2} \log 2\pi,
\end{equation}
where $\mathrm{RSS}_0$ and $\mathrm{RSS}_d$ are residual sums of
squares under the null and disruption models, $\sigma_\lambda$ is the
scale of a half-normal prior on~$\lambda$, and
$I_{\lambda\lambda} = \dd^\top \dd / \hat\sigma^2$ is the observed
Fisher information for~$\lambda$.  The three terms have distinct roles:
the first captures the likelihood improvement from adding the disruption
component; the second imposes the prior cost of the estimated intensity;
and the third is the Occam factor, penalizing models where $\lambda$ is
precisely determined.  Unlike the fixed $\log p$ penalty in BIC, the
Occam factor scales with the \emph{shape} of the likelihood surface,
providing better calibration for the disruption detection problem.

The $\ezero$ gap $\Delta_{\mathrm{gap}} = \estar(d{>}0) - \estar(d{=}0)$
is also computed: the difference in fitted life expectancy between the
disruption and null models for the same cluster.  A large gap indicates
that the null model had to shift $\estar$ downward to accommodate excess
mortality that the disruption model explains at the correct $\estar$.
An optional gap threshold can be applied as an additional filter.

\subsection{Identifiability analysis}
\label{sec:fitting:identifiability}

The key challenge in life table fitting is \emph{partial
non-identifiability} between the disruption profiles and the trajectory
tangent.  When $\dd$ is approximately parallel to $\tk$ at some~$\estar$,
a shift in life expectancy can mimic the disruption signal.

We quantify this using the correlation between $\tk$ and $\dd$, computed
across all $(k, \estar)$ pairs.  The \emph{orthogonal fraction}
$\|\dd^\perp\| / \|\dd\|$, where
$\dd^\perp = \dd - (\dd \cdot \hat{\tk}) \hat{\tk}$, measures the
identifiable component.

{\setlength\LTleft{0pt}\setlength\LTright{0pt}
\begin{longtable}{@{\extracolsep{\fill}}lrrr@{}}
\caption{Identifiability of disruption types.  Median values across all $(k, \ezero)$ pairs.} \label{tab:s8_identifiability} \\
\toprule
Type & Median $|\rho(\mathbf{t}_k, \boldsymbol{\delta}_d)|$ & Median VIF & Orthogonal fraction \\
\midrule
\endfirsthead
\caption[]{Identifiability of disruption types.  Median values across all $(k, \ezero)$ pairs. (continued)} \\
\toprule
Type & Median $|\rho(\mathbf{t}_k, \boldsymbol{\delta}_d)|$ & Median VIF & Orthogonal fraction \\
\midrule
\endhead
\midrule
\multicolumn{4}{r}{\textit{Continued on next page}} \\
\endfoot
\bottomrule
\endlastfoot
War & 0.81 & 2.9 & 0.58 \\
Respiratory & 0.84 & 3.4 & 0.55 \\
Enteric & 0.91 & 5.8 & 0.41 \\
\end{longtable}}

All three disruption types show substantial correlation with the
trajectory tangent ($|\rho|$ ranging from 0.74 for war to 0.85 for
respiratory), reflecting a fundamental property
of mortality space: any vector that increases mortality across ages (as
all disruption profiles do) correlates with the tangent direction, which
also represents a general mortality increase as $\estar$ decreases.
War profiles retain the largest identifiable component (58\%) because
the distinctive young-adult male mortality hump has no analog in the
smooth trajectory.  Respiratory and enteric profiles are more difficult
to distinguish from trajectory shifts, with orthogonal fractions of 55\%
and 41\% respectively.  This is an intrinsic limitation of the single-schedule
problem, not a fitting failure.

\subsection{Cross-validation and evaluation}
\label{sec:fitting:evaluation}

\subsubsection{Data}

The fitter is evaluated on HMD ground truth: 500 randomly sampled
non-exceptional country-years (for false positive estimation) and all
529 exceptional country-years (192~war, 303~respiratory, 34~enteric).
Among the exceptional cases, 263 have strong disruptions ($\lambda > 1$):
124~war, 116~respiratory, 23~enteric.

\subsubsection{Cross-validation protocol}
\label{sec:fitting:cv}

The two tuning parameters -- $\sigma_\lambda$ (prior scale) and the
$\ezero$ gap threshold -- are selected via 5-fold stratified
cross-validation.  On each fold, a grid of $(\sigma_\lambda,
\text{gap})$ values is swept on the training set (maximizing strong
accuracy subject to a false positive budget), then evaluated on the
held-out fold.  The consensus selection across folds determines the
final parameters.

The fitting itself (Stages 1--3) is performed once on all data; only the
decision thresholds are cross-validated.  This is valid because the
thresholds affect only how the stored Bayes factors are compared, not
the fitted $\estar$ or $\lhat$ values.

\subsubsection{Results}
\label{sec:fitting:results}

\Cref{tab:s8_results} summarizes performance at three operating points
on the sensitivity/false-positive Pareto frontier.

{\setlength\LTleft{0pt}\setlength\LTright{0pt}
\begin{longtable}{@{\extracolsep{\fill}}lccc@{}}
\caption{Fitting performance at three operating points on the sensitivity/false-positive Pareto frontier.  ``Strong'' refers to exceptional country-years with $\lambda > 1$.  FP = false positives out of 500 non-exceptional schedules.  ``Conservative'': tightest $(\sigma_\lambda, \text{gap})$ from the parameter sweep with FP $\leq 10$.  ``BIC'': model selection by minimum BIC.  ``Laplace BF + CV'': cross-validated Laplace Bayes factor (the recommended operating point).} \label{tab:s8_results} \\
\toprule
Metric & Conservative & BIC & Laplace BF + CV \\
\midrule
\endfirsthead
\caption[]{Fitting performance at three operating points on the sensitivity/false-positive Pareto frontier.  ``Strong'' refers to exceptional country-years with $\lambda > 1$.  FP = false positives out of 500 non-exceptional schedules.  ``Conservative'': tightest $(\sigma_\lambda, \text{gap})$ from the parameter sweep with FP $\leq 10$.  ``BIC'': model selection by minimum BIC.  ``Laplace BF + CV'': cross-validated Laplace Bayes factor (the recommended operating point). (continued)} \\
\toprule
Metric & Conservative & BIC & Laplace BF + CV \\
\midrule
\endhead
\midrule
\multicolumn{4}{r}{\textit{Continued on next page}} \\
\endfoot
\bottomrule
\endlastfoot
Strong accuracy (correct type) & 30.4\% & 40.7\% & 39.5\% \\
War (strong, correct type) & 39.5\% & 67.7\% & 47.6\% \\
Respiratory (strong) & 23.3\% & 12.9\% & 29.3\% \\
Enteric (strong) & 17.4\% & 34.8\% & 47.8\% \\
Detection (any type, strong) & 66.5\% & 79.5\% & 97.7\% \\
\addlinespace
False positives & 179 & 294 & 476 \\
Speed (ms/fit) & 9 & 9 & 9 \\
\end{longtable}}

\begin{figure}[!htbp]
\centering
\includefigifexists{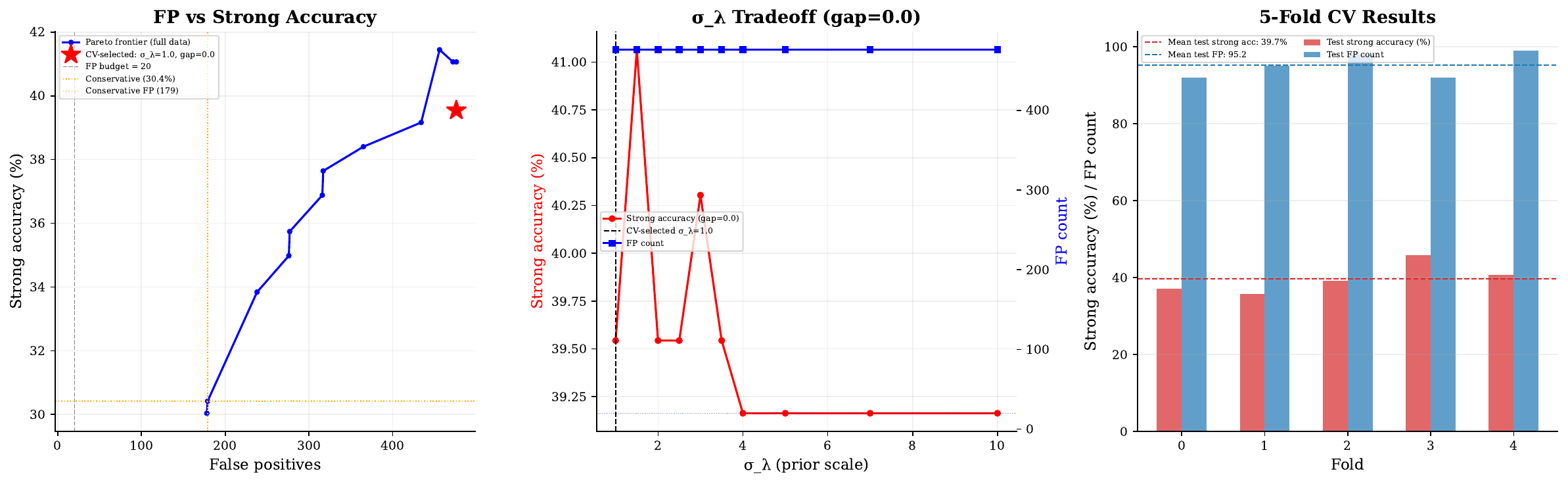}
\caption{Sensitivity/false-positive tradeoff.  Left: Pareto frontier
of false positives vs.\ strong accuracy across all tuning
configurations.  Center: metrics as a function of the prior scale
$\sigma_\lambda$.  Right: 5-fold cross-validation stability -- all folds
independently select $\sigma_\lambda = 1.0$, gap $= 0.0$.}
\label{fig:s8_tradeoff}
\end{figure}

The results reveal a characteristic \emph{Pareto frontier} between
sensitivity and false positive rate.  At the conservative operating
point (tight $\sigma_\lambda$ and gap threshold), the fitter achieves
modest strong accuracy with very few false positives.  At the
permissive operating point (cross-validated Laplace Bayes factor), strong
accuracy improves substantially but at the cost of many more false
positives.  No configuration tested -- across BIC,
Laplace Bayes factors, penalty multiplier sweeps, and gap
thresholds -- simultaneously achieves $>30\%$ strong accuracy with $<20$
false positives.  \Cref{fig:s8_tradeoff} documents this frontier.

This frontier reflects the geometric identifiability constraints
(\cref{sec:fitting:identifiability}): the disruption profiles correlate
with structured mortality variation in normal schedules, so any
criterion permissive enough to detect moderate disruptions will also
fire on a subset of normals.

\subsubsection{Per-type performance}

War disruptions are the best identified, with 48\% correct type
classification and 100\% detection among strong events.  The distinctive
young-adult male mortality hump is geometrically unique and cannot be
produced by trajectory shifts alone.

Respiratory disruptions show genuine partial identifiability.  The broad age
distribution of respiratory excess mortality overlaps with the trajectory
tangent direction, limiting identification, but the Bayes factor captures
enough orthogonal signal for meaningful detection.

Enteric disruptions (48\% correct type) suffer from a small sample
($n = 23$ strong) and an age pattern concentrated in infants and young
children that partially overlaps with the trajectory tangent at low
$\estar$ values.

\subsubsection{Cross-validation stability}
\label{sec:fitting:cv-stability}

The 5-fold CV showed remarkable stability: all five folds independently
selected $\sigma_\lambda = 1.0$ and gap threshold $= 0.0$.  The
per-fold test strong accuracy ranged from 35.7\% to 45.8\% (mean
$39.7\% \pm 3.9\%$), with per-fold false positive counts stable at
92--99.  The variance is driven by which exceptional cases fall in each
fold, not by instability in the parameter selection.

\subsubsection{Confusion matrices}

\begin{figure}[!htbp]
\centering
\includefigifexists{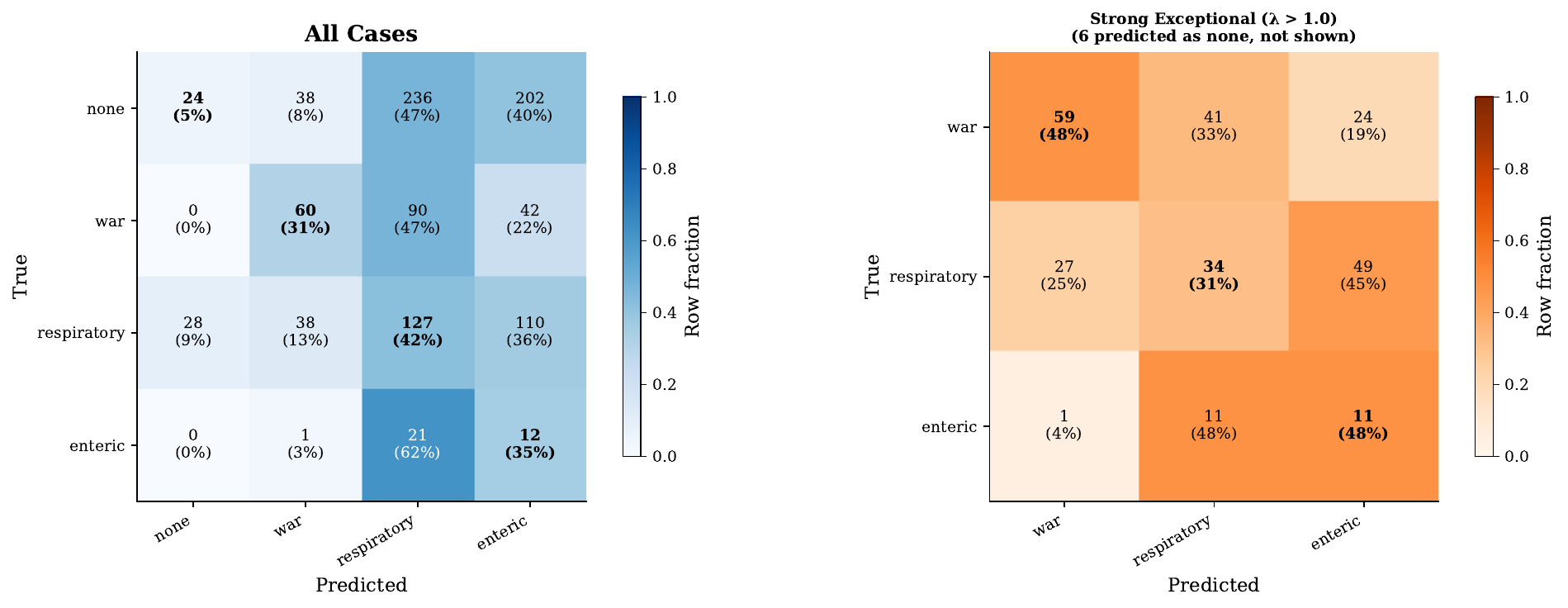}
\caption{Confusion matrices for the life table fitter at the
CV-selected operating point.  Left: all exceptional country-years.
Right: strong disruptions only ($\lambda > 1$).}
\label{fig:s8_confusion}
\end{figure}

\Cref{fig:s8_confusion} shows the confusion matrices for the
cross-validated fitter operating at the aggressive ($\sigma_\lambda =
1.0$) point: one for all exceptional cases and one restricted to strong
disruptions ($\lambda > 1$).  Among strong events, 97.7\% are detected
as having some disruption; the principal classification challenge is
distinguishing which \emph{type} of disruption is present.

\subsubsection{Example fits}

\begin{figure}[!htbp]
\centering
\includefigifexists{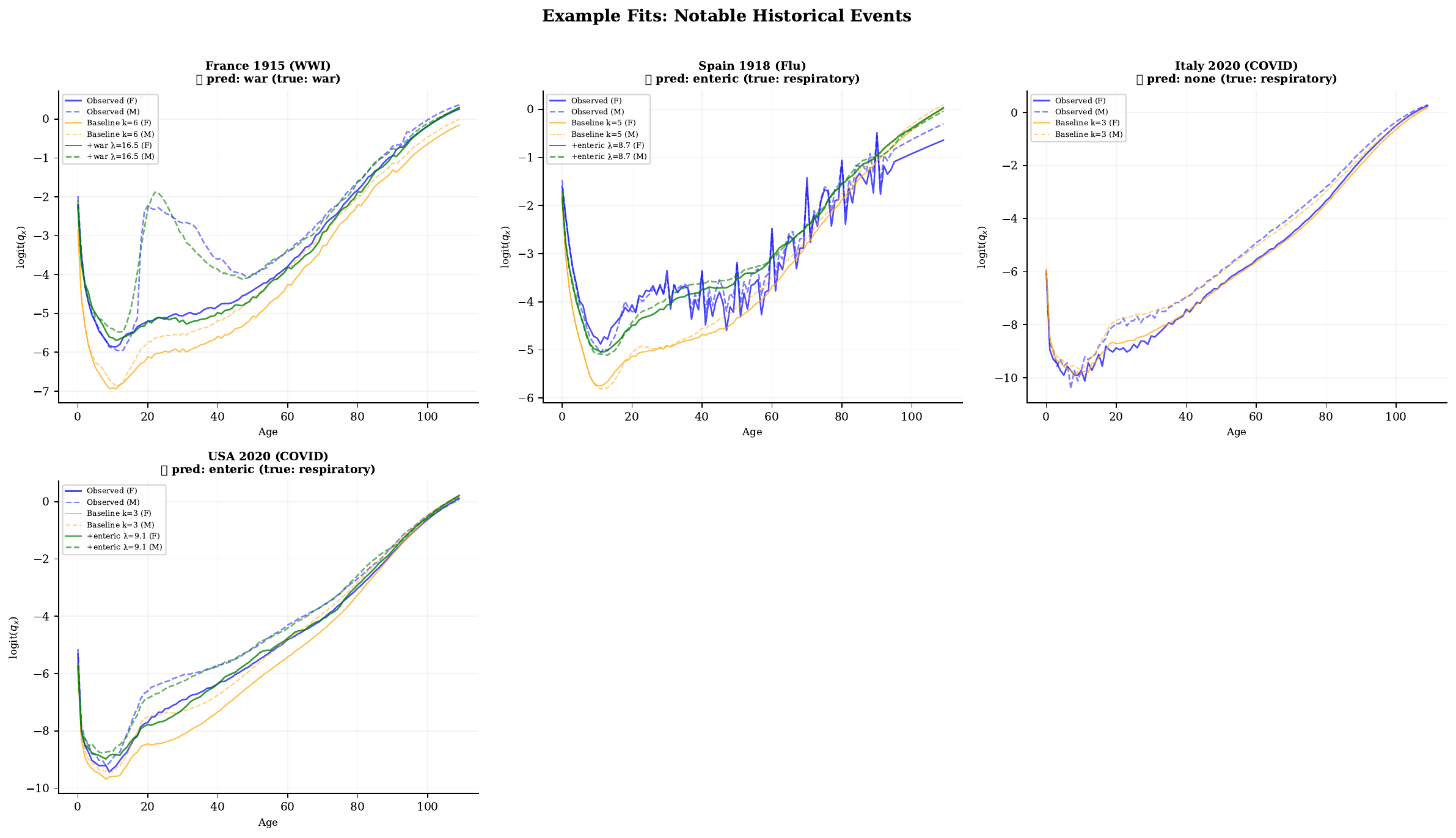}
\caption{Example life table fits for selected historical events.
Each panel shows the observed mortality schedule (gray), the
fitted baseline (blue), and the full fitted model including the
disruption component (red dashed).}
\label{fig:s8_examples}
\end{figure}

\Cref{fig:s8_examples} presents example fits for historically
notable events spanning the three disruption types.  Each panel shows
the observed schedule, the fitted baseline $\zk(\estar)$, and the
decomposition into baseline plus disruption.  The fitter correctly
identifies the iconic wartime pattern in France~1918 and Russia~1943,
the respiratory signature in the 1918 pandemic, and correctly estimates
life expectancy at birth in each case.

\subsection{Computational details}
\label{sec:fitting:computation}

The fitter is implemented in Python with NumPy for vectorized linear
algebra.  The batch regression functions process all 150 $\ezero$ grid
points simultaneously using array broadcasting and
\texttt{searchsorted}-based interpolation, avoiding Python loops over the
$p = 2A = 220$ dimensions.  The full evaluation set (1,029 schedules
$\times$ $K$~clusters $\times$ 150 grid points $\times$ 4 disruption
types $=$ 4.3~million regressions) completes in approximately 9~seconds
on a single core.  The $(\sigma_\lambda, \text{gap})$ sweep is
instantaneous because it only recomputes the Bayes factor from stored
RSS values.

\section{Predicting Schedules from Summary Indicators}
\label{sec:svdcomp}

The preceding sections build a generative model: given a cluster, life
expectancy, and optional disruption, the system produces a mortality
schedule.  We now address a different task -- predicting a
complete sex-age mortality schedule from summary demographic indicators
such as child mortality ($_5q_0$) or child and adult mortality
($_5q_0, {}_{45}q_{15}$).  This is the task addressed by the
SVD-Comp model \citep{Clark2019}, the Log-Quad model
\citep{Wilmoth2012}, and the traditional Coale--Demeny regional model
life tables \citep{CoaleDemeny1966}.

The key innovation here is to reformulate SVD-Comp within the Tucker
framework.  In the original SVD-Comp, separate per-sex SVDs are
computed, and the right singular vector elements (weights) are modeled
as functions of $_5q_0$ via polynomial regression -- independently for
each sex and each component.  In the Tucker formulation, both sexes
share the decomposition, the weights are elements of the effective core
matrix $G_{ct}$, and a single neural network jointly predicts all
weights from the summary indicators.  The sex differential is enforced
by the shared factor matrices $\bS$ and $\bA$ rather than by
independent models.

\subsection{From SVD-Comp to Tucker core prediction}
\label{sec:svdcomp:framework}

Recall from \cref{sec:tucker:reconstruction} that any country-year
schedule can be written as
$M_{:,:,c,t} = \bS \, G_{ct} \, \bA^\top$, where
$G_{ct} \in \R^{r_1 \times r_2}$ is the effective core matrix.
The reconstruction is linear in $\operatorname{vec}(G_{ct})$: using the
row-major vectorization convention in which sex varies before age --
matching the concatenation $\bm{z}_{c,t} = (\hat{y}_{1,1},\ldots,
\hat{y}_{1,A},\, \hat{y}_{2,1},\ldots,\hat{y}_{2,A})^\top$ of
\cref{eq:cluster_feature} -- and defining the
$(2A) \times (r_1 r_2)$ reconstruction matrix
\begin{equation}
  \label{eq:svdcomp:recon}
  \mathbf{R} = \bS \otimes \bA\,,
\end{equation}
where $\otimes$ denotes the Kronecker product and the same row-major
ordering is applied to $\operatorname{vec}(G_{ct})$, we have
$\operatorname{vec}(M_{:,:,c,t}) = \mathbf{R} \, \operatorname{vec}(G_{ct})$.

Using all $r_2$ age components produces noisy predictions because the
higher-order components capture fine-grained structure that cannot be
meaningfully predicted from one or two summary indicators.  Following
the original SVD-Comp \citep{Clark2019}, which uses $c_{\text{age}} = 4$
components, we truncate the age dimension to the first $c_{\text{age}}$ components
(here $c_{\text{age}} = 6$, because the Tucker decomposition separates the sex
dimension into $\bS$, leaving the age basis to capture only
age-specific variation).  The truncated reconstruction matrix
$\mathbf{R}_{c_{\text{age}}} \in \R^{2A \times r_1 c_{\text{age}}}$ uses only the first $c_{\text{age}}$
columns of $\bA$, and the prediction target becomes
$\operatorname{vec}(G_{ct}^{(c_{\text{age}})}) \in \R^{r_1 c_{\text{age}}}$ -- the truncated
core weights.  With $r_1 = 2$ and $c_{\text{age}} = 6$, this is a 12-dimensional
target that captures $> 99.9\%$ of the age-specific variance.

\subsection{Neural network with reconstruction loss}
\label{sec:svdcomp:nn}

A multilayer perceptron maps the summary indicators to the truncated
core weights:
\begin{equation}
  \label{eq:svdcomp:nn}
  \operatorname{vec}(\hat{G}_{ct}^{(c)})
    = f_\theta\bigl(\logit({}_5q_0^F),\, \logit({}_5q_0^M)
      [,\, \logit({}_{45}q_{15}^F),\, \logit({}_{45}q_{15}^M)]\bigr),
\end{equation}
where $f_\theta$ is the network with parameters $\theta$ and the
bracketed inputs are included only in the two-parameter model.  The
architecture is two hidden layers of 64 units each with ReLU
activations.

The loss function operates on the \emph{reconstructed schedule}, not on
the core weights directly:
\begin{equation}
  \label{eq:svdcomp:loss}
  \mathcal{L}
    = \underbrace{\frac{1}{2A} \lVert
        \mathbf{R}_{c_{\text{age}}} \operatorname{vec}(\hat{G}_{ct}^{(c_{\text{age}})})
        - \mathbf{z} \rVert^2}_{\text{full-schedule MSE}}
    + \;\alpha \underbrace{\frac{1}{2 \cdot 5}\sum_{s \in \{F,M\}}
        \sum_{a=0}^{4}
        \bigl(\hat{z}_{s,a} - z_{s,a}\bigr)^2}_{\text{age 0--4 MSE}},
\end{equation}
where $\mathbf{z} = \operatorname{vec}(M_{:,:,c,t})$ is the observed
schedule on the logit scale and $\alpha = 10$.  The first term averages
over all $2A = 220$ age-sex cells; the second averages over the 10
age 0--4 cells (5~ages $\times$ 2~sexes) and then upweights by
$\alpha$.  Because $1/(2A) \approx 0.005$ while $\alpha/(2 \cdot 5) = 1$,
each age 0--4 cell receives roughly 220 times the gradient of other
cells.

The rationale for this upweighting is that the network's primary input is
$_5q_0$ -- a summary measure that aggregates mortality across the five
single-year ages 0--4 -- but its output is a complete schedule with
age-specific mortality at \emph{each} of those ages.  Many different
age-specific patterns within ages 0--4 can produce the same $_5q_0$: a
schedule with very high neonatal mortality and low mortality at ages 1--4
is indistinguishable in terms of $_5q_0$ from one with more evenly
distributed infant and child mortality.  Without the age-specific
upweighting, the network could learn to reproduce the correct $_5q_0$ in
aggregate while distributing mortality across ages 0--4 in a way that
does not match the observed within-group structure.  The upweighting
forces the network to get each individual age right within the input age
range, not just the summary -- ensuring that the predicted schedule is
faithful to the observed age pattern at the ages that carry the most
information about the input.

The gradient flows through $\mathbf{R}_{c_{\text{age}}}$
into the network via the chain rule:
\[
\frac{\partial \mathcal{L}}{\partial \theta}
  = \frac{\partial \mathcal{L}}{\partial \hat{\mathbf{z}}}
    \; \mathbf{R}_{c_{\text{age}}}
    \; \frac{\partial \hat{G}}{\partial \theta}\,.
\]

Training uses Adam optimization with weight decay ($\ell_2 = 10^{-5}$),
batch size 256, and early stopping on a 10\% held-out validation set
(patience 30 epochs).

\subsection{Two models}
\label{sec:svdcomp:models}

Two models are trained on the ${\sim}4{,}000$ non-exceptional
country-years in the HMD tensor:

\paragraph{Model~1 (one-parameter).}
Input: $[\logit({}_5q_0^F),\, \logit({}_5q_0^M)]$.
This serves the same purpose as the original SVD-Comp
\citep{Clark2019} and Log-Quad \citep{Wilmoth2012}: predicting
full schedules when only child mortality is available.

\paragraph{Model~2 (two-parameter).}
Input: $[\logit({}_5q_0^F),\, \logit({}_5q_0^M),\,
  \logit({}_{45}q_{15}^F),\, \logit({}_{45}q_{15}^M)]$.
Adding adult mortality provides direct information about working-age
mortality, substantially improving predictions at ages 15--60.

\subsection{Evaluation and cross-validation}
\label{sec:svdcomp:eval}

Both models are evaluated on the full training set (in-sample) and via
5-fold cross-validation (out-of-sample).  Metrics include: RMSE on
the logit scale, mean absolute error (MAE) in $_5q_0$,
${}_{45}q_{15}$, and $\ezero$ on the probability/years scale, and
MAE in single-year $q_x$ at ages 0--4.

\begin{figure}[!htbp]
\centering
\includefigifexists{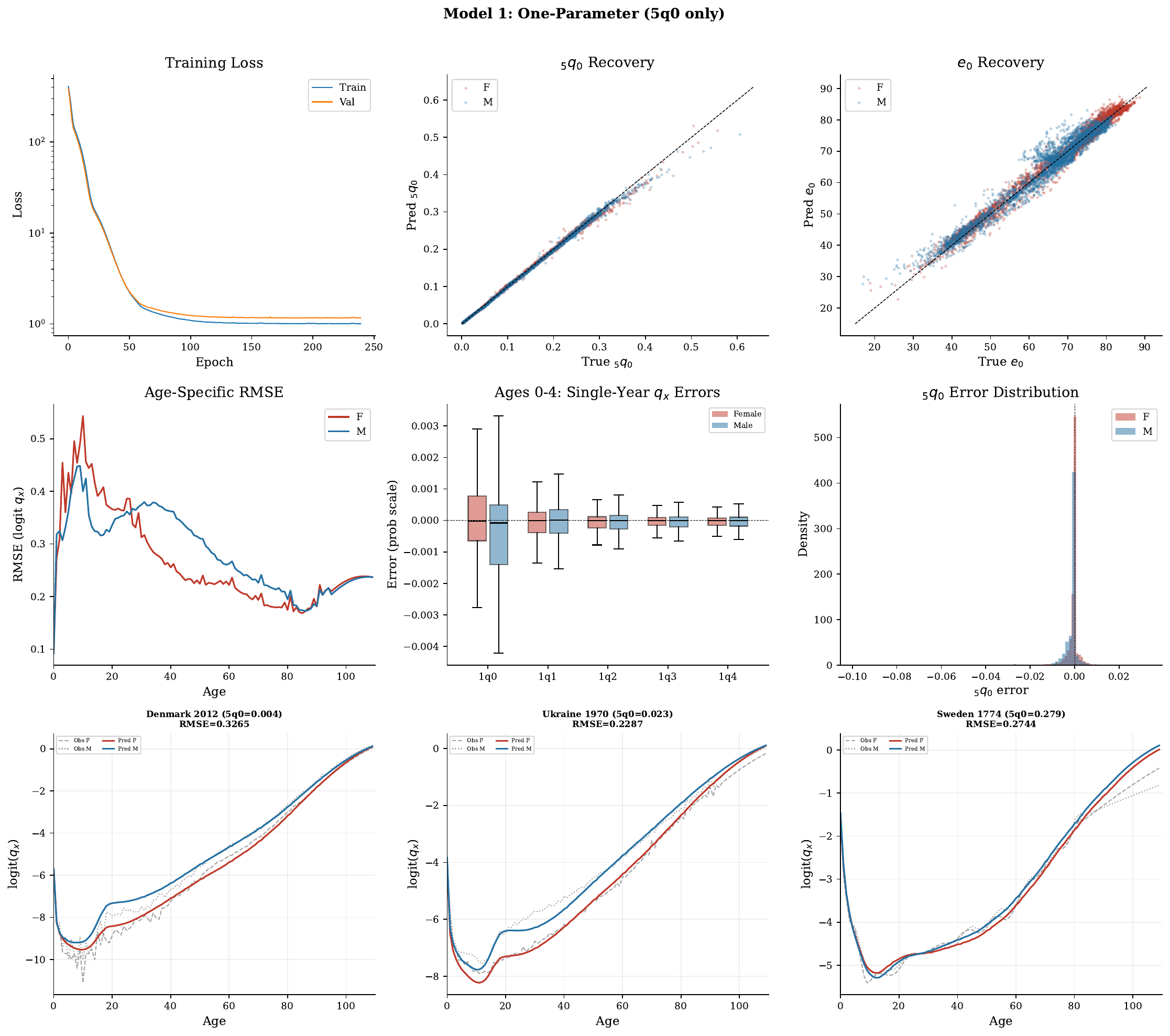}
\caption{Model~1 (one-parameter, $_5q_0$ only) diagnostics: training
loss, $_5q_0$ and $\ezero$ recovery, age-specific RMSE, single-year
age 0--4 errors, and example predictions at low, medium, and high
mortality levels.}
\label{fig:s9_model1_diag}
\end{figure}

\begin{figure}[!htbp]
\centering
\includefigifexists{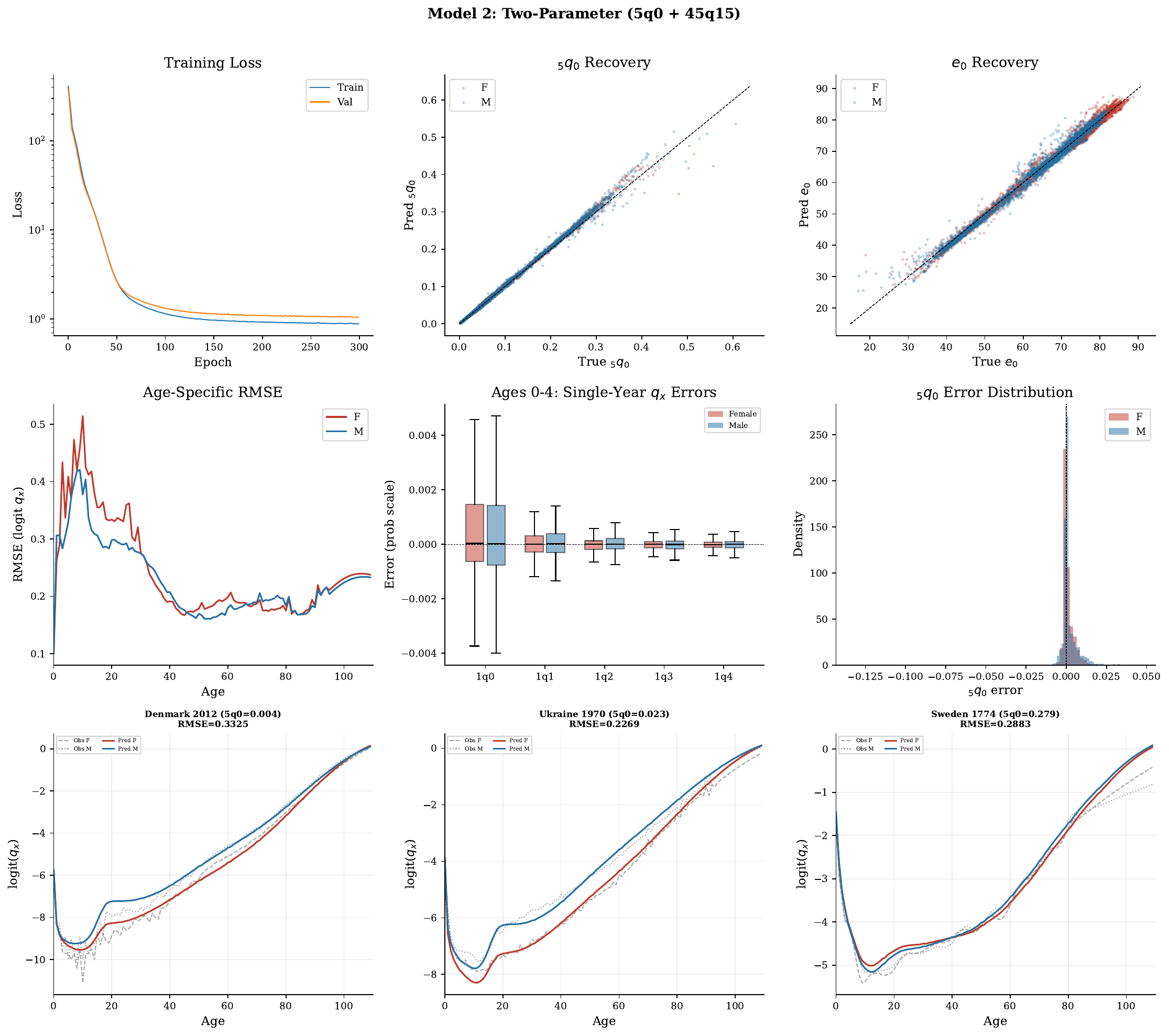}
\caption{Model~2 (two-parameter, $_5q_0 + {}_{45}q_{15}$) diagnostics.
The additional adult mortality input substantially reduces RMSE at
working ages (15--60) compared to Model~1.}
\label{fig:s9_model2_diag}
\end{figure}

\begin{figure}[!htbp]
\centering
\includefigifexists{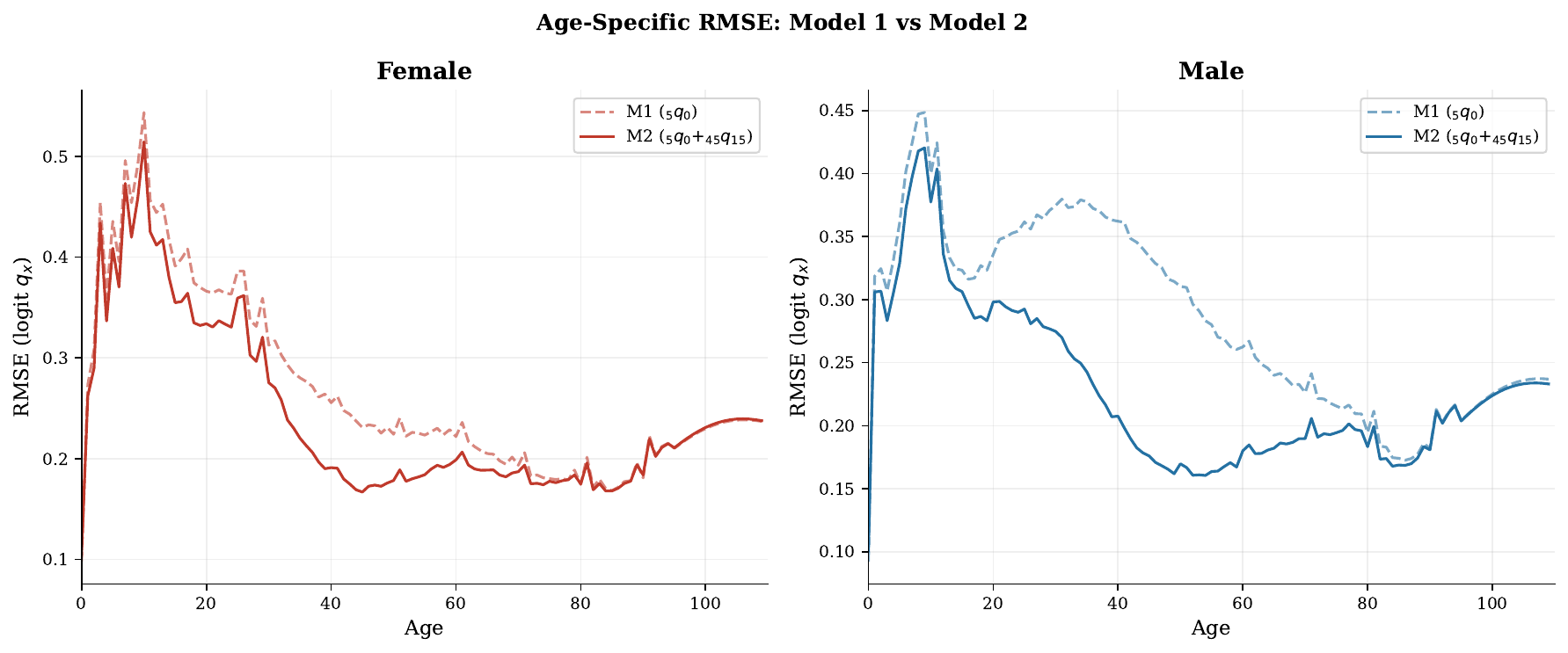}
\caption{Age-specific RMSE comparison: Model~1 (dashed) vs.\ Model~2
(solid) for female (left) and male (right).  The two-parameter model
improves predictions at all ages, with the largest gains at working
ages.}
\label{fig:s9_comparison}
\end{figure}

\Cref{fig:s9_model1_diag} presents the one-parameter model diagnostics.
The $_5q_0$ recovery panel confirms that the age 0--4 penalty produces
faithful reproduction of the input indicator.  The age-specific RMSE
panel shows that reconstruction error is concentrated at the oldest
ages, where the truncated basis necessarily loses resolution.
\Cref{fig:s9_model2_diag} shows analogous results for the
two-parameter model, with substantially lower RMSE at working ages.
\Cref{fig:s9_comparison} compares the two models directly, showing
the age-specific RMSE improvement from adding the adult mortality input.

\subsection{Advantages over SVD-Comp}
\label{sec:svdcomp:advantages}

The Tucker formulation improves on the original SVD-Comp in several
ways.  First, both sexes are predicted by a single model through the
shared factor matrices $\bS$ and $\bA$, ensuring that the sex
differential is demographically consistent without requiring separate
per-sex models.  Second, the neural network learns nonlinear
input--weight relationships automatically, replacing the hand-crafted
polynomial features of the original SVD-Comp (eight terms per
component per sex, Equation~12 of \citealt{Clark2019}).  Third,
the reconstruction loss ensures that the optimization directly targets
schedule accuracy rather than core weight accuracy -- higher-order
components with small contributions to the reconstruction are
automatically downweighted.  Fourth, extending the model to additional
inputs (e.g.\ ${}_{45}q_{15}$, cluster membership, or epidemiological
covariates) requires only changing the input dimension of the network,
with no structural modifications.

\section{Forecasting Mortality Schedules}
\label{sec:forecasting}

The preceding sections develop the MDMx system as a \emph{generative}
model: given a cluster label, life expectancy, and optional disruption
specification, it produces a complete sex-specific mortality schedule.
We now develop a \emph{forecasting} framework that projects the
system's low-dimensional parameters forward in time, producing
probabilistic forecasts of the full sex-age mortality surface.

The central insight is that the Tucker decomposition provides the
forecasting target directly.  Every country-year's mortality schedule is
fully determined by its effective core matrix
$G_{ct} \in \R^{r_1 \times r_2}$ (\cref{eq:Gct}), and the
reconstruction
$\hat{M}_{:,:,c,t} = \bS \, G_{ct} \, \bA^\top$
(\cref{sec:tucker:reconstruction}) is exact within the Tucker subspace.
Forecasting reduces to projecting the $r_1 \times r_2 = 84$
elements of $G_{ct}$ forward in time -- a drastically
lower-dimensional problem than forecasting the $2A = 220$-dimensional
$\logit(\qx)$ vector directly.  Moreover, because the reconstruction
uses the shared factor matrices $\bS$ and $\bA$, the resulting female
and male schedules are \emph{structurally coherent}: they are linked
through the Tucker basis by construction, not by post hoc adjustment.

This structural coherence distinguishes the approach from the dominant
family of mortality forecasting methods that originate with the
Lee--Carter model \citep{LeeCarter1992}.  Lee--Carter projects a
single temporal index $k_t$ via random walk with drift -- essentially
a rank-one SVD extrapolation.  Extensions add higher-order components
\citep{BoothMaindonaldSmith2002,HyndmanUllah2007}, flexible time-series
models \citep{deJongTickle2006}, coherent multi-population structure
\citep{LiLee2005,HyndmanBoothYasmeen2013}, and more nuanced treatment of
the rotation of the age pattern of mortality decline
\citep{LiLeeGerland2013}.  Recent reviews
\citep{BoothTickle2008,ShangBoothHyndman2011,BaselliniCamardaBooth2023}
document persistent challenges: the fixed age pattern in Lee--Carter
tends to underpredict life expectancy
\citep{LeeMillerLeeCarter2001,BoothEtAl2006}, independent sex-specific
fits produce divergent forecasts requiring post hoc adjustment, and
prediction intervals are sensitive to the choice of time-series model.
The present approach addresses all three issues through the Tucker
architecture: rank elevation from 1 to 5 effective components allows the
age pattern to rotate, structural sex coherence is automatic, and a
drift-constrained Kalman filter with hierarchical target provides both
regularisation and principled uncertainty quantification.

\subsection{PCA reduction of $G_{ct}$}
\label{sec:forecasting:pca}

Although 84 parameters is far fewer than 220, running a multivariate
Kalman filter with a $168$-dimensional state vector ($84 \times 2$ for
level and drift) is computationally demanding because each country
requires maximum likelihood estimation of the Kalman hyperparameters.  A
PCA on all observed $\text{vec}(G_{ct})$ vectors reveals that
five components capture 97.1\% of the total variance, with the first
component alone accounting for 91.8\%.  The PCA projection is
\begin{equation}
\label{eq:pca_scores}
\bm{s}_{c,t} = V^\top \bigl(\text{vec}(G_{ct}) - \bar{g}\bigr)
\;\in\; \R^{N_{PC}}\,,
\qquad N_{PC} = 5\,,
\end{equation}
where $V \in \R^{84 \times 5}$ contains the first five principal
component loadings and $\bar{g}$ is the HMD-wide mean $G_{ct}$.

The combined map from PCA scores to the full $\logit(\qx)$ schedule is
linear:
\begin{equation}
\label{eq:scores_to_z}
\bm{z} = \bm{z}_{\text{mean}} + L \, \bm{s}\,,
\qquad
L \in \R^{2A \times N_{PC}}\,,
\end{equation}
where column $j$ of $L$ is the vectorised Tucker reconstruction of the
$j$-th principal component.  This linearity has important consequences
for prediction intervals (\cref{sec:forecasting:intervals}).

\subsection{Drift-constrained hierarchical Kalman filter}
\label{sec:forecasting:kalman}

The state vector for each country is
$\bm{x}_t = [\bm{\ell}_t^\top, \bm{\delta}_t^\top]^\top
\in \R^{2 N_{PC}}$,
where $\bm{\ell}_t \in \R^{N_{PC}}$ is the level (PCA scores) and
$\bm{\delta}_t \in \R^{N_{PC}}$ is the drift.

The standard damped local linear trend \citep{Harvey1989,DurbinKoopman2012},
building on the Kalman filter \citep{Kalman1960},
mean-reverts the drift toward zero as $\rho < 1$.  We modify this so
that the drift mean-reverts toward a hierarchical target
$\bm{\delta}_{\text{hier}}$ rather than toward zero:
\begin{align}
\label{eq:fc_level}
\bm{\ell}_t &= \bm{\ell}_{t-1} + \bm{\delta}_{t-1}
               + \bm{\eta}_t^{\ell}\,, \\
\label{eq:fc_drift}
\bm{\delta}_t &= \rho \, \bm{\delta}_{t-1}
                 + (1 - \rho) \, \bm{\delta}_{\text{hier}}
                 + \bm{\eta}_t^{\delta}\,,
\end{align}
with observation equation
$\bm{y}_t = H \, \bm{x}_t + \bm{\varepsilon}_t$,
where $H = [I_{N_{PC}} \;\; 0]$ selects the level.  This is
implemented as a state intercept $\bm{b}$ in the Kalman predict step:
$\bm{x}_{t|t-1} = F \, \bm{x}_{t-1|t-1} + \bm{b}$,
where $\bm{b} = [\bm{0},\; (1-\rho) \bm{\delta}_{\text{hier}}]^\top$
and $F$ has the standard DLLT structure with $\rho$ in the drift block.

When $\rho < 1$, the forecast drift converges to
$\bm{\delta}_{\text{hier}}$ at rate $\rho^h$, so the long-run
forecast approaches the hierarchical consensus regardless of the
country's recent trajectory.  The Kalman hyperparameters -- the
$N_{PC}$ diagonal elements each of $Q^{\ell}$ (level innovation
variance), $Q^{\delta}$ (drift innovation variance), and $R$
(observation noise), plus the scalar $\rho$ (damping) -- total
$3 N_{PC} + 1 = 16$ per country and are estimated by maximum likelihood
via the Kalman prediction error decomposition, optimised using L-BFGS-B
with $\rho \in [0.80, 0.999]$.

\subsection{Two-level hierarchy: an empirical discovery}
\label{sec:forecasting:hierarchy}

The original architecture specified a three-level hierarchy:
\begin{equation}
\label{eq:hier3}
\bm{\delta}_{\text{hier}} =
    w_1 \, \bm{\delta}_{\text{HMD}}
  + w_2 \, \bm{\delta}_{\text{cluster}}
  + w_3 \, \bm{\delta}_{\text{country}}\,,
\qquad w_1 + w_2 + w_3 = 1\,,
\end{equation}
where each component drift is estimated from the last $W = 20$ years of
training data at the appropriate level.  The clusters are those
identified in \cref{sec:clustering}.  A grid search over the full
three-element simplex at step size 0.05 (231 points) found that
\textbf{the optimal cluster weight is zero}: every top-ten
configuration had $w_{\text{cluster}} \leq 0.05$, and the best was
$(0.80, 0.00, 0.20)$.

The interpretation is that the clusters from \cref{sec:clustering:gmm}
group countries by mortality \emph{pattern} (level and age-structure
shape), not by \emph{trajectory} (the pace and direction of recent
change).  Countries within a structural cluster can have very different
recent trends -- for example, the Czech Republic improving rapidly and
Russia stagnating, both in the same post-Soviet cluster.  The hierarchy
simplifies to two levels:
\begin{equation}
\label{eq:hier2}
\bm{\delta}_{\text{hier}} =
    0.80 \, \bm{\delta}_{\text{HMD}}
  + 0.20 \, \bm{\delta}_{\text{country}}\,.
\end{equation}
The clusters remain central to the decomposition
(\cref{sec:clustering}), trajectory construction
(\cref{sec:reconstruction}), and reconstruction pipeline -- they are
simply not informative for the forecast drift target.

\subsection{Prediction intervals}
\label{sec:forecasting:intervals}

\subsubsection{Schedule-level intervals (exact)}

Because the map from PCA scores to $\logit(\qx)$ is linear
(\cref{eq:scores_to_z}), the Kalman forecast covariance propagates
exactly:
$\text{Cov}(\bm{z}_h) = L \, H \, P_{h|T} \, H^\top \, L^\top$,
where $P_{h|T}$ is the $h$-step-ahead state covariance.

\subsubsection{Life expectancy intervals (delta method)}

Life expectancy $\ezero$ is a nonlinear function of the schedule.
The delta method linearises the $\bm{s} \to \ezero$ mapping:
\begin{equation}
\label{eq:delta_method}
\sigma^2_{\ezero}(h) = J_h \; \text{Cov}_{\text{scores}}(h) \; J_h^\top\,,
\end{equation}
where $J_h = \partial \ezero / \partial \bm{s}$ is the Jacobian
evaluated numerically at the $h$-step forecast.  Monte Carlo validation
at all 1{,}934 cross-validation test points confirmed that the
$\expit$ nonlinearity adds only 5\% to the uncertainty (MC/delta
$\sigma$ ratio = 1.051), so the delta method is essentially exact for
this problem.  A scalar calibration factor $\kappa =
\text{SD}(\text{CV z-scores})$ is applied to account for remaining
miscalibration of the Kalman covariance.

\subsection{Cross-validation results}
\label{sec:forecasting:results}

The system is evaluated by rolling-origin cross-validation using six
origins (1960, 1970, 1980, 1990, 2000, 2010), each with a 15-year
horizon and a minimum of 30 training years, producing 1{,}934
non-exceptional test points.

\subsubsection{Point forecast accuracy}
\label{sec:forecasting:accuracy}

{\setlength\LTleft{0pt}\setlength\LTright{0pt}
\begin{longtable}{@{\extracolsep{\fill}}rrrr@{}}
\caption{Rolling-origin cross-validation results (two-level hierarchy, 80/20 HMD-wide/country, per-country MLE).} \label{tab:fc_results} \\
\toprule
Origin & n & MAE & Bias \\
\midrule
\endfirsthead
\caption[]{Rolling-origin cross-validation results (two-level hierarchy, 80/20 HMD-wide/country, per-country MLE). (continued)} \\
\toprule
Origin & n & MAE & Bias \\
\midrule
\endhead
\midrule
\multicolumn{4}{r}{\textit{Continued on next page}} \\
\endfoot
\bottomrule
\endlastfoot
1960 & 194 & 1.34 & 0.18 \\
1970 & 251 & 1.91 & 1.69 \\
1980 & 312 & 1.94 & 1.88 \\
1990 & 483 & 1.38 & -0.16 \\
2000 & 443 & 1.31 & -0.96 \\
2010 & 251 & 0.73 & 0.53 \\
Overall & 1934 & 1.44 & 0.35 \\
\end{longtable}}

\Cref{tab:fc_results} summarises the point forecast accuracy.  The
overall $\ezero$ MAE is 1.44~years with a bias of $+0.35$~years.
The 1970 and 1980 origins have the largest errors because they require
forecasting through the fastest phase of the mortality transition --
a drift-based model cannot anticipate inflection points.  At modern
origins (1990--2010), the MAE is 0.71--1.41~years.  The fitted $\rho$
distribution has a median of 0.80, with 108 of 164 country-origin fits
at the lower bound, indicating that most countries benefit from
substantial damping -- consistent with the deceleration of mortality
improvement at high $\ezero$ \citep{OeppenVaupel2002}.

\subsubsection{Benchmark comparison}
\label{sec:forecasting:benchmark}

{\setlength\LTleft{0pt}\setlength\LTright{0pt}
\begin{longtable}{@{\extracolsep{\fill}}lrrrr@{}}
\caption{Benchmark comparison. Lee--Carter (R \texttt{demography::lca}, adjust=none) and Hyndman--Ullah (R \texttt{demography::fdm}, order=6, ARIMA scores) on original HMD $m_x$ schedules.} \label{tab:fc_benchmark} \\
\toprule
Method & n & MAE & RMSE & Bias \\
\midrule
\endfirsthead
\caption[]{Benchmark comparison. Lee--Carter (R \texttt{demography::lca}, adjust=none) and Hyndman--Ullah (R \texttt{demography::fdm}, order=6, ARIMA scores) on original HMD $m_x$ schedules. (continued)} \\
\toprule
Method & n & MAE & RMSE & Bias \\
\midrule
\endhead
\midrule
\multicolumn{5}{r}{\textit{Continued on next page}} \\
\endfoot
\bottomrule
\endlastfoot
MDMx & 1934 & \textbf{1.436} & \textbf{1.802} & \textbf{0.349} \\
LC & 1914 & 1.738 & 2.143 & -1.184 \\
HU & 1914 & 1.439 & 1.855 & -0.662 \\
\end{longtable}}

\begin{figure}[!htbp]
\centering
\includegraphics[width=\textwidth]{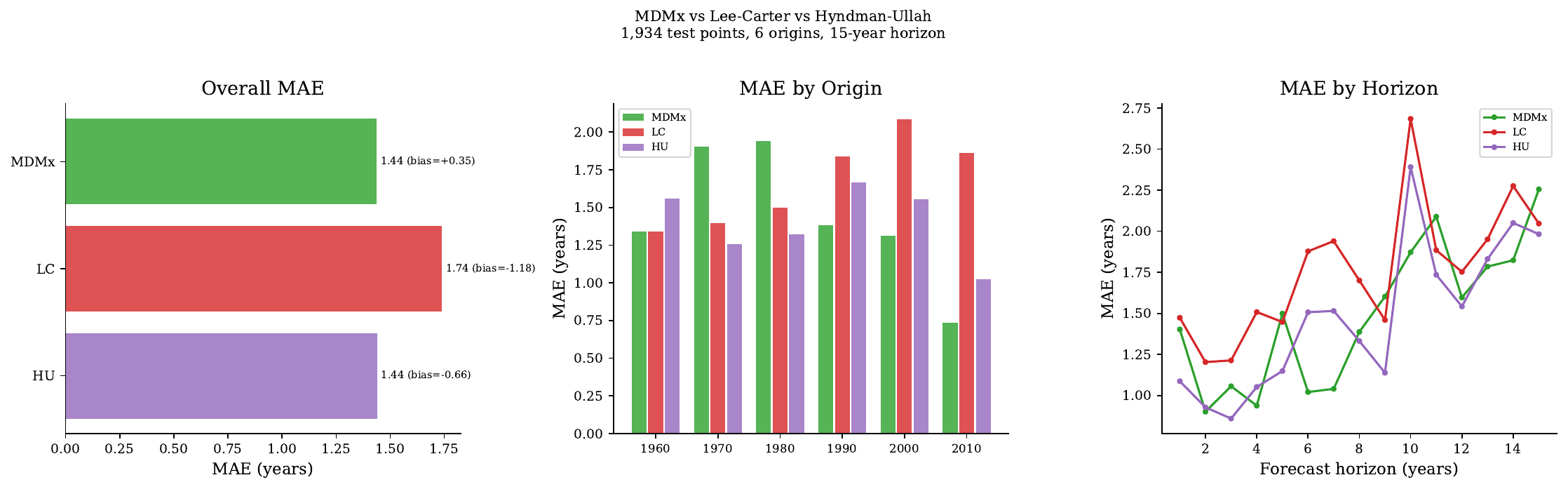}
\caption{Benchmark comparison on 1{,}934 test points (R
\texttt{demography} package).  Left: overall MAE.  Centre: MAE by
forecast origin.  Right: MAE by forecast horizon.  MDMx (green)
outperforms Lee--Carter (red) across all origins and horizons;
Hyndman--Ullah (purple) matches MDMx on average but with larger
negative bias.}
\label{fig:fc_benchmark}
\end{figure}

\Cref{tab:fc_benchmark} compares MDMx against Lee--Carter
\citep{LeeCarter1992} and Hyndman--Ullah \citep{HyndmanUllah2007},
both computed by the R \texttt{demography} package on the HMD's own
graduated $m_x$ schedules -- the same cross-validation origins,
horizons, and $\ezero$ computation apply to all three methods.  MDMx
achieves 17\% lower MAE than Lee--Carter (1.44 vs.\ 1.74~years) and
matches Hyndman--Ullah (1.44~years each), though the methods differ
markedly in bias ($+0.35$ for MDMx vs.\ $-1.18$ for Lee--Carter and
$-0.66$ for Hyndman--Ullah) and in sex-gap coherence
(\cref{sec:forecasting:sex}).
\Cref{fig:fc_benchmark} shows the comparison by origin and horizon.

Lee--Carter's rank-1 SVD imposes a fixed age pattern of decline
($b_x$) that cannot rotate -- a well-documented limitation
\citep{LiLeeGerland2013,BoothMaindonaldSmith2002,
BaselliniCamardaBooth2023}.  Hyndman--Ullah's rank-6 SVD with
ARIMA-forecasted scores provides enough flexibility to match MDMx on
$\ezero$ MAE, though it carries a larger negative bias ($-0.66$ vs.\
$+0.35$~years) -- underpredicting both female and male $\ezero$, with
the male underprediction more severe.  The per-origin breakdown reveals
complementary strengths: Hyndman--Ullah performs better at early
origins (1970--1980), while MDMx excels at later origins (1990--2010)
where the hierarchical drift constraint anchors the forecast to the
recent HMD-wide consensus.

\subsubsection{Sex-specific accuracy and coherence}
\label{sec:forecasting:sex}

{\setlength\LTleft{0pt}\setlength\LTright{0pt}
\begin{longtable}{@{\extracolsep{\fill}}llrrlr@{}}
\caption{Sex-specific forecast accuracy and sex-gap coherence.} \label{tab:fc_sex} \\
\toprule
Method & Sex & n & MAE & RMSE & Bias \\
\midrule
\endfirsthead
\caption[]{Sex-specific forecast accuracy and sex-gap coherence. (continued)} \\
\toprule
Method & Sex & n & MAE & RMSE & Bias \\
\midrule
\endhead
\midrule
\multicolumn{6}{r}{\textit{Continued on next page}} \\
\endfoot
\bottomrule
\endlastfoot
MDMx & Female & 1934 & \textbf{1.281} & \textbf{1.610} & \textbf{0.383} \\
MDMx & Male & 1934 & \textbf{1.631} & \textbf{2.049} & \textbf{0.316} \\
MDMx & Average & 1934 & \textbf{1.436} & \textbf{1.802} & \textbf{0.349} \\
LC & Female & 1914 & 1.551 & 1.879 & -1.053 \\
LC & Male & 1914 & 2.081 & 2.595 & -1.315 \\
LC & Average & 1914 & 1.738 & 2.143 & -1.184 \\
HU & Female & 1914 & 1.332 & 1.700 & -0.523 \\
HU & Male & 1914 & 1.654 & 2.163 & -0.800 \\
HU & Average & 1914 & 1.439 & 1.855 & -0.662 \\
MDMx & Gap (F$-$M) & 1934 & \textbf{0.595} & -- & \textbf{0.067} \\
LC & Gap (F$-$M) & 1914 & 1.112 & -- & 0.261 \\
HU & Gap (F$-$M) & 1914 & 0.836 & -- & 0.277 \\
\end{longtable}}

\begin{figure}[!htbp]
\centering
\includegraphics[width=\textwidth]{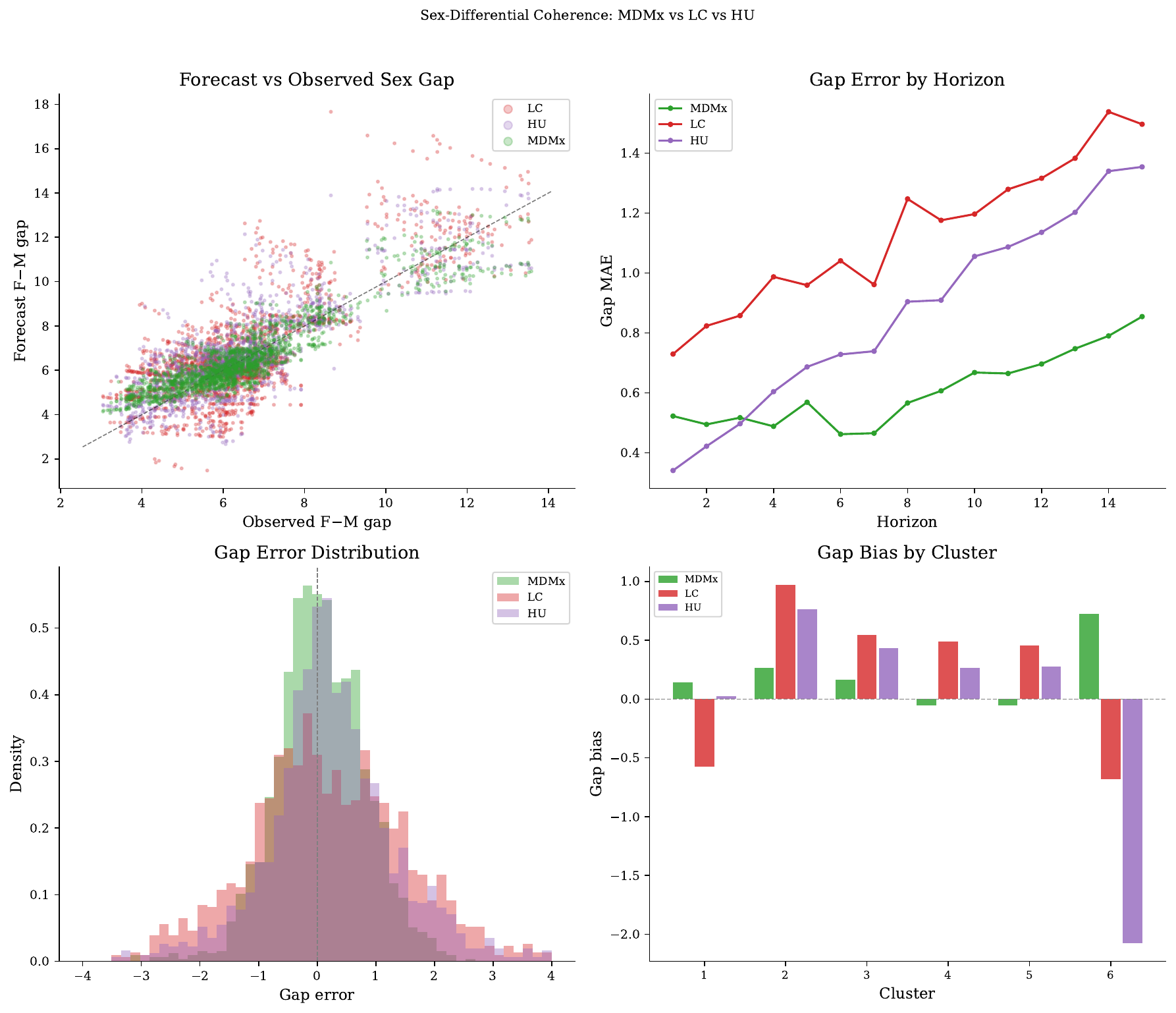}
\caption{Sex-differential coherence across methods.  Top left: forecast
vs.\ observed F--M gap -- MDMx clusters tightly on the diagonal.  Top
right: gap error by horizon -- Lee--Carter diverges.  Bottom left: gap
error distribution.  Bottom right: gap bias by cluster.}
\label{fig:fc_sex_coherence}
\end{figure}

\Cref{tab:fc_sex} shows that MDMx wins on female $\ezero$ and
matches Hyndman--Ullah on male $\ezero$.  The sex-gap
($\ezero^F - \ezero^M$) coherence reveals the structural advantage
most clearly: MDMx forecasts the gap with MAE of 0.60~years and
near-zero bias ($+0.067$~years), whereas Lee--Carter's independent
fits produce a gap MAE of 1.11~years (bias $+0.261$) and
Hyndman--Ullah's a gap MAE of 0.84~years (bias $+0.277$) --
both overestimating the female advantage because their independent
sex-specific fits are unconstrained.
\Cref{fig:fc_sex_coherence} visualises this.

\subsubsection{Prediction interval calibration}
\label{sec:forecasting:pi}

\begin{figure}[!htbp]
\centering
\includegraphics[width=\textwidth]{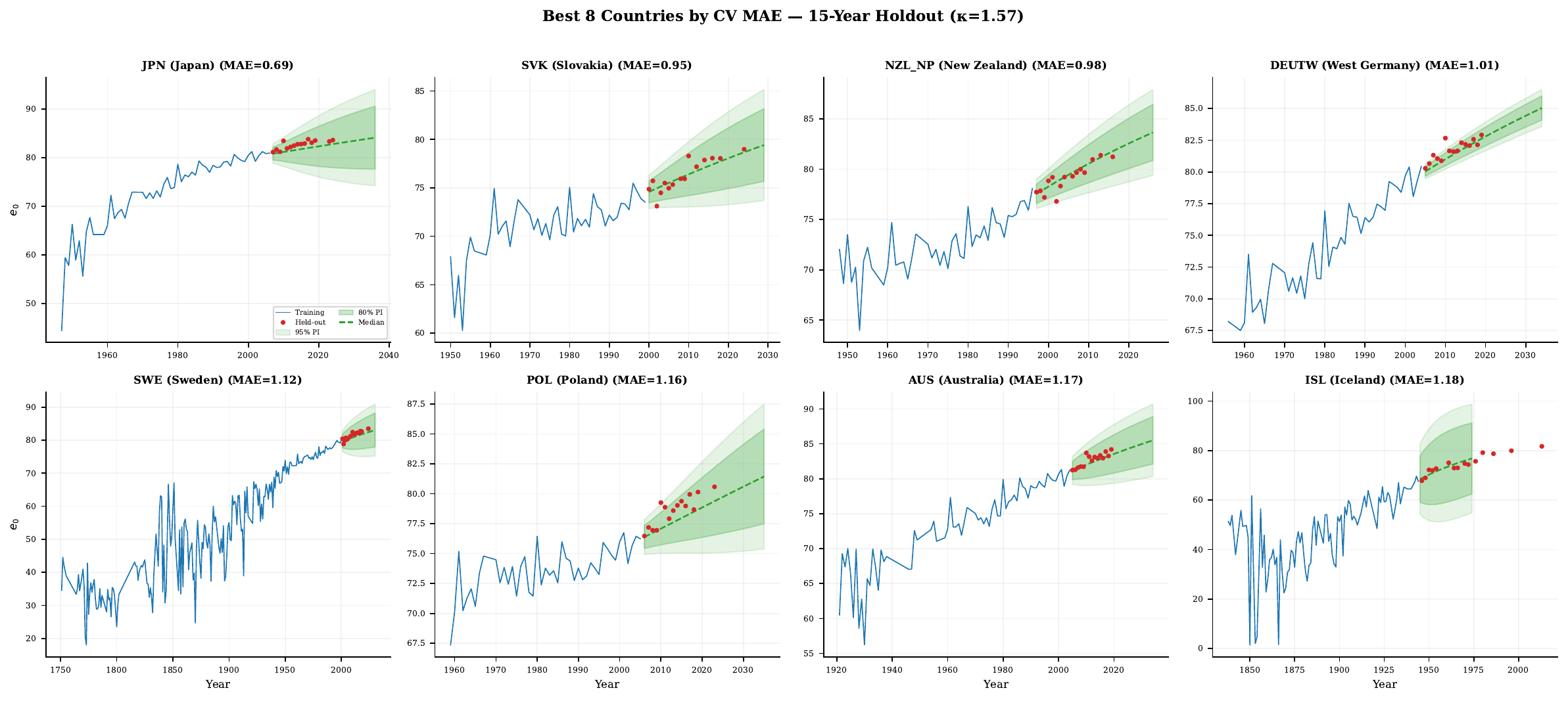}
\caption{Best-forecast countries: the eight populations with the lowest
average $\ezero$ MAE across all CV origins and horizons.  Blue: training
data.  Red dots: held-out observations.  Green dashed: median forecast
with 80\% (dark shading) and 95\% (light shading) prediction intervals.
Title shows per-country MAE.}
\label{fig:fc_best_countries}
\end{figure}

\begin{figure}[!htbp]
\centering
\includegraphics[width=\textwidth]{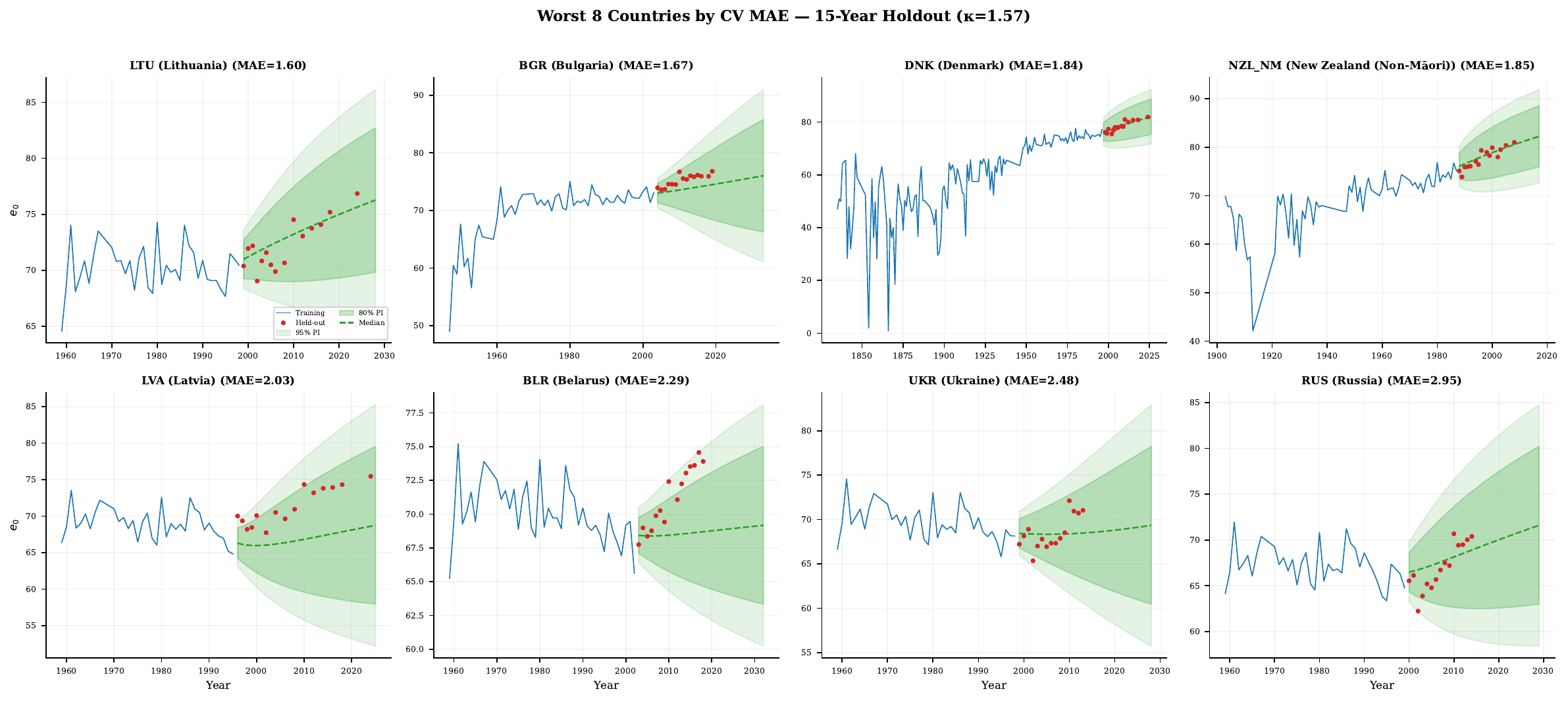}
\caption{Worst-forecast countries: the eight populations with the highest
average $\ezero$ MAE.  Layout as in \cref{fig:fc_best_countries}.  These
countries typically exhibit volatile or non-monotonic trajectories that
are difficult for any extrapolative method.}
\label{fig:fc_worst_countries}
\end{figure}

The raw delta-method 95\% interval achieves 86.9\% coverage; the 80\%
interval achieves 75.7\%.  After calibration with $\kappa = 1.57$, the
95\% coverage reaches 93.7\% and the 80\% coverage 87.3\%.
\Cref{fig:fc_best_countries,fig:fc_worst_countries} display the
countries with, respectively, the lowest and highest average $\ezero$
MAE across all CV origins and horizons.  The best-performing countries
tend to have long, smooth historical trajectories with steady improvement
-- conditions under which the hierarchical drift and cluster structure
work well.  The worst-performing countries include those with volatile
recent histories (stagnation or reversal), short time series, or
atypical trajectories that the hierarchical drift target cannot capture.
Even for the worst countries, the prediction intervals generally contain
the held-out observations, confirming that the calibrated intervals
are appropriately wide.

\subsubsection{Schedule-level evaluation}
\label{sec:forecasting:schedule}

The $l_x$-weighted RMSE by age group reveals that youth ages (5--14) are
the hardest to forecast (wRMSE = 0.549), followed by child (1--4,
0.443) and working ages (15--49, 0.398).  Mature (50--69, 0.195) and
old (70+, 0.120) ages are the easiest, because old-age mortality
improvements follow more universal trajectories across countries.

\subsubsection{Sex-differential dynamics}
\label{sec:forecasting:sexdiff}

The forecast sex differential is nearly stationary, evolving much more
slowly than the overall level.  The level is captured primarily by PC~1
(91.8\% of variance), which has a strong, consistent drift.  The
sex-differential dynamics live in PCs~2--5, which have weaker,
more heterogeneous drifts that the 80\% HMD-wide weighting pulls to near
zero -- reflecting that HMD-wide sex-gap convergence has been slow and
geographically uneven.  The scalar $\rho$ applies identical damping to
all five PCA components, so the same mechanism that prevents level
overshoot also prevents the sex differential from evolving.  A natural
extension would be component-specific $\rho$ or hierarchy weights;
the current conservative forecasts are defensible because the
alternative -- unconstrained sex-gap drift -- risks the implausible
divergence that plagues independent-sex methods.

\subsubsection{Origin continuity and forecast surfaces}
\label{sec:forecasting:surfaces}

\begin{figure}[!htbp]
\centering
\includegraphics[width=\textwidth]{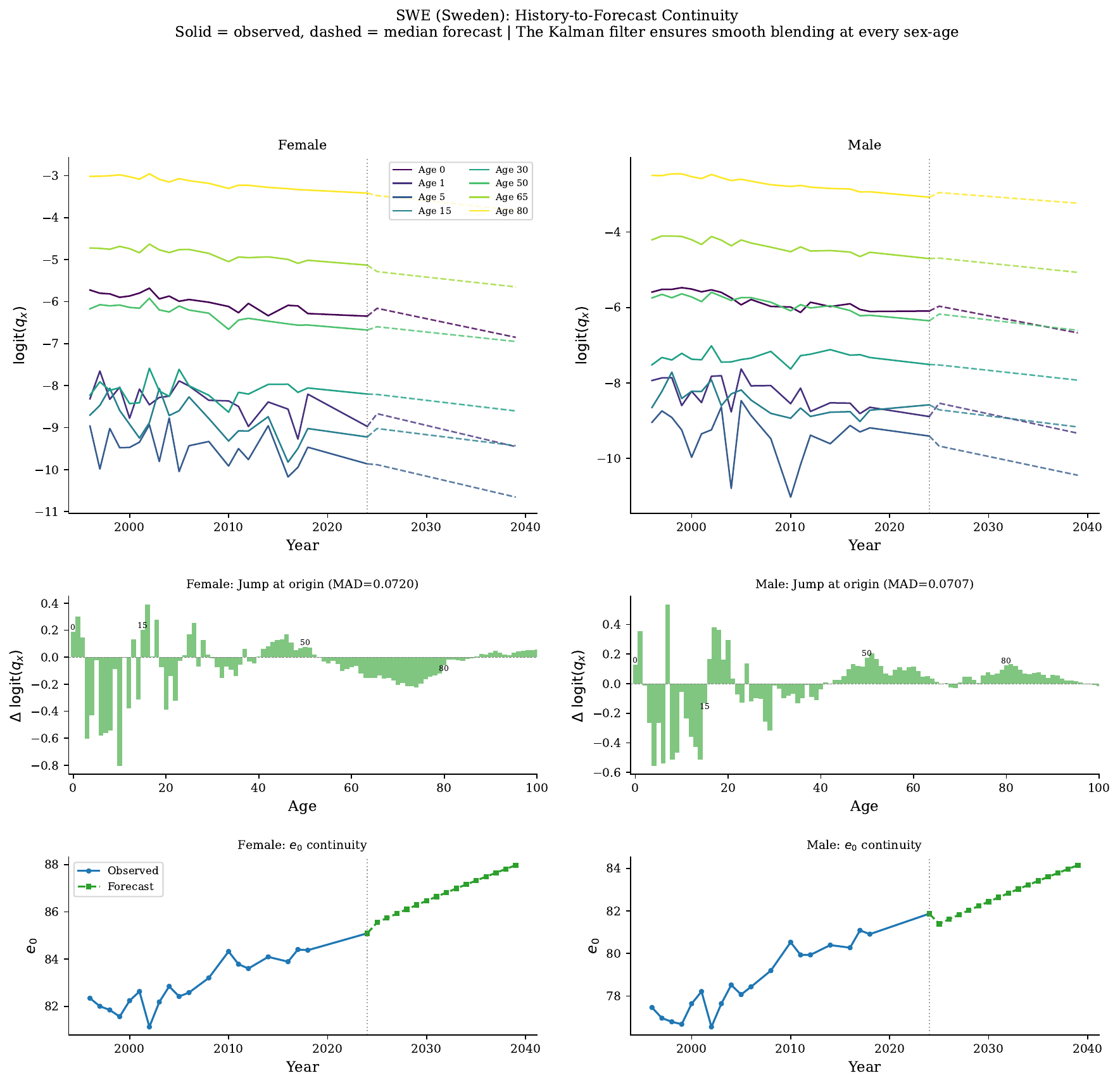}
\caption{Origin continuity for Sweden.  Top: $\logit(\qx)$ time
series at selected ages flowing smoothly from observed (solid) to
forecast (dashed).  Centre: jump at origin across all ages -- near zero
everywhere.  Bottom: $\ezero$ continuity.}
\label{fig:fc_continuity}
\end{figure}

The Kalman filter's last filtered state serves as the forecast
jump-off, so the median forecast joins the observed data smoothly at
every sex-age combination by construction.  The median absolute jump
at the origin is below $10^{-3}$ in $\logit(\qx)$ for essentially all
ages (\cref{fig:fc_continuity}).

\begin{figure}[!htbp]
\centering
\includegraphics[width=\textwidth]{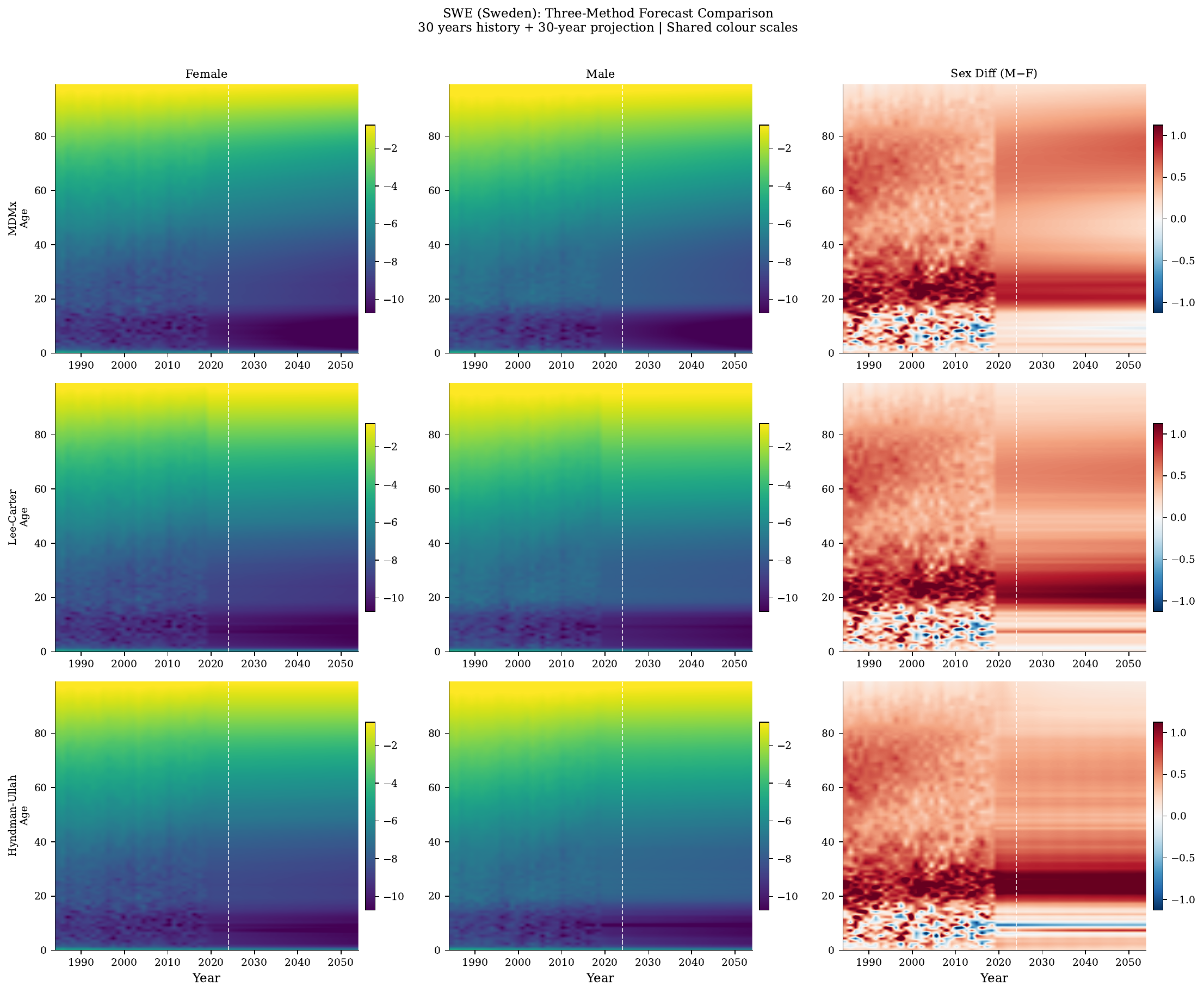}
\caption{Three-method forecast comparison for Sweden.  Rows: MDMx,
Lee--Carter, Hyndman--Ullah.  Columns: female $\logit(\qx)$, male
$\logit(\qx)$, sex differential (M$-$F).  All panels share
colour scales.  White dashed line marks the forecast origin.}
\label{fig:fc_method_surface}
\end{figure}

\Cref{fig:fc_method_surface} provides a side-by-side comparison of
the three methods' forecast mortality surfaces.  MDMx produces smooth
age-pattern evolution with coherent sex structure; Lee--Carter's
fixed $b_x$ produces a rigid, non-rotating age pattern; and
Hyndman--Ullah shows more flexibility but can produce age-pattern
distortions at long horizons.

\subsection{Relation to existing methods}
\label{sec:forecasting:comparison}

The MDMx forecasting system can be understood as a generalised
Lee--Carter in Tucker coordinates.  \Cref{tab:fc_comparison} summarises
the structural correspondence.

{\setlength\LTleft{0pt}\setlength\LTright{0pt}
\begin{longtable}{@{\extracolsep{\fill}}l>{\raggedright\arraybackslash}p{3.5cm}>{\raggedright\arraybackslash}p{3.5cm}>{\raggedright\arraybackslash}p{3.5cm}@{}}
\caption{Structural comparison of MDMx forecasting with Lee--Carter and
         Hyndman--Ullah.} \label{tab:fc_comparison} \\
\toprule
Feature & Lee--Carter & Hyndman--Ullah & MDMx \\
\midrule
\endfirsthead
\caption[]{Structural comparison (continued)} \\
\toprule
Feature & Lee--Carter & Hyndman--Ullah & MDMx \\
\midrule
\endhead
\midrule
\multicolumn{4}{r}{\textit{Continued on next page}} \\
\endfoot
\bottomrule
\endlastfoot
Decomposition & Rank-1 SVD of $\log(m_{x,t})$
  & Rank-6 SVD of smoothed $\log(m_{x,t})$
  & Tucker of $\mathrm{logit}(q_x)$ tensor \\
\addlinespace
Age pattern & Fixed $b_x$
  & 6 basis functions, fixed
  & 41 basis functions via $\bm{A}$, shared \\
\addlinespace
Sex handling & Independent fits
  & Independent fits
  & Structural coherence via $\bm{S}$ \\
\addlinespace
Time series & RW + drift on $k_t$
  & RW + drift on each score
  & Damped LLT Kalman on 5~PCA scores \\
\addlinespace
Multi-population & None (or Li--Lee)
  & Product-ratio
  & Hierarchical drift (HMD-wide + country) \\
\addlinespace
PI method & MC on $k_t$
  & MC on scores
  & Delta method ($\kappa$-calibrated) \\
\end{longtable}}

The key advances are the rank elevation (from 1 to 5 effective
components), the hierarchical drift constraint (which provides natural
regularisation for short-series countries), and the structural sex
coherence.  The rank elevation captures the rotation of the age pattern
of mortality decline -- the primary source of Lee--Carter's documented
tendency to underpredict life expectancy
\citep{LeeMillerLeeCarter2001,BoothEtAl2006,BaselliniCamardaBooth2023}.
The hierarchical drift provides small-population regularisation:
countries with limited data stay close to the HMD-wide consensus, while
long-series countries express their individual trends.  And the
structural sex coherence eliminates the divergence problem that
requires ad hoc adjustment in independent-sex methods
\citep{LiLee2005,HyndmanBoothYasmeen2013}.

\begin{figure}[!htbp]
\centering
\includegraphics[width=\textwidth]{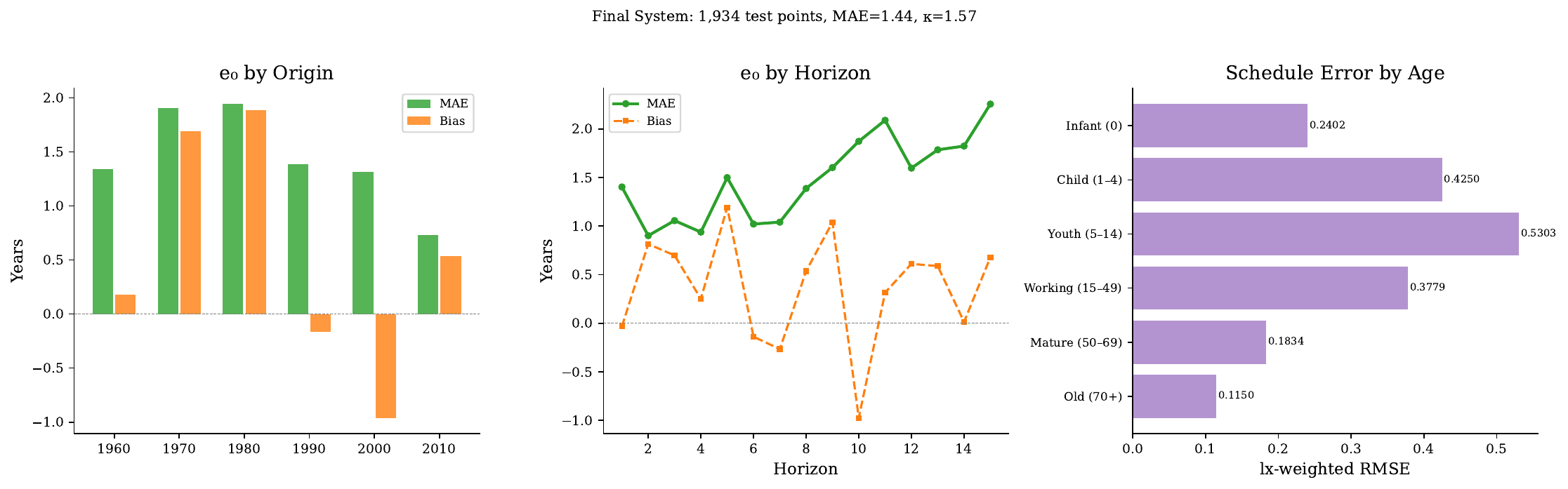}
\caption{Final system evaluation.  Left: $\ezero$ MAE and bias by
forecast origin.  Centre: MAE and bias by forecast horizon.  Right:
schedule-level $l_x$-weighted RMSE by age group.}
\label{fig:fc_summary}
\end{figure}

\Cref{fig:fc_summary} presents a three-panel summary of the final
system's performance.

\section{Discussion}
\label{sec:discussion}

We have presented the theory, methods, and empirical results for
a unified mortality modeling framework built from a Tucker tensor
decomposition of the Human Mortality Database.  The framework serves
simultaneously as a model life table system, a life table fitter, a
summary-indicator prediction engine, and a mortality forecasting system.
We organize this discussion by component, summarizing each stage's
architecture and findings, placing them in the context of existing
methods, and noting limitations and directions for further work.

\subsection{Decomposition and structural sex coherence}

The most consequential design decision is to treat the HMD as a single
four-dimensional object and decompose it jointly across sex, age,
country, and year, rather than fitting separate models to individual
countries or stacking dimensions into matrices.  The Tucker decomposition
(HOSVD) extracts shared factor matrices for sex ($\bS$) and age
($\bA$), together with country and year loadings and a core tensor
encoding their interactions.  The result is a compact representation
that captures $>$99.99\% of the per-mode variance in the data
(reconstruction RMSE $<$ 0.08 in $\logit(q_x)$) with Tucker ranks
$r_1 = 2$ (sex, full rank) and $r_2 = 42$ (age).

The practical payoff is structural sex coherence.  Because $\bS$ and
$\bA$ mediate every reconstruction, female and male schedules are linked
through the same basis at every stage of the framework -- in the
generative model, in the life table fitter, in the summary-indicator
predictor, and in the forecaster.  This coherence is not imposed post hoc
(as in product-ratio methods; \citealt{HyndmanBoothYasmeen2013}) or
through common-factor constraints (as in \citealt{LiLee2005}), but is an
automatic consequence of the decomposition itself.

The system is also nonparametric with respect to both age and time.
Unlike the Lee--Carter model, which assumes a fixed age pattern of
mortality change ($b_x$), or the log-quadratic model, which assumes a
specific functional form for the age curve, the Tucker decomposition
learns the age basis functions and their temporal evolution directly from
the data.  This flexibility is essential for capturing the rotation of
the age pattern of mortality decline -- a phenomenon that a rank-one
decomposition cannot represent.

\subsection{Clustering, trajectories, and the generative model}

The level-controlled clustering strategy -- operating on Tucker core
features with the grand-mean age component removed
(\cref{sec:clustering:features}) -- cleanly separates the age-pattern
signal from the mortality-level signal.  This is a structural advantage
over methods that cluster mortality schedules directly (whether on raw
$\qx$ or on PCA scores of reconstructed schedules), which inevitably
conflate level and shape.  By removing the level contribution within the
Tucker framework itself, rather than through ad hoc mean-centering, the
clustering identifies canonical age-pattern families that are stable
over time and that draw members from diverse geographies and periods
\citep{Omran1971,Mesle2004}.  A complementary epoch classification
characterizes each country's direction and speed of change -- rapid
improvement, stagnation, worsening -- providing a two-layer description
of each country-year that separates \emph{what kind} of mortality regime
from \emph{how fast} it is changing.

The cluster-specific trajectories provide a form of heterogeneity that
existing systems handle only partially.  The Coale--Demeny families
capture regional variation but with fixed functional forms; the Wilmoth
log-quadratic model allows for variation in two parameters but not in the
shape of the age curve beyond what two parameters can express.  The
present system allows each cluster to have its own trajectory through the
full $2A$-dimensional space, and the rotation of the age pattern of
decline \citep{LiLeeGerland2013,Vekas2019} emerges naturally from these
trajectories rather than being imposed as a model extension.  A neural
trajectory model (\cref{sec:reconstruction:neural_trajectory}) provides
smooth interpolation across cluster boundaries and safe extrapolation
beyond observed ranges, closely matching the LOWESS estimates within each
cluster's observed $\ezero$ range while maintaining demographic
plausibility at extremes.

The generative model (\cref{eq:full_model}) combines these components:
given a mortality regime (cluster), a target mortality level ($\ezero$),
and an optional disruption specification (type and intensity), it
produces a complete, sex-specific, single-year-of-age schedule of $\qx$
from birth through age~109.

\subsection{Exceptional mortality}

The explicit treatment of exceptional mortality -- identifying it
formally, removing it from the baseline decomposition, and modeling it
separately -- is, to our knowledge, new in the model life table
literature.  Existing systems either exclude exceptional years informally
or allow them to influence the model implicitly.  The separable
disruption model represents excess mortality as the product of a
type-specific age-sex profile and a scalar intensity, producing both a
clean baseline and a principled way to add disruptions back at arbitrary
intensity.

The three disruption types produce distinctive, interpretable profiles.
War disruptions concentrate excess mortality in young-adult males, with a
pattern geometrically distinct from any secular trajectory shift.
Respiratory pandemics show broad age elevation with variable young-adult
selectivity, consistent with known epidemiological differences between
pathogen strains.  The enteric/infectious disease profile shows
broader-than-expected elevation across young and adult ages, a finding
that warrants further investigation given the heterogeneous events
grouped under this label.  Sub-clustering within each type reveals
historically recognizable subtypes -- trench warfare versus total war,
pandemics with different age selectivity -- with improved explained
variance relative to the single-profile model.

The baselines for exceptional years are estimated primarily by a neural
core model (\cref{sec:exceptional:neural_core}) that predicts
uncontaminated core tensor slices from country loadings and year features,
avoiding the systematic bias that na\"{\i}ve projection introduces
(\cref{sec:exceptional:naive}).  Temporal interpolation and penalized
projection are developed as alternatives and cross-checks, and a
comparison of all four baseline approaches (na\"{\i}ve, temporal,
penalized, neural core) documents the neural core's advantage.

\subsection{Life table fitting}

The life table fitter (\cref{sec:fitting}) inverts the generative model:
given a single observed schedule with no temporal or geographic context,
it discovers whether a disruption is present, identifies which type best
explains the residual pattern, and estimates its intensity -- all through
a three-stage algorithm (linearized grid search, Gauss--Newton
refinement, exact Bayesian evaluation) that runs in ${\sim}8$~ms per
schedule.  The regression framing is central: the
Frisch--Waugh--Lovell decomposition ensures that disruption detection and
life expectancy estimation are properly orthogonalized, while the Laplace
Bayes factor provides a model comparison criterion that scales with the
shape of the likelihood surface rather than imposing a fixed penalty.

Cross-validation characterizes the sensitivity/false-positive Pareto
frontier inherent in the single-schedule setting: 40\% correct
disruption-type classification among strong events at the cross-validated
operating point, with war events best identified (48\% type accuracy,
100\% detection among strong events).  The fundamental limitation is the
high correlation between disruption profiles and the trajectory tangent
($|\rho| > 0.8$), which makes the distinction between disruption and
normal variation inherently noisy when only one schedule is available.
This is a geometric property of the problem, not a deficiency of the
algorithm; for applications with temporal context -- monitoring a
country's mortality over time, for instance -- anchoring $\estar$ from
neighboring non-exceptional years would break the absorption mechanism
and should dramatically improve respiratory and enteric detection.

\subsection{Summary-indicator prediction}

The summary-indicator prediction models (\cref{sec:svdcomp}) demonstrate
that the Tucker framework provides a natural generalization of the
SVD-Comp approach \citep{Clark2019}.  By predicting the truncated
core weights rather than the schedule directly, the models inherit the
smoothness of the Tucker age basis functions and the internally
consistent sex differential from the shared factor matrices.  The
truncation to $c_{\text{age}} = 6$ age components is essential: using the
full $r_2$ components produces noisy reconstructions because the
higher-order weights are effectively unpredictable from one or two
summary indicators.

Two models are trained: a one-parameter model using $_5q_0$ alone, and a
two-parameter model using both $_5q_0$ and ${}_{45}q_{15}$.  The
two-parameter model substantially outperforms the one-parameter model at
working ages (15--60), confirming the practical value of adult mortality
information when it is available.  Both models can be extended to
additional inputs -- cluster membership, epidemiological covariates, or
temporal information -- by simply changing the input dimension, with no
structural modifications to the architecture.  This extensibility,
combined with the joint sex prediction and automatic nonlinear feature
learning of the neural network, represents a meaningful advance over the
polynomial regression approach of the original SVD-Comp.

\subsection{Forecasting}

The forecasting framework (\cref{sec:forecasting}) operates in the
Tucker parameter space rather than on age-specific rates directly,
projecting PCA-reduced effective core matrices forward through a damped
local linear trend Kalman filter with the drift state constrained toward
a two-level hierarchical target blending HMD-wide (80\%) and
country-level (20\%) trends.  The Tucker reconstruction then delivers
complete sex-specific schedules at every horizon, structurally coherent
by construction.

The empirical finding that cluster-level drift is redundant -- despite
clusters being central to the decomposition and reconstruction --
illustrates a broader principle: structures useful for pattern
description need not be the same structures useful for temporal
extrapolation.  The two-level hierarchy is both simpler and more
accurate than the three-level alternative.

Rolling-origin cross-validation (six origins, 15-year horizon) yields an
$\ezero$ MAE of 1.44~years -- 17\% better than Lee--Carter and
matching Hyndman--Ullah (both computed by the R \texttt{demography}
package on HMD $m_x$ schedules; \cref{tab:fc_benchmark}).  Where MDMx
distinguishes itself from Hyndman--Ullah is in sex-gap coherence:
the structural Tucker architecture produces a gap MAE of 0.60~years
with near-zero bias, versus 0.84~years for Hyndman--Ullah and
1.11~years for Lee--Carter.  The architecture addresses the three
persistent challenges of the Lee--Carter family: age-pattern rotation
emerges naturally from the joint evolution of five PCA scores (versus a
fixed $b_x$); sex coherence is structural (versus independent fits);
and prediction intervals, calibrated from cross-validation z-scores,
achieve 93.7\% coverage at the nominal 95\% level.  The best-performing
countries tend to have long, smooth historical trajectories; the worst
performers exhibit volatile or non-monotonic recent histories that are
difficult for any extrapolative method.

The system exhibits a modest positive bias ($+0.35$~years), suggesting
that the hierarchical drift target is slightly optimistic on average.
The sex differential is nearly stationary, because the sex-gap dynamics
live in secondary PCA components with weak drifts that the scalar $\rho$
and HMD-wide pooling damp toward zero.  Both features could be addressed
by allowing $\rho$ and hierarchy weights to vary by PCA component,
permitting the sex-differential dynamics to evolve more freely while
keeping the level heavily damped.

\subsection{Neural components as targeted extensions}

The system integrates neural network components at specific points where
the linear framework encounters limitations -- not as replacements for
the interpretable tensor decomposition, but as targeted extensions
trained on its outputs.  The neural core predicts uncontaminated
baselines at exceptional years; the neural trajectory enables smooth
inter-cluster interpolation and extrapolation; the neural sub-cluster
profiles capture within-type disruption heterogeneity; and the
summary-indicator networks predict Tucker core weights from one or two
mortality measures.  In each case, the neural component operates within
the interpretive framework the decomposition provides, and the
all-important age basis and sex structure come from the Tucker factor
matrices, not from the network.

These components do add complexity: each introduces hyperparameters
(network architecture, learning rate, regularization strength) that
require tuning, and their predictions are less transparent than the
corresponding linear methods.  All neural components are implemented from
scratch in pure NumPy, keeping the system free of deep learning framework
dependencies.  The neural trajectory validation
(\cref{sec:results:neural_trajectory}) and disruption sub-clustering
results (\cref{sec:results:profile_results}) evaluate whether the
improved fit justifies the additional complexity.

\subsection{Limitations}

The framework has several important limitations.

The HMD is the foundation of the system, and the HMD is not globally
representative.  Its roughly 50 populations are predominantly European,
with limited representation from East Asia, the Americas, and Oceania,
and no representation from Africa, South Asia, or most of the
developing world.  The mortality regimes, trajectories, and disruption
profiles estimated here reflect the experience of high-income and
upper-middle-income countries.  Applying the system to populations
outside this domain -- particularly to sub-Saharan Africa, where
HIV/AIDS has produced mortality patterns with no analogue in the
HMD -- requires caution and likely extension with additional data sources.

The imputation of missing entries in the ragged tensor introduces bias,
as discussed in \cref{sec:preprocessing:imputation_bias}.  The weighted
decomposition approach outlined there is our preferred mitigation, but
its effectiveness remains to be demonstrated empirically.

The separable disruption model assumes a single canonical profile per
disruption type.  Events within a type may differ substantially (the
1918 influenza pandemic had a distinctive young-adult excess not seen
in other influenza pandemics), and the single-profile model captures
only the average pattern.  The sub-clustering extension
(\cref{sec:exceptional:subclustering}) mitigates this by identifying
distinct sub-profiles, but is constrained by the number of events
available for each type.

The life table fitter faces an inherent identifiability constraint in the
single-schedule setting: because disruption profiles correlate with
structured mortality variation in normal schedules, any criterion
permissive enough to detect moderate disruptions will also produce
false positives.  This is a geometric property of the problem, not a
deficiency of the algorithm, and temporal context would largely resolve
it.

The forecasting system exhibits a modest positive bias ($+0.35$~years in
$\ezero$) and produces nearly stationary sex differentials, both
consequences of the uniform damping and HMD-wide pooling applied across
all PCA components.  The 95\% prediction interval coverage is 93.7\% --
close to nominal but slightly undercovered, suggesting room for improved
calibration.

\subsection{Future directions}

Several extensions suggest themselves.

\emph{Broader data.}  Incorporating mortality data from demographic
surveillance systems and civil registration systems in low- and
middle-income countries would broaden the range of mortality regimes
represented, potentially adding clusters that capture the epidemiological
profiles of sub-Saharan Africa and South Asia.  Extending the
decomposition to include cohort life tables would add a fifth dimension
and allow the system to capture cohort effects directly.

\emph{Uncertainty in the generative model.}  Developing a probabilistic
version of the trajectory model -- replacing the point estimate
$\hat{\bm{\mu}}_k(\ezero^{\ast})$ with a predictive distribution --
would provide uncertainty quantification for the reconstructed schedules
and the life table fitter, not just for the forecasts.

\emph{Richer neural architectures.}  The neural components use simple
multi-layer perceptrons; architectures that encode demographic structure
(monotonicity constraints at older ages, attention mechanisms across
clusters) could improve both accuracy and interpretability.  The
sub-clustering of disruption profiles could be extended to a
hierarchical model that shares information across disruption types.

\emph{Forecasting refinements.}  Allowing $\rho$ and hierarchy weights
to vary by PCA component would permit the sex-differential dynamics to
evolve more freely while keeping the level heavily damped.
Incorporating time-varying covariates (GDP, healthcare expenditure,
cause-of-death composition) into the Kalman drift target, developing
covariate-driven disruption activation models, and integrating the
disruption overlay from \cref{sec:exceptional} into the forecast
prediction intervals are all promising directions.

\emph{Anchored fitting.}  For applications with temporal context, fixing
$\estar$ from neighboring non-exceptional years would break the
absorption mechanism that drives the sensitivity/false-positive tradeoff,
dramatically improving respiratory and enteric detection.

\emph{Population forecasting.}  Coupling the system with a demographic
projection framework would allow it to serve not only as a model life
table system but as a component of population forecasting, with the
Tucker architecture providing the mortality input.

\emph{Extending the framework.} A similar approach could be
applied to fertility and perhaps population projection more generally
by tying mortality and fertility together. It may also be possible to include
covariates in the tensor to encourage consistency across an array of
demographic and related dimensions.


\newpage
\section{Notation}
\label{sec:notation}

Scalars are denoted by lowercase italic letters ($a$, $s$), vectors by
lowercase bold ($\bm{v}$), matrices by uppercase bold ($\bm{U}$), and
tensors of order three or higher by calligraphic uppercase ($\M$, $\G$).
Component indices within the decomposition are written in lowercase
fraktur ($\fs$, $\fa$, $\fc$, $\ft$) to distinguish them from the
dimension indices ($s$, $a$, $c$, $t$) they parallel.
A subscript indexes an element: $\M_{s,a,c,t}$ is the element of the
tensor~$\M$ at sex~$s$, age~$a$, country~$c$, and year~$t$.
The same convention applies to vectors extracted from factor matrices:
a fraktur subscript selects a \emph{component} (column), as in
$\bm{a}_\fa$ (the $\fa$-th age basis vector), while an italic
subscript selects a \emph{data unit} (row), as in $\bm{s}_s$ (the
loading vector for sex~$s$) or $\bm{u}_c$ (the loading vector for
country~$c$).

\Cref{tab:notation} collects the principal symbols used throughout.

\begin{longtable}{@{}l@{\hspace{1.5em}}l@{\hspace{1.5em}}>{\raggedright\arraybackslash}p{0.62\textwidth}@{}}
\caption{Principal notation.}
\label{tab:notation} \\
\toprule
Symbol & Dimension & Meaning \\
\midrule
\endfirsthead
\caption[]{Principal notation. (continued)} \\
\toprule
Symbol & Dimension & Meaning \\
\midrule
\endhead
\midrule
\multicolumn{3}{r@{}}{\textit{Continued on next page}} \\
\endfoot
\bottomrule
\endlastfoot
\multicolumn{3}{@{}l}{\textit{Dimensions and indices}} \\
$S$ & -- & number of sex categories ($S=2$: female, male) \\
$A$ & -- & number of single-year age groups \\
$C$ & -- & number of countries (or populations) \\
$T$ & -- & number of calendar years \\
$s, a, c, t$ & -- & indices for sex, age, country, year \\
$\fs, \fa, \fc, \ft$ & -- & component indices for sex, age, country, year \\
\addlinespace
\multicolumn{3}{@{}l}{\textit{Demographic quantities}} \\
$\qx$ & -- & probability of dying between exact ages $x$ and $x{+}1$;
             in tensor equations $x = a$ \\
$\ezero$ & -- & life expectancy at birth \\
${}_5q_0$ & -- & under-five mortality (probability of dying before age~5) \\
${}_{45}q_{15}$ & -- & adult mortality (probability of dying between ages 15 and 60) \\
\addlinespace
\multicolumn{3}{@{}l}{\textit{Tensors and factor matrices}} \\
$\M$ & $S \times A \times C \times T$
      & mortality data tensor (in logit-transformed $\qx$) \\
$y_{s,a,c,t}$ & --
      & element of $\M$: $\logit(\qx)$ at sex $s$, age $a$, country $c$, year $t$ \\
$\G$ & $r_1 \times r_2 \times r_3 \times r_4$
      & core tensor \\
$\bS$ & $S \times r_1$ & sex factor matrix \\
$\bA$ & $A \times r_2$ & age factor matrix (columns are age basis vectors) \\
$\bC$ & $C \times r_3$ & country factor matrix (loadings) \\
$\bT$ & $T \times r_4$ & year factor matrix (loadings) \\
\addlinespace
\multicolumn{3}{@{}l}{\textit{Basis vectors (columns of factor matrices; fraktur subscript = component)}} \\
$\bm{a}_\fa$ & $A \times 1$
      & column $\fa$ of $\bA$: the $\fa$-th age basis vector \\
$\bm{c}_\fc$ & $C \times 1$
      & column $\fc$ of $\bC$: the $\fc$-th country basis vector \\
$\bm{t}_\ft$ & $T \times 1$
      & column $\ft$ of $\bT$: the $\ft$-th year basis vector \\
\addlinespace
\multicolumn{3}{@{}l}{\textit{Loading vectors (rows of factor matrices; italic subscript = data unit)}} \\
$\bm{s}_s$ & $1 \times r_1$
      & row $s$ of $\bS$: sex-$s$ position in component space \\
$\bm{u}_c$ & $1 \times r_3$
      & row $c$ of $\bC$: country-$c$ position in component space \\
$\bm{w}_t$ & $1 \times r_4$
      & row $t$ of $\bT$: year-$t$ position in component space \\
$r_1, r_2, r_3, r_4$ &  -- 
      & Tucker ranks (number of retained components per dimension) \\
$\tau$ & --
      & cumulative variance threshold for rank selection \\
\addlinespace
\multicolumn{3}{@{}l}{\textit{Transforms}} \\
$\logit(p)$ & -- & $\log\!\bigl(p\,/\,(1-p)\bigr)$ \\
$\expit(x)$ & -- & $1\,/\,(1 + \exp(-x))$ \\
$q_{\min}$ & -- & floor for $\qx$ before logit (e.g., $10^{-8}$) \\
\addlinespace
\multicolumn{3}{@{}l}{\textit{Operations}} \\
$\nmode{n}$ & -- & $n$-mode tensor--matrix product \\
$\bm{M}_{(n)}$ & $I_n \times \prod_{m \neq n} I_m$
      & mode-$n$ unfolding (matricization) of~$\M$ \\
$\bm{\sigma}^{(n)}$ &  -- 
      & singular values of the mode-$n$ unfolding \\
$\hat{\M}$ & $S \times A \times C \times T$
      & reconstructed (approximated) tensor \\
$\bm{O}$ & $C \times T$
      & observed mask ($O_{c,t} = 1$ if country-year is observed) \\
\addlinespace
\multicolumn{3}{@{}l}{\textit{Clustering and reconstruction}} \\
$\bm{z}_{c,t}$ & $2A \times 1$
      & sex-age feature vector for country-year $(c,t)$ \\
$G_{c,t}$ & $r_1 \times r_2$
      & effective core matrix for country-year $(c,t)$ \\
$\bm{f}_{c,t}$ & $r_1(r_2 - 1) \times 1$
      & age-structure feature (level-controlled) \\
$K$ & -- & number of clusters \\
$g_{c,t}$ & -- & cluster label for country-year $(c,t)$ \\
$\hat{\bm{\mu}}_k$ & $\R \to \R^{2A}$
      & mean trajectory for cluster~$k$ \\
\addlinespace
\multicolumn{3}{@{}l}{\textit{Exceptional mortality}} \\
$\mathcal{E}$ &  -- 
      & set of exceptional country-years \\
$d_{c,t}$ & -- & disruption label ($0$: none; $1$: war; $2$: resp.; $3$: enteric) \\
$\bm{\delta}_d$ & $2A \times 1$
      & disruption profile for type~$d$ (smoothed unit vector) \\
$\lambda_{c,t}$ & -- & disruption intensity for country-year $(c,t)$ \\
$\bm{r}_{c,t}$ & $2A \times 1$
      & residual vector (observed minus baseline) \\
$\alpha$ & -- & penalty parameter (penalized projection);
        age 0--4 upweight in \cref{sec:svdcomp} \\
\addlinespace
\multicolumn{3}{@{}l}{\textit{Neural extensions}} \\
$\tilde{G}_{c,t}$ & $r_1 \times r_2$
      & direct projection: $\bS^\top \M_{:,:,c,t} \bA$ (training target) \\
$\hat{G}_{c,t}$ & $r_1 \times r_2$
      & neural core: MLP-predicted core slice for country-year $(c,t)$ \\
$\mathcal{O}$ & --
      & set of observed non-exceptional country-years \\
$\bm{\phi}(t)$ & $\R^9$ & year feature encoding (polynomial + Fourier);
        not to be confused with activation function $\phi(\cdot)$ \\
$\bm{\psi}(\ezero^{\ast})$ & $\R^7$ & $\ezero$ feature encoding (polynomial + Fourier) \\
$\bm{e}_k$ & $\R^{d_k}$ & cluster embedding: PCA of cluster centroids \\
$d_k$ & -- & cluster embedding dimension, $\min(8, K)$ \\
$W_\ell, \bm{b}_\ell$ & varies & weight matrix and bias vector of network layer $\ell$ \\
$\bm{h}_\ell$ & varies & hidden-layer activation (output of layer $\ell$) \\
$\phi(\cdot)$ & -- & element-wise activation function (ReLU: $\max(0, x)$) \\
$L$ & -- & number of network layers (neural extensions only;
        distinct from linear map $L$ in forecasting) \\
$\eta$ & -- & learning rate (gradient descent step size) \\
$\lambda_{\mathrm{reg}}$ & -- & weight decay ($L_2$ regularization) strength \\
$\hat{\bm{z}}_{\mathrm{NN}}$ & $2A \times 1$
      & neural trajectory: MLP-predicted schedule at $(k, {\ezero}^{\ast})$ \\
$\bm{\delta}_{d,k}$ & $2A \times 1$
      & disruption sub-profile for sub-cluster $k$ within type $d$ \\
$K_d$ & -- & number of sub-clusters within disruption type $d$ \\
\addlinespace
\multicolumn{3}{@{}l}{\textit{Life table fitting}} \\
$\yobs$ & $2A \times 1$
      & observed logit($\qx$) schedule (input to fitter) \\
$\zk(\eref)$ & $2A \times 1$
      & cluster-$k$ trajectory evaluated at reference $\ezero$ \\
$\tk(\eref)$ & $2A \times 1$
      & trajectory tangent $\partial \zk / \partial \ezero$ at $\eref$ \\
$\De$ & -- & fitted $\ezero$ shift: $\estar = \eref + \De$ \\
$\estar$ & -- & fitted life expectancy at birth \\
$\lhat$ & -- & estimated disruption intensity \\
$\sigma_\lambda$ & -- & prior scale on $\lambda$ (Bayes factor penalty) \\
$\boldsymbol{\varepsilon}$ & $2A \times 1$
      & residual (noise) vector \\
$\log \mathrm{BF}$ & -- & Laplace-approximated log Bayes factor ($d$ vs.\ null) \\
\addlinespace
\multicolumn{3}{@{}l}{\textit{Summary-indicator prediction (\cref{sec:svdcomp})}} \\
$\mathbf{R}$ & $2A \times r_1 r_2$
      & reconstruction matrix: $\bS \otimes \bA$ (Kronecker product) \\
$\mathbf{R}_{c_{\text{age}}}$ & $2A \times r_1 c_{\text{age}}$
      & truncated reconstruction matrix (first $c_{\text{age}}$ age components) \\
$c_{\text{age}}$ & --
      & number of retained age components in summary-indicator models
        (distinct from country index $c$) \\
\addlinespace
\multicolumn{3}{@{}l}{\textit{Forecasting}} \\
$N_{PC}$ & --
      & number of PCA components (5) \\
$V$ & $r_1 r_2 \times N_{PC}$
      & PCA loadings matrix (columns are principal components of
        $\text{vec}(G_{ct})$) \\
$\bar{g}$ & $r_1 r_2 \times 1$
      & HMD-wide mean of $\text{vec}(G_{ct})$ (PCA centering) \\
$\bm{s}_{c,t}$ & $N_{PC} \times 1$
      & PCA scores for country $c$, year $t$ \\
$L$ & $2A \times N_{PC}$
      & linear map from PCA scores to $\logit(\qx)$ schedule
        (distinct from layer count $L$ in neural extensions) \\
$\bm{x}_{c,t}$ & $2 N_{PC} \times 1$
      & Kalman state: $[\bm{\ell}_{c,t}^\top,\, \bm{\delta}_{c,t}^\top]^\top$ \\
$\bm{\ell}_{c,t}$ & $N_{PC} \times 1$
      & Kalman level state (PCA scores) \\
$\bm{\delta}_{c,t}$ & $N_{PC} \times 1$
      & Kalman drift state (distinct from disruption profile $\bm{\delta}_d$) \\
$\bm{\delta}_{\text{HMD}}$ & $N_{PC} \times 1$
      & HMD-wide mean drift (average across all countries) \\
$\bm{\delta}_{\text{country}}$ & $N_{PC} \times 1$
      & country-specific drift (from last $W$ training years) \\
$\bm{\delta}_{\text{hier}}$ & $N_{PC} \times 1$
      & hierarchical drift target:
        $0.80\,\bm{\delta}_{\text{HMD}} + 0.20\,\bm{\delta}_{\text{country}}$ \\
$w_1, w_2, w_3$ & --
      & simplex weights for HMD-wide, cluster, and country drifts
        ($w_1 + w_2 + w_3 = 1$); grid search finds $w_2 = 0$ \\
$\rho_c$ & --
      & damping parameter for country $c$ ($\rho_c \in [0.80, 0.999]$) \\
$F$ & $2 N_{PC} \times 2 N_{PC}$
      & Kalman transition matrix (DLLT structure with $\rho$ in drift block) \\
$\bm{b}$ & $2 N_{PC} \times 1$
      & state intercept: $[\bm{0},\; (1-\rho)\,\bm{\delta}_{\text{hier}}]^\top$ \\
$Q^{\ell}, Q^{\delta}$ & $N_{PC} \times N_{PC}$
      & level and drift innovation covariances (diagonal) \\
$R$ & $N_{PC} \times N_{PC}$
      & observation noise covariance (diagonal);
        context distinguishes from reconstruction matrix $\mathbf{R}$
        in \cref{sec:svdcomp} \\
$H$ & $N_{PC} \times 2 N_{PC}$
      & observation matrix: $[I_{N_{PC}} \;\; 0]$ \\
$P_{h|T}$ & $2 N_{PC} \times 2 N_{PC}$
      & $h$-step-ahead state covariance \\
$J_h$ & $1 \times N_{PC}$
      & Jacobian $\partial \ezero / \partial \bm{s}$ at horizon $h$ \\
$\kappa$ & --
      & PI calibration factor: $\text{SD}(\text{CV z-scores})$ \\
$\sigma_{\ezero}(h)$ & --
      & delta-method standard deviation of $\ezero$ at horizon $h$ \\
$W$ & -- & drift estimation window (years; default 20) \\
\end{longtable}

Additional notation -- including smoother parameters (Gaussian kernel
bandwidth $\sigma$, ramp age, scale factor), Gaussian mixture model
specification ($\pi_k$, $\bm{\Sigma}_k$), and BIC formula -- is
introduced where it is first needed.


\newpage
\section{Computational Environment}
\label{sec:computational}

All computations were performed on an Apple MacBook Pro with an
Apple M1~Max processor (10 cores: 8 performance, 2 efficiency) and
64\,GB unified memory, running macOS.

The analysis pipeline is implemented in Python~3.14 within a Quarto
notebook environment, managed by \texttt{uv} (package installer) and
\texttt{pyenv} (Python version management), with Positron as the IDE.
The Python dependencies, as specified in the project's
\texttt{pyproject.toml}, are:

\begin{itemize}
    \item \textbf{Core scientific computing:} NumPy, SciPy, pandas
    \item \textbf{Machine learning and statistics:} scikit-learn,
        statsmodels, PyTorch (available but not used for the neural
        components, which are implemented from scratch in NumPy)
    \item \textbf{Visualization:} matplotlib, seaborn
    \item \textbf{Data storage and retrieval:} DuckDB
        \citep{MuehleisenetAl2019}, requests (for HMD data download)
    \item \textbf{Interactive application:} Shiny for Python
    \item \textbf{Signal processing:} ruptures (change-point detection
        for epoch classification), supersmoother (adaptive smoothing)
    \item \textbf{Utilities:} openpyxl (Excel I/O), python-dotenv
        (credential management), Jupyter (notebook kernel)
\end{itemize}

\noindent
The forecasting benchmarks (\cref{sec:forecasting:benchmark}) additionally
require R~($\geq$~4.1) with the following packages:

\begin{itemize}
    \item \textbf{Mortality modeling:} \texttt{demography}%
        \footnote{R.~J. Hyndman with contributions from H.~Booth,
        L.~Tickle, and J.~Maindonald,
        \texttt{demography}: Forecasting Mortality, Fertility,
        Migration and Population Data. R~package,
        \url{https://pkg.robjhyndman.com/demography/}.}
        -- implements the Lee--Carter (\texttt{lca}) and functional
        demographic (\texttt{fdm}) models used as benchmark comparators
    \item \textbf{Time series forecasting:} \texttt{forecast}%
        \footnote{R.~J. Hyndman and Y.~Khandakar,
        Automatic time series forecasting: the \texttt{forecast} package
        for~R. \textit{Journal of Statistical Software}, 27(3), 2008.}
        -- provides the ARIMA engine used by \texttt{fdm} for score
        extrapolation and the random walk with drift used by \texttt{lca}
    \item \textbf{Functional time series:} \texttt{ftsa},
        \texttt{rainbow} -- supporting libraries for the functional
        data decomposition in \texttt{fdm}
\end{itemize}

\noindent
All neural network components -- the neural core, neural trajectory
model, disruption sub-cluster profiles, and summary-indicator prediction
networks -- are implemented from scratch in pure NumPy without any deep
learning framework.  The \texttt{SimpleMLP} class and
\texttt{AdamOptimizer} \citep{KingmaBa2015} are defined in the
project's shared utility module (\texttt{py/\_mdmlt\_utils.py}) and
support forward pass, backpropagation with He initialization
\citep{HeZhangRenSun2015}, and serialization for caching trained
weights.

Document preparation uses \LaTeX\ via KOMA-Script (\texttt{scrartcl})
with Palatino/\texttt{mathpazo} typography; the computational notebook
pipeline comprises nine Quarto chapters (\texttt{qmd/00-setup.qmd}
through \texttt{qmd/08-forecast.qmd}).  The Lee--Carter and
Hyndman--Ullah benchmarks in \cref{sec:forecasting:benchmark} are
computed by the R \texttt{demography} package via a
subprocess bridge, using the HMD's own graduated $m_x$ schedules and
person-year exposures.

The Python codebase totals approximately 19{,}000 lines across the nine
Quarto notebooks and two shared utility modules.  The notebooks follow
the manuscript's section structure: data ingest and preprocessing
(ch.~01), Tucker decomposition (ch.~02), clustering and epoch
classification (ch.~03), trajectory estimation and reconstruction
(ch.~04), exceptional mortality modeling (ch.~05), life table fitting
(ch.~06), summary-indicator prediction (ch.~07), and forecasting with
cross-validation and benchmarking (ch.~08).  A shared environment module
(\texttt{py/\_mdmlt\_env.py}, 363~lines) provides bootstrap
configuration, figure and table export utilities, and chapter-level
logging; a shared utility module (\texttt{py/\_mdmlt\_utils.py},
399~lines) implements the \texttt{SimpleMLP} neural network class,
\texttt{AdamOptimizer}, Tucker reconstruction helpers, and the life
table forward model.  Intermediate results are cached in a DuckDB
database and pickle files, allowing individual chapters to be re-rendered
independently.

An interactive Shiny web application demonstrating the life table
generator, fitter, and summary-indicator prediction is deployed at
\url{https://samclark.shinyapps.io/mdmx/}.

The complete source code is available from the author.

\paragraph{AI research assistance.}
Claude (Anthropic, Claude Opus 4) served as a research assistant
throughout the development of this project.  Its contributions included
writing and debugging Python code for the computational pipeline,
drafting and editing \LaTeX\ manuscript text, performing literature
searches, conducting numerical cross-checks between the Quarto output and
manuscript claims, and iterating on architectural decisions for all
modules through interactive empirical experimentation.  All
substantive scientific decisions -- including defining and framing the questions; designing the analytical approach; choosing the specific methods; optimizing and fine-tuning each method; validating and interpreting results; and organizing and creating the manuscript -- were made by the author.  The AI
assistant's outputs were reviewed, verified, and revised by the author
before incorporation.


\newpage
\appendix
\section{Event Dictionaries}
\label{app:events}

This appendix provides the complete event dictionaries used to classify
exceptional country-years in the HMD mortality tensor.  Three dictionaries
are defined, one for each disruption type: armed conflicts
(\cref{app:wars}), respiratory pandemics (\cref{app:respiratory}), and
enteric pandemics (\cref{app:enteric}).  Each entry records the event name,
the calendar years of impact, and the HMD countries affected.  A
country-year $(c,t)$ is flagged as exceptional if and only if it appears in
at least one dictionary.  When a country-year appears in more than one
dictionary (e.g., belligerent countries during 1918 appear under both war
and respiratory pandemic), the label assigned is that of the event expected
to dominate the mortality signal, as discussed in
\cref{sec:exceptional:events}.

The dictionaries are compiled from standard historical and demographic
sources, including \citet{Urlanis1971}, \citet{Clodfelter2017},
\citet{Patterson1986}, \citet{Hays2005}, and \citet{KarlinskyKobak2021}.
Only events expected to produce a discernible signature in national
age-sex mortality schedules are included; minor skirmishes or localized
outbreaks that affected only a small fraction of the national population
are omitted.  The dictionaries list every affected country regardless of
whether the HMD contains data for that country-year; the intersection with
actual data availability is handled during preprocessing
(\cref{sec:preprocessing:exceptional}).

Country names follow current HMD conventions.  Where historical boundaries
differ from modern ones (e.g., Austria-Hungary, the Russian Empire, the
Ottoman Empire), affected territory is mapped to the modern HMD
country or countries that inherited the relevant population.

\subsection{Wars}
\label{app:wars}

\Cref{tab:wars} lists the armed conflicts included in the war event
dictionary.  The table covers conflicts from the mid-eighteenth century
through the late twentieth century that caused substantial military or
civilian mortality in at least one HMD country.

{\setlength\LTleft{0pt}\setlength\LTright{0pt}
\begin{longtable}{@{}>{\raggedright\arraybackslash}p{3.8cm} >{\raggedright\arraybackslash}p{2.2cm} >{\raggedright\arraybackslash}p{\dimexpr\textwidth-6cm-6\tabcolsep\relax}@{}}
\caption{War event dictionary.  Each row records an armed conflict, the
calendar years during which it produced substantial mortality, and the HMD
countries affected.}
\label{tab:wars} \\
\toprule
Event & Years & HMD countries affected \\
\midrule
\endfirsthead
\caption[]{War event dictionary. (continued)} \\
\toprule
Event & Years & HMD countries affected \\
\midrule
\endhead
\midrule
\multicolumn{3}{r@{}}{\textit{Continued on next page}} \\
\endfoot
\bottomrule
\endlastfoot
Seven Years' War
    & 1756--1763
    & Sweden \\
\addlinespace
Gustav III's Russian War
    & 1788--1790
    & Sweden, Finland \\
\addlinespace
French Revolutionary Wars
    & 1792--1802
    & France, Netherlands, Belgium, Switzerland, Italy, Austria \\
\addlinespace
Napoleonic Wars
    & 1803--1815
    & France, Netherlands, Belgium, Denmark, Norway, Sweden,
      Switzerland, Italy, Spain, Austria \\
\addlinespace
Finnish War
    & 1808--1809
    & Sweden, Finland \\
\addlinespace
First Schleswig War
    & 1848--1851
    & Denmark \\
\addlinespace
Crimean War
    & 1853--1856
    & France, England \& Wales, Scotland, Italy \\
\addlinespace
Second Italian War of Independence
    & 1859
    & France, Italy \\
\addlinespace
Second Schleswig War
    & 1864
    & Denmark \\
\addlinespace
Franco-Prussian War
    & 1870--1871
    & France \\
\addlinespace
Russo-Turkish War
    & 1877--1878
    & Russia, Bulgaria \\
\addlinespace
Boer War
    & 1899--1902
    & England \& Wales, Scotland, Australia,
      New Zealand, Canada \\
\addlinespace
Russo-Japanese War
    & 1904--1905
    & Russia, Japan \\
\addlinespace
Balkan Wars
    & 1912--1913
    & Bulgaria, Greece \\
\addlinespace
World War I
    & 1914--1918
    & France, Belgium, Italy, England \& Wales, Scotland,
      Australia, New Zealand, Canada, USA, Austria, Hungary,
      Czech Republic, Slovakia, Bulgaria, Russia, Ukraine,
      Estonia, Latvia, Lithuania, Poland \\
\addlinespace
Finnish Civil War
    & 1918
    & Finland \\
\addlinespace
Russian Civil War
    & 1918--1922
    & Russia, Ukraine, Belarus, Estonia, Latvia, Lithuania \\
\addlinespace
Irish War of Independence \& Civil War
    & 1919--1923
    & Ireland \\
\addlinespace
Greco-Turkish War
    & 1919--1922
    & Greece \\
\addlinespace
Spanish Civil War
    & 1936--1939
    & Spain \\
\addlinespace
World War II
    & 1939--1945
    & France, Belgium, Netherlands, Luxembourg, Denmark,
      Norway, Finland, England \& Wales, Scotland,
      Northern Ireland, Italy, Austria, Germany,
      Czech Republic, Slovakia, Hungary, Poland, Bulgaria,
      Greece, Estonia, Latvia, Lithuania, Russia, Ukraine,
      Belarus, Australia, New Zealand, Canada, USA, Japan \\
\addlinespace
Korean War
    & 1950--1953
    & South Korea, USA, Australia, Canada \\
\end{longtable}}

\subsection{Respiratory pandemics}
\label{app:respiratory}

\Cref{tab:respiratory} lists the respiratory pandemic events included in the
dictionary.  Because airborne pandemics typically affect all countries within
one or two years, the country lists are extensive.  The years recorded are
those in which excess mortality attributable to the pandemic is expected to
appear in national vital statistics.

{\setlength\LTleft{0pt}\setlength\LTright{0pt}
\begin{longtable}{@{}>{\raggedright\arraybackslash}p{3.8cm} >{\raggedright\arraybackslash}p{2.2cm} >{\raggedright\arraybackslash}p{\dimexpr\textwidth-6cm-6\tabcolsep\relax}@{}}
\caption{Respiratory pandemic event dictionary.}
\label{tab:respiratory} \\
\toprule
Event & Years & HMD countries affected \\
\midrule
\endfirsthead
\caption[]{Respiratory pandemic event dictionary. (continued)} \\
\toprule
Event & Years & HMD countries affected \\
\midrule
\endhead
\midrule
\multicolumn{3}{r@{}}{\textit{Continued on next page}} \\
\endfoot
\bottomrule
\endlastfoot
Russian (Asiatic) influenza
    & 1889--1890
    & All HMD countries with data during the period:
      Sweden, Denmark, Norway, Iceland, Finland, Netherlands,
      Belgium, France, Switzerland, England \& Wales,
      Scotland, Italy, Spain \\
\addlinespace
Spanish influenza
    & 1918--1920
    & All HMD countries with data during the period:
      Sweden, Denmark, Norway, Iceland, Finland, Netherlands,
      Belgium, France, Switzerland, England \& Wales,
      Scotland, Italy, Spain, Australia, New Zealand,
      Canada, USA, Japan \\
\addlinespace
Asian influenza
    & 1957--1958
    & All HMD countries with data during the period \\
\addlinespace
Hong Kong influenza
    & 1968--1969
    & All HMD countries with data during the period \\
\addlinespace
COVID-19
    & 2020--2022
    & All HMD countries with data during the period \\
\end{longtable}}

\subsection{Enteric pandemics}
\label{app:enteric}

\Cref{tab:enteric} lists the enteric pandemic events included in the
dictionary.  These are principally cholera pandemics of the nineteenth
century.  Unlike respiratory pandemics, cholera outbreaks did not strike all
countries simultaneously; the years listed for each country reflect the
period during which a given pandemic wave produced significant mortality in
that specific population.

{\setlength\LTleft{0pt}\setlength\LTright{0pt}
\begin{longtable}{@{}>{\raggedright\arraybackslash}p{3.8cm} >{\raggedright\arraybackslash}p{2.2cm} >{\raggedright\arraybackslash}p{\dimexpr\textwidth-6cm-6\tabcolsep\relax}@{}}
\caption{Enteric pandemic event dictionary.}
\label{tab:enteric} \\
\toprule
Event & Years & HMD countries affected (with peak years) \\
\midrule
\endfirsthead
\caption[]{Enteric pandemic event dictionary. (continued)} \\
\toprule
Event & Years & HMD countries affected (with peak years) \\
\midrule
\endhead
\midrule
\multicolumn{3}{r@{}}{\textit{Continued on next page}} \\
\endfoot
\bottomrule
\endlastfoot
Second cholera pandemic
    & 1826--1837
    & Sweden (1834), Finland (1831), France (1832),
      England \& Wales (1832), Scotland (1832),
      Netherlands (1832), Belgium (1832--1833),
      Italy (1835--1837), Spain (1833--1835) \\
\addlinespace
Third cholera pandemic
    & 1846--1860
    & Sweden (1850, 1853--1854), Norway (1848--1849, 1853),
      Denmark (1853), Finland (1853),
      Netherlands (1848--1849, 1853--1855),
      Belgium (1848--1849, 1853--1854),
      France (1849, 1853--1854),
      Switzerland (1855),
      England \& Wales (1848--1849, 1853--1854),
      Scotland (1848--1849, 1853--1854),
      Italy (1854--1855), Spain (1854--1855) \\
\addlinespace
Fourth cholera pandemic
    & 1863--1875
    & Sweden (1866), Norway (1866),
      Netherlands (1866--1867), Belgium (1866),
      France (1865--1866),
      England \& Wales (1866), Scotland (1866),
      Italy (1865--1867), Spain (1865) \\
\addlinespace
Fifth cholera pandemic
    & 1881--1896
    & France (1884), Italy (1884--1885, 1893),
      Spain (1885, 1890) \\
\addlinespace
Sixth cholera pandemic
    & 1899--1923
    & Italy (1910--1911), Russia (1910, 1921--1922) \\
\end{longtable}}

\newpage

\end{document}